\documentclass[11pt]{article}
\pdfoutput=1
\usepackage[margin=1in]{geometry} 

\usepackage{etoolbox}

\usepackage[round]{natbib}
\usepackage{array,amsmath,amsthm,amssymb,amsfonts,titling,bbm, titlesec, bm, physics, enumitem, accents, setspace, fancyhdr, natbib, geometry, pdflscape}
\usepackage{graphicx}
\usepackage{float}
\usepackage[font=footnotesize,labelfont=bf]{caption}
\usepackage{array}
\usepackage[title]{appendix}
\usepackage{afterpage}
\usepackage{listings}
\usepackage{prodint}
\usepackage{algorithm}
 \usepackage{booktabs}
\usepackage{multirow,booktabs}
\usepackage{subcaption}
\newlength{\continueindent}
\usepackage{etoolbox}
\usepackage{setspace}
\usepackage[dvipsnames]{xcolor}
\definecolor{mydarkgray}{rgb}{0.25,0.25,0.25}
\definecolor{Bleu}{RGB}{0,0,204}
\usepackage[hypertexnames=false]{hyperref}
\hypersetup{
colorlinks,
  citecolor=Bleu,
  linkcolor=Bleu,
  urlcolor=Bleu}

\usepackage{makecell}
\usepackage{tikz}
\def\checkmark{\tikz\fill[scale=0.4](0,.35) -- (.25,0) -- (1,.7) -- (.25,.15) -- cycle;}

\usepackage{tikz}
\usetikzlibrary{positioning, arrows.meta}

\usepackage{amssymb}
\usepackage{pifont}
\usepackage{booktabs}
\usepackage{multirow}

\lstdefinestyle{compact}{
    basicstyle=\ttfamily\small,%
    lineskip=-0.5pt,
    aboveskip=0pt,
    belowskip=0pt,
    keepspaces=true,
    breaklines=true,
    columns=fullflexible
}

\newcommand{\INPUT}{\item[\textbf{Input:}]}

\newcolumntype{P}[1]{>{\centering\arraybackslash}p{#1}}

\usepackage{amsmath}
\usepackage{amssymb}
\usepackage{amsthm,thmtools}
\usepackage{enumitem}
\usepackage{amsfonts}
\RequirePackage{algorithm}
\RequirePackage{algorithmic}
\usepackage{bm,upgreek}
\usepackage{mathrsfs}  
\usepackage{calc}%
\usepackage{subcaption}
\usepackage{authblk}

\theoremstyle{plain}
\newtheorem{theorem}{Theorem}
\newtheorem{corollary}{Corollary}

\newtheorem{lemma}{Lemma}

\theoremstyle{definition}
\newtheorem{example}{Example}

\providecommand{\keywords}[1]
{
  \small
  \textbf{\textit{Keywords:}} #1
}
\usepackage[utf8]{inputenc}
\usepackage{textcomp}
\DeclareMathOperator*{\argmin}{argmin}

\DeclareMathOperator*{\esssup}{ess\,sup}
\DeclareMathOperator*{\essinf}{ess\,inf}
\usepackage[capitalize,noabbrev]{cleveref}

\DeclareFontFamily{U}{jkpmia}{}
\DeclareFontShape{U}{jkpmia}{m}{it}{<->s*jkpmia}{}
\DeclareFontShape{U}{jkpmia}{bx}{it}{<->s*jkpbmia}{}
\DeclareMathAlphabet{\mathfrak}{U}{jkpmia}{m}{it}
\SetMathAlphabet{\mathfrak}{bold}{U}{jkpmia}{bx}{it}

\usepackage[hang,flushmargin]{footmisc}

\title{Doubly robust inference via calibration}
\author[1,5]{Lars van der Laan}
\author[2,3]{Alex Luedtke}
\author[4,1]{Marco Carone}

\affil[1]{\footnotesize Department of Statistics, University of Washington, Seattle, WA, USA}
\affil[2]{\footnotesize Department of Health Care Policy, Harvard Medical School, Boston, MA, USA}
\affil[3]{\footnotesize Department of Statistics, Harvard University, Cambridge, MA, USA}
\affil[4]{\footnotesize Department of Biostatistics, University of Washington, Seattle, WA, USA}
\affil[5]{\footnotesize Current affiliation: Department of Management Science and Engineering, Stanford University, Stanford, CA, USA; \texttt{vdlaan@stanford.edu}}

 \AtBeginEnvironment{algorithmic}{\normalsize}
 \usepackage{tikz}
 \usepackage{pgfplots}
 \pgfplotsset{compat=1.18}

 \usepackage{setspace}

\date{}

\begin{document}

\allowdisplaybreaks
\maketitle

 \singlespacing

\begin{abstract}

Doubly robust estimators are widely used for estimating average treatment effects and other linear summaries of regression functions. While consistency requires only one of two nuisance functions to be estimated consistently, asymptotic normality for linear functionals typically requires sufficiently fast convergence of both. We address this mismatch by showing that calibrating the nuisance
estimators within a doubly robust procedure can yield doubly robust asymptotic normality. We introduce the idea of \emph{calibrated debiased machine learning (DML)} and propose a specific implementation in which standard DML is augmented with a simple isotonic regression adjustment.  We show that, under a partial orthogonality condition, a calibrated DML estimator remains asymptotically normal if either the regression function or Riesz representer of the functional is estimated sufficiently well, allowing the other to converge arbitrarily slowly or even inconsistently. We also propose a bootstrap-assisted method for constructing confidence intervals, enabling doubly robust inference without additional nuisance estimation. In a range of semi-synthetic benchmark datasets, calibrated DML reduces bias and improves coverage relative to standard DML. Our method can be integrated into existing DML pipelines by adding just a few lines of code to calibrate cross-fitted estimates via isotonic regression.

\end{abstract}


\doublespacing

\section{Introduction}

\subsection{Motivation}

 Nonparametric inference on the causal effects of dynamic and stochastic treatment interventions using flexible statistical learning techniques is a fundamental problem in causal inference. These inferential targets fall within a broader class of target parameters that can be expressed as linear functionals of the outcome regression, that is, the conditional mean of the outcome given treatment assignment and baseline covariates \citep{chernozhukov2018double, rotnitzky2021characterization, hirshberg2021augmented}. To achieve inference for such parameters, various debiased machine learning (DML) frameworks are available, including one-step estimation \citep{pfanzagl1985contributions, bickel1993efficient}, estimating equations and double machine learning \citep{robins1994estimation,robins1995analysis, robinsCausal, vanderlaanunified, DoubleML}, and targeted minimum loss estimation (TMLE) \citep{van2006targeted, vanderLaanRose2011}, and sieve-based estimators \citep{qiu2021universal, cho2024kernel}. When performing inference on linear functionals, debiased machine learning methods frequently produce estimators that demonstrate robustness against inconsistent or slow estimation of certain nuisance functions, a property known as double robustness \citep{bang2005doubly, smucler2019unifying}. For example, when estimating the average treatment effect (ATE),  doubly robust estimators remain consistent as long as either the outcome regression or the propensity score is consistently estimated, even if the other is inconsistently estimated. This robustness property extends more generally to linear functionals of the outcome regression, necessitating consistent estimation of only the outcome regression itself or the Riesz representer of the linear functional \citep{chernozhukov2018double, rotnitzky2021characterization}.

Extending the double robustness property of debiased machine learning estimators
to statistical inference is challenging.  By \emph{double robustness for
statistical inference}, we mean that valid root-\(n\) inference remains possible
when one nuisance function is estimated consistently at a sufficiently fast
rate, even if the other is not. Standard debiased machine learning procedures,
however, typically require both the outcome regression and the Riesz representer
to be estimated consistently and sufficiently quickly, so that the product of
their estimation errors is \(o_p(n^{-1/2})\). If this condition fails,
asymptotic linearity and normality can be lost, invalidating standard confidence
intervals and \(p\)-values \citep{benkeser2017doubly}.   In practice,
inconsistency may arise from model misspecification, such as omitted
interactions or nonlinearities, or from over-regularization of flexible learners;
slow convergence may arise even under consistency when the nuisance functions
are intrinsically complex, for example, due to nonsparse high-dimensional
interactions or limited smoothness. 

One approach to achieving doubly robust inference is to use parametric methods,
such as maximum likelihood estimation within a parametric model, to estimate the nuisance parameters.
These methods exhibit model double robustness, maintaining asymptotic linearity
when at least one nuisance model is correctly specified
\citep{vermeulen2015bias,vermeulen2016data,shook2025double}. However, their
usefulness for nonparametric inference is limited by their reliance on strong
modeling assumptions, since in practice both models are often misspecified
\citep{kang2007demystifying}.
 
To overcome the limitations of parametric methods, several estimators exhibiting
\emph{sparsity double robustness}---a sparsity-based analogue of model double
robustness---have been proposed \citep{avagyan2017honest,athey2018approximate,
bradic2019sparsity,bradic2019minimax,smucler2019unifying,tan2020model,
zhang2021dynamic,ghosh2022doubly,liu2023root,bradic2024high,baybutt2026doubly}.
In high-dimensional generalized linear models, \cite{avagyan2017honest} and
\cite{bradic2019sparsity} introduced doubly robust \(\ell_1\)-regularized
estimators of the ATE that are asymptotically normal when either the outcome
regression or propensity score is sufficiently sparse, allowing misspecification
of the other nuisance model. Related sparse methods include the singly robust
estimators of \cite{athey2018approximate} and \cite{tan2020model}, which remain
valid under outcome-regression misspecification when the propensity score is
sufficiently sparse, and the approximate-sparsity framework of
\cite{bradic2019minimax}. However, these guarantees rely on high-dimensional
sparsity or dictionary-based approximations and tailored regularized estimators
\citep{bradic2019minimax,smucler2019unifying,liu2023root}. These methods are
therefore not learner-agnostic and have limited ability to leverage arbitrary
modern machine learning methods.

 By contrast, the line of work initiated by \cite{van2014targeted} and further
developed by \cite{benkeser2017doubly} provides a learner-agnostic debiasing
framework---based on targeted maximum likelihood estimation (TMLE)
\citep{vanderLaanRose2011}---for doubly robust inference with nuisance functions
estimated using generic machine learning methods. Under suitable conditions, the
resulting estimators are \emph{doubly robust asymptotically linear} (DRAL): they
remain asymptotically linear if one nuisance estimator is inconsistent, provided
the other is estimated sufficiently well, typically at rate \(o_p(n^{-1/4})\).
Concretely, for the ATE, if \(\pi_n\) and \(\mu_n\) estimate the propensity score
\(\pi_0\) and outcome regression \(\mu_0\), respectively, then the DRAL condition
is the \emph{best-rate} condition
\(\min\{\|\mu_n-\mu_0\|,\|\pi_n-\pi_0\|\}=o_p(n^{-1/4})\), whereas rate double robustness requires the
product-rate condition
\(\|\mu_n-\mu_0\|\,\|\pi_n-\pi_0\|=o_p(n^{-1/2})\). For the ATE, these methods
iteratively update initial propensity score and outcome regression estimates
using low-dimensional (often one-dimensional) regressions that enforce moment
conditions and debias the cross-product remainder in the asymptotic expansion.
This approach has since been extended to mean outcomes under informative
missingness \citep{diaz2017doubly}, counterfactual survival curves
\citep{diaz2019statistical}, and semiparametric inference for the ATE in
partially linear regression models \citep{dukes2024doubly}.

The TMLE-based framework for doubly robust inference has some limitations.
First, the iterative debiasing procedure is difficult to study theoretically
and can be computationally demanding to implement, posing challenges for its
application to large-scale data or complex data structures. While \cite{bonvini2024doubly} proposed a non-iterative, nonparametric DRAL
estimator of the ATE based on higher-order influence functions, their procedure
requires careful tuning of bivariate kernel-smoothing parameters, for which
practical guidance remains limited. Second, when considering new
parameters of interest, a new debiasing algorithm must be derived on a
case-by-case basis, requiring careful analysis of the relevant remainder terms
and novel targeting constructions. To date, there is no unified, learner-agnostic approach to doubly robust
inference for a generic linear functional of the outcome regression. Finally, the available guarantees are mainly designed for settings in which one nuisance estimator is inconsistent. Existing results do not strengthen inference when both nuisance
estimators are consistent but one converges too slowly for the product-rate
condition to hold. In fact, they typically assume that both nuisance estimators converge to their
(possibly incorrect) limits at rate \(o_p(n^{-1/4})\).

\subsection{Contributions of this work}

  \begin{figure}[tb]
\centering
\begin{tikzpicture}[node distance=0.8cm and 1.5cm, every node/.style={draw, rounded corners, align=center, minimum height=1cm, minimum width=2.5cm}, >=Stealth]

\node (estimate_cal) {Estimate Nuisances};
\node (calibrate_cal) [right=of estimate_cal, draw=blue, ultra thick] {Calibrate};
\node (debias_cal) [right=of calibrate_cal] {Debias};

\draw[->] (estimate_cal) -- (calibrate_cal);
\draw[->] (calibrate_cal) -- (debias_cal);

\end{tikzpicture}
\caption{We show \textbf{calibrating the nuisances} before debiasing ensures doubly robust asymptotic normality.}
\label{fig:dml_flowchart}
\end{figure}

We develop new methods for doubly robust inference on linear functionals of the
outcome regression. Our goal is to construct DRAL estimators that permit valid
inference when one of the two nuisance functions is estimated arbitrarily slowly
or even inconsistently, provided that the other converges at an
\(o_p(n^{-1/4})\) rate. We first establish a link between calibration, a
technique typically used in prediction and classification tasks
\citep{zadrozny2001obtaining,niculescu2005predicting}, and doubly robust
asymptotic linearity. We then introduce a general framework, \emph{calibrated
debiased machine learning} (calibrated DML). As outlined in
Figure~\ref{fig:dml_flowchart}, calibrated DML begins with standard cross-fitted
estimates of the outcome regression and Riesz representer, obtained using any
machine learning algorithm. It then adds a single post-hoc step: each nuisance
estimator is \emph{calibrated} by fitting a one-dimensional map from its fitted
values to its target, for example, using isotonic regression
\citep{barlow1972isotonic}. We show that this calibration step enforces a particular empirical orthogonality between nuisance errors, which enables a standard one-step debiased estimator to achieve doubly robust asymptotic linearity under a partial orthogonality condition comparable to those assumed in prior work on learner-agnostic doubly robust inference (see, e.g., \citealp{van2014targeted,benkeser2017doubly,diaz2017doubly,dukes2024doubly,bonvini2024doubly}).

Confidence intervals can be constructed using the closed-form influence function of the calibrated DML estimator, without requiring knowledge of which nuisance estimators converge sufficiently quickly. We show that no correction is needed to the standard influence function-based variance estimator when both nuisance estimators are consistent for their respective targets, even if one converges arbitrarily slowly.   To handle cases in which one nuisance may be inconsistently estimated, we propose a computationally efficient bootstrap-assisted procedure that avoids the additional nuisance estimation required in \cite{van2014targeted} and related approaches, and remains valid under the same nuisance-rate conditions.

\begin{table}[tb]
\centering
\caption{
Summary of properties (consistency, $n^{1/2}$-rate asymptotic normality, and nonparametric efficiency) of G-computation plug-in, IPW, AIPW and calibrated DML (proposal) estimators of the ATE when nuisances are estimated using machine learning,  in settings where only the propensity score vs. only the outcome regression vs. both nuisances are estimated well (i.e., consistently at a rate faster than $n^{-1/4}$). ($\ast$) Our proposed estimator is efficient if one nuisance is estimated well, provided the other is estimated consistently, regardless of rate.
}
\label{tab:calibration_summary}
\resizebox{\textwidth}{!}{%
\begin{tabular}{l@{\hskip 6pt}!{\vrule width 0.8pt}@{\hskip 6pt}
>{\centering\arraybackslash}m{1.7cm}
>{\centering\arraybackslash}m{1.8cm}
>{\centering\arraybackslash}m{1.7cm}|
>{\centering\arraybackslash}m{1.7cm}
>{\centering\arraybackslash}m{1.8cm}
>{\centering\arraybackslash}m{1.7cm}|
>{\centering\arraybackslash}m{1.7cm}
>{\centering\arraybackslash}m{1.8cm}
>{\centering\arraybackslash}m{1.7cm}
}
\multicolumn{1}{c}{} & \multicolumn{3}{c}{\makecell{only propensity score \\ estimated well}} 
& \multicolumn{3}{c}{\makecell{only outcome regression \\ estimated well}} 
& \multicolumn{3}{c}{\makecell{both nuisance functions \\ estimated well}} \\[.175in]
\textbf{estimator} 
& \textbf{consistent} & \textbf{normal} & \textbf{efficient}
& \textbf{consistent} & \textbf{normal} & \textbf{efficient}
& \textbf{consistent} & \textbf{normal} & \textbf{efficient} \\
\midrule\midrule
G-computation      
& {\color{darkgray}--} & {\color{darkgray}--} & {\color{darkgray}--} 
& \checkmark & {\color{mydarkgray}\texttimes} & {\color{mydarkgray}\texttimes} 
& {\color{darkgray}--} & {\color{darkgray}--} & {\color{darkgray}--} \\
IPW          
& \checkmark & {\color{mydarkgray}\texttimes} & {\color{mydarkgray}\texttimes} 
& {\color{darkgray}--} & {\color{darkgray}--} & {\color{darkgray}--} 
& {\color{darkgray}--} & {\color{darkgray}--} & {\color{darkgray}--} 
\\
AIPW                  
& \checkmark & {\color{mydarkgray}\texttimes} & {\color{mydarkgray}\texttimes}
& \checkmark & {\color{mydarkgray}\texttimes} & {\color{mydarkgray}\texttimes} 
& \checkmark & \checkmark & \checkmark \\
\textbf{calibrated DML} (ours)              
& \checkmark & \checkmark  & \checkmark {\small $\ast$}
& \checkmark & \checkmark  & \checkmark {\small $\ast$}
& \checkmark & \checkmark & \checkmark \\
\bottomrule
\end{tabular}%
}
\end{table}

   Table~\ref{tab:calibration_summary} summarizes the properties of a specific implementation of calibrated DML in the well-studied setting of ATE estimation. This implementation employs isotonic regression for calibration and a cross-fitted augmented inverse probability weighted (AIPW) estimator for debiasing. Compared to the standard AIPW estimator, calibrated DML remains asymptotically normal under weaker convergence requirements: it suffices that one nuisance be consistently estimated at an $o_p(n^{-1/4})$ rate; the other nuisance may be inconsistently estimated. It is efficient when \emph{both} nuisances are consistently estimated, at least one at an $o_p(n^{-1/4})$ rate. Calibrated DML can yield even larger improvements over non-doubly robust estimators based on the G-computation formula or inverse probability weighting (IPW), since these estimators are neither doubly robust nor generally asymptotically normal under flexible nuisance estimation.

Our approach is motivated by the DRAL estimators of the ATE proposed by
\citet{van2014targeted} and further developed in subsequent work
\citep{benkeser2017doubly,diaz2017doubly,diaz2019statistical,dukes2024doubly}.
We share the goal of constructing nuisance estimators that solve certain score equations
and thereby yield doubly robust asymptotic linearity, but our strategy differs
substantially, even for the ATE. In particular, our approach is a non-iterative,
computationally efficient two-step procedure that can be implemented using
standard isotonic regression software. Moreover, it does not depend on the
specific form of the linear functional; it requires only initial nuisance
estimators and black-box access to an empirical estimator of the target
functional.   As in \cite{benkeser2017doubly}, our estimator naturally facilitates  Wald-type
inference based on a closed-form influence function, without requiring the
analyst to know which nuisance is correctly specified. However, unlike that work,
our proposed bootstrap-assisted inference procedure avoids estimating the
additional nuisance components that appear in the closed-form influence
function, significantly simplifying implementation.  Our theoretical results also refine prior work by showing that asymptotic
linearity holds not only when one nuisance function is consistently estimated,
but also when both nuisance functions are consistent but one converges too
slowly. Finally, we study doubly robust inference for generic continuous linear
functionals, which requires a new analysis of the doubly robust remainder that
leverages the Riesz representation property.

\subsection{Related work}
 
\label{sec:relatedwork}

\medskip  \noindent \emph{Calibration in causal inference.}  Prior work has studied the finite-sample benefits of calibrating propensity score estimates in IPW and AIPW estimators of the ATE \citep{gutman2022propensity, deshpande2023calibrated, ballinari2024improving, van_der_Laan2024stabilized, rabenseifner2025calibrationstrategiesrobustcausal}. These studies focus primarily on calibrating a single nuisance function---the propensity score---for a single parameter (the ATE), and do not provide theoretical justification beyond the observation that calibration can reduce nuisance estimation error. We extend this line of work by showing that, for general linear functionals, calibrating \emph{both} the outcome regression and the Riesz representer is key for further debiasing (in finite samples and asymptotically) and for constructing DRAL estimators. More broadly, our work shows that calibration can yield asymptotic guarantees and relax nuisance-rate requirements, rather than serving only to improve the finite-sample quality of nuisance estimators. In the special case of the ATE, we find that calibrating the outcome regression can reduce finite-sample bias and remove certain asymptotic bias terms.

\medskip\noindent \emph{Sparsity double robustness.}
Our calibration-based approach to doubly robust inference is closely related to
the literature on (approximate) sparsity-based doubly robust inference
mentioned above \citep{athey2018approximate,bradic2019minimax,
smucler2019unifying,tan2020model,liu2023root}. Methodologically, our approach
enforces one-dimensional orthogonality conditions on nuisance errors via
calibration (see \eqref{eqn::orthogonaloutcome}--\eqref{eqn::orthogonalRiesz}),
whereas sparsity-based methods typically enforce high-dimensional moment
conditions. In particular, when the nuisances are estimated by
\(\ell_1\)-regularized empirical risk minimization over linear classes, the
Karush--Kuhn--Tucker conditions yield approximate empirical orthogonality
over suitable linear subspaces (see, e.g., \citealp[Eq.~15]{tan2020model};
\citealp[Sec.~2.1]{bradic2019sparsity}; \citealp[Sec.~4.1]{smucler2019unifying}).
\cite{smucler2019unifying} develop a unifying \(\ell_1\)-regularization
framework for sparse doubly robust estimation of mixed-bias parameters,
establishing rate and model double robustness under approximate-sparsity
conditions. For the ATE, \cite{liu2023root} relax sparsity requirements on
misspecified nuisance limits by imposing sparsity only on the consistently
estimated nuisance, although their construction is not generally
semiparametrically efficient. We also note a difference in terminology:
\cite{tan2020model}, \cite{liu2023root}, and related work use ``calibration''
as in survey sampling, referring to adjustments that enforce moment
conditions \citep{deville1992calibration}, whereas we use ``calibration'' as in
machine learning to mean mean-unbiasedness conditional on the model's
own prediction, a one-dimensional objective. The analogous high-dimensional
goal is closely related to \emph{multi-accuracy} \citep{roth2022uncertain}.
We discuss these differences in more detail in
Appendix~\ref{appendix:relatedwork}.

A salient precedent to this work is \cite{liu2023root}, which also makes use of a post-hoc
correction. However, while \cite{liu2023root} solve a high-dimensional moment-matching
problem built on preliminary \(\ell_1\)-regularized generalized linear models,
we use a one-dimensional calibration step and allow arbitrary
user-supplied first-stage learners. The guarantees differ accordingly: we show
that one-dimensional calibration suffices for DRAL under a
partial-orthogonality condition on the nuisances (Condition~\ref{cond:ortho}),
whereas \cite{liu2023root} rely on sparsity conditions tailored to their
high-dimensional setting and require an auxiliary sample (e.g.,  via
sample splitting) and careful tuning. Our calibrated DML estimators are
nonparametric efficient when both nuisances are consistent and at least
one converges faster than \(n^{-1/4}\)
(Table~\ref{tab:calibration_summary}), whereas \cite{liu2023root} do not
generally establish efficiency.

\medskip\noindent \emph{Higher-order influence functions and undersmoothing.}
Higher-order influence function (HOIF) estimators \citep{robins2008higher, liu2017semiparametric}
use higher-order bias corrections to relax the nuisance-rate requirements of
first-order doubly robust methods, and can yield asymptotically valid and
efficient inference even when nuisance estimators fail to satisfy the usual
\(o_p(n^{-1/2})\) product-rate condition. Related robustness to slow nuisance rates can also be obtained without explicit HOIF-based corrections by combining sample splitting with undersmoothing
\citep{newey2018cross, mcgrath2022nuisance, mcclean2024double}. While HOIF methods can accommodate flexible first-stage machine learning
estimators, their higher-order bias corrections typically rely on smoothness and
approximability assumptions, often of H\"older type, to control higher-order
remainders. They also introduce additional tuning parameters---such as the
choice of basis functions, truncation order, sieve dimension, or bandwidth---for
which practical guidance can be limited. Consequently, HOIF estimators can be
more complex to implement and tune, and their performance may be sensitive to
these assumptions and choices, especially in high-dimensional settings.

For the expected conditional covariance parameter, \cite{bonvini2024doubly}
show that modified HOIF-style estimators can attain guarantees similar to those
of the TMLE-style doubly robust inference approach of
\cite{benkeser2017doubly}. Their HOIF-based DRAL estimators use the outcome
regression and propensity score as a bivariate dimension reduction and achieve
validity under conditions comparable to those imposed in our work and related
TMLE-based approaches \citep{benkeser2017doubly}. In contrast, calibrated DML
uses a simple post-hoc calibration step to control the relevant
bias, requiring no additional nuisance estimation beyond standard first-order
methods. When implemented with isotonic calibration, this post-hoc step is also
tuning-free.

\subsection{Organization} This paper is organized as follows. In Section \ref{section::startsetup}, we introduce the class of regression functionals we consider and provide background on a standard rate doubly robust debiased estimator. In Section \ref{section:prelimDRinference}, we formulate the problem of doubly robust inference, introduce a model-agnostic linear expansion for the bias of one-step debiased estimators, and demonstrate how further debiasing of nuisances can be employed to construct DRAL estimators. In Section \ref{section::proposedestimator}, we present a novel link between nuisance estimator calibration and the doubly robust inference properties of debiased machine learning estimators, use this link to build our general calibrated DML framework, and propose a calibrated DML estimator based on isotonic regression with accompanying bootstrap-based confidence intervals. In Section \ref{section::theory}, we establish the theoretical properties of our proposed calibrated DML estimators and confidence intervals. Finally, we empirically investigate the performance of calibrated DML in Section \ref{section::experiments} and provide concluding remarks in Section \ref{section::conclusion}.

 \section{Problem setup}
 
 \label{section::startsetup}
 
\subsection{Data structure and target parameter}
\label{section::setup}

Suppose we observe an i.i.d.\ sample $Z_1,\ldots,Z_n$ of
$Z:=(W,A,Y)\sim P_0$, where $W\in\mathcal W\subset\mathbb R^d$ are baseline
covariates, $A\in\mathcal A\subset\mathbb R^p$ is a treatment, and $Y\in\mathbb R$
is a bounded outcome. For any distribution $P$, let
$\mu_P(a,w) := E_P(Y\,|\, A=a, W=w)$ denote the outcome regression and
$P f := \int f(z)\,dP(z)$ for any $P$-integrable function $f$. We denote by
$P_{0,A,W}$ the marginal distribution of $(A,W)$ under $P_0$, by
$\mathcal H := L^2(P_{0,A,W})$ the corresponding Hilbert space of
square-integrable functions on $\mathcal A \times \mathcal W$ with inner product
$\langle \mu_1,\mu_2\rangle := P_{0,A,W}(\mu_1\mu_2)$, and
by $\|\cdot\|$ the induced norm. For general $P$, we write
$\|\mu\|_P := \{P_{A,W}(\mu^2)\}^{1/2}$, where $P_{A,W}$ is the
marginal law of $(A,W)$ under $P$. Let $\mathcal{M}$ be a nonparametric model containing $P_0$.
To simplify notation, we write \(S_0\) for any summary \(S_{P_0}\) of the true
distribution \(P_0\).

Our objective is to develop doubly robust nonparametric inference for a
linear summary \(\tau_0 := \psi_0(\mu_0)\) of the outcome regression \(\mu_0\),
where \(\psi_0 : \mathcal{H} \to \mathbb{R}\) is an \(L^2\)-continuous linear
functional. We allow the functional of interest to depend on the unknown
data-generating distribution \(P_0\), viewing \(\psi_0\) as the evaluation at
\(P_0\) of a smooth mapping \(P \mapsto \psi_P\) defined on the statistical
model \(\mathcal{M}\). As illustrated at the end of this subsection, many
parameters take this form, including average causal effects; in particular, this framework
includes the class of continuous linear functionals studied in
\cite{chernozhukov2022automatic} and \cite{bradic2019minimax}.

To facilitate $\sqrt{n}$-inference on $\tau_0$, we impose the following smoothness conditions on the mapping
\((P,\mu)\mapsto \psi_P(\mu)\), which are sufficient for pathwise
differentiability of the composite parameter \(\Psi : P \mapsto \psi_P(\mu_P)\)
at \(P_0\). For each \(P\in\mathcal{M}\), we view \(\psi_P\) as a bounded linear
functional on the normed space \((\mathcal{H}, \|\cdot\|_{L^2(P_{0,A,W})})\),
equipped with the corresponding operator norm.

\begin{enumerate}[label=\bf{A\arabic*)}, ref={A\arabic*}, series = introcond]
    \item \textit{(linearity and boundedness)}\label{cond::linear}  it holds that $\psi_0(c\mu_1 + \mu_2) = c\psi_0(\mu_1) + \psi_0(\mu_2)$ for each $\mu_1,\mu_2\in \mathcal{H}$ and $c\in\mathbb{R}$ and that $\sup_{\mu \in \mathcal{H}\backslash \{0\}} |\psi_0(\mu)|/\|\mu\| < \infty$;
    \item \textit{(Hellinger continuity)} \label{cond::continuous}  $P \mapsto \psi_P$ is a continuous map at $P_0$ with domain $\mathcal{M}$ equipped with the Hellinger distance and codomain $\{\psi_P: P \in \mathcal{M}\}$ equipped with the operator norm;  
    \item \textit{(pathwise differentiability)}\label{cond::pathwise} for each $\mu \in \mathcal{H}$, $P \mapsto \psi_P(\mu)$ is pathwise differentiable at $P_0$ relative to $\mathcal{M}$ with efficient influence function $\phi_{0, \mu} \in L^2_0(P_0)$.
\end{enumerate} 

A key object associated with a continuous linear functional is its Riesz
representer, which plays a central role in debiasing and in the construction of
asymptotically linear estimators of \(\tau_0\); see \cite{williams2025riesz} for an
accessible introduction. Since \(\psi_0\) is a continuous linear functional on
\(\mathcal{H}\), the Riesz representation theorem guarantees the existence of a
unique \(\alpha_0 \in \mathcal{H}\) such that
\begin{equation}
\label{eqn::reiszrep}
\psi_0(\mu) = \langle \alpha_0, \mu \rangle
\quad \text{for all } \mu \in \mathcal{H}.
\end{equation}
Section~\ref{section::prelimDR} shows how~\eqref{eqn::reiszrep} motivates
automatic debiased machine learning estimators of \(\tau_0\). Under our
conditions, the same \(\alpha_0\) also appears in the nonparametric efficient
influence function (EIF) for the parameter mapping \(\Psi\) and hence in the
associated semiparametric efficiency bound \citep{bickel1993efficient}; see
\cite{van2000asymptotic} for a review. Letting \(D_{\mu,\alpha}\) denote the
map \(z\mapsto \alpha(a,w)\{y-\mu(a,w)\}\), the following theorem states
this EIF.

\begin{theorem}[Pathwise differentiability]
\label{theorem::EIF}
Under \ref{cond::linear}--\ref{cond::pathwise}, the parameter
\(\Psi:\mathcal{M} \rightarrow \mathbb{R}\) is pathwise differentiable at \(P_0\)
and has nonparametric efficient influence function
\(D_0 := \phi_{0,\mu_0} + D_{\mu_0,\alpha_0}\), where \(\phi_{0,\mu_0}\), the component arising from the
dependence of \(\psi_0\) on \(P_0\), is defined in \ref{cond::pathwise}.
\end{theorem}

\noindent Following \cite{van2000asymptotic}, we call an estimator efficient for
\(\tau_0\) relative to \(\mathcal M\) if it is asymptotically linear with
influence function equal to \(D_0\). By the
convolution theorem, for any regular estimator $\tau_n$ of \(\tau_0\), the asymptotic variance of $n^{1/2}(\tau_n-\tau_0)$ is no smaller than \(P_0D_0^2\), and this lower bound is attained by efficient
estimators.

A key class of parameters covered by our framework consists
of linear functionals that can be represented as expectations of a possibly unknown
linear map.

\begin{example}[Expectation-representable linear functionals]
\label{example::expectfunctional}
Many parameters of interest in causal inference, such as counterfactual means and the ATE, can be written as functionals of the form \citep{chernozhukov2022automatic, bradic2019minimax, rotnitzky2021characterization}
\begin{equation}
    \mu \mapsto \psi_0(\mu) = E_0\{m(Z,\mu)\}\label{eqn::simplefunctional}
\end{equation}
for a known map $m:\mathcal{Z}\times\mathcal{H}\to\mathbb{R}$ that is linear in its
last argument. For example, the counterfactual mean outcome under a dynamic intervention that
sets treatment to $d(w)\in\mathcal{A}$ for individuals with covariate value
$w\in\mathcal{W}$ is obtained by taking $m:(z,\mu)\mapsto \mu(d(w),w)$, whereas
the ATE corresponds to $m:(z,\mu)\mapsto \mu(1,w)-\mu(0,w)$. Doubly robust inference for the ATE parameter was considered in \cite{van2014targeted} and \cite{benkeser2017doubly}. For such functionals, Lemma~\ref{lem:hellinger_continuity} in
Appendix~\ref{appendix::conditions} verifies
Conditions~\ref{cond::linear}--\ref{cond::pathwise} when $\mu \mapsto m(\cdot, \mu)$ is $L^2$-continuous and $\mathcal{M}$ is suitably dominated by $P_0$, and it is shown that $\phi_{0,\mu}:=m(\cdot,\mu)-\psi_0(\mu)$. More generally, our framework
allows for nuisance-dependent functionals of the form
$\psi_0(\mu)=E_0\{m_0(Z,\mu)\}$, where $m_0$ is a map that may depend on
$P_0$. One such case is the causal effect under a stochastic intervention, for
which $m_0(z,\mu) := \int \mu(a,w)\,\omega_0(a\,|\, w)\,da$ with a $P_0$-specific 
conditional density $\omega_0$. \qed
\end{example}

In Appendix~\ref{sec::mixedbias}, we show that the expected conditional covariance parameter
arising in semiparametric ATE estimation under partially linear models
\citep{robinson1988root,li2011higher}, as a measure of ``causal influence''
\citep{diaz2024non}, and in conditional independence testing
\citep{shah2020hardness} also falls
within our framework; doubly robust inference for this parameter was previously
studied by \cite{dukes2024doubly} and \cite{bonvini2024doubly}. We further show
that the average treatment effect under covariate-dependent outcome missingness
likewise fits within our framework; doubly robust inference for this estimand
was previously considered by \cite{diaz2017doubly}.

\subsection{One-step debiased estimation and rate double robustness}
\label{section::prelimDR}

Suppose we have an estimator \(\mu_n\) of the outcome regression \(\mu_0\) and an
estimator \(\psi_n\) of the target functional \(\psi_0\). For the
functionals described in
Example~\ref{example::expectfunctional}, a natural choice of $\psi_n$ is the empirical
plug-in map
\[
\psi_n(\mu):=\frac{1}{n}\sum_{i=1}^n m(Z_i,\mu).
\]
If the parameter involves a nuisance-dependent functional, 
\(\psi_n\) must satisfy Condition~\ref{cond::DRremainder}
in Appendix~\ref{appendix::mainconditions}. A natural estimator of the target parameter
\(\tau_0\) is then the plug-in estimator \(\psi_n(\mu_n)\). However, when
\(\mu_n\) is obtained using flexible statistical learning methods,
\(\psi_n(\mu_n)\) is typically not \(n^{1/2}\)-consistent and may fail to be
asymptotically normal, precluding valid inference, due to excess bias, as the
error \(\psi_0(\mu_n)-\psi_0(\mu_0)\) typically depends linearly on
\(\mu_n-\mu_0\) \citep{vanderLaanRose2011}. Debiasing methods aim to remove this
leading bias and thereby restore root-\(n\) inference.

A widely used approach is \emph{automatic debiased machine learning} (autoDML)
\citep{chernozhukov2018double,chernozhukov2022automatic,van2025automaticDML},
which augments the plug-in estimator with a bias correction based on an estimate
of the Riesz representer \(\alpha_0\) of \(\psi_0\). Indeed, by the Riesz
representation property~\eqref{eqn::reiszrep},
\[
\psi_0(\mu_n)-\psi_0(\mu_0)
=
\langle \alpha_0,\mu_n-\mu_0\rangle,
\]
so the leading plug-in error is linear in \(\mu_n-\mu_0\). Given an estimator
\(\alpha_n\) of \(\alpha_0\), this expression motivates the \textit{one-step debiased estimator}
\[
\tau_n
:=
\psi_n(\mu_n)
+
\frac{1}{n}\sum_{i=1}^n \alpha_n(A_i,W_i)\{Y_i-\mu_n(A_i,W_i)\},
\]
where the second term provides an empirical bias correction. This construction
generalizes the augmented inverse-probability-weighted estimator
\citep{robinsCausal,robins1995analysis,bang2005doubly,bruns2023augmented}. The nuisance estimators \(\mu_n\) and \(\alpha_n\)
may be constructed using flexible supervised learning methods. For example, \(\mu_n\) and \(\alpha_n\) may be obtained using empirical risk minimization based, respectively, on 
the least-squares loss \((z,\mu)\mapsto\{y-\mu(a,w)\}^2\) and empirical Riesz loss \((z,\alpha)\mapsto \alpha^2(a,w)-2\psi_n(\alpha)\)
\citep{chernozhukov2018double,chernozhukov2022riesznet}. Here, ``automatic'' refers to the fact that both the form of \(\tau_n\) and the
losses used to learn \(\mu_n\) and \(\alpha_n\) are generic, and do not require
tailoring to the specific functional \(\psi_n\).

A key property of the one-step debiased estimator \(\tau_n\) is \emph{rate double
robustness} \citep{bradic2019minimax, rotnitzky2021characterization}. Under mild regularity
conditions, if \(\|\alpha_n-\alpha_0\|=o_p(1)\), \(\|\mu_n-\mu_0\|=o_p(1)\), and
\(\|\alpha_n-\alpha_0\|\,\|\mu_n-\mu_0\|=o_p(n^{-1/2})\), then \(\tau_n\) is
asymptotically linear for \(\tau_0\) with influence function \(D_0\), i.e.,
\(\tau_n-\tau_0 = P_nD_0 + o_p(n^{-1/2})\). Hence, by the Central Limit Theorem, \(n^{1/2}(\tau_n-\tau_0)\) is  asymptotically
normal, and, by Theorem~\ref{theorem::EIF}, $\tau_n$ is efficient. Additionally,  \(\tau_n\) is doubly
robust for consistency: it converges in probability to \(\tau_0\) provided that
either \(\|\alpha_n-\alpha_0\|=o_p(1)\) or \(\|\mu_n-\mu_0\|=o_p(1)\). The
required regularity conditions amount to negligibility of certain
empirical-process remainder terms, which can be ensured, for example, by sample
splitting or cross-fitting \citep{schick1986asymptotically,
vanderLaanRose2011, chernozhukov2018double}.

Rate double robustness does \emph{not} guarantee doubly robust inference: valid
Wald-type inference based on standard rate doubly robust estimators requires that \emph{both} nuisance estimators be consistent
and satisfy the product-rate condition above (e.g., by both converging at rate
\(o_p(n^{-1/4})\)). By contrast, when  one
nuisance estimator is inconsistent or converges too slowly, the product-rate condition may fail for such estimators, and non-negligible asymptotic bias may then preclude asymptotic
linearity, invalidating Wald-type confidence intervals and typically causing
undercoverage. This gap motivates the stronger objective of \emph{doubly robust
asymptotic linearity}: estimators that remain asymptotically linear---and not only consistent---whenever \emph{either} nuisance estimator is sufficiently accurate, allowing the
other nuisance estimator to be inconsistent or converge arbitrarily slowly.

\section{Preliminaries on doubly robust inference}
\label{section:prelimDRinference}

\subsection{Statistical objective: doubly robust asymptotic linearity}
\label{section:prelimDRinference:sec1}

We aim to construct estimators of \( \tau_0 \) that are doubly robust
asymptotically linear (DRAL), meaning they remain asymptotically linear whenever
either
\begin{enumerate}
    \item[(i)] $\|\alpha_n-\alpha_0\| = o_p(n^{-1/4})$ and $\mu_n$ converges in
    probability to some $\overline{\mu}_0$, or
    \item[(ii)] $\|\mu_n - \mu_0\|= o_p(n^{-1/4})$ and $\alpha_n$ converges in
    probability to some $\overline{\alpha}_0$.
\end{enumerate}
The DRAL property strengthens rate double robustness by allowing one nuisance
estimator to converge arbitrarily slowly or even be inconsistent. In
particular, it yields \emph{best-rate double robustness}, replacing the usual
product-rate condition \( \| \alpha_n - \alpha_0 \| \cdot \| \mu_n - \mu_0 \| =
o_p(n^{-1/2}) \) with the weaker one-sided requirement \( \| \alpha_n - \alpha_0 \|^2
\wedge \| \mu_n - \mu_0 \|^2 = o_p(n^{-1/2}) \). We refer to this condition as `one-sided' because, unlike the product-rate
condition, which couples nuisance error rates so that the faster-converging nuisance
estimator can compensate for the slower-converging nuisance estimator, the best-rate double robustness condition does not. Without additional
assumptions on the data-generating distribution, rate double robustness cannot
be improved upon \citep{balakrishnan2023fundamental, jin2024structure,
jin2025sharp}. We therefore seek estimators that are DRAL under additional
assumptions, while retaining rate double robustness when these conditions fail.

To illustrate the improvement over rate double robustness, suppose that \(\mu_n\)
and \(\alpha_n\) converge to \(\mu_0\) and \(\alpha_0\) at rates
\(n^{-\beta/(2\beta+d)}\) and \(n^{-\gamma/(2\gamma+d)}\), where \(d\) is the
dimension of \(W\) and \(\beta,\gamma>0\) are smoothness exponents for the
nuisance functions. Such rates are attainable, for example, with local
polynomial and random forest-type methods when \(\mu_0\) and \(\alpha_0\) are
H\"older-smooth
\citep{stone1982optimal,wainwright2019high,biau2012analysis,mourtada2020minimax}.
To satisfy the product-rate condition
underlying rate double robustness, one typically requires \(\sqrt{\beta \gamma}>d/2\)
\citep{robins2008higher}; thus, if one nuisance is rough
\((\beta\wedge\gamma \ll d/2)\), the other must be sufficiently smooth to compensate, with greater smoothness required for greater roughness of the first nuisance.
By contrast, DRAL requires only \(\beta\vee\gamma>d/2\), so asymptotic
linearity can hold even when one nuisance is rough, provided the other is
sufficiently smooth to a degree independent of how rough the first nuisance is. An analogous separation holds under sparsity: if
\(\alpha_0\) and \(\mu_0\) have sparsities \(s_\alpha\) and \(s_\mu\) in ambient
dimension \(p\), then rate double robustness typically requires
\(\sqrt{s_\alpha s_\mu}\,\log p=o(\sqrt{n})\), whereas DRAL requires only
\((s_\alpha\wedge s_\mu)\,\log p=o(\sqrt{n})\)
\citep{bradic2019minimax,smucler2019unifying,liu2023root}. More generally, known convergence guarantees for the nuisance estimators translate
the estimator-level DRAL condition into primitive conditions on the nuisance
classes. In this sense, DRAL provides an infinite-dimensional analogue of model
double robustness.

 To elucidate how doubly robust asymptotic linearity can arise, we study the
one-step debiased estimator \(\tau_n\) under possible nuisance
misspecification. Suppose that \(\alpha_n\) and
\(\mu_n\) tend, respectively, to $\overline{\alpha}_0$ and $\overline{\mu}_0$ in probability, in the sense that
\(\|\alpha_n-\overline{\alpha}_0\|=o_p(1)\) and
\(\|\mu_n-\overline{\mu}_0\|=o_p(1)\), where at least one limit is correct,
i.e., either \(\overline{\mu}_0=\mu_0\) or \(\overline{\alpha}_0=\alpha_0\).
Under sample splitting, or under constraints on nuisance estimator complexity, we
obtain the linear approximation: 
\begin{equation}
  \tau_n - \tau_0
  = (P_n - P_0)\,\overline{D}_0
    + o_p(n^{-1/2})
    + \langle \alpha_0 - \alpha_n, \mu_n - \mu_0 \rangle,
  \label{eqn::drbias2}
\end{equation}
where \(\overline{D}_0 := \phi_{0,\overline{\mu}_0}
+ D_{\overline{\mu}_0,\overline{\alpha}_0}\). Thus,  the only obstruction to asymptotic linearity of
\(\tau_n\) is the cross-product remainder
\(\langle \alpha_0-\alpha_n,\,\mu_n-\mu_0\rangle\). Most of the existing
literature controls this remainder via rate double robustness, imposing the
product-rate condition \(\|\alpha_n-\alpha_0\|\,\|\mu_n-\mu_0\|=o_p(n^{-1/2})\),
which typically requires both nuisance estimators to be consistent and to
converge sufficiently fast \citep{robinsCausal,robins1995analysis,
vanderLaanRose2011,DoubleML}. If both limits are correct, so that
\(\overline{\mu}_0=\mu_0\) and \(\overline{\alpha}_0=\alpha_0\), then
\(\overline{D}_0\) coincides with the efficient influence function \(D_0\) in
Theorem~\ref{theorem::EIF}, yielding efficiency.

When the product-rate condition fails---because one nuisance estimator is
inconsistent or converges too slowly---the cross-product remainder in
\eqref{eqn::drbias2} is no longer negligible and can drive the behavior of
\(n^{1/2}(\tau_n-\tau_0)\), precluding asymptotic linearity of the one-step
estimator. Achieving DRAL therefore requires
additional control or debiasing of \(\langle \alpha_0-\alpha_n,\,\mu_n-\mu_0\rangle\)
so that it is negligible, or itself admits an asymptotically linear representation,
even when \(\|\alpha_n-\alpha_0\|\,\|\mu_n-\mu_0\|=o_p(n^{-1/2})\) does not hold. In the
remainder of this section, we analyze this term and show how, under suitable conditions, additional
debiasing yields DRAL whenever either (i) or (ii) holds. Readers primarily interested in the estimator may wish to proceed
directly to Section~\ref{section::proposedestimator}.

\subsection{General strategy: debiasing the cross-product remainder}
\label{sec::genstrat}

\subsubsection{A learner-agnostic bias expansion under calibration and partial orthogonality}
 \label{sec::genstrat:sub1}
We now describe a general strategy for debiasing the cross-product remainder
\(\langle \alpha_0-\alpha_n,\mu_n-\mu_0\rangle\) by calibrating the nuisance
estimators \(\alpha_n\) and \(\mu_n\). The goal is to construct these estimators
so that the cross-product remainder admits an asymptotically linear
representation, behaving like the empirical mean of a mean-zero, possibly
degenerate, influence function. This strategy builds on
\citet{van2014targeted} and subsequent work
\citep{benkeser2017doubly,diaz2017doubly,diaz2019statistical,dukes2024doubly},
which construct DRAL estimators of average treatment effects when one nuisance
estimator may be inconsistent. We extend this line of work in two directions:
we consider a broad class of statistical parameters beyond the ATE, and we analyze regimes in which both nuisance estimators are
consistent but one may converge arbitrarily slowly. In these settings, we
establish asymptotic linearity of the cross-product remainder and give refined
conditions under which the remaining higher-order terms are negligible.

 We use the following notation throughout this subsection. The nuisance
errors are denoted by \(\Delta_{\alpha,n}:=\alpha_n-\alpha_0\) and
\(\Delta_{\mu,n}:=\mu_n-\mu_0\). For any map
\(v:\mathcal A\times\mathcal W\to\mathbb R^k\), \(\Pi_v\) denotes the
\(L^2(P_0)\)-projection onto \(\{h\circ v:h:\mathbb R^k\to\mathbb R\}\). When
\(v\) and \(f\) are nonrandom,
\((\Pi_v f)(a,w)=E_0\{f(A,W)\,|\, v(A,W)=v(a,w)\}\). Define the projected Riesz
and regression errors by
\[
s_{n,0}:=\Pi_{\mu_n}(\alpha_0-\alpha_n)
        =-\Pi_{\mu_n}\Delta_{\alpha,n},
\qquad
r_{n,0}:=\Pi_{\alpha_n}(\mu_0-\mu_n)
        =-\Pi_{\alpha_n}\Delta_{\mu,n}.
\]
Thus, \(s_{n,0}\) is the component of the Riesz error
\(\alpha_0-\alpha_n\) explained by \(\mu_n\), while \(r_{n,0}\) is the
component of the regression error \(\mu_0-\mu_n\) explained by
\(\alpha_n\).

Our debiasing strategy calibrates the nuisance estimators so that their errors
are empirically orthogonal along the projected directions \(s_{n,0}\) and
\(r_{n,0}\). Specifically, we construct calibrated estimators \(\mu_n\) and
\(\alpha_n\) satisfying
\begin{equation}
\frac{1}{n}\sum_{i=1}^n s_{n,0}(A_i,W_i)\{Y_i-\mu_n(A_i,W_i)\}
=0 ,
\label{eqn::orthogonaloutcome}
\end{equation}
\begin{equation}
\psi_n(r_{n,0})
-
\frac{1}{n}\sum_{i=1}^n r_{n,0}(A_i,W_i)\alpha_n(A_i,W_i)
=0 .
\label{eqn::orthogonalRiesz}
\end{equation}
The first condition orthogonalizes the regression residuals against the
projected Riesz error \(s_{n,0}\). The second enforces the empirical Riesz
representation $\psi_n(r_{n,0})=\langle \alpha_n,r_{n,0}\rangle_{P_n}$ along the projected regression error \(r_{n,0}\). If
\(\psi_n(\cdot)=\langle\cdot,\alpha_n^\star\rangle_{P_n}\) for an empirical
representer \(\alpha_n^\star\), then the second condition amounts to $\langle r_{n,0},\alpha_n^\star-\alpha_n\rangle_{P_n}=0.$ For the average treatment effect, \citet{benkeser2017doubly} approximately
satisfied analogous conditions using parameter-specific TMLE-style procedures
based on estimates of \(s_{n,0}\) and \(r_{n,0}\); see
Appendix~\ref{sec::drtmle}. In the next section, we show that isotonic
calibration provides a simple way to enforce these conditions exactly. For now, we
treat \(\mu_n\) and \(\alpha_n\) satisfying
\eqref{eqn::orthogonaloutcome}--\eqref{eqn::orthogonalRiesz} as given. 

To build intuition, we first study the cross-product remainder under \eqref{eqn::orthogonaloutcome} and show how this condition reduces sensitivity to estimation error in \(\alpha_n\) when \(\mu_n\) is consistent for \(\mu_0\). Following \cite{van2014targeted} and related work, we begin with the observation that the cross-product remainder depends only on the component of the Riesz estimation error \(\alpha_n(A,W)-\alpha_0(A,W)\) lying in the linear span of \(\mu_n(A,W)\) and \(\mu_0(A,W)\). Specifically, by the law of iterated expectations, conditioning on \((\mu_n(A,W),\mu_0(A,W))\) under \(P_0\) yields
\begin{align*}
\langle \alpha_0 - \alpha_n , \mu_n - \mu_0 \rangle
&= \langle \Pi_{(\mu_n, \mu_0)}(\alpha_0 - \alpha_n), \mu_n - \mu_0\rangle.
\end{align*}
The orthogonality condition \eqref{eqn::orthogonaloutcome} then applies a minimal correction to the identifiable component \(\Pi_{\mu_n}(\alpha_0-\alpha_n)\) of the projected error \(\Pi_{(\mu_n,\mu_0)}(\alpha_0-\alpha_n)\), that is, the part determined by \(\mu_n(A,W)\). Specifically, following \cite{benkeser2017doubly}, \cite{dukes2024doubly}, and \cite{bonvini2024doubly}, we consider the decomposition
\begin{equation}
\label{eqn::DRexpansion}
\begin{aligned}
\langle \alpha_0 - \alpha_n , \mu_n - \mu_0 \rangle
&= \langle \Pi_{\mu_n}(\alpha_0 - \alpha_n), \mu_n - \mu_0\rangle
  + \langle (\Pi_{(\mu_n , \mu_0)} - \Pi_{\mu_n})(\alpha_0 - \alpha_n),
           \mu_n - \mu_0\rangle \\
&= \langle s_{n,0}, \mu_n - \mu_0\rangle
  + \langle (\Pi_{(\mu_n , \mu_0)} - \Pi_{\mu_n}) \Delta_{\alpha,n},
           \mu_0 - \Pi_{\mu_n} \mu_0 \rangle,
\end{aligned}
\end{equation}
where the second term is negligible when \(\mu_0\) provides little information about the residual \(\Delta_{\alpha,n}\) beyond what is already captured by \(\mu_n\). The first term is the leading bias component and can prevent asymptotic linearity of \(\tau_n\) when \(\alpha_0\) is estimated too slowly or inconsistently. The second term is what we refer to below as the higher-order cross-product remainder. The key role of \eqref{eqn::orthogonaloutcome} is to facilitate the linearization of this leading term up to an empirical process fluctuation,
\[
\langle s_{n,0}, \mu_n - \mu_0\rangle
= - P_0 A_{n,0}
= (P_n - P_0) A_{n,0}
= (P_n-P_0) A_0 + (P_n - P_0) (A_{n,0} - A_0),
\]
where   $A_{n,0}(z)
:= \Pi_{\mu_n}(\alpha_0-\alpha_n)(a,w)\{y-\mu_n(a,w)\}$ and $A_0(z)
:= \Pi_{\mu_0}(\alpha_0-\overline\alpha_0)(a,w)\{y-\mu_0(a,w)\}.$
The second equality uses that \(P_n A_{n,0}=0\), as implied by
\eqref{eqn::orthogonaloutcome}. Provided that
\(\|A_{n,0}-A_0\|=o_p(1)\) and \(A_{n,0}\) satisfies an appropriate empirical
process condition, the final term on the right is \(o_p(n^{-1/2})\). We verify
this condition for isotonic calibration in our proofs; see also
Appendix~\ref{appendix::EPhandle}.

The following theorem formalizes the above by deriving a bias expansion showing
how \eqref{eqn::orthogonaloutcome} and \eqref{eqn::orthogonalRiesz} debias the
cross-product remainder. To control the higher-order terms in
\eqref{eqn::DRexpansion} and related expansions, we impose the following
high-level condition, with sufficient conditions given later. Informally, we
require that the estimation error of the slower or inconsistent nuisance be
asymptotically orthogonal after partialling out the other nuisance:
\begin{enumerate}[label=\bf{B\arabic*)}, ref={B\arabic*},series=sketch]
  \item \textit{Asymptotic partial orthogonality:} At least one of the following holds: \label{cond:ortho} 
    \begin{enumerate}[label={\roman*)}, ref={\ref{cond:ortho}\roman*}]\vspace{-.05in}
      \item \label{cond:orthoA} \(\langle (\Pi_{(\mu_n , \mu_0)}-\Pi_{\mu_n})\Delta_{\alpha,n}
        \,,\, \mu_0 - \Pi_{\mu_n} \mu_0 \rangle = O_p(\|\mu_n -  \mu_0\|^2)\),\vspace{-.05in}
      \item \label{cond:orthoB} \(\langle (\Pi_{(\alpha_n , \alpha_0)}-\Pi_{\alpha_n})\Delta_{\mu,n}
        \,,\, \alpha_0 - \Pi_{\alpha_n} \alpha_0 \rangle = O_p(\|\alpha_n -  \alpha_0\|^2)\).
    \end{enumerate}
\end{enumerate}
The term in \ref{cond:orthoA} is a covariance between the
component of \(\Pi_{(\mu_n,\mu_0)}\Delta_{\alpha,n}\) that remains after
partialling out functions of \(\mu_n\) and the approximation error
\(\mu_0-\Pi_{\mu_n}\mu_0\). When this condition holds, the higher-order term in
\eqref{eqn::DRexpansion} is \(O_p(\|\mu_n-\mu_0\|^2)\); it is therefore
asymptotically negligible when \(\|\mu_n-\mu_0\| = o_P(n^{-1/4})\), regardless
of the quality of \(\alpha_n\). Similarly, Condition~\ref{cond:orthoB} ensures a related higher-order term is negligible with the roles of $\alpha$ and $\mu$ reversed. We provide sufficient conditions in Section \ref{sec::sufficientconds}. As a concrete example, Condition~\ref{cond:orthoA} holds if \(\Pi_{(\mu_n,\mu_0)}\Delta_{\alpha,n}\) admits the approximate partially linear representation
\begin{equation}
\label{eqn::PLM}
    \Pi_{(\mu_n,\mu_0)}\Delta_{\alpha,n} = g_n \circ \mu_n + \beta_n \circ (\mu_n,\mu_0)\cdot(\mu_n-\mu_0) + o_p(\|\mu_n-\mu_0\|)
\end{equation}
in \(L^2(P_0)\) for some \(g_n\colon \mathbb{R}\to\mathbb{R}\) and \(\beta_n\colon \mathbb{R}^2\to\mathbb{R}\) satisfying \(\|\beta_n\|_{\infty}=O_p(1)\). In particular, this condition holds if \((\Pi_{(\mu_n,\mu_0)}\Delta_{\alpha,n})(A,W)\) is sufficiently smooth in \(\mu_n(A,W)\) and \(\mu_0(A,W)\).

The following theorem involves two additional nuisance functions:
\begin{equation}
\label{eqn::nuisnew}
\begin{aligned}
s_0(a,w)
&:= E_0\!\left\{\alpha_0(A,W)-\overline{\alpha}_0(A,W)
    \mid \mu_0(A,W)=\mu_0(a,w)\right\},\\
r_0(a,w)
&:= E_0\!\left\{\mu_0(A,W)-\overline{\mu}_0(A,W)
    \mid \alpha_0(A,W)=\alpha_0(a,w)\right\}.
\end{aligned}
\end{equation}
In the short-hand notation already introduced, this can be written as
\(s_0=\Pi_{\mu_0}(\alpha_0-\overline\alpha_0)\) and
\(r_0=\Pi_{\alpha_0}(\mu_0-\overline\mu_0)\). Define $B_{n,0}(z)
:=
\phi_{r_{n,0}}(z)+\psi_0(r_{n,0})
-r_{n,0}(a,w)\alpha_n(a,w)$ and $
B_0(z)
:=
\phi_{r_0}(z)+\psi_0(r_0)
-r_0(a,w)\alpha_0(a,w).$
 The theorem does not require \(\mu_n\) and \(\alpha_n\) to converge to
\(\overline{\mu}_0\) and \(\overline{\alpha}_0\), but for simplicity we impose this convergence
in all subsequent results and discussion.
 
\begin{theorem}[Expansion of cross-product remainder under orthogonality]
\label{theorem::DRbiasexpansion} Both of the following are valid:
\begin{enumerate}[label=(\roman*)]
    \item \emph{Regression-based decomposition:} If $\mu_n$ satisfies \eqref{eqn::orthogonaloutcome}, 
\begin{align*}
\langle \alpha_0 - \alpha_n, \mu_n - \mu_0 \rangle  &= P_n A_0 +  (P_n - P_0)(A_{n,0} - A_0)  + \langle (\Pi_{(\mu_n, \mu_0)} - \Pi_{\mu_n})\Delta_{\alpha,n}, \,\mu_0 - \Pi_{\mu_n} \mu_0 \rangle .
\end{align*}
 If \ref{cond:orthoA} holds, then
$ \langle (\Pi_{(\mu_n, \mu_0)} - \Pi_{\mu_n})\Delta_{\alpha,n}, \,\mu_0 - \Pi_{\mu_n} \mu_0 \rangle
= O_p\left(\|\mu_n - \mu_0 \|^2\wedge\|\Pi_{\mu_n}\mu_0 - \mu_0 \|\|\alpha_n - \alpha_0\| \right)$. 
\item \emph{Riesz representer-based decomposition:} If $\alpha_n$ satisfies \eqref{eqn::orthogonalRiesz}, 
\begin{align*}
\langle \alpha_0 - \alpha_n, \mu_n - \mu_0 \rangle  &=  P_n B_0 +  (P_n - P_0)(B_{n,0} - B_0) + \langle (\Pi_{(\alpha_n, \alpha_0)} - \Pi_{\alpha_n})\Delta_{\mu,n}, \,\alpha_0 - \Pi_{\alpha_n} \alpha_0 \rangle   + Rem_{\psi_n}(r_{n,0})
\end{align*}
 with
$ Rem_{\psi_n}(r_{n,0}) := \psi_n(r_{n,0}) - \psi_0(r_{n,0}) - P_n \phi_{r_{n,0}}$. If \ref{cond:orthoB} holds, then
$\langle (\Pi_{(\alpha_n, \alpha_0)} - \Pi_{\alpha_n})\Delta_{\mu,n}, \,\Delta_{\alpha,n} \rangle
=  O_p\left(\|\alpha_n - \alpha_0 \|^2 \wedge\|\mu_n - \mu_0 \|\|\Pi_{\alpha_n}\alpha_0 - \alpha_0\| \right)$.
 
\end{enumerate}

\end{theorem}
When the nuisance estimators satisfy the empirical orthogonality conditions in \eqref{eqn::orthogonaloutcome} and \eqref{eqn::orthogonalRiesz}, Theorem~\ref{theorem::DRbiasexpansion} decomposes the cross-product remainder into an asymptotically linear term, an empirical process remainder, and a higher-order cross-product remainder. Under the regression-based decomposition, these terms are, respectively, \(P_nA_0\), \((P_n-P_0)(A_{n,0}-A_0)\), and \(\langle (\Pi_{(\mu_n,\mu_0)}-\Pi_{\mu_n})\Delta_{\alpha,n},\, \mu_0-\Pi_{\mu_n}\mu_0\rangle\).  Recalling \eqref{eqn::drbias2}, if the remainder terms are \(o_p(n^{-1/2})\), then the one-step debiased estimator \(\tau_n\), constructed from the orthogonalized nuisance estimators \(\mu_n\) and \(\alpha_n\), is DRAL with influence function \(\overline{D}_0 + 1(\overline{\mu}_0 \neq \mu_0)\,A_0 + 1(\overline{\alpha}_0 \neq \alpha_0)\,B_0\).  Moreover, if both nuisance estimators are consistent, so that \(\overline{\alpha}_0=\alpha_0\) and \(\overline{\mu}_0=\mu_0\), then \(A_0=0\) and \(B_0=0\). Hence, \(\tau_n\) is asymptotically linear with influence function \( D_0\), provided the cross-product remainder is indeed asymptotically negligible.

The traditional cross-product remainder is often bounded using the
Cauchy--Schwarz inequality, yielding $|\langle \alpha_0-\alpha_n,\mu_n-\mu_0\rangle|
\le
\|\alpha_n-\alpha_0\|\,\|\mu_n-\mu_0\|$ \citep{vanderLaanRose2011,chernozhukov2018double,bonvini2024doubly}.
Similarly, for the higher-order cross-product remainder, an application of the
Cauchy--Schwarz inequality yields
\begin{equation}
 \big|
 \langle (\Pi_{(\mu_n,\mu_0)}-\Pi_{\mu_n})\Delta_{\alpha,n},
  \,\mu_0-\Pi_{\mu_n}\mu_0\rangle
 \big|
 \le
 \|(\Pi_{(\mu_n,\mu_0)}-\Pi_{\mu_n})(\alpha_n-\alpha_0)\|\,
 \|\mu_0-\Pi_{\mu_n}\mu_0\|.
 \label{eqn::worstcasebound}
\end{equation}
This bound is no larger than the usual product bound because it projects out the components of the nuisance errors explained by \(\mu_n\).  Thus, in view of \eqref{eqn::drbias2}, even when the partial orthogonality condition fails, calibration yields a linear approximation to $\tau_n-\tau_0$ whose remainder, up to higher-order terms, admits a worst-case bound that is generally sharper (and never worse) than the classical worst-case bound $\|\alpha_n-\alpha_0\|\,\|\mu_n-\mu_0\|$ for the remainder of the linear approximation available without calibration. When the partial orthogonality condition holds, the bound in
\eqref{eqn::worstcasebound} translates into improved rates: under \ref{cond:orthoA}, it converges to zero at least as fast as both $\|\mu_n - \mu_0 \|^2$ and $\|\alpha_n - \alpha_0\|\|\Pi_{\mu_n}\mu_0 - \mu_0 \|=O_p(\|\alpha_n-\alpha_0\|\|\mu_n - \mu_0\|)$. An analogous argument can instead be made using the Riesz-based decomposition, in which \(\alpha_n\) is projected out from both errors.

\subsubsection{Sufficient conditions for partial orthogonality}
\label{sec::sufficientconds}

A sufficient condition for \ref{cond:ortho} is that the projected Riesz error \(\Pi_{(\mu_n,\mu_0)}\Delta_{\alpha,n}\) admits the partially linear expansion in \eqref{eqn::PLM}. At a high level, \ref{cond:orthoA} rules out near-perfect alignment between the projected-out regression error and the Riesz error. For example, if \(\mu_0 - \Pi_{\mu_n}\mu_0 = \varepsilon_n \Delta_{\alpha,n}/\|\Delta_{\alpha,n}\|\) for some \(\varepsilon_n=o(1)\), then \(\langle (\Pi_{(\mu_n,\mu_0)}-\Pi_{\mu_n})\Delta_{\alpha,n}, \mu_0-\Pi_{\mu_n}\mu_0\rangle\) is of order \(\|\Delta_{\alpha,n}\|\,\|\mu_0-\Pi_{\mu_n}\mu_0\|\), which need not be \(o_p(\|\mu_n-\mu_0\|^2)\), especially when \(\|\Delta_{\alpha,n}\|\) does not converge to zero in probability. Conditions excluding this type of behavior already appear in related work; see, for example, Condition~(v) in Appendix~F of \citet{benkeser2020nonparametric}, Theorem~1 of \citet{benkeser2017doubly}, Assumption~3 of \citet{dukes2024doubly}, Lemma~I.1 of \citet{wang2023super}, and Section~3.1 of \citet{bonvini2024doubly}. We next adapt these ideas to our setting and present sufficient conditions in increasing order of strength.

  \begin{enumerate}[label=\bf{\ref{cond:ortho}$^{*}$}, ref={\ref{cond:ortho}$^{*}$}]
  \item \textit{Projection error coupling}  \citep{van2014targeted}: \label{cond:projectioncoupling1}  
      \begin{enumerate}[label={\roman*)}, ref={\ref{cond:projectioncoupling1}\roman*}]\vspace{-.05in}
     \item \label{cond:projectioncoupling1A}   $  \|(\Pi_{\mu_n, \mu_0} - \Pi_{\mu_n})\Delta_{\alpha,n}\|  =   O_p(\|\mu_n - \mu_0 \|)$\,; \vspace{-.05in}
     \item \label{cond:projectioncoupling1B}   $\|(\Pi_{\alpha_n, \alpha_0} - \Pi_{\alpha_n})\Delta_{\mu,n}\|  = O_p( \|\alpha_n - \alpha_0 \|)$\,.
 
    \end{enumerate}

\end{enumerate}
The above condition is also assumed in \citet[][Theorem~1]{benkeser2017doubly} and \citet[][Assumption~3]{dukes2024doubly}; related conditions appear in \cite{diaz2017doubly} and \cite{diaz2019statistical}.
  
  \begin{enumerate}[label=\bf{\ref{cond:ortho}$^{**}$}, ref={\ref{cond:ortho}$^{**}$}]
    \item\textit{Lipschitz continuity} 
\citep{bonvini2024doubly}: Denoting\ $\mathcal{D}_n := \{Z_1,\ldots,Z_n\}$, it holds that, for all $n$ large enough:\label{cond:projectioncoupling2}  
    \begin{enumerate}[
  label={\roman*)},
  ref=\ref{cond:projectioncoupling2}\roman*,
  leftmargin=-10pt,
  labelsep=-0.7em,
  labelindent=0pt,
  align=left
]\item $(\widehat{m}, m) \mapsto E_0\{\Delta_{\alpha,n}(A,W) \,|\,  \mu_n(A,W) = \widehat{m}, \mu_0(A,W) = m, \mathcal{D}_n\}$ is $P_0$--almost surely Lipschitz continuous with uniformly bounded constant;\label{cond:projectioncoupling2A}  
    \item  $(\widehat{a}, a) \mapsto E_0\{\Delta_{\mu,n}(A,W) \,|\,  \alpha_n(A,W) = \widehat{a}, \alpha_0(A,W) = a, \mathcal{D}_n\}$ is $P_0$--almost surely Lipschitz continuous with uniformly bounded constant.\label{cond:projectioncoupling2B}  
    \end{enumerate}
    
\end{enumerate}
The above condition is used in \citet[][Section~3.1]{bonvini2024doubly} and related bounded derivative conditions are also imposed in \citet[][Condition~v in Appendix~F]{benkeser2020nonparametric}, \citet[][Lemma~1]{van2023adaptive}, and \citet[][Lemma~I.1]{wang2023super} to analyze regressions involving estimated features.

 \begin{lemma}[Sufficient conditions for \ref{cond:ortho}]
 \label{lemma::sufficientcoupling}
Conditions \ref{cond:orthoA} and \ref{cond:orthoB} are implied by
\ref{cond:projectioncoupling1A} and \ref{cond:projectioncoupling1B},
respectively. Moreover, \ref{cond:projectioncoupling1A} and
\ref{cond:projectioncoupling1B} are implied by
\ref{cond:projectioncoupling2A} and \ref{cond:projectioncoupling2B},
respectively.
\end{lemma}

Roughly, Condition~\ref{cond:projectioncoupling1A} requires that \(\Pi_{(\mu_n,\mu_0)}\Delta_{\alpha,n}\) be approximated well by \(\Pi_{\mu_n}\Delta_{\alpha,n}\) whenever \(\mu_n\) converges to \(\mu_0\) in norm. Heuristically, it asserts that, as \(\|\mu_n-\mu_0\|\to 0\), the information in \[\{\mu_n(A,W),\mu_0(A,W),Z_1,\ldots,Z_n\}\] for predicting \(\Delta_{\alpha,n}\) is asymptotically equivalent to that in \(\{\mu_n(A,W),Z_1,\ldots,Z_n\}\). While such equivalences need not hold in general, a sufficient condition is Condition~\ref{cond:projectioncoupling2}, which ensures that \(\Pi_{(\mu_n,\mu_0)}\Delta_{\alpha,n}\) and \(\Pi_{(\alpha_n,\alpha_0)}\Delta_{\mu,n}\) are almost surely Lipschitz continuous as bivariate transformations of \((\mu_n,\mu_0)\) and \((\alpha_n,\alpha_0)\), respectively. By symmetry between \(\mu_n\) and \(\mu_0\), Condition~\ref{cond:projectioncoupling2A} also implies \(\|(\Pi_{(\mu_n,\mu_0)}-\Pi_{\mu_0})\Delta_{\alpha,n}\| = O_p(\|\mu_n-\mu_0\|)\) and hence, by the triangle inequality, \(\|(\Pi_{\mu_0}-\Pi_{\mu_n})\Delta_{\alpha,n}\| = O_p(\|\mu_n-\mu_0\|)\). An analogous conclusion holds under
Condition~\ref{cond:projectioncoupling2B}. Conditions~\ref{cond:ortho}--\ref{cond:projectioncoupling2}
may also be weakened to H\"older-type continuity conditions, as in
\citet{bonvini2024doubly}. In that case, Theorem~\ref{theorem::DRbiasexpansion}
yields a corresponding weaker bias expansion, with rate determined by the
H\"older exponent; for instance, the quadratic exponent in
Condition~\ref{cond:ortho} could be replaced by \(1+\varepsilon\) for some
\(\varepsilon>0\).

More broadly, some form of structural assumption is unavoidable for
meaningfully improving upon standard doubly robust estimators, since
first-order methods are optimal in structure-agnostic settings
\citep{balakrishnan2023fundamental, jin2025sharp}. The conditions
above are one such set of
assumptions. For comparison, analyses of higher-order influence function
estimators \citep{robins2008higher} typically impose H\"older-type smoothness
assumptions on the potentially high-dimensional error functions
\(\Delta_{\mu,n}\) and \(\Delta_{\alpha,n}\), whereas sparsity-based methods
require approximate sparsity and linearity of the nuisance models
\citep{liu2023root}. Our guarantees instead rely on the \emph{asymptotic
partial orthogonality} condition in \ref{cond:ortho}, which is qualitatively
different from the smoothness and approximability assumptions used in these
approaches. In particular, the sufficient Condition~\ref{cond:projectioncoupling2}
requires only smoothness of certain bivariate transformations of the true and
estimated nuisances. Informally, Condition~\ref{cond:projectioncoupling2A} may be
plausible when the joint level sets
\(\{(a,w): \mu_n(a,w)=\widehat m,\ \mu_0(a,w)=m\}\) vary smoothly with
\((\widehat m,m)\), and the conditional mean of \(\Delta_{\alpha,n}\) over
these level sets depends continuously on \((\widehat m,m)\). Importantly, even when these additional conditions do not hold,  \eqref{eqn::worstcasebound} shows that satisfaction of the empirical orthogonality conditions typically results in a one-step debiased estimator with attenuated bias.  

\section{Calibrated debiased machine learning}
\label{section::proposedestimator}

\subsection{Enforcing orthogonality via calibration}
\label{section::calibration}

 We now establish a key insight of this paper: the
empirical orthogonality equations \(\eqref{eqn::orthogonaloutcome}\) and
\(\eqref{eqn::orthogonalRiesz}\) can be enforced by calibrating the nuisance
estimators \(\mu_n\) and \(\alpha_n\).  Combined with Theorem~\ref{theorem::DRbiasexpansion} of the previous subsection, this
reformulation shows that calibration enables the construction of DRAL debiased
machine learning estimators, provided the nuisance estimators satisfy the
partial orthogonality conditions in \ref{cond:ortho}. In particular,
calibration implicitly performs the bias correction needed for the error
decomposition in Theorem~\ref{theorem::DRbiasexpansion} to apply. We now
formalize this connection and introduce our calibrated debiased machine
learning framework. 

Calibration is a widely used post-hoc technique in predictive modeling that
transforms a model's fitted values to better align with its prediction target \citep{lichtenstein1977calibration, zadrozny2002transforming,
niculescu2005predicting, bella2010calibration, guo2017calibration,
gupta2020distribution}. Motivated by this idea, we say that \(\mu_n\) and \(\alpha_n\) are
\emph{(perfectly) empirically calibrated} with respect to the least-squares and
Riesz losses if
\begin{align}
\sum_{i=1}^n \{Y_i - \mu_n(A_i, W_i)\}^2
&= \min_{\theta}\sum_{i=1}^n \{Y_i - \theta(\mu_n(A_i, W_i))\}^2,
\label{eqn::calpropertyoutcome} \\
\sum_{i=1}^n \{\alpha_n(A_i, W_i)\}^2 - 2 \psi_n(\alpha_n)
&= \min_{\theta}\Bigg\{
    \sum_{i=1}^n \{\theta(\alpha_n(A_i, W_i))\}^2
    - 2 \psi_n(\theta \circ \alpha_n)
  \Bigg\},
\label{eqn::calpropertyRiesz}
\end{align}
where the minima are taken over all one-dimensional functions
\(\theta:\mathbb{R}\to\mathbb{R}\). Equivalently, the empirical risk of each nuisance estimator cannot be reduced by
post-composition with an arbitrary one-dimensional mapping
\citep{whitehouse2024orthogonal}.  Beyond enabling doubly robust inference, empirical calibration can improve
estimator stability and predictive accuracy. Calibration for the outcome regression ensures that $\mu_n(a,w)$ equals the empirical mean outcome
among individuals with the same predicted value, whereas calibration for the propensity
score enforces covariate balance across strata of the estimated propensity score
\citep{deshpande2023calibrated,van_der_Laan2024stabilized}.

Empirical calibration of the nuisance estimators implies the orthogonality
conditions \eqref{eqn::orthogonaloutcome} and \eqref{eqn::orthogonalRiesz}. To
ensure that this implication holds for \(\alpha_n\), we assume that the
estimated functional \(\psi_n\) is linear, as holds automatically for
\(\psi_n(\mu) := \frac{1}{n}\sum_{i=1}^n m(Z_i,\mu)\) in
Example~\ref{example::expectfunctional}.
\begin{enumerate}[label=\bf{B\arabic*)}, ref={B\arabic*},resume=sketch]
 \item \textit{(Linearity of the estimated functional)}
 \label{cond::estboundedlinear}
 \(P_0\)--almost surely, $\psi_n:\mathcal{H} \rightarrow \mathbb{R}$ satisfies \(\psi_n(c\mu_1 + \mu_2) = c\,\psi_n(\mu_1) + \psi_n(\mu_2)\)
 for each \(\mu_1,\mu_2\in \mathcal{H}\) and \(c\in\mathbb{R}\).
\end{enumerate}
 Under \ref{cond::estboundedlinear}, the first-order optimality conditions for
\eqref{eqn::calpropertyoutcome} and \eqref{eqn::calpropertyRiesz} yield the
equivalent moment conditions: for each transformation
\(\theta:\mathbb{R}\to\mathbb{R}\),
\begin{align}
    \frac{1}{n}\sum_{i=1}^n \theta\left(\mu_n(A_i,W_i)\right)\{Y_i - \mu_n(A_i,W_i)\}
    &= 0, \label{eqn::calpropertyoutcome2}\\
    \frac{1}{n}\sum_{i=1}^n \theta\left(\alpha_n(A_i,W_i)\right)\alpha_n(A_i,W_i)
    - \psi_n(\theta \circ \alpha_n)
    &= 0. \label{eqn::calpropertyRiesz2}
\end{align}
In other words, calibration requires the residuals of \(\mu_n\) and \(\alpha_n\) to be
orthogonal to \emph{every} one-dimensional transformation of the corresponding
fitted values. In particular, they are orthogonal to the specific
transformations \(s_{n,0}=\Pi_{\mu_n}(\alpha_0 - \alpha_n)\) and
\(r_{n,0}=\Pi_{\alpha_n}(\mu_0 - \mu_n)\) appearing in the orthogonality conditions. We summarize this implication in the
following lemma.

\begin{lemma}[Empirical calibration debiases the cross-product remainder]
\label{theorem:empiricalcalibration}
If $\mu_n$ is empirically calibrated so that \eqref{eqn::calpropertyoutcome}
holds, then $\mu_n$ satisfies \eqref{eqn::orthogonaloutcome}. If, in addition, \ref{cond::estboundedlinear} holds and $\alpha_n$ satisfies
\eqref{eqn::calpropertyRiesz}, then $\alpha_n$ satisfies
\eqref{eqn::orthogonalRiesz}.
\end{lemma}
\noindent We refer to debiased machine learning with calibrated nuisance estimators as
\emph{calibrated DML}. 

In practice, calibration is implemented by applying a post-hoc one-dimensional map to an
initial nuisance estimator, chosen to reduce the empirical loss over a flexible class \citep{whitehouse2024orthogonal}.
A simple option is histogram (binning) calibration: partition the fitted values and assign a constant calibrated prediction within each bin, chosen to minimize the loss within that bin (e.g., an
empirical mean) \citep{zadrozny2001obtaining,gupta2021distribution,van2025generalized}. Since histogram-calibrated estimators minimize the loss within each bin, they
also minimize the loss within each level set of the calibrated fit, and
therefore satisfy \eqref{eqn::calpropertyoutcome} and
\eqref{eqn::calpropertyRiesz} exactly. A widely used alternative is \emph{isotonic calibration} \citep{zadrozny2002transforming,niculescu2005predicting},
which fits a monotone map and can be viewed as adaptive binning regularized by a
monotonicity constraint \citep{van2023causal,van2025generalized}. It is tuning-free \citep{guntuboyina2018nonparametric}  and, unlike histogram binning, can only improve or leave unchanged the empirical risk, since the identity map is monotone. Moreover, the calibration properties in \eqref{eqn::calpropertyoutcome} and
\eqref{eqn::calpropertyRiesz} follow because isotonic regression is piecewise
constant on the blocks produced by the pooled adjacent violators algorithm, and
within each block the fitted value minimizes the corresponding empirical loss
\citep{barlow1972isotonic}.

\subsection{Proposed estimator using isotonic calibration}
\label{sec:proposedest}
   We now introduce \emph{isocalibrated DML}, a concrete instantiation of calibrated DML that enforces \eqref{eqn::calpropertyoutcome} and \eqref{eqn::calpropertyRiesz} exactly through isotonic calibration. The procedure is automatic: it requires no knowledge of the specific form of
the linear functional \(\psi_0\), only black-box access to an asymptotically
linear estimator \(\psi_n(\mu)\) of \(\psi_0(\mu)\) for any fixed \(\mu\).%

Our post-hoc procedure uses isotonic regression \citep{barlow1972isotonic} to
calibrate an initial outcome regression \(\mu_n\) and Riesz representer
estimator \(\alpha_n\).
Specifically, we define the isotonic-calibrated (`isocalibrated') nuisance
estimators \(\mu_n^*\) and \(\alpha_n^*\) by
\begin{equation}
\begin{aligned}
\mu_n^* := f_n \circ \mu_n, \quad \textnormal{where}\quad
f_n &\in \argmin_{f \in \mathcal{F}_{\text{iso}}}
\sum_{i=1}^n \bigl\{ Y_i - f(\mu_n(A_i, W_i))\bigr\}^2, \\
\alpha_n^* := g_n \circ \alpha_n, \quad \textnormal{where}\quad
g_n &\in \argmin_{g \in \mathcal{F}_{\text{iso}}}
\sum_{i=1}^n \bigl\{ g(\alpha_n(A_i, W_i))\bigr\}^2
- 2\psi_n(g \circ \alpha_n),
\end{aligned}
\label{eqn::metastep}
\end{equation}
where \(\mathcal{F}_{\text{iso}}\) denotes the class of real-valued monotone
nondecreasing functions on \(\mathbb{R}\). Our proposed isocalibrated DML estimator is then
\begin{equation}
\tau_n^* := \psi_n(\mu_n^*) + \frac{1}{n}\sum_{i=1}^n
\alpha_n^*(A_i,W_i)\{Y_i - \mu_n^*(A_i,W_i)\},
\label{eqn::ICDRnot}
\end{equation}
that is, the one-step debiased machine learning estimator formed using the calibrated
nuisance estimators.  

 To obtain theoretical guarantees without empirical process conditions, we use cross-fitting to construct the initial
nuisance estimators \citep{van2011cross, DoubleML}. Algorithm~\ref{alg:DR} displays the cross-fitted version of \eqref{eqn::ICDRnot}; the only change is that cross-fitted initial nuisance estimates are used for calibration and to compute $\tau_n^*$. The calibration step is still performed on the full sample so that
the orthogonality conditions hold in sample. This does not undermine the
cross-fitting guarantees: monotone function classes have finite uniform entropy integrals, which we use to control empirical process terms arising in our analysis
\citep{rabenseifner2025calibrationstrategiesrobustcausal}. Cross-fitting can produce nuisance estimates that are poorly empirically calibrated on holdout data, and full-sample calibration corrects this.

\begin{algorithm}[!htb]
\begin{algorithmic}[1]
{\small
\caption{Calibrated DML using isotonic calibration} \label{alg:DR}
 \vspace{.1in}
\INPUT dataset $\mathcal{D}_n = \{O_i: i=1,\ldots,n\}$, number $J$ of cross-fitting splits
\vspace{.05in}
\STATE partition $\mathcal{D}_n$ into datasets $\mathcal{C}^{(1)},\mathcal{C}^{(2)},\ldots,\mathcal{C}^{(J)}$;
\FOR {$s = 1,\ldots,J$}
\STATE get initial estimators $\alpha_{n,s}$ of $\alpha_0$ and $\mu_{n,s}$ of $\mu_0$ from $\mathcal{E}^{(s)} := \mathcal{D}_n \backslash \mathcal{C}^{(s)}$;
\STATE get initial estimators $\mu \mapsto \psi_{n,s}(\mu)$ of $\mu \mapsto \psi_0(\mu)$ from $\mathcal{D}_n$, where sample-splitting is performed as needed to satisfy Condition \ref{cond::DRremainder};
\ENDFOR
\STATE for {$s = 1,\ldots,J$}, set $j(i):=s$ for each $i\in \mathcal{C}^{(s)}$;
\STATE compute calibrators $f_n$ and $g_n$ using pooled out-of-fold estimates as\vspace{-.05in}
\begin{align*}
    f_n &\in \argmin_{f \in \mathcal{F}_{\text{iso}}} \sum_{i=1}^n  \big{\{} Y_i - f(\mu_{n,j(i)}(A_i, W_i))\big{\}}^2\\
    g_n &\in \argmin_{g \in \mathcal{F}_{\text{iso}}} \sum_{i=1}^n\big{\{}g(\alpha_{n,j(i)}(A_i, W_i))^2 - 2\psi_{n,j(i)}(g \circ \alpha_{n,j(i)}) \big{\}};
\end{align*}
\vspace{-.1in}
\STATE for {$s = 1,\ldots,J$}, set $ \mu_{n,s}^* := f_n \circ \mu_{n,s}$ and $ \alpha_{n,s}^* := g_n \circ \alpha_{n,s} $;
 \STATE compute calibrated one-step debiased estimator \vspace{-.05in}
 \begin{equation}
   \label{eqn::crossfitOnestep}  \tau_n^* := \frac{1}{J}\sum_{j=1}^J\left[\psi_{n,j}(\mu_{n,j}^*)  +   \frac{1}{n}\sum_{i=1}^n \alpha_{n,j(i)}^*(A_i,W_i)\left\{Y_i - \mu_{n,j(i)}^*(A_i,W_i)\right\}\right].
\end{equation}\vspace{-.2in}
\RETURN $\tau_n^*$
}
\end{algorithmic}
\vspace{.05in}
\end{algorithm}

The isotonic regression solutions \(f_n\) and \(g_n\) are uniquely defined only at the observed values of \(\mu_n\) and \(\alpha_n\), and any such solution guarantees empirical calibration. Following \citet{groeneboom1993isotonic}, we take the unique c\`adl\`ag piecewise-constant versions with jumps only at these observed values. Computing \(g_n\) in \eqref{eqn::metastep} may involve a nonstandard loss, but it can be obtained as the empirical risk minimizer over monotone univariate regression trees of unbounded depth. Thus, \eqref{eqn::metastep} can be solved using regression-tree software that supports monotonicity constraints and custom losses, such as the \texttt{R} and \texttt{Python} implementations of \texttt{xgboost} \citep{chen2016xgboost}; see \citet{lee2025rieszboost} for fitting trees with Riesz losses. Equivalently, \(g_n\) can be computed by empirical risk minimization over a basis of threshold indicators with a nonnegative coefficient constraint. The next example shows that, for many causal parameters, this reduces to standard isotonic regression.

\setcounter{example}{0}

\begin{example}[continued]
\label{example::ATE}
Consider the $G$-computation identification $\tau_0(a_0) := E_0\{\mu_0(a_0, W)\}$
of the mean of the counterfactual outcome $Y^{a_0}$ for $a_0 \in \mathcal{A}$. In this
case, the Riesz representer $\alpha_0$, given by $(a,w) \mapsto
1(a = a_0)/\pi_0(a_0 \mid w)$, is determined by the propensity score $\pi_0$.
Given estimators $\mu_n$ of $\mu_0$ and $\pi_n$ of $\pi_0$, an isocalibrated DML
estimator of $\tau_0(a_0)$ (without cross-fitting) is
\[
\tau_n^*(a_0) := \frac{1}{n}\sum_{i=1}^n \left[\mu_n^*(a_0, W_i) +
\alpha_n^*(A_i,W_i)\{Y_i - \mu_n^*(A_i, W_i)\}\right],
\]
where, for each $w\in\mathcal{W}$, we set $\mu_n^*(a_0, w) :=
f_{n,a_0}(\mu_n(a_0 ,w))$ and $\alpha_n^*(a_0 \mid w) :=
g_{n,a_0}(\pi_n^{-1}(a_0 \mid w))$ with
\begin{align*}
   f_{n,a_0} &\in \argmin_{f \in \mathcal{F}_{\textnormal{iso}}}
   \sum_{i=1}^n 1(A_i=a_0)\left\{ Y_i - f(\mu_n(a_0, W_i))\right\}^2;\\
   g_{n,a_0} &\in \argmin_{g \in \mathcal{F}_{\textnormal{iso}}}
   \sum_{i=1}^n\left[\left\{1(A_i=a_0)\, g(\pi_n^{-1}(A_i \mid W_i))\right\}^2
   - 2 g(\pi_n^{-1}(a_0 \mid W_i)) \right]. 
\end{align*}
Here, we optionally stratify the outcome-regression calibration step
in~\eqref{eqn::metastep} by treatment, ensuring empirical calibration within
each treatment level and hence marginal empirical calibration as well. A useful property of isotonic regression is that, for a broad class of convex
losses---namely, Bregman losses---the isotonic fit is \emph{universal}: the
solution is invariant to the particular loss used within this class
\citep{jordan2022characterizing}. Bregman loss functions for Riesz
representers and density ratios are provided in \cite{hines2025learning}, and
these losses can therefore be used for calibration. In particular, the second
step can be implemented via logistic balancing-weight regression
\citep{sverdrup2026balnet}.

A similar invariance holds for isotonic calibration of the propensity score, which implicitly also calibrates the corresponding inverse weights, since \(x \mapsto 1/x\) is monotone on \((0,\infty)\). In particular,
$g_{n,a_0}$ equals $t \mapsto 1/\widetilde{g}_{n,a_0}(t)$, where
\begin{align*}
   \widetilde{g}_{n,a_0} \in \argmin_{g \in \mathcal{F}_{\textnormal{iso}}}
   \sum_{i=1}^n \left\{ 1(A_i = a_0) - g\bigl(\pi_n(a_0 \mid W_i)\bigr) \right\}^2.
\end{align*}
This equivalence is shown in \cite{van_der_Laan2024stabilized} using the
first-order conditions that characterize isotonic regression solutions
$g_{n,a_0}$ and $\widetilde{g}_{n,a_0}$. The calibrated propensities
\(\{\widetilde{g}_{n,a_0}(\pi_n(A_i \mid W_i))\}_{i=1}^n\) may be zero for
observations with \(A_i \neq a_0\), and thus induce infinite weights; however,
this is not problematic, since the AIPW estimator only requires evaluating the
corresponding weights for observations with \(A_i = a_0\). \texttt{Python} and
\texttt{R} implementations of this calibration procedure are provided in
Appendix~\ref{sec:code}. \qed
\end{example}

\setcounter{example}{3}

Our estimator is related to the automatic debiased machine learning
(autoDML) estimators of
\cite{chernozhukov2018double,chernozhukov2022riesznet,chernozhukov2022automatic},
which use tailored loss functions to learn cross-fitted estimators of the Riesz
representer and outcome regression. In particular, autoDML uses the Riesz loss
to estimate the Riesz representer, which may improve finite-sample performance
and nuisance estimator quality.
By itself, however, autoDML does not remove the usual product-rate requirement
for asymptotic linearity. In contrast, our calibrated DML framework uses these
loss functions in a second stage to \emph{calibrate} user-supplied cross-fitted
nuisance estimators to enable doubly robust inference. The two approaches are therefore complementary:
calibrated DML can be incorporated into an autoDML pipeline by isotonically
calibrating the cross-fitted nuisance estimates before forming the one-step
estimator.

 \subsection{Bootstrap-assisted confidence intervals}
\label{sec:bootalgo}
In previous works, doubly robust confidence intervals were obtained through a standard Wald construction based on the asymptotic normality of the corrected estimator, using adjusted estimators of the asymptotic variance \citep{benkeser2017doubly, dukes2024doubly, bonvini2024doubly}. Such confidence intervals are valid if either nuisance estimator is consistent at a sufficiently fast rate. A drawback of previous approaches is that they require consistently estimating additional nuisance functions not required by the estimator itself. Here, we propose a new method that avoids this requirement. It uses a variant of the bootstrap to construct doubly robust confidence intervals associated with the estimator described in Algorithm~\ref{alg:DR} \citep{efron1994introduction}.  In Appendix~\ref{appendix::ERMnuis}, we also propose a complementary Wald-type confidence interval that exploits the piecewise-constant structure of the isocalibrated nuisance estimators by estimating the required nuisance quantities over their fitted level sets.

Besides avoiding estimating new nuisances, our proposed bootstrap-assisted inference procedure is easy to describe and computationally efficient. Holding the initial cross-fitted nuisance estimators fixed, our procedure recomputes the calibrated DML estimator on bootstrap samples, only refitting the isotonic regressions and recomputing the empirical mean used to evaluate the DML estimator. In this way, each bootstrap replicate can be solved in near-linear time using standard software \citep{barlow1972isotonic}. Consequently, computing our bootstrap procedure is typically much less expensive than fitting the original DML estimator.

We focus our description of the automated algorithm on the case where the linear functional \( \psi_0 \) is of the form \( \mu \mapsto E_0\{m(Z, \mu)\} \) for a known map \( m: \mathcal{Z} \times \mathcal{H} \rightarrow \mathbb{R} \). In such cases, we consider the empirical plug-in estimator \( \psi_n \) of \( \psi_0 \), defined as \( \mu \mapsto \frac{1}{n} \sum_{i=1}^n m(Z_i, \mu) \). A similar algorithm can be devised for the general case by appropriately bootstrapping the estimation of the functional \( \psi_n \). As in the previous subsection, we first describe our approach in the simpler case where the initial estimators are not cross-fitted. The general procedure allowing for cross-fitting is provided in Algorithm \ref{alg::bootstrap}. To this end, let $\alpha_n$ and $\mu_n$ be initial estimators of $\alpha_0$ and $\mu_0$ computed once using the original sample. Our approach involves bootstrapping the meta isotonic fitting step of \eqref{eqn::metastep} used originally to obtain the calibrators $f_n$ and $g_n$. Specifically, we sample $\{Z_i^{\#}: i=1,\ldots,n\}$ with replacement from the original sample $\{Z_i: i=1,\ldots,n\}$ and compute the bootstrapped calibrators
\begin{align*}
     f_n^{\#} &\in \argmin_{f \in \mathcal{F}_{\text{iso}}} \sum_{i=1}^n  \big{\{} Y_{i}^{\#} - f(\mu_{n}(A_i^{\#}, W_i^{\#}))\big{\}}^2;\\
    g_n^{\#} &\in \argmin_{g \in \mathcal{F}_{\text{iso}}}  \sum_{i=1}^n \big{\{}g(\alpha_{n}(A_i^{\#}, W_i^{\#}))^2 - 2m(Z_i^{\#},  g \circ \alpha_{n}) \big{\}}\,.
\end{align*}  
Next, we obtain the bootstrap-calibrated nuisance estimators $\alpha_{n}^{\#} := g_n^{\#} \circ \alpha_{n}$ and $\mu_{n}^{\#} := f_n^{\#} \circ \mu_{n}$, and compute the bootstrapped estimator $\tau_n^{\#}$ of $\tau_0$ from $\mu_{n}^{\#}$ and $\alpha_{n}^{\#}$ following the same steps as in \eqref{eqn::ICDRnot}. This procedure is repeated $K$ times to obtain $K$ bootstrap estimators $\tau_{n,1}^{\#},\ldots,\tau_{n,K}^{\#}$ of $\tau_0$. The choice $K = 10{,}000$ has been recommended for use in practice \citep{efron1994introduction}.
 
To construct a bootstrap-assisted confidence interval, we first compute the bootstrap average $\overline{\tau}_n^{\#} := \frac{1}{K}\sum_{k=1}^K \tau_{n,k}^{\#}$. For given confidence level $\rho \in (0,1)$, we use the interval \[\text{CI}_n(\rho) := \left(\tau_n^*-q(\rho) \sigma_n^{\#},\,\tau_n^*+q(\rho) \sigma_n^{\#}\right),\] where $\sigma_n^{\#}$ is the empirical standard deviation of $\tau_{n,1}^{\#},\ldots,\tau_{n,K}^{\#}$ and, for each $\gamma\in(0,1)$, $q(\gamma)$ denotes the $\gamma/2$--quantile of the standard normal distribution. 
Alternatively, we could consider a percentile bootstrap interval $\left(\tau_n^*-q_{n}(\rho),\,\tau_n^*-q_n(2-\rho)\right)$, where for each $\gamma\in(0,1)$ $q_{n}(\gamma)$ denotes the empirical $\gamma/2$--quantile of the centered bootstrapped estimates $\tau_{n,1}^{\#} - \overline{\tau}_n^{\#},\ldots,\tau_{n,K}^{\#} - \overline{\tau}_n^{\#}$; this interval may be more robust to finite sample deviations from normality. We refer to Algorithm \ref{alg::bootstrap} in Appendix \ref{appendix::bootinfalgo} for the pseudo-code implementing this approach with cross-fitting.

\section{Theoretical properties}
 \label{section::theory}

\subsection{Notation for cross-fitting and fold-averaging}
\label{section::theory::crossfit}

We extend the notation of Section~\ref{sec::genstrat} to accommodate
cross-fitting. For fold-specific functions \(f_{\diamond}=\{f_j:j\in[J]\}\), with
\(f_j\in L^2(P_0)\), we write $ \overline P_0 f_{\diamond}:=\frac1J\sum_{j=1}^J P_0 f_j$ and $\|f_{\diamond}\|_{\overline P_0}
    :=\{\overline P_0(f_{\diamond}^2)\}^{1/2}.$
Equivalently, \(f_{\diamond}\) is viewed as a function
\((z,j)\mapsto f_j(z)\) on \(\mathcal Z\times[J]\), with \(j\) averaged
uniformly. For fold-specific feature maps
\(v_{\diamond}=\{v_j:\mathcal A\times\mathcal W\to\mathbb R^k\}_{j=1}^J\), we define
\[
    (\overline\Pi_{v_{\diamond}} f_{\diamond})_j
    :=\theta^\star\circ v_j,
    \qquad
    \theta^\star\in
    \arg\min_{\theta}
    \|f_{\diamond}-\theta\circ v_{\diamond}\|_{\overline P_0}^2,
\]
where the minimization is over measurable
\(\theta:\mathbb R^k\to\mathbb R\) with
\(\|\theta\circ v_{\diamond}\|_{\overline P_0}<\infty\). Thus,
\(\overline\Pi_{v_{\diamond}} f_{\diamond}\) is the fold-averaged
\(L^2(\overline P_0)\)-projection onto functions of \(v_{\diamond}\), with a
single map \(\theta\) shared across folds.

\subsection{Doubly robust asymptotic linearity and inference}
\label{Section::DR}

\subsubsection{Main result}

We now state our main theorem, which establishes doubly robust asymptotic
linearity of the isocalibrated DML estimator in Algorithm~\ref{alg:DR}. We first
introduce the required conditions. Let
\(\mu_{n,\diamond}^* := \{\mu_{n,j}^*: j \in [J]\}\) and
\(\alpha_{n,\diamond}^* := \{\alpha_{n,j}^*: j \in [J]\}\) denote the
isocalibrated nuisance estimators, and define
\(\Delta_{\alpha,n,j} := \alpha_{n,j}^* - \alpha_0\) and
\(\Delta_{\mu,n,j} := \mu_{n,j}^* - \mu_0\). We use the following
fold-averaged analogues of the higher-order cross-product remainder terms in
Condition~\ref{cond:ortho}:
\begin{align*}
    \overline{\gamma}_{\mu,0}(\mu_{n, \diamond}^*, \alpha_{n, \diamond}^*)
    &=
    \left\langle
    \overline{\Pi}_{(\mu_{n,\diamond}^*,\, \mu_0)}\Delta_{\alpha,n,\diamond}
    - \overline{\Pi}_{\mu_{n,\diamond}^*}\Delta_{\alpha,n,\diamond}
    \,,\,
    \mu_0 - \overline{\Pi}_{\mu_{n,\diamond}^*}\mu_0
    \right\rangle_{\overline{P}_0}
     ,\\
    \overline{\gamma}_{\alpha,0}(\mu_{n, \diamond}^*, \alpha_{n, \diamond}^*)
    &=
    \left\langle
    \overline{\Pi}_{(\alpha_{n,\diamond}^*,\, \alpha_0)}\Delta_{\mu,n,\diamond}
    - \overline{\Pi}_{\alpha_{n,\diamond}^*}\Delta_{\mu,n,\diamond}
    \,,\,
    \alpha_0 - \overline{\Pi}_{\alpha_{n,\diamond}^*}\alpha_0
    \right\rangle_{\overline{P}_0}.
\end{align*}
The theorem relies on the following key conditions, together with the technical
regularity conditions stated in
Appendix~\ref{appendix::mainconditions}
(Conditions~\ref{cond::estboundedlinearcross}--\ref{cond::variation}). These
conditions are commonly used in the debiased machine learning literature
\citep{van2023causal,rabenseifner2025calibrationstrategiesrobustcausal}. They
include asymptotic linearity of the estimated linear functional
\(\psi_{n,j}\) for nuisance-dependent targets, uniform boundedness of the
outcome and nuisance functions, and uniform boundedness of the total variation
of certain one-dimensional calibration functions, as in
\citet{van2023causal}, to control the empirical process remainders in
Theorem~\ref{theorem::DRbiasexpansion}. When verifying the asymptotic linearity
condition requires additional rate conditions on \(\mu_n\), \(\alpha_n\), or
other nuisance estimators, the DRAL guarantee holds under those additional
prerequisites.

\begin{enumerate}[label=\bf{C\arabic*)}, ref={C\arabic*}, series=cond1]
  \item \textit{At least one nuisance is estimated well:}
  \label{cond::DRlimits}\label{cond::DRmisDRconsist}
  For some $(\overline{\alpha}_0,\overline{\mu}_0) \in \mathcal{H}^2$, either of the following holds:
  \begin{enumerate}[label={\roman*)}, ref={\ref{cond::DRlimits}\roman*}]
    \item \label{cond::DRmisDRconsist::1}
    $\|\alpha_{n,\diamond}^* - \alpha_0 \|_{\overline{P}_0} = o_p(n^{-1/4})$ and $\|\mu_{n,\diamond}^* - \overline{\mu}_0\|_{\overline{P}_0} = o_p(1)$;
    \item \label{cond::DRmisDRconsist::2}
    $\|\mu_{n,\diamond}^* - \mu_0\|_{\overline{P}_0} = o_p(n^{-1/4})$ and $\|\alpha_{n,\diamond}^* - \overline{\alpha}_0\|_{\overline{P}_0} = o_p(1)$.
  \end{enumerate}

\item \textit{Partial orthogonality and consistency:}
\label{cond::mainprojectioncoupling1}
The following holds jointly with whichever
part of \ref{cond::DRmisDRconsist} itself holds:
\begin{enumerate}[label={\roman*)}, ref={\ref{cond::mainprojectioncoupling1}\roman*}]\vspace{-.05in}
  \item \label{cond::mainprojectioncoupling1:i}
  Under \ref{cond::DRmisDRconsist::1},
  $ \overline{\gamma}_{\alpha,0}(\mu_{n, \diamond}^*, \alpha_{n, \diamond}^*) = O_p(  \|\alpha_{n,\diamond}^* - \alpha_0\|_{\overline{P}_0}^2) $ and
  $\|(\overline{\Pi}_{\alpha_{n,\diamond}^*} - \overline{\Pi}_{\alpha_0})\Delta_{\mu,n,\diamond}\|_{\overline{P}_0}
  = o_p(1)$. 
  \item \label{cond::mainprojectioncoupling1:ii}
  Under \ref{cond::DRmisDRconsist::2}, 
  $\overline{\gamma}_{\mu,0}(\mu_{n, \diamond}^*, \alpha_{n, \diamond}^*) =  O_p( \|\mu_{n,\diamond}^* - \mu_0\|_{\overline{P}_0}^2)$ and
  $\|(\overline{\Pi}_{\mu_{n,\diamond}^*} - \overline{\Pi}_{\mu_0})\Delta_{\alpha,n,\diamond}\|_{\overline{P}_0}
  = o_p(1)$.
\end{enumerate}
\end{enumerate}
Condition~\ref{cond::DRmisDRconsist} is a DRAL-type rate condition. Under
\ref{cond::DRmisDRconsist::1} we have $\overline{\alpha}_0=\alpha_0$, whereas
under \ref{cond::DRmisDRconsist::2} we have $\overline{\mu}_0=\mu_0$.
The first part of \ref{cond::mainprojectioncoupling1} ensures that a cross-averaged
version of the expansion in Theorem~\ref{theorem::DRbiasexpansion} applies to
the cross-fitted calibrated nuisance estimators; sufficient conditions akin to
Lemma~\ref{lemma::sufficientcoupling} are given in Appendix \ref{appendix::sufficientCondsCross}.
The second part is used to control cross-fitted versions of the empirical process
remainders appearing in Theorem \ref{theorem::DRbiasexpansion}. When
both parts of \ref{cond::DRmisDRconsist} hold, \ref{cond::mainprojectioncoupling1}
is unnecessary, since the usual product-rate condition then holds.

  The limiting
distribution of the isocalibrated DML estimator is governed by the influence function
\[
\chi_0
:= \phi_{0,\overline{\mu}_0}
+ D_{\overline{\mu}_0,\overline{\alpha}_0}
+ A_0
+ \Bigl\{B_0 + \psi_0(\overline{\mu}_0-\mu_0)\Bigr\},
\]
where \(A_0(z) = s_0(a,w)\{y-\mu_0(a,w)\}\) and
\(B_0(z) = \phi_{0,r_0}(z) + \psi_0(r_0) - r_0(a,w)\alpha_0(w,a)\). When \(\psi_0(\mu)=E_0\{m(Z,\mu)\}\), as in
\eqref{eqn::simplefunctional}, $B_0$ simplifies to
\(m(\cdot,r_0) - r_0\alpha_0\). 
 The first two terms have the same form as
the efficient influence function \(D_0\) in Theorem~\ref{theorem::EIF}, with
nuisance components replaced by their in-probability limits. The remaining terms are additional influence function
components induced by inconsistent nuisance estimation. Under \ref{cond::DRlimits}, at
most one of these components is nonzero: if \(\overline{\alpha}_0=\alpha_0\),
then \(A_0=0\), whereas if \(\overline{\mu}_0=\mu_0\), then
\(B_0+\psi_0(\overline{\mu}_0-\mu_0)=0\).

\begin{theorem}[Main result: DRAL of \(\tau_n^*\)]
\label{theorem::DRinference}
Suppose Conditions~\ref{cond::linear}--\ref{cond::pathwise},
\ref{cond::estboundedlinearcross}--\ref{cond::variation}, and that at least
one nuisance is estimated well (\ref{cond::DRlimits}) with partially orthogonal
errors (\ref{cond::mainprojectioncoupling1}). Then:
\begin{enumerate}[label=\roman*)]
\item The isocalibrated DML estimator \(\tau_n^*\) of \(\tau_0\) is
asymptotically linear with influence function \(\chi_0\).
\item If, in addition, \((\overline{\alpha}_0,\overline{\mu}_0)=(\alpha_0,\mu_0)\),
then \(\chi_0 = D_0\), so \(\tau_n^*\) is regular and efficient.
\end{enumerate}
Consequently, \(n^{1/2}(\tau_n^*-\tau_0)\) converges in distribution to a
mean-zero normal random variable with variance
\(\sigma_0^2 := \mathrm{var}_0\{\chi_0(Z)\}\).
\end{theorem}

Inference may be conducted using the bootstrap, as described in
Section~\ref{section::bootstrap}, or, given a consistent estimator of
\(\sigma_0^2\), using Wald-type confidence intervals and tests based on the
influence function \(\chi_0\). For
Wald-type inference, the unknown quantities \(r_0\) and \(s_0\) in \(\chi_0\) can
be estimated via empirical risk minimization, as discussed in
Appendix~\ref{appendix::ERMnuis}. If both nuisance
estimators are consistent, then \(r_0\) and \(s_0\) are identically zero, and
standard confidence intervals based on the calibrated nuisances are valid without
additional correction. For parameters that can be expressed as smooth transformations of linear
summaries---such as relative differences between counterfactual means---DRAL
estimators can be obtained by first constructing DRAL estimators of each summary
and then applying the corresponding plug-in map. Inference may then be based on
either the delta method or the bootstrap. For instance, in
Example~\ref{example::ATE}, DRAL estimators of the ATE \(\tau_0(1)-\tau_0(0)\) and
relative ATE \(\tau_0(1)/\tau_0(0)\) are \(\tau_n^*(1)-\tau_n^*(0)\) and
\(\tau_n^*(1)/\tau_n^*(0)\), respectively.

When both nuisance estimators are consistent, Theorem~\ref{theorem::DRinference}
shows that the isocalibrated DML estimator is regular, asymptotically linear,
and efficient for \(\tau_0\) under weaker rate conditions than standard doubly robust
estimators and related DRAL estimators. In particular, it avoids product-rate
conditions and allows one nuisance function to be estimated arbitrarily
slowly. By contrast, the TMLE-based DRAL estimators of the ATE proposed in
\cite{van2014targeted} and related work \citep{benkeser2017doubly, dukes2024doubly}
typically require the nuisance estimators to converge sufficiently quickly to
their (possibly incorrect) limits, imposing $\|\mu_{n,\diamond}^* - \overline{\mu}_0\|_{\overline{P}_0}\,
\|\alpha_{n,\diamond}^* - \overline{\alpha}_0\|_{\overline{P}_0}
= o_P(n^{-1/2}).$
When both nuisance estimators are consistent, this offers no improvement over
rate double robustness. We view this requirement as an artifact of existing
theoretical analyses rather than an inherent limitation of TMLE-based
estimators, and our analysis can be adapted to relax it. For the expected conditional covariance parameter, \cite{bonvini2024doubly}
recently showed that the estimation rate achieved by the isocalibrated DML
estimator is minimax optimal under a hybrid model that combines smoothness and
structure-agnostic assumptions on the nuisance functions, similar to the
conditions of Lemma~\ref{lemma::sufficientcoupling}. Consequently, the rate
conditions for asymptotic normality imposed by \ref{cond::DRmisDRconsist}
cannot be meaningfully improved without additional assumptions.

 The convergence-rate conditions in \ref{cond::DRlimits} are stated
in terms of the isocalibrated nuisance estimators, which raises the
question of whether the calibration step can degrade the rates achieved by the
first-stage learners. Appendix~\ref{appendix::calnuisances} addresses this by
showing that isotonic calibration preserves the \(L^2\) convergence rates of
the first-stage nuisance estimators, up to an additional \(O_p(n^{-1/3})\)
term. Since \(n^{-1/3}=o(n^{-1/4})\), this calibration error is negligible
relative to the \(o_p(n^{-1/4})\) rate required in \ref{cond::DRlimits}, and
thus calibration imposes no additional rate restrictions. Moreover, in the
high-dimensional and nonparametric regimes that motivate our work, the
one-dimensional \(n^{-1/3}\) calibration term is typically dominated by
first-stage estimation error. This term reflects estimation of a monotone
function in one dimension and therefore depends favorably on problem constants,
independent of covariate dimension \citep{chatterjee2013improved}. By contrast,
rates faster than \(n^{-1/3}\) for high-dimensional first-stage learning
typically require strong smoothness or extreme sparsity assumptions, and can
involve constants that scale poorly with dimension and other structural
parameters. Lemma~\ref{lemma::Rates} in
Appendix~\ref{appendix::calnuisances} further shows that the isocalibrated nuisance
estimators achieve the same mean-squared-error rate as the best monotone
transformation of the original nuisance estimators, where ``best'' is defined as
the minimizer of the relevant population risk: least-squares risk for the
outcome regression and Riesz risk for the Riesz representer. Thus, isotonic
calibration relaxes the consistency requirement by allowing the initial nuisance
estimators to be misspecified up to an unknown monotone transformation.

When \(\tau_0 = E_0\{\mu_0(1,W)\}\) is a counterfactual mean, the DRAL properties
in Lemma~\ref{theorem:empiricalcalibration} have a simple interpretation:
confounding bias can be removed by adjusting either for the outcome regression
\(\mu_0(A,W)\) or for the inverse-propensity weight
\(\alpha_0(A,W)=1\{A=1\}/P_0(A=1\mid W)\). The latter amounts to adjusting for
the propensity score \(P_0(A=1\mid W)\), a classical balancing score
\citep{rosenbaum1983central,imai2014covariate}, whereas \(\mu_0(A,W)\) is a
deconfounding score \citep{benkeser2020nonparametric,d2021deconfounding}. More formally, recalling the notation of Section~\ref{section::calibration},
the calibration property in \eqref{eqn::calpropertyoutcome2} debiases the
plug-in estimator \(\psi_n(\mu_n)\) under models in which \(\mu_n(A,W)\)
provides a dimension reduction of \((A,W)\) \citep{benkeser2020nonparametric}.
Similarly, the calibration property in \eqref{eqn::calpropertyRiesz2} reduces
to covariate balance conditional on the estimated propensity score
\citep{van_der_Laan2024stabilized}, and thus debiases the weighted-outcome
estimator \(\frac{1}{n}\sum_{i=1}^n \alpha_n(A_i,W_i)Y_i\) under models in which
\(\alpha_n(A,W)\) provides a dimension reduction \citep{rosenbaum1983central}.
Intuitively, the calibrated DML estimator combines these two debiasing
mechanisms, yielding bias correction whenever at least one nuisance estimator
is consistent.

The calibrated IPW and plug-in estimators are special cases of our general framework, obtained by setting \(\mu_{n,j}\) or \(\alpha_{n,j}\) to zero, respectively. By Theorem~\ref{theorem::DRinference}, each is asymptotically linear. However, unless the nuisance estimator they build upon is obtained from parametric methods, these singly robust estimators are usually not regular, let alone efficient, and we do not generally recommend their use in practice.  

\subsubsection{Irregularity under nuisance misspecification}
\label{sec:irregularmain}

 Under the stated conditions, Theorem~\ref{theorem::DRinference} shows that calibrated DML is regular whenever both nuisance estimators are consistent and at least one converges faster than \(n^{-1/4}\). When flexible learners are used, this consistency requirement is mild. Indeed, many machine learning methods---including neural networks, random forests, and gradient boosting---are universal approximators and can consistently estimate smooth nuisance functions under suitable conditions \citep{hornik1989multilayer, anthony2009neural, zhang2005boosting, biau2008consistency}.  Calibrated DML and related DRAL estimators \citep{benkeser2017doubly,dukes2024doubly,bonvini2024doubly} may be \emph{irregular} when one nuisance estimator is inconsistent. Regularity is desirable because it is needed for locally uniformly valid asymptotic inference to be possible; without it, there may exist arbitrarily close distributions for which the asymptotic approximation requires arbitrarily large sample sizes to be accurate \citep{LeebModelSelect2005}. Nevertheless, when one nuisance function is inconsistent, the estimator remains asymptotically normal and continues to support pointwise valid inference, with the resulting irregularity quantified in Appendix~\ref{sec::irregular}. There, we show that under any fixed local perturbation \(P_{0,tn^{-1/2}}\) of \(P_0\), the rescaled estimation error \(\sqrt{n}\{\tau_n^*-\Psi(P_{0,tn^{-1/2}})\}\) is asymptotically normal with a nonzero mean of order \(t(\|\overline{\mu}_0-\mu_0\| \,\vee\, \|\overline{\alpha}_0-\alpha_0\|)\). Consequently, confidence intervals for calibrated DML estimators may still attain close-to-nominal coverage provided that either the magnitude of perturbation (\(|t|\)) or  inconsistency \(\bigl(\|\overline{\mu}_0-\mu_0\| \,\vee\, \|\overline{\alpha}_0-\alpha_0\|\bigr)\) is sufficiently small. This underscores the need to estimate both nuisance functions accurately, since
irregularity scales with the degree of inconsistency.

\subsection{ Doubly robust validity of bootstrap-assisted confidence intervals }
\label{section::bootstrap}

We now show the bootstrap-assisted confidence intervals from Algorithm \ref{alg::bootstrap} provide  doubly robust asymptotically valid inference for $\tau_0$. Like the Wald interval based on Theorem \ref{theorem::DRinference}, constructing our bootstrap-assisted intervals does not require having prior knowledge of which nuisance estimator is consistent, or whether both are. 
However, unlike the Wald approach, it does not require estimating the unknown quantities $r_0$ and $s_0$ appearing in the influence function of $\tau_n^*$. Hence, this bootstrap procedure is fully automated, only requiring computing the isocalibrated DML estimator within each bootstrap sample.  

We recall that Algorithm~\ref{alg::bootstrap} assumes that \(\psi_0\) has the form \(\mu \mapsto E_0\{m(Z,\mu)\}\), and that \(J^{-1}\sum_{j=1}^J \psi_{n,j}\) is taken to be the empirical plug-in estimator \(\mu \mapsto n^{-1}\sum_{i=1}^n m(Z_i,\mu)\). Below, we let \(P_{n,j}^\#\) denote the empirical distribution of an arbitrary bootstrap sample \(\mathcal{C}^{\#(j)}\) drawn from the empirical distribution \(P_{n,j}\) of \(\mathcal{C}^{(j)}\). Let \(\tau_n^{\#}\) denote the corresponding bootstrapped isocalibrated DML estimator obtained from the bootstrap-calibrated nuisance estimators \(\mu_{n,1}^{\#},\ldots,\mu_{n,J}^{\#}\) and \(\alpha_{n,1}^{\#},\ldots,\alpha_{n,J}^{\#}\), as in Algorithm~\ref{alg::bootstrap}. Denoting by \(\Delta_{\alpha,n,j}^\# := \alpha_{n,j}^{\#}-\alpha_0\) and \(\Delta_{\mu,n,j}^\# := \mu_{n,j}^{\#}-\mu_0\), for \(j=1,\ldots,J\), the nuisance estimation errors, the following condition is used in the upcoming theorem. The result also assumes two standard technical conditions (\ref{cond::estboundboot}--\ref{cond::lipschitzisotonic}) stated in Appendix~\ref{appendix::bootconditions}.

\begin{enumerate}[label=\bf{D\arabic*)}, ref={D\arabic*}]
\item \textit{Partial orthogonality and projection error bounds:}
\label{cond::mainprojectioncoupling1::boot}
The following holds jointly with whichever
part of \ref{cond::DRmisDRconsist} itself holds:
\begin{enumerate}[label={\roman*)}, ref={\ref{cond::mainprojectioncoupling1::boot}\roman*}]\vspace{-.05in}
  \item \label{cond::mainprojectioncoupling1:i::boot}
  Under \ref{cond::DRmisDRconsist::1},
  $\overline{\gamma}_{\alpha,0}(\mu_{n,\diamond}^{\#}, \alpha_{n,\diamond}^{\#}) = O_p(\|\alpha_{n,\diamond}^{\#} - \alpha_0\|_{\overline{P}_0}^2)$ and
  $\|(\overline{\Pi}_{\alpha_{n,\diamond}^{\#}} - \overline{\Pi}_{\alpha_0})\Delta_{\mu,n,\diamond}^\#\|_{\overline{P}_0}
  = o_p(1)$. 
  \item \label{cond::mainprojectioncoupling1:ii::boot}
Under \ref{cond::DRmisDRconsist::2}, 
  $\overline{\gamma}_{\mu,0}(\mu_{n,\diamond}^{\#}, \alpha_{n,\diamond}^{\#}) = O_p(\|\mu_{n,\diamond}^{\#} - \mu_0\|_{\overline{P}_0}^2)$ and
  $\|(\overline{\Pi}_{\mu_{n,\diamond}^{\#}} - \overline{\Pi}_{\mu_0})\Delta_{\alpha,n,\diamond}^\#\|_{\overline{P}_0}
  = o_p(1)$.
\end{enumerate}
 \end{enumerate}

\begin{theorem}[Double robustness of confidence intervals]
\label{theorem::DRboot}
Assume the conditions of Theorem \ref{theorem::DRinference}, so that at least
one nuisance is estimated well (\ref{cond::DRlimits}) with partially orthogonal
errors (\ref{cond::mainprojectioncoupling1}).
Under \ref{cond::mainprojectioncoupling1::boot}, and \ref{cond::estboundboot}--\ref{cond::lipschitzisotonic}, the conditional law of $n^{\frac{1}{2}}(\tau_n^{\#} - \tau_n^*)$ given $ \{P_{n,j}: j=1,\ldots,J\}$ converges marginally in $P_0$--probability to the law of a mean-zero normal random variable with variance $\sigma_0^2 := var_0\{\chi_0(Z)\}$.
\end{theorem}
Under regularity conditions, as the number of bootstrap samples $K \rightarrow \infty$, the scaled version $n^{\frac{1}{2}}\sigma_n^{\#}$ of the bootstrap standard error of Algorithm \ref{alg::bootstrap} is a consistent estimator of the limiting standard deviation $\sigma_0$ of $n^{\frac{1}{2}}(\tau_n^*-\tau_0)$. Hence, by Theorem \ref{theorem::DRboot}, the bootstrap-assisted confidence interval $\text{CI}_n(\rho)$ from Algorithm \ref{alg::bootstrap} has doubly robust asymptotically exact coverage, in the sense that $P_0\{\tau_0 \in \text{CI}_n(\rho)\} \rightarrow 1 - \rho$ as $n \rightarrow \infty$, provided $K \rightarrow \infty$ sufficiently quickly.

\section{Numerical experiments}
\label{section::experiments}

\subsection{Experiment 1: performance under inconsistent nuisance estimation}

\label{section::experiments1}
We evaluate the numerical performance of the calibrated DML estimator through simulation, focusing on inference for the ATE \( \tau_0 = E_0\{\mu_0(1,W) - \mu_0(0,W)\} \). The estimator is defined as \( \tau_n^* := \tau_n^*(1) - \tau_n^*(0) \), where \( \tau_n^*(a) \) is the calibrated DML estimator of the counterfactual mean \( E_0\{\mu_0(a,W)\} \), constructed as in Example~\ref{example::ATE} using Algorithm~\ref{alg:DR}. We compare its performance to the standard cross-fitted AIPW estimator from Section~\ref{section::setup} and the DR-TMLE estimator from \cite{benkeser2017doubly}, described in Appendix~\ref{sec::drtmle} and implemented in the \texttt{drtmle} \texttt{R} package \citep{benkeser2023doubly}. To quantify uncertainty, we use bootstrap confidence intervals (Algorithm~\ref{alg::bootstrap}) for calibrated DML and Wald-type intervals based on influence function variance estimates for AIPW and DR-TMLE. For calibrated DML, we also report the corrected Wald-type confidence interval from Appendix~\ref{appendix::ERMnuis}, which uses level-set estimators for the additional nuisance quantities, along with the corresponding uncorrected Wald interval, which is generally invalid under nuisance inconsistency.

We consider the same data-generating process used in the experiments of \cite{benkeser2017doubly}. 
 Covariates \( W_1 \sim \text{Uniform}(-2, 2) \) and \( W_2 \sim \text{Bernoulli}(0.5) \) are drawn independently. The treatment and outcome are binary, with propensity \( \pi_0(1 \mid (w_1, w_2)) := \text{expit}(-w_1 + 2w_1 w_2) \) and outcome regression \( \mu_0(a, w_1, w_2) := \text{expit}(0.2a - w_1 + 2w_1 w_2) \).  This simple, low-dimensional design is intended mainly as an illustrative check of performance under controlled nuisance model misspecification. Experiment~2 provides a more challenging benchmark with higher-dimensional covariates and fully data-adaptive nuisance estimation. To evaluate the doubly robust inference property, we examine three scenarios of inconsistent nuisance estimation for each estimator:
\begin{itemize}
\item[(a)] when both the outcome regression and treatment mechanism are estimated consistently;\vspace{-.15in}
\item[(b)] when only the treatment mechanism is estimated consistently;\vspace{-.15in}
\item[(c)] when only the outcome regression is estimated consistently.
\end{itemize}
In all settings, any consistently estimated nuisance function is obtained using a 5-fold cross-fitted Nadaraya–Watson kernel smoother with bandwidth selected via cross-validation and stratified by both \( A \) and \( W_2 \)  \citep{hofmeyr2020fast}. In settings (b) and (c), to induce inconsistency in the estimation of the outcome regression and treatment mechanisms, we estimate the propensity score \( \pi_0 \) and outcome regressions \( \mu_0 \) using misspecified main-term logistic regression models.  For DR-TMLE, we apply the same kernel method to estimate the nuisance functions \( r_0 \) and \( s_0 \).   Further details and code for reproducing these analyses are available in the GitHub repository linked in Appendix \ref{sec:code}.  

For each scenario and sample sizes ranging from 250 to 10{,}000, we generated and analyzed 5{,}000 datasets. Empirical estimates of bias, standard error, and 95\% confidence interval coverage are presented in Figure~\ref{fig::exp1}. Results for the setting in which both nuisance functions are consistently estimated appear in Appendix~\ref{appendix::exp1}. In scenarios with inconsistent nuisance estimation, both calibrated DML and DR-TMLE exhibited DRAL behavior, showing lower bias and better coverage than AIPW. Relative to DR-TMLE, calibrated DML exhibited less bias, achieved better coverage, and had smaller or comparable variance. In settings where both nuisance functions were consistently estimated, all estimators performed similarly, as expected. For calibrated DML, the bootstrap and corrected Wald-type confidence intervals performed qualitatively similarly and appeared asymptotically equivalent, as predicted by the theory. By contrast, the uncorrected Wald intervals did not attain nominal coverage under nuisance inconsistency: they either undercovered or overcovered, depending on which nuisance component was misspecified.

\begin{figure}[tb]
     \centering 
\captionsetup{skip=5pt}
\captionsetup[subfigure]{skip=1pt}
\includegraphics[width=0.45\linewidth]{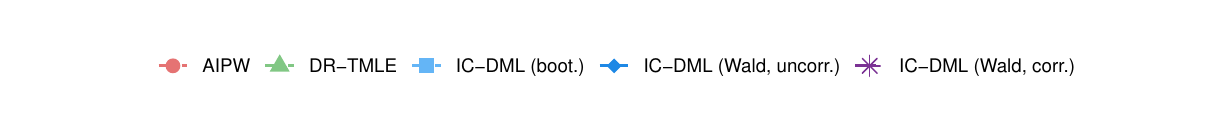}
\vspace{-0.6em}

    \begin{subfigure}[b]{0.95\linewidth}  
    \centering 
    \includegraphics[width=\linewidth]{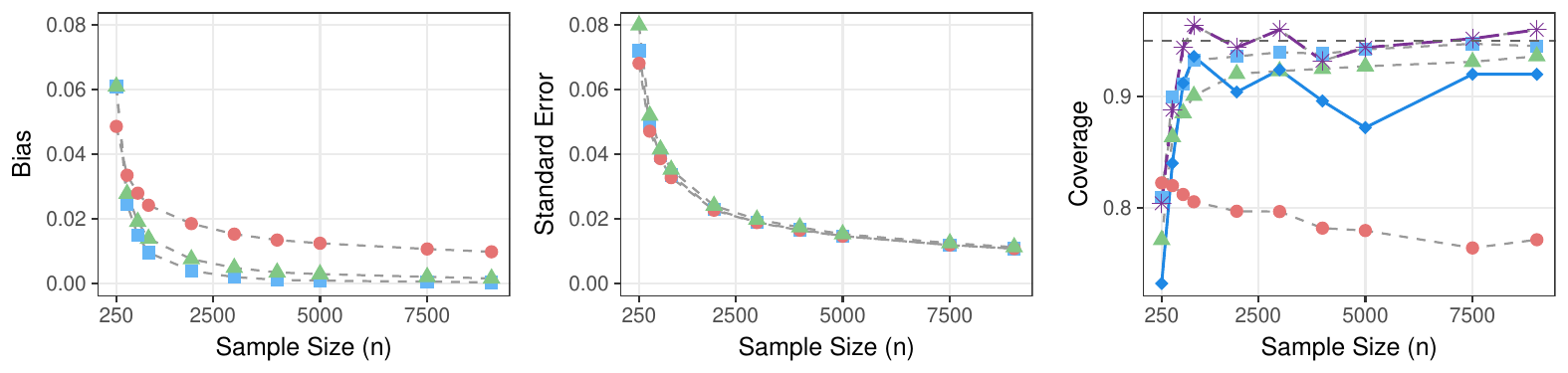}
    \subcaption{Only outcome regression is estimated consistently.}
    \end{subfigure}  
    \vspace{-0.45em}
    \begin{subfigure}[b]{0.95\linewidth}  
    \centering 
    \includegraphics[width=\linewidth]{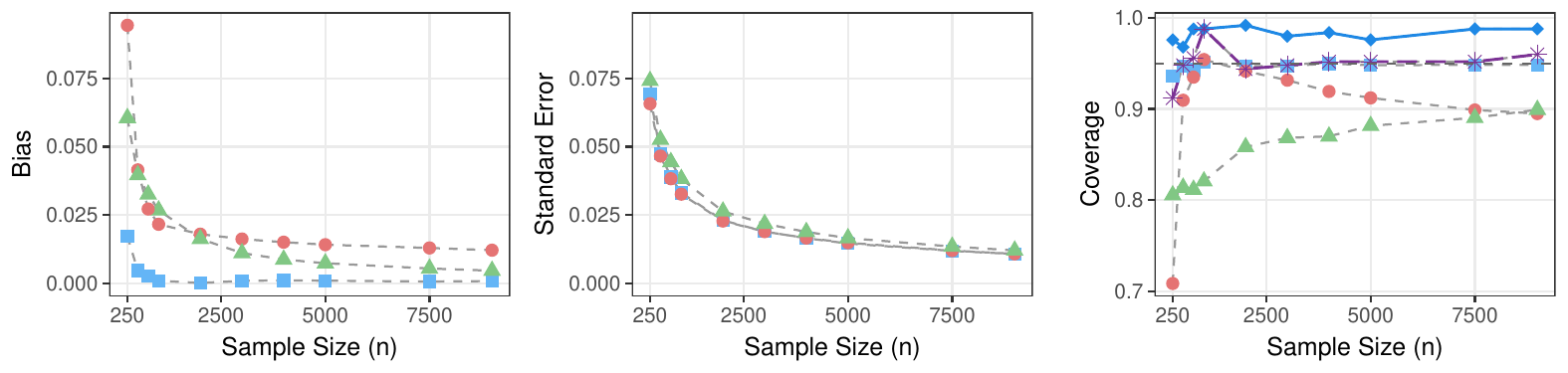}
   \subcaption{Only propensity score is estimated consistently.}
    \end{subfigure}  
\vspace{0.15em}
\caption{Empirical bias, standard error, and 95\% confidence interval coverage for calibrated DML (IC-DML), DR-TMLE and AIPW estimators under singly consistent estimation of the outcome regression and treatment mechanism. }
       \label{fig::exp1}  

\end{figure}

\subsection{Experiment 2: structured outcome regression and high-dimensional propensity}
\label{section::experiments2}

Experiment~1 deliberately misspecifies one nuisance estimator. We next consider a setting in which both nuisance functions are estimated flexibly, but the two first-stage tasks differ in difficulty. The target is again the ATE, and we compare AIPW with calibrated DML throughout. The outcome regression is additive in the covariates on the logit scale and is estimated using a gradient-boosted tree learner that can exploit this structure. In contrast, the propensity score is a rough, high-dimensional function estimated by random forests, and its estimation becomes increasingly difficult as the dimension grows. The design therefore isolates a finite-sample regime in which propensity estimation is difficult while outcome regression is comparatively stable.

In each repetition, \(W=(W_1,\ldots,W_d)\) has independent \(U[-1,1]\)
coordinates, and the treatment and outcome are Bernoulli with a smooth logit-additive outcome regression and a rough nonadditive
propensity score. We estimate the outcome regressions separately by treatment
arm with depth-one, additive \texttt{xgboost} fits constrained to univariate trees. We
estimate the propensity score with \texttt{ranger} probability forests using 500
trees, maximum depth 8, and \(mtry=\lfloor \sqrt d \rfloor\). The dimension
curve fixes \(n=10{,}000\) and varies \(d\in\{1,2,4,8,12,16\}\). The
sample-size grid varies \(n\in\{2500,5000,10000,15000\}\) and
\(d\in\{4,8,12,16\}\), with 500 Monte Carlo repetitions per setting.
Calibrated DML uses isotonic calibration and the level-set corrected Wald
interval throughout.
Additional implementation details are given in
Appendix~\ref{appendix::exp2_holder}.

\begin{figure}[th!b]
     \centering
    \begin{subfigure}[t]{0.51\linewidth}
    \centering
    \includegraphics[width=\linewidth]{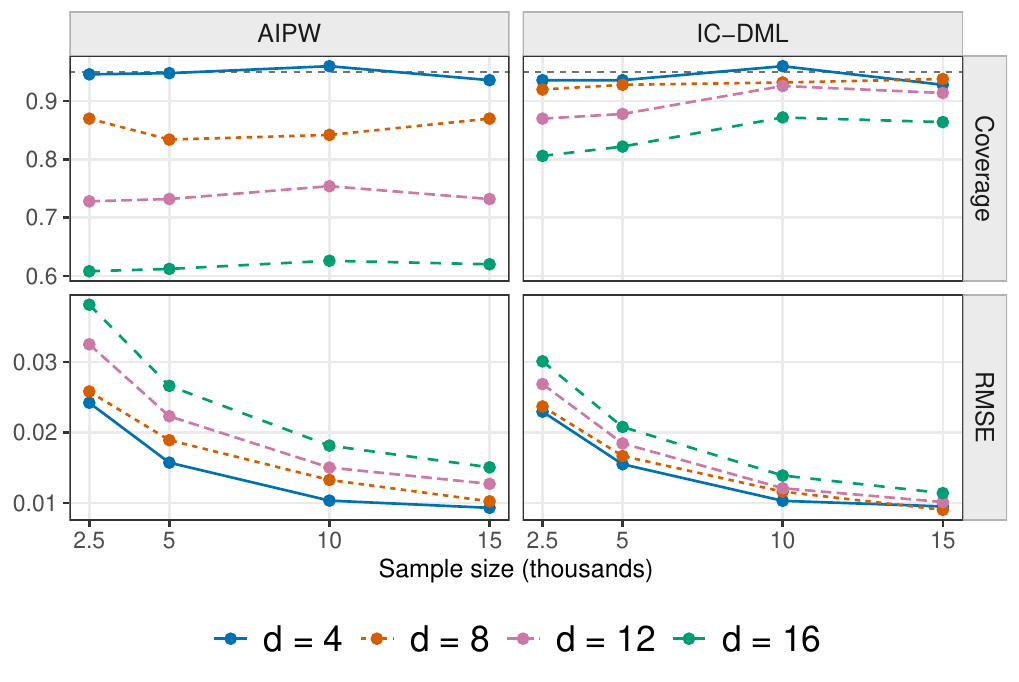}
    \subcaption{Sample-size grid.}
    \end{subfigure}
    \hfill
    \begin{subfigure}[t]{0.47\linewidth}
    \centering
    \includegraphics[width=\linewidth]{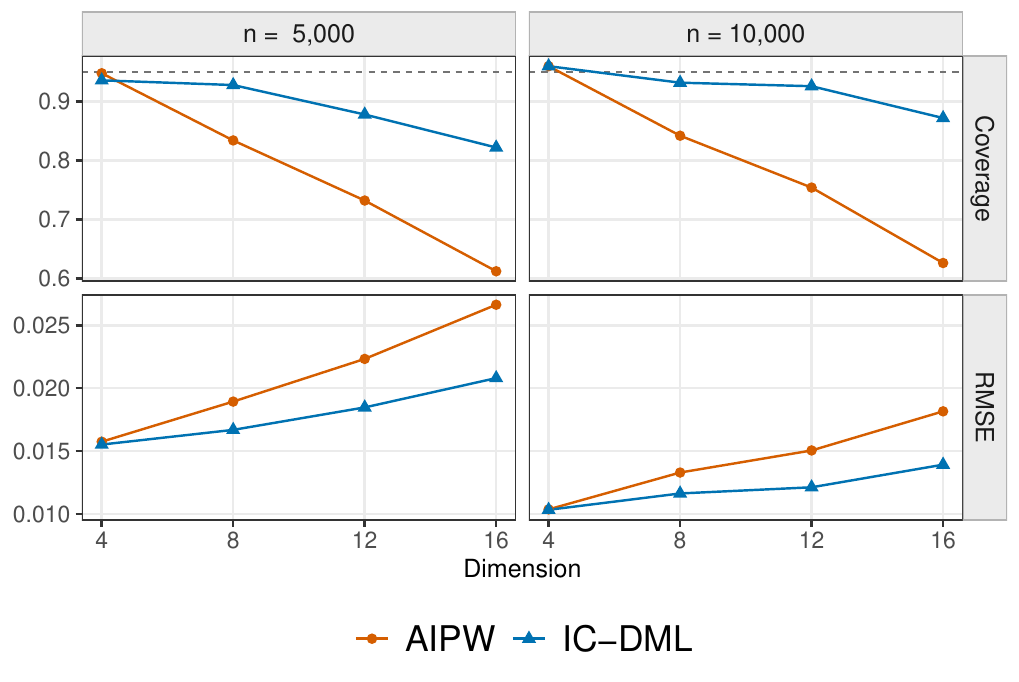}
    \subcaption{\(n=5{,}000\) and \(n=10{,}000\), varying \(d\).}
    \end{subfigure}
\caption{Varying sample size and dimension in the Hölder-smooth simulation. Empirical coverage and RMSE are shown for logit-additive outcome regression estimated by depth-one \texttt{xgboost} and a high-dimensional propensity score estimated by random forests. The left panel varies sample size within dimensions, and the right panel fixes \(n=5{,}000\) or \(n=10{,}000\) and varies dimension. The dashed line marks nominal 95\% coverage.}
       \label{fig::exp2_holder}
\end{figure}

Figure~\ref{fig::exp2_holder} shows that AIPW and calibrated DML have similar
coverage and RMSE in low dimensions. As dimension increases in the fixed-\(n\)
display, AIPW coverage deteriorates sharply and its RMSE increases more rapidly,
while calibrated DML remains closer to nominal coverage with lower RMSE in the
moderate- and high-dimensional cells. In the sample-size grid, the two
methods are comparable for \(d=4\). For \(d=8,12,\) and \(16\), the
random-forest propensity error remains visible through AIPW undercoverage.
RMSE generally decreases with \(n\) for both methods, but the coverage pattern
is consistent with finite-sample sensitivity to the high-dimensional propensity
estimation problem. The depth-one \texttt{xgboost} outcome learner continues to
benefit from the logit-additive structure, although its error can still
increase with \(d\) as the number of additive components grows.

\subsection{Experiment 3: benchmarking on semi-synthetic datasets}
\label{section::experiments3}

In this experiment, we evaluate the performance of our method on a collection of semi-synthetic datasets commonly used to benchmark causal inference methods. These datasets feature moderately high-dimensional covariates, limited overlap, confounding, and treatment imbalance (Table~\ref{tab:dataset_summary}). We include the Twins, Lalonde PSID, and Lalonde CPS datasets from the \textit{realCause} generative modeling framework \citep{neal2020realcause}, which are derived from original data sources \citep{louizos2017causal, lalonde1986evaluating}. We also use several datasets from the 2017 Atlantic Causal Inference Challenge (ACIC-2017, settings 17–24) \citep{hahn2019atlantic}, which are based on covariates from the Infant Health and Development Program (IHDP) \citep{brooks1992effects}, and 680 datasets from the scaling component of ACIC-2018 \citep{shimoni2018benchmarking}, with sample sizes of 1{,}000, 2{,}500, 5{,}000, and 10{,}000. The final dataset (IHDP) is a semi-synthetic version of the IHDP data introduced in \cite{hill2011bayesian} (Setting B) and further analyzed in \cite{shalit2017estimating}. In these benchmarks, the covariates are fixed and derived from real data, and the inferential target is the sample average treatment effect, defined as $ \tau_{n,0}:= \frac{1}{n}\sum_{i=1}^n (Y_{1,i} - Y_{0,i}),$
the empirical mean of the individual treatment effects in the observed sample.

\begin{table}[tb]
\centering
\footnotesize
\caption{Sample size of total study ($n_{total}$), expected size of treatment arm $(n_{treated})$, and covariate dimension $(d)$ for the semi-synthetic datasets of the third experiment. For the ACIC-2017 synthetic dataset, we employed simulation setting 24, characterized by additive errors, a large signal, large noise, and strong confounding.}
\label{tab:dataset_summary}
\begin{tabular}{|c|c|c|c|c|}
\hline 
Dataset & $n_{total}$ & $n_{treated}$ & $d$ &  source\\
\hline
\hline
ACIC-2017  & 4302 & $2233$ &  58 & \cite{hahn2019atlantic}\\
\hline
ACIC-2018  & varied & varied &  varied & \cite{shimoni2018benchmarking}\\
\hline
IHDP & 672 & 123 &  25 & \cite{shalit2017estimating}\\
\hline
Lalonde CPS & 16177 & $167$ & 9 & \cite{neal2020realcause} \\
\hline
Lalonde PSID & 2675 & $175$ &  9 & \cite{neal2020realcause}\\
\hline
Twins & 11984 & 8935 &  76 & \cite{neal2020realcause}\\
\hline  
\end{tabular}
\end{table}

We compare the isocalibrated DML (IC-DML) estimator to the AIPW estimator, with both defined as in Experiment~1. DR-TMLE was excluded from this experiment due to the computational expense of the iterative debiasing step, as discussed in the Introduction. Cross-fitted nuisance functions (propensity score and outcome regression) are estimated using gradient-boosted regression trees via \texttt{xgboost} \citep{chen2015xgboost}, with the maximum tree depth selected by 10-fold cross-validation. Under smoothness conditions, these nuisance estimators will be consistent, though the rate of convergence may be slow in moderate dimensions \citep{zhang2005boosting}. For the AIPW estimator, we truncate the cross-fitted propensity score estimates
to lie in \([c_n,1-c_n]\), where \(c_n :=  \min\{0.05, 25/(n^{1/2}\log n)\}\), following
\citet{gruber2022data}. We do not apply additional truncation for
calibrated DML, since isotonic calibration induces automatic truncation
\citep{van_der_Laan2024stabilized}.

 For each collection of $k$ dataset realizations, Table~\ref{fig::exp3} reports scaled absolute bias, scaled root mean square error (RMSE), and nominal 95\% confidence interval coverage. In the ACIC-2017 settings where AIPW exhibits substantial bias, calibration noticeably reduces bias: in settings 18, 20, and 24, scaled bias decreases from 0.371 to 0.169 (55\% reduction), 1.82 to 0.534 (71\% reduction), and 0.316 to 0.087 (73\% reduction), respectively. Coverage improves in the same settings from 0.21 to 0.71, 0.29 to 0.89, and 0.30 to 0.90. In settings where AIPW bias is already small, the point estimates are generally similar.

\begin{table}[tb]
    \centering
    \caption{Semi-synthetic benchmark results. Bias and RMSE are scaled by the average absolute effect size; coverage uses the stored primary confidence interval for each method. For clarity, we report only sample size 10000 results for ACIC-2018 and the settings with strong confounding for ACIC-2017. The full results can be found in Appendix \ref{appendix::simRealFull}.}\label{fig::exp3}
    \begin{subtable}{\textwidth}
        \centering
        \footnotesize
        \setlength{\tabcolsep}{2.5pt}
\begin{tabular}{|c||P{1.15cm}|P{1.15cm}||P{1.25cm}|P{1.25cm}||P{1.25cm}|P{1.25cm}|}
            \hline
            \multirow{2}{*}{\textbf{Dataset}} &
              \multicolumn{2}{c||}{\textbf{Coverage}} &
              \multicolumn{2}{c||}{\textbf{RMSE}} &
              \multicolumn{2}{c|}{\textbf{Absolute Bias}} \\
            \cline{2-7}
            & AIPW & IC-DML & AIPW & IC-DML & AIPW & IC-DML \\
            \hline\hline
             ACIC-2017 (18) & 0.21 & \textbf{0.71} & 1.01 & \textbf{0.607} & 0.371 & \textbf{0.169} \\
            \hline
             ACIC-2017 (20) & 0.29 & \textbf{0.89} & 2.16 & \textbf{1.48} & 1.82 & \textbf{0.534} \\
            \hline
             ACIC-2017 (22) & 0.40 & \textbf{0.80} & 0.147 & \textbf{0.109} & 0.0395 & \textbf{0.0216} \\
            \hline
             ACIC-2017 (24) & 0.30 & \textbf{0.90} & 0.37 & \textbf{0.269} & 0.316 & \textbf{0.087} \\
            \hline
             ACIC-2018 (10000) & 0.63 & \textbf{0.73} & 16.7 & 16.7 & \textbf{1.47} & 1.5 \\
            \hline
             IHDP & \textbf{0.58} & 0.55 & \textbf{0.461} & 0.47 & \textbf{0.129} & 0.132 \\
            \hline
             Lalonde CPS & 0.14 & \textbf{0.65} & \textbf{0.193} & 0.33 & 0.105 & \textbf{0.0696} \\
            \hline
             Lalonde PSID & 0.48 & \textbf{0.83} & \textbf{0.19} & 0.432 & 0.0424 & \textbf{0.0236} \\
            \hline
             Twins & \textbf{0.96} & \textbf{0.96} & 0.112 & \textbf{0.11} & 0.0603 & \textbf{0.0565} \\
            \hline
        \end{tabular}
    \end{subtable}
\end{table}

\section{Conclusion}\label{section::conclusion}

In this work, we introduced a unified debiased machine learning framework for doubly robust inference on linear functionals of the outcome regression. These functionals form an important subclass of the mixed-bias parameters characterized by \cite{rotnitzky2021characterization}. For linear functionals, our results provide an affirmative answer to the open question raised by \cite{smucler2019unifying}: learner-agnostic doubly robust inference with black-box nuisance estimators is possible, provided asymptotic partial orthogonality holds. A second contribution is to clarify the role of calibration in the doubly robust inferential properties of debiased machine learning estimators. Our arguments rely only on the calibrated nuisance estimators satisfying the calibration score equations in Section~\ref{Section::DR}. Although our implementation uses isotonic calibration, these equations can also be satisfied by other histogram-type or step-function calibration procedures, including honest regression trees \citep{wager2018estimation}, the highly adaptive lasso \citep{van2017generally,benkeser2016highly,fang2021multivariate}, and Venn--Abers calibration \citep{vovk2012venn,van2024self,van2025generalized}. Overall, our theoretical and empirical results support the use of calibrated nuisance estimators in causal inference as a general strategy for reducing bias from inconsistent nuisance estimation and enabling valid doubly robust inference.  

Several open questions remain. First, although Appendix~\ref{sec::mixedbias} outlines an extension of our framework to more general parameters characterized by the mixed bias property \citep{rotnitzky2021characterization}, a complete development of this approach is left to future work. As an illustrative proof-of-concept of this direction, we proposed calibrated DML estimators for the expected conditional covariance parameter arising in partially linear regression models, in the spirit of the estimators proposed by \cite{dukes2024doubly} and \cite{bonvini2024doubly}. Second, it would be useful to develop an isotonic-calibrated targeted minimum loss estimator that inherits both the plug-in property and doubly robust asymptotic linearity, as in \cite{van2014targeted} and \cite{benkeser2017doubly}. The main challenge is that isotonic calibration and TMLE targeting enforce different, and generally incompatible, score equations, so that applying one step may undo the constraints imposed by the other. One possible approach is to alternate between isotonic calibration and targeting until the score equations needed for both DRAL and the plug-in property are solved simultaneously. Alternatively, the score-preserving TMLE of \cite{pimentel2025score} could be applied to the calibrated nuisance estimates so that the targeting step preserves the empirical calibration scores. Finally, extending our framework to accommodate multiple and sequential robustness in longitudinal data structures presents another promising direction \citep{rotnitzky2017multiply,luedtke2017sequential}.

 \hspace{0.5cm}

\noindent\textbf{Acknowledgements.} Research reported in this publication was supported by NIH grants DP2-LM013340 and  R01-HL137808, and NSF grant DMS-2210216. The content is solely the responsibility of the authors and does not necessarily represent the official views of the funding agencies. 

\singlespacing

\clearpage
\onehalfspacing
\appendix

\setcounter{equation}{0}
\renewcommand{\theequation}{S\arabic{equation}}
\setcounter{theorem}{0}
\setcounter{figure}{0}
\setcounter{table}{0}
\setcounter{lemma}{0}
\setcounter{corollary}{0}
\renewcommand{\thetheorem}{S\arabic{theorem}}
\renewcommand{\thecorollary}{S\arabic{corollary}}
\renewcommand{\thelemma}{S\arabic{lemma}}
\renewcommand{\thefigure}{S\arabic{figure}}
\renewcommand{\thetable}{S\arabic{table}}
\renewcommand{\thealgorithm}{S\arabic{algorithm}}

Sections~A--H collect supporting discussion, examples, calibration background,
and implementation details. Sections~I--L contain the proofs of the main
theoretical results and their supporting lemmas. Section~M reports additional
simulation results.

\section{Background on semiparametric efficiency}
\label{appendix::efficiencybackground}

\noindent Following \cite{van2000asymptotic}, we call an estimator
asymptotically linear for \(\tau_0=\Psi(P_0)\), relative to
\(\mathcal M\), with influence function \(D\in L^2_0(P_0)\) if
\[
    \widehat\tau_n-\tau_0
    =
    P_n D + o_p(n^{-1/2}).
\]
It is efficient if \(D=D_0\), the efficient influence function. We say that an
estimator is regular at \(P_0\) with respect to \(\mathcal M\) if, along every
regular parametric submodel
\(\{P_\varepsilon:\varepsilon\}\subset\mathcal M\) through \(P_0\) and every
fixed \(h\), the law under \(P_{h/\sqrt n}\) of
\[
    \sqrt n\{\widehat\tau_n-\Psi(P_{h/\sqrt n})\}
\]
converges to a limit that does not depend on \(h\) or on the submodel. By the
convolution theorem (e.g., \citealp[Ch.~25]{van2000asymptotic}; see also
\citealp{van1996weak}), any regular estimator of \(\tau_0\) has asymptotic
variance at least \(P_0D_0^2\), and this lower bound is attained by efficient
estimators. Moreover, efficient estimators are locally asymptotically minimax
among all estimators \citep{hajek1972local,le2012asymptotic}.

\section{Conditions for main text}

\subsection{Additional conditions for Theorem \ref{theorem::DRinference}}
 \label{appendix::mainconditions}

Theorem \ref{theorem::DRinference} further relies on the following regularity conditions, versions of which are commonly assumed in the debiased machine learning literature \citep{van2023causal,rabenseifner2025calibrationstrategiesrobustcausal}. 
We set
$r_{n,j}^* := (\overline{\Pi}_{\alpha_{n,\diamond}^*}(\mu_0 - \mu_{n,\diamond}^*))_j$
and
$s_{n,j}^* := (\overline{\Pi}_{\mu_{n,\diamond}^*}(\alpha_0 - \alpha_{n,\diamond}^*))_j$ for $j=1,\ldots,J$,
which respectively approximate \(r_0 := \Pi_{\alpha_0} (\mu_0 - \overline{\mu}_0)\) and \(s_0 := \Pi_{\mu_0} (\alpha_0 - \overline{\alpha}_0)\), defined analogously to \eqref{eqn::nuisnew}. Let \( P_{n,j} \) denote the empirical measure corresponding to the \( j \)-th fold \( \mathcal{C}^{(j)} \) in the cross-fitting procedure. 

\begin{enumerate}[label=\textbf{C\arabic*)}, ref={C\arabic*}, start=3]

  \item \textit{Linearity of estimated functional:}\label{cond::estboundedlinearcross}
For each \(j \in [J]\), the map \(\psi_{n,j}:\mathcal{H}\to\mathbb{R}\) is linear
\(P_0\)-almost surely; that is,
\(\psi_{n,j}(c\mu_1+\mu_2)=c\,\psi_{n,j}(\mu_1)+\psi_{n,j}(\mu_2)\) for all
\(\mu_1,\mu_2\in\mathcal{H}\) and \(c\in\mathbb{R}\).
            \item \textit{Asymptotic linearity of the functional estimator:} \label{cond::DRremainder}
For each $j \in [J]$, $\psi_{n,j}(\mu_{n,j}^*) - \psi_0(\mu_{n,j}^*) - P_{n,j} \phi_{0,\overline{\mu}_0} = o_p(n^{-1/2})$ and, under \ref{cond::DRmisDRconsist::1}, also $\psi_{n,j}(r_{n,j}^*) - \psi_0(r_{n,j}^*) - P_{n,j}\phi_{0,r_0} = o_p(n^{-1/2})$.
     \item \textit{Boundedness of nuisance functions:} \label{cond::estnuisbound} for some $M\in(0, \infty)$, $Y$ and the functions $\mu_{n,j}^*$, $\alpha_{n,j}^*$, $\mu_0$, $\alpha_0$, $\overline{\mu}_0$ and $\overline{\alpha}_0$ are uniformly bounded by $M$ with $P_0$--probability tending to one for each $j=1,\ldots,J$; \label{cond::bound}
   \item \textit{Calibration functions have finite total variation:} \label{cond::variation} 
There exists \(M<\infty\) such that, for all \(n\in\mathbb N\), there are
measurable representatives
\(\theta_{n,\mu}:\mathbb R\to\mathbb R\) and
\(\theta_{n,\alpha}:\mathbb R\to\mathbb R\) satisfying
\(\theta_{n,\mu}\circ\mu_{n,\diamond}
=
\overline{\Pi}_{\mu_{n,\diamond}}\Delta_{\alpha,n,\diamond}\) and
\(\theta_{n,\alpha}\circ\alpha_{n,\diamond}
=
\overline{\Pi}_{\alpha_{n,\diamond}}\Delta_{\mu,n,\diamond}\) almost surely,
with total variation norms bounded above by $M$ almost surely.

\end{enumerate}
Condition~\ref{cond::estboundedlinearcross} requires the estimated functional to
be linear; in most applications, this linearity is inherited from \(\psi_0\)
and is therefore satisfied by standard plug-in estimators. 
Condition~\ref{cond::DRremainder} is automatic when \(\psi_0\) is known and
\(\psi_{n,j}=\psi_0\). It also holds, under standard conditions, for
expectation-representable functionals
\(\psi_0(f)=P_0 m(\cdot,f)\) satisfying \eqref{eqn::simplefunctional}, with
known \(m\), when estimated by the empirical plug-in map
\(\psi_{n,j}(f)=P_{n,j}m(\cdot,f)\); see
Lemma~\ref{lemma::remainderlipschitzsmall} of
Appendix~\ref{appendix::dealingwithEP}. The condition is included mainly to
allow more general nuisance-dependent linear functionals, for which \(\psi_0\)
may itself depend on unknown components of \(P_0\). For example, for a
distribution-shift functional
\(\psi_0(\mu)=E_{P_0^{\mathrm{shift}}}\{m(Z,\mu)\}\), with
\(P_0^{\mathrm{shift}}\neq P_0\), the condition holds for
\(\psi_n(f)=P_n^{\mathrm{shift}}m(\cdot,f)\) under similar conditions
\citep{chernozhukov2023automatic}. When verifying
Condition~\ref{cond::DRremainder} requires additional rate conditions on
\(\mu_n\), \(\alpha_n\), or other nuisance estimators, the DRAL guarantee applies
conditional on those prerequisites.

Conditions~\ref{cond::bound} and \ref{cond::variation} are used to control
empirical-process remainders. Condition~\ref{cond::bound} imposes standard
boundedness assumptions and could potentially be relaxed to suitable tail
conditions (e.g., sub-Gaussianity). Condition \ref{cond::variation} rules out settings in which the optimal
\(L^2(\overline{P}_0)\) predictors of the estimation errors
\(\Delta_{\alpha,n,\diamond}\) and \(\Delta_{\mu,n,\diamond}\), given only the
initial nuisance estimators \(\mu_{n,\diamond}\) and \(\alpha_{n,\diamond}\), are
pathological---in the sense that they have infinite variation norm when viewed
as univariate functions of the nuisance estimators \citep{van2023causal}.  

\subsection{Additional conditions for Theorem \ref{theorem::DRboot}}
 \label{appendix::bootconditions}

We recall that Algorithm \ref{alg::bootstrap} assumes $\psi_0$ has the form $\mu \mapsto E_0\{m(Z, \mu)\}$ and $\frac{1}{J}\sum_{j=1}^J\psi_{n,j}$ is taken to be the empirical plug-in estimator $\mu \mapsto \frac{1}{n}\sum_{i=1}^n m(Z_i, \mu)$. Below, we define $P_{n,j}^\#$ as the empirical distribution of an arbitrary bootstrap sample $\mathcal{C}^{\#(j)}$ drawn from the empirical distribution $P_{n,j}$ of $\mathcal{C}^{(j)}$. Moreover, we denote the bootstrapped calibrated nuisance estimators corresponding to $P_{n,j}^\#$ as $\mu_{n,j}^{\#}$ and $\alpha_{n,j}^{\#}$, which are obtained as in Algorithm \ref{alg::bootstrap}. Finally, we define $\tau_n^{\#}$ as the corresponding bootstrapped isocalibrated DML estimator obtained from $\mu_{n,1}^{\#},\ldots,\mu_{n,J}^{\#}$ and $\alpha_{n,1}^{\#},\ldots,\alpha_{n,J}^{\#}$. Denoting by $\Delta_{\alpha,n,j}^\# := \alpha_{n,j}^{\#} - \alpha_0$ and $\Delta_{\mu,n,j}^\# := \mu_{n,j}^{\#} - \mu_0$ for $j=1,\ldots,J$ the nuisance estimation errors, the following additional conditions are used in Theorem \ref{theorem::DRboot}.

 \begin{enumerate}[label=\bf{D\arabic*)}, ref={D\arabic*}, start=2]
 \item \textit{Boundedness of bootstrap nuisance estimators:} \label{cond::estboundboot} The functions $\mu_{n,j}^{\#}$ and $\alpha_{n,j}^{\#}$ are uniformly bounded by $M$ with $P_0$--probability tending to one for each $j=1,\ldots,J$;
    \item \textit{Lipschitz continuity of the functional:} \label{cond::lipschitzisotonic} 
$\mathcal{A}$ is finite, and there exists a constant $L \in (0,\infty)$ such that, for all $\mu \in \mathcal{H}$, $\|m(\cdot,\mu)\| \leq L\|\mu\|$ and $|m(Z,\mu)| \leq L \max_{a \in \mathcal{A}} |\mu(a,W)|$ $P_0$-almost surely.
    \end{enumerate}
Condition~\ref{cond::estboundboot} is a bootstrap variant of
Condition~\ref{cond::estnuisbound}. Condition~\ref{cond::lipschitzisotonic} is
used to formally verify \ref{cond::DRremainder} and the analogous condition for
the bootstrapped estimators. The \(L^2(P_0)\) Lipschitz condition \(\|m(\cdot,\mu)\|\leq L\|\mu\|\) is standard in the autoDML literature \citep{chernozhukov2022automatic}; in particular, it ensures that the influence-function component \(\varphi_{\mu_0}=m(\cdot,\mu_0)-E_0\{m(Z,\mu_0)\}\) has finite variance whenever \(\|\mu_0\|<\infty\). The pointwise Lipschitz condition is automatic for the mean potential outcome at treatment level \(a_0\), for which \(m(Z,\mu)=\mu(a_0,W)\), since
\[
    |m(Z,\mu)| = |\mu(a_0,W)|
    \leq \max_{a\in\mathcal A}|\mu(a,W)|.
\]
The \(L^2(P_0)\) Lipschitz bound then follows from positivity: if \(P_0(A=a\mid W)\geq \epsilon>0\) for all \(a\in\mathcal A\), then \(\|m(\cdot,\mu)\|^2=E_0\{\mu(a_0,W)^2\}\leq \epsilon^{-1}E_0\{\mu(A,W)^2\}=\epsilon^{-1}\|\mu\|^2\).

\subsection{Verification of \ref{cond::linear}--\ref{cond::pathwise} for expectation-representable functionals}
\label{appendix::conditions}

The expectation-representable linear functionals in
Example~\ref{example::expectfunctional} satisfy
Conditions~\ref{cond::linear}--\ref{cond::pathwise} under a \(P_0\)-operator bound and a local domination condition that transfers this bound to nearby distributions. For \(P\in\mathcal M\), define $\psi_P(\mu):=E_P\{m(Z,\mu)\}.$ Denote the operator norm of $\psi_P-\psi_{P_0}$ on \((\mathcal H,\|\cdot\|)\) by
\[
\|\psi_P-\psi_{P_0}\|_{\mathrm{op}}
:=
\sup_{\mu\in\mathcal H:\|\mu\|\le 1}
\left|\psi_P(\mu)-\psi_{P_0}(\mu)\right|.
\]
We assume:
\begin{enumerate}[label=\textnormal{(E\arabic*)}, ref={E\arabic*}]
    \item \label{cond:expectlinear} \(m\) is linear in its second argument;
    \item \label{cond:expectl2} $ B_0:=
    \sup_{\mu\in\mathcal H:\|\mu\|\le 1}
    \left(E_0[m(Z,\mu)^2]\right)^{1/2}<\infty ;$
    \item \label{cond:expectdom} for some \(\eta>0\) and \(C_\eta<\infty\),
    each \(P\in\mathcal M\) satisfying \(H(P,P_0)<\eta\) obeys
    \(P\ll P_0\) and \(dP/dP_0\le C_\eta\), \(P_0\)-almost surely.
\end{enumerate}

\begin{lemma}[Verification for expectation functionals]
\label{lem:hellinger_continuity}
Under \ref{cond:expectlinear}--\ref{cond:expectdom}, \(\psi_0\) satisfies
Conditions~\ref{cond::linear}--\ref{cond::pathwise}. Specifically,
\[
\sup_{\mu\in\mathcal H:\mu\ne0}
\frac{|\psi_0(\mu)|}{\|\mu\|}
\le B_0 .
\]
Moreover, if \(H(P,P_0)<\eta\), then
\[
\|\psi_P-\psi_{P_0}\|_{\mathrm{op}}
\le 2\sqrt{C_\eta+1}\,B_0\,H(P,P_0).
\]
Finally, for each \(\mu\in\mathcal H\), \(P\mapsto\psi_P(\mu)\) is pathwise
differentiable at \(P_0\) with efficient influence function
\[
\phi_{0,\mu}(Z)=m(Z,\mu)-\psi_0(\mu).
\]
\end{lemma}

 \section{Additional details on DRAL and calibrated DML}

This section collects examples and auxiliary discussion that support the
general framework of Sections~\ref{sec::genstrat} and
\ref{section::calibration}. The subsections below provide additional examples,
conditions used in the main theory, and implementation details for the
calibrated DML procedure.

\subsection{Additional examples of parameters}

\begin{example}[counterfactual means under general interventions]
    For simplicity, suppose \(\mathcal{A}\) is discrete, and denote the propensity score by \(\pi_0(a \mid w) := P_0(A = a \mid W = w)\). Let $Y^{a_0}$ denote the potential outcome that would be observed if treatment $a_0 \in \mathcal{A}$ were administered. Under causal conditions \citep{rubin1974}, the mean of $Y^{a_0}$ is identified by $\tau_0 := E_0\{\mu_0(a_0,W)\}$, which corresponds to the linear functional $\psi_0: \mu \mapsto E_0\{\mu(a_0,W)\}$. Doubly robust inference for this parameter was considered in \cite{van2014targeted} and \cite{benkeser2017doubly}. More generally, the mean of the counterfactual outcome under a stochastic intervention that, given $W=w$, draws  a treatment value $A^*$ at random from the distribution with probability mass function $a\mapsto \omega_0(a |w)$ depending only on $P_{0,A,W}$ can be identified by $\tau_0 := \iint \mu_0(a,w) \omega_0(a  |  w) P_{0,A,W}(da,dw)$. If $(a,w) \mapsto \alpha_0(a,w):= \omega_0(a  |  w)/\pi_0(a |  w)$ has finite $\mathcal{H}$--norm, then conditions \ref{cond::linear}--\ref{cond::pathwise} can be verified and $\alpha_0$ is the nonparametric Riesz representer of $\tau_0$.
\end{example}

\begin{example}[Expected conditional covariance]
Consider the expected conditional covariance between $A$ and $Y$ adjusted for
$W$,
\[
\tau_0 = E_0\!\left[\{A-\pi_0(W)\}\{Y-m_0(W)\}\right],
\]
where $\pi_0(w) := E_0[A \mid W=w]$ and $m_0(w) := E_0[Y \mid W=w]$. This
parameter arises in semiparametric ATE estimation under partially linear models
\citep{robinson1988root,li2011higher}, as a measure of ``causal influence''
\citep{diaz2024non}, and in conditional independence testing
\citep{shah2020hardness}. Note that
$\tau_0 = E_0[AY] - E_0\{A\,m_0(W)\}$. The first term is efficiently estimated
by the empirical plug-in $n^{-1}\sum_{i=1}^n A_iY_i$. The second is a linear
functional of $m_0$, namely $\psi_0: m \mapsto E_0\{A\,m(W)\}$, with Riesz
representer $\pi_0$, and can be estimated by its empirical plug-in
\(\psi_n: m \mapsto n^{-1}\sum_{i=1}^n A_i m(W_i)\). Consequently,
$E_0\{A\,m_0(W)\}$ is an estimand within our framework, and our methods apply to
doubly robust inference for $\tau_0$.

Technically, $\psi_0$ depends on the marginalized conditional mean $m_0$ rather
than the treatment-specific conditional mean $\mu_0$; however, this does not
affect the applicability of our approach, since $m_0$ can be calibrated under
squared error loss in the same way (see Appendix~\ref{sec::mixedbias}). Related
work on doubly robust inference for this parameter includes
\cite{dukes2024doubly} and \cite{bonvini2024doubly}.
\end{example}

 \begin{example}[average treatment effect under covariate-dependent outcome missingness]
Suppose that the observed data unit consists of $(W, S, \Delta, \Delta Y)$, where $S \in \{0,1\}$ is a binary treatment, $\Delta \in \{0,1\}$ is a missingness indicator taking value $1$ if $Y$ is observed, and $\Delta Y$ is the observed outcome, equal to $Y$ if $\Delta = 1$ and $0$ otherwise. Under causal conditions, the ATE under covariate-dependent outcome missingness is identified by the estimand $\tau_0 :=   E_0\{E_0(\Delta Y  \,|\,  S = 1, \Delta = 1, W) - E_0(\Delta Y \,|\,  S = 0, \Delta = 1, W)\}$. Doubly robust inference for this estimand was considered in \cite{diaz2017doubly}. Using the notation of this section, we can write $Z := (W,A,Y')$, where $A := (S, \Delta)$ is a multi-valued interventional variable and $Y' := \Delta Y$ is an outcome with missingness. The linear functional corresponding to $\tau_0$ is $\psi_0:\mu \mapsto E_0\{\mu((1, 1), W) - \mu((0, 1), W)\}$. The nonparametric Riesz representer of $\psi_0$ is given by $\alpha_0:(s, \delta, w ) \mapsto (2s-1)\delta/P_0(\Delta =1,S =s\,|\, W=w)$.

\end{example}

\subsection{Controlling empirical process remainders in Theorem \ref{theorem::DRbiasexpansion}}
\label{appendix::EPhandle}
In this section, we provide sufficient conditions for the asymptotic
negligibility of the empirical-process remainders $(P_n-P_0)(A_{n,0}-A_0)$ and
$(P_n-P_0)(B_{n,0}-B_0)$ that appear in the regression-based and Riesz-based
decompositions in Theorem~\ref{theorem::DRbiasexpansion}, respectively.
Importantly, the lemmas show that these remainders are $o_p(n^{-1/2})$ under
conditions requiring only that the nuisance associated with the decomposition
is consistent, while the potentially inconsistent nuisance merely converges in
probability to a limit. The proofs of the lemmas are given in Appendix~\ref{appendix::proofEP}.

The next lemma establishes negligibility of $(P_n-P_0)(A_{n,0}-A_0)$. Recall that $A_0(z) = \Pi_{\mu_0}(\alpha_0 - \overline{\alpha}_0)\{y - \mu_0(z)\}$ and $A_{n,0}(z) = \Pi_{\mu_n}(\alpha_0 - \alpha_n)\{y - \mu_n(z)\}$.

\begin{enumerate}[label=\textbf{(G\arabic*)},ref=G\arabic*,series=EP]
    \item \emph{Boundedness.} $\mu_n$, $\alpha_n$, $\mu_0$, $\alpha_0$, and $Y$ are uniformly bounded. \label{controlEP::bounded}
    \item \emph{Consistency and projection coupling.} $\|\overline{\alpha}_0 - \alpha_n\| = o_p(1)$, $\|\mu_n - \mu_0\| = o_p(1)$, and
    $\|\Pi_{\mu_n}(\overline{\alpha}_0 - \alpha_0)- \Pi_{\mu_0}(\overline{\alpha}_0 - \alpha_0)\|
    = O_p(\|\mu_n - \mu_0\|^{\beta})$
    for some $\beta \in (0,1]$. \label{controlEP::couplingmu}
    \item \emph{Donsker condition.} With probability tending to one, $A_{n,0} - A_0$ belongs to a fixed Donsker function class. \label{controlEP::donsker}
\end{enumerate}

\begin{lemma}
\label{lemma::negEPOutcome}
   Assume \ref{controlEP::bounded}-\ref{controlEP::couplingmu}. Then, $\|A_{n,0} - A_0\| = o_p(1)$. If, in addition, \ref{controlEP::donsker} holds, then $(P_n - P_0)(A_{n,0} - A_0) = o_p(n^{-1/2})$.
\end{lemma}

In our proof of Theorem~\ref{theorem::DRinference}, to establish DRAL for the
isocalibrated DML estimator with cross-fitting, we avoid Donsker assumptions
such as \ref{controlEP::donsker}. Instead, we use cross-fitting to remove any
complexity conditions on the initial nuisance estimators $\alpha_{n,j}$ and
$\mu_{n,j}$. We then exploit that the calibrated nuisance estimators
$\alpha_{n,j}^*$ and $\mu_{n,j}^*$ differ from $\alpha_{n,j}$ and $\mu_{n,j}$ only through
monotone transformations. Since the class of monotone functions has finite
uniform entropy integral (and is therefore Donsker), this controls the
complexity of the calibrated nuisance estimators. With cross-fitting, to
control the complexity of functions of the form $s_{n,0}=\Pi_{\mu_n}
\Delta_{\alpha,n}$, it suffices that the transformation
$\theta_{\alpha,n}:\mathbb{R}\to\mathbb{R}$ defined by
$s_{n,0}=\theta_{\alpha,n}\circ \mu_n$ has bounded variation (as in
\ref{cond::variation}), which likewise yields controlled complexity. See our proofs in Appendix \ref{appendix::dealingwithEP} for details.

The next lemma establishes negligibility of $(P_n-P_0)(B_{n,0}-B_0)$. Recall
that $B_0 := \phi_{r_0} + \psi_0(r_0) - r_0 \alpha_0$, and
$B_{n,0} := \phi_{r_{n,0}} + \psi_0(r_{n,0}) - r_{n,0} \alpha_n$. Note that
\[
(P_n - P_0) (B_{n,0} - B_0)
=
(P_n - P_0)\{\phi_{r_{n,0}} - \phi_{r_0} -  r_{n,0} \alpha_n + r_0 \alpha_0\}.
\]

\begin{enumerate}[label=\textbf{(G\arabic*)},ref=G\arabic*, resume = EP]
    \item \emph{Consistency and projection coupling.} $\|\overline{\mu}_0 - \mu_n\| = o_p(1)$, $\|\alpha_n - \alpha_0\| = o_p(1)$, and
    $\|\Pi_{\alpha_n}(\overline{\mu}_0 - \mu_0)- \Pi_{\alpha_0}(\overline{\mu}_0 - \mu_0)\|
    = O_p(\|\alpha_n - \alpha_0\|^{\beta})$
    for some $\beta \in (0,1]$. \label{controlEP::couplingriesz}
    \item \textit{Lipschitz continuity}: $\|\phi_{r_{n,0}} - \phi_{r_0} \|  = O_p(\|r_{n,0} - r_0 \| )$. \label{controlEP::lip}
    \item \emph{Donsker condition.} With probability tending to one, $\phi_{r_{n,0}} - \phi_{r_0} -  r_{n,0} \alpha_n + r_0 \alpha_0$ belongs to a fixed Donsker function class. \label{controlEP::donskerriesz}
\end{enumerate}

\begin{lemma}
\label{lemma::negEPRiesz}
Assume Conditions~\ref{controlEP::bounded} and~\ref{controlEP::couplingriesz}--\ref{controlEP::lip}. Then,
$\|\phi_{r_{n,0}} - \phi_{r_0} -  r_{n,0} \alpha_n + r_0 \alpha_0 \|=o_p(1)$. If, in addition, \ref{controlEP::donskerriesz} holds, then
$(P_n-P_0)(B_{n,0}-B_0)=o_p(n^{-1/2})$.
\end{lemma}

\subsection{Bootstrap-assisted confidence intervals with cross-fitting}
\label{appendix::bootinfalgo}

 \begin{algorithm}[H]
\caption{Bootstrap-assisted doubly robust CIs}
\label{alg::bootstrap}
\begin{algorithmic}[1]
\small 
\INPUT cross-fitted estimators $(\alpha_{n,j}, \mu_{n,j})$ excluding fold $\mathcal{C}^{(j)}$ for $j=1,\ldots,J$,  isocalibrated DML estimator $\tau_n^*$; confidence level $1-\rho$; bootstrap replication number $K$;
\FOR {$k = 1,\ldots,K$}
\STATE for {$s = 1,\ldots,J$}, draw bootstrap sample $\mathcal{C}_{k}^{(s)\#}$ from $\mathcal{C}_k^{(s)}$;
\STATE set $\mathcal{D}_{n,k}^{\#} :=  \{Z_1^{\#}, \dots, Z_n^{\#}\} = \{\mathcal{C}_{k}^{(s)\#}: s=1,\ldots,J\} $;
\STATE for {$s = 1,\ldots,J$}, set $j(i)=s$ for each $i\in\{1,\ldots,n\}$ such that $Z_i^{\#} \in \mathcal{C}_k^{(s)\#}$;
\STATE compute calibrators $f^{\#(k)}_n$ and $g^{\#(k)}_n$ using pooled out-of-fold estimates as\vspace{-.05in}
\begin{align*}
f_n^{\#(k)} &\in \argmin_{f \in \mathcal{F}_{\text{iso}}} \sum_{i =1}^n \bigl\{ Y_{i}^{\#} - f(\mu_{n,j(i)}(A_i^{\#}, W_i^{\#}))\bigr\}^2;\\
g_n^{\#(k)} &\in \argmin_{g \in \mathcal{F}_{\text{iso}}} \sum_{i =1}^n \bigl\{g (\alpha_{n,j(i)}(A_i^{\#}, W_i^{\#}))^2 - 2m(Z_i^{\#},  g \circ \alpha_{n,j(i)})\bigr\}\,;
\end{align*}\vspace{-.05in}
\STATE for {$s = 1,\ldots,J$}, set $  \mu_{n,s}^{(k)\#}  := f_n^{(k)\#} \circ \mu_{n,s} $ and $  \alpha_{n,s}^{(k)\#} := g_n^{(k)\#} \circ \alpha_{n,s} $;
\STATE compute calibrated one-step debiased estimator
$$\tau_{n,k}^{\#} := \frac{1}{n} \sum_{i =1}^n \left[ m(Z_i^{\#}, \mu_{n,j(i)}^{(k)\#})  +  \alpha_{n,j(i)}^{(k)\#}(A_i^{\#}, W_i^{\#})\left\{Y_i^{\#} - \mu_{n,j(i)}^{(k)\#}(A_i^{\#},W_i^{\#})\right\}\right];$$
\ENDFOR 
\STATE set $\sigma_n^{\# 2} := \frac{1}{K}\sum_{k=1}^K (\tau_{n,k}^{\#} - \overline{\tau}_n^{\#})^2$ with $\overline{\tau}_n^{\#} := \frac{1}{K}\sum_{k=1}^K \tau_{n,k}^{\#};$
\STATE construct interval $\text{CI}_n(\rho) := (\tau_n^* - q(\rho)\sigma_n^{\#},\tau_n^* + q(\rho)\sigma_n^{\#});$  
\RETURN  \textit{bootstrap-assisted confidence interval} $\text{CI}_n(\rho)$ for $\tau_0$.
\end{algorithmic}
\end{algorithm}

\subsection{Convergence rates of isotonic calibrated nuisances}

\label{appendix::calnuisances}

The convergence-rate conditions in \ref{cond::DRlimits} are stated
in terms of the isotonic-calibrated nuisance estimators, which raises the
question of whether the calibration step can degrade the rates achieved by the
first-stage learners. This Appendix addresses this by
showing that isotonic calibration preserves the \(L^2\) convergence rates of
the first-stage nuisance estimators, up to an additional \(O_p(n^{-1/3})\)
term. Since \(n^{-1/3}=o(n^{-1/4})\), this calibration error is negligible
relative to the \(o_p(n^{-1/4})\) rate required in \ref{cond::DRlimits}, and
thus calibration imposes no additional rate restrictions. Moreover, in the
high-dimensional and nonparametric regimes that motivate our work, the
one-dimensional \(n^{-1/3}\) calibration term is typically dominated by
first-stage estimation error. This term reflects estimation of a monotone
function in one dimension and therefore depends favorably on problem constants,
independent of covariate dimension \citep{chatterjee2013improved}. By contrast,
rates faster than \(n^{-1/3}\) for high-dimensional first-stage learning
typically require strong smoothness or extreme sparsity assumptions, and can
involve constants that scale poorly with dimension and other structural
parameters. Lemma~\ref{lemma::Rates} below further shows that the calibrated nuisance
estimators achieve the same mean-squared-error rate as the best monotone
transformation of the original nuisance estimators, where ``best'' is defined as
the minimizer of the relevant population risk. Thus, isotonic
calibration relaxes the consistency requirement by allowing the initial nuisance
estimators to be misspecified up to an unknown monotone transformation.

The following lemma is a direct application of Theorem~\ref{theorem::MSE}, stated and proven in Appendix \ref{appendix::calrates}.

\begin{lemma}
\label{lemma::Rates}
    Suppose that $\psi_0$ has the form \eqref{eqn::simplefunctional} with mapping $m:\mathcal{A}\times\mathcal{W}\times \mathcal{H}\rightarrow\mathbb{R}$, and that $\frac{1}{J}\sum_{j=1}^J\psi_{n,j}$ is taken to be the plug-in estimator $\mu \mapsto \frac{1}{n}\sum_{i=1}^n m(Z_i, \mu)$. If Conditions~\ref{cond::linear}, \ref{cond::estboundedlinearcross}, \ref{cond::estnuisbound}, and~\ref{cond::lipschitzisotonic} hold, then
    \begin{align*}
        \sum_{j=1}^J\|\mu_{n,j}^* - \mu_0 \|^2\,&=\,\min_{f \in \mathcal{F}_{\textnormal{iso}}} \sum_{j=1}^J\|f \circ \mu_{n,j} - \mu_0 \|^2 + O_p(n^{-\frac{2}{3}})\,;\\
       \sum_{j=1}^J \|\alpha_{n,j}^* - \alpha_0 \|^2\,&=\,\min_{g \in \mathcal{F}_{\textnormal{iso}}} \sum_{j=1}^J\|g \circ \alpha_{n,j} - \alpha_0 \|^2 + O_p(n^{-\frac{2}{3}})\,.
    \end{align*}
\end{lemma}

Since the identity transformation is itself monotone non-decreasing, the above
lemma implies that the isotonic-calibrated nuisance estimators typically
perform no worse than the uncalibrated ones, and may perform better when the
uncalibrated estimators are poorly calibrated or only consistent up to a
monotone transformation. Thus, beyond enabling doubly robust asymptotic
linearity, isotonic calibration relaxes the consistency requirements on the
nuisance estimators: it suffices that the initial cross-fitted nuisance
estimators be consistent up to a monotone transformation.

\subsection{Estimation of nuisance functions in the limiting distribution}
\label{appendix::ERMnuis}

When calibrated DML is used together with the bootstrap-assisted inference
procedure of Algorithm~\ref{alg::bootstrap}, it is not necessary to estimate
the projected nuisance-error functions
\[
\begin{aligned}
s_0(a,w)
&:= E_0\!\left\{\alpha_0(A,W)-\overline{\alpha}_0(A,W)
    \mid \mu_0(A,W)=\mu_0(a,w)\right\},\\
r_0(a,w)
&:= E_0\!\left\{\mu_0(A,W)-\overline{\mu}_0(A,W)
    \mid \alpha_0(A,W)=\alpha_0(a,w)\right\},
\end{aligned}
\]
which appear in the limiting distribution of
Theorem~\ref{theorem::DRinference}. Nevertheless, these functions can be
estimated by univariate empirical risk minimization if desired. For example,
one may wish to estimate them when implementing the TMLE approach to doubly
robust inference discussed below.

Although \(s_0\) and \(r_0\) are functions on \(\mathcal A\times\mathcal W\),
they are identified by one-dimensional regression functions. Specifically, let
\[
\begin{aligned}
f_{0,\mu_0}(t)
&:= E_0\!\left\{\alpha_0(A,W)-\overline{\alpha}_0(A,W)
    \mid \mu_0(A,W)=t\right\},\\
g_{0,\alpha_0}(t)
&:= E_0\!\left\{\mu_0(A,W)-\overline{\mu}_0(A,W)
    \mid \alpha_0(A,W)=t\right\},
\end{aligned}
\]
where the conditional expectations denote square-integrable versions on the
supports of \(\mu_0(A,W)\) and \(\alpha_0(A,W)\), respectively. Then
\[
    s_0 = f_{0,\mu_0}\circ \mu_0,
    \qquad
    r_0 = g_{0,\alpha_0}\circ \alpha_0 .
\]
Thus, estimating \(s_0\) and \(r_0\) reduces to estimating the univariate
nuisance functions \(f_{0,\mu_0}\) and \(g_{0,\alpha_0}\).

 To estimate \(s_0\), first observe that
\(s_0=f_{0,\overline{\mu}_0}\circ\overline{\mu}_0\), where
\(f_{0,\overline{\mu}_0}\circ\overline{\mu}_0\) is the Riesz representer of the
linear functional
\[
    h \mapsto
    \langle \alpha_0-\overline{\alpha}_0,h\rangle_{P_0}
    =
    \psi_0(h)-\langle \overline{\alpha}_0,h\rangle_{P_0}
\]
with respect to the linear space
\[
    \mathcal H_{\overline{\mu}_0}
    :=
    \left\{f\circ\overline{\mu}_0:
    f:\mathbb R\to\mathbb R,\ 
    \|f\circ\overline{\mu}_0\|<\infty\right\}.
\]
Thus, by the Riesz representation stated above
Theorem~\ref{theorem::EIF}, \(f_{0,\overline{\mu}_0}\) can be estimated by
univariate empirical risk minimization after replacing
\(\overline{\mu}_0\) and \(\overline{\alpha}_0\) with their estimators. For
example, over a univariate function class \(\mathcal F\), one may minimize
\[
    f \mapsto
    \frac{1}{n}\sum_{i=1}^n
    \left[
    \{f\circ\mu_{n,j(i)}^*\}^2(A_i,W_i)
    -
    2\left\{
    \psi_{n,j(i)}(f\circ\mu_{n,j(i)}^*)
    -
    (\alpha_{n,j(i)}^*\, f\circ\mu_{n,j(i)}^*)(A_i,W_i)
    \right\}
    \right].
\]
Similarly, \(r_0=g_{0,\overline{\alpha}_0}\circ\overline{\alpha}_0\), where
\[
    g_{0,\overline{\alpha}_0}(t)
    =
    E_0\{Y-\overline{\mu}_0(A,W)
    \mid \overline{\alpha}_0(A,W)=t\}.
\]
Therefore, \(g_{0,\overline{\alpha}_0}\) can be estimated by regressing the
residuals \(\{Y_i-\mu_{n,j(i)}^*(A_i,W_i)\}_{i\in[n]}\) on
\(\{\alpha_{n,j(i)}^*(A_i,W_i)\}_{i\in[n]}\), using, for example, locally
adaptive regression splines \citep{mammen1997locally} or random forests
\citep{breiman2001random}.

 For implementation, one may exploit that the isotonic calibration maps
\(\mu_{n,j}^*\) and \(\alpha_{n,j}^*\) are piecewise constant. Their level sets
therefore define finite bins, so the correction functions can be estimated
without choosing an additional smoothing basis. Write
\[
    m_i=\mu_{n,j(i)}^*(A_i,W_i),\qquad
    a_i=\alpha_{n,j(i)}^*(A_i,W_i),\qquad
    e_i=Y_i-\mu_{n,j(i)}^*(A_i,W_i).
\]
For \(t\in\{m_i:i\in[n]\}\), let \(I_{\mu,t}:=\{i:m_i=t\}\) and
\(n_{\mu,t}:=|I_{\mu,t}|\). Empirical Riesz minimization over functions that are
constant on the bins induced by \(\mu_{n,\diamond}^*\) gives
\[
    \widehat f_\mu(t)
    =
    \frac{
    \sum_{i=1}^n
    \psi_{n,j(i)}
    \left\{\mathbf 1\big(\mu_{n,j(i)}^*=t\big)\right\}
    -
    \sum_{i\in I_{\mu,t}} a_i
    }{n_{\mu,t}},
    \qquad
    \widehat s(A_i,W_i)=\widehat f_\mu(m_i).
\]
Similarly, for \(u\in\{a_i:i\in[n]\}\), let
\(I_{\alpha,u}:=\{i:a_i=u\}\) and \(n_{\alpha,u}:=|I_{\alpha,u}|\). Least
squares over the bins induced by \(\alpha_{n,\diamond}^*\) gives the residual
mean estimator
\[
    \widehat g_\alpha(u)
    =
    \frac1{n_{\alpha,u}}\sum_{i\in I_{\alpha,u}} e_i,
    \qquad
    \widehat r(A_i,W_i)=\widehat g_\alpha(a_i).
\]
Thus, \(\widehat s\) is a level-set empirical Riesz estimator, whereas
\(\widehat r\) is the corresponding level-set residual mean estimator.

\subsection{Doubly robust inference via TMLE}
 \label{sec::drtmle}

For the special case of doubly robust inference for the average treatment
effect, \cite{van2014targeted} and \cite{benkeser2017doubly} propose a
targeted minimum loss estimation (TMLE) approach that refines given nuisance
estimators so that they asymptotically satisfy
\eqref{eqn::orthogonaloutcome} and \eqref{eqn::orthogonalRiesz}. The
TMLE-based approach iteratively updates the initial nuisance estimators
$\mu_n$ and $\alpha_n$ using a reweighted generalized linear regression
algorithm with offset to produce \textit{targeted} estimators $\mu_n^*$ and
$\alpha_n^*$ satisfying the empirical score equations
\begin{align}
   &\frac{1}{n}\sum_{i=1}^{n} s_n^*(A_i,W_i)\{Y_i-\mu_n^*(A_i,W_i)\}\,=\,   o(n^{-\frac{1}{2}})\,; \label{eqn::tmle1}\\ 
    &\frac{1}{n}\sum_{i=1}^{n} r_n^*(A_i,W_i) \alpha_n^*(A_i,W_i) - \psi_n(r_n^*) \,=\,o(n^{-\frac{1}{2}})\, . \label{eqn::tmle2}
\end{align} 
Here, $s_n^*$ and $r_n^*$ estimate the data-dependent nuisance functions
$s_{n,0}^* := \Pi_{\mu_n^*}(\alpha_0 - \alpha_n^*)$ and
$r_{n,0}^* := \Pi_{\alpha_n^*}(\mu_0 - \mu_n^*)$. They can be obtained, for
example, by univariate empirical risk minimization with features
$\mu_n^*(A_1,W_1),\ldots,\mu_n^*(A_n,W_n)$ and
$\alpha_n^*(A_1,W_1),\ldots,\alpha_n^*(A_n,W_n)$, respectively; we provide
details in Appendix~\ref{appendix::ERMnuis}. Under suitable regularity
conditions, \eqref{eqn::tmle1} and \eqref{eqn::tmle2} imply that
\eqref{eqn::orthogonaloutcome} and \eqref{eqn::orthogonalRiesz} hold
asymptotically, so the one-step debiased estimator based on $\mu_n^*$ and
$\alpha_n^*$ is indeed DRAL.

As discussed in the Introduction, the TMLE-based procedure for doubly robust
inference can be computationally intensive because it requires estimating two
univariate nuisance functions at each iteration, including $r_{n,0}^*$ and
$s_{n,0}^*$ in the final step. In addition, the final estimation of
$r_{n,0}^*$ and $s_{n,0}^*$ may introduce unnecessary finite-sample bias in
the one-step debiased estimator $\tau_n$, since the empirical moment
equations \eqref{eqn::orthogonaloutcome} and \eqref{eqn::orthogonalRiesz} are
then solved only approximately. We also note that neither
\cite{van2014targeted} nor \cite{benkeser2017doubly} formally show that their
iterative refinement procedures preserve the convergence properties of the
original nuisance estimators while enforcing the empirical orthogonality
equations. Such a result would be needed to ensure that, under consistent
estimation of both nuisance functions, the refinement step does not degrade
the resulting estimator or inference. In the following section, we propose a
novel procedure for constructing one-step debiased estimators that avoids
these issues and, unlike prior work, simultaneously applies to a whole class
of target parameters.

 One might ask whether the bias correction achieved by calibration in
Theorem~\ref{theorem::DRbiasexpansion}, or by a TMLE that solves
\eqref{eqn::tmle1} and \eqref{eqn::tmle2}, could instead be obtained through a
one-step adjustment that subtracts an estimate of
\(\langle \Pi_{\mu_n}(\alpha_0-\alpha_n), \mu_n-\mu_0\rangle\),
rather than by enforcing \eqref{eqn::orthogonaloutcome}, together with the
analogous correction for the Riesz-representer bias term. However, as argued by
\citet{benkeser2017doubly}, such an approach requires knowing which nuisance
component contributes the dominant error. Nuisance targeting and nuisance calibration avoid this difficulty by debiasing
the nuisance estimators themselves, rather than by applying a direct correction
to the parameter estimate. This limitation is not intrinsic to all one-step-type estimators: it can be
avoided by targeting different bias terms and adding U-statistic corrections
motivated by higher-order influence functions; see
\citet{bonvini2024doubly}.

\section{Estimator irregularity under inconsistent nuisance estimation}
\label{sec::irregular}

As emphasized in the main text, Theorem~\ref{theorem::DRinference} shows that
the isocalibrated DML estimator is DRAL and is regular and efficient when both
nuisance functions are consistently estimated, even if one converges at an
insufficient rate. Recall that an asymptotically linear estimator
\(\widehat{\psi}_n\) of \(\psi_0\), with influence function \(f_0\), is
\emph{regular} at \(P_0\) for the functional
\(\Psi:\mathcal{M}\to\mathbb{R}\) if, for every one-dimensional regular
parametric submodel
\(\{P_{0,t}:t\in(-\epsilon,\epsilon)\}\subset\mathcal{M}\) through \(P_0\) at
\(t=0\),
\[
n^{1/2}\{\widehat{\psi}_n-\Psi(P_{0,t_n})\}
\stackrel{d}{\longrightarrow} N(0,P_0 f_0^2)
\]
under sampling from \(P_{0,t_n}\), where \(t_n:=tn^{-1/2}\) for any fixed
\(t\in\mathbb{R}\). Regularity means that the centered limiting distribution of
the estimator remains stable under local perturbations of the data-generating
distribution.

When one nuisance function is inconsistently estimated, the calibrated
estimator is typically irregular under a nonparametric model. In this case, it
may remain asymptotically linear at \(P_0\), but with an influence function that
differs from the nonparametric efficient influence function for the original
target parameter \citep{van2000asymptotic,dukes2024doubly}. This distinction is
important because irregularity means that the limiting distribution need not be
stable under \(n^{-1/2}\)-local perturbations of \(P_0\). Consequently, the
asymptotic approximation at \(P_0\) may be unreliable at nearby distributions,
and very large samples may be needed before confidence intervals attain their
nominal coverage \citep{LeebModelSelect2005}. Regularity, by contrast, provides
the stability needed for locally uniform asymptotic inference over local
neighborhoods of \(P_0\).

This irregularity under nuisance misspecification is not unique to calibrated
DML. For standard one-step debiased estimators, inconsistency of one nuisance
typically leaves a non-negligible product-bias term, so the estimator may fail
to be root-\(n\) consistent and hence may not be asymptotically linear
\citep{dukes2024doubly}. Calibrated DML, by contrast, can remain
asymptotically linear and provide \emph{pointwise}, rather than locally uniform,
root-\(n\) inference in such settings. Nevertheless, regularity remains
desirable, and in practice one should still aim to estimate both nuisance
functions as accurately as possible.

We now characterize this irregularity more precisely. The following results
describe the local asymptotic behavior of the isocalibrated DML estimator under
one-sided nuisance misspecification; their proofs are given in
Appendix~\ref{appendix::regularproof}. We first define a projection-based
oracle parameter \(\Psi_0\), for which the estimator remains regular. We then
identify submodels on which \(\Psi_0\) coincides with the original target
parameter \(P\mapsto\Psi(P):=\psi_P(\mu_P)\), so that regularity for
\(\Psi_0\) transfers to regularity for \(\Psi\). Finally, we quantify the local
asymptotic bias that arises for inference on \(\Psi\) under general local
perturbations.

At a high level, when one nuisance function is inconsistently estimated, the
oracle parameter replaces the corresponding population nuisance by its best
one-dimensional projection onto the model generated by the consistently
estimated nuisance. To define this parameter, we first introduce some notation.
For each distribution \(P\), define the affine spaces
\[
\overline{\Theta}_{\alpha_0,P}
:=
\{\overline{\mu}_0 + f \circ \alpha_0: f:\mathbb R\to\mathbb R\}
\cap L^2(P),
\qquad
\overline{\Theta}_{\mu_0,P}
:=
\{\overline{\alpha}_0 + g \circ \mu_0: g:\mathbb R\to\mathbb R\}
\cap L^2(P),
\]
where \(f\) and \(g\) range over all one-dimensional real-valued
transformations. Let the corresponding \(L^2(P)\)-orthogonal projections be
\[
\overline{\Pi}_{\alpha_0,P}\mu_P
:=
\argmin_{\mu\in\overline{\Theta}_{\alpha_0,P}}
\|\mu_P-\mu\|_P,
\qquad
\overline{\Pi}_{\mu_0,P}\alpha_P
:=
\argmin_{\alpha\in\overline{\Theta}_{\mu_0,P}}
\|\alpha_P-\alpha\|_P .
\]
We then define the projection-based oracle parameter \(\Psi_0\), with the
definition depending on which nuisance function is consistently estimated, by
\[
    \Psi_{0}(P)
    :=
    \begin{cases}
        \Psi(P),
        & \text{if } (\overline{\alpha}_0,\overline{\mu}_0)
        =(\alpha_0,\mu_0), \\
        \overline{\psi}_{\mu_0,P}(\mu_P),
        & \text{if } \overline{\mu}_0 = \mu_0
        \text{ and } \overline{\alpha}_0 \neq \alpha_0, \\
        \psi_P(\overline{\Pi}_{\alpha_0,P}\mu_P),
        & \text{if } \overline{\alpha}_0 = \alpha_0
        \text{ and } \overline{\mu}_0 \neq \mu_0 .
    \end{cases}
\]
Here,
\[
\overline{\psi}_{\mu_0,P}
:
h\mapsto
\langle \overline{\Pi}_{\mu_0,P}\alpha_P,h\rangle_P
\]
can be viewed as an oracle approximation of the linear functional
\(\psi_P:h\mapsto \langle\alpha_P,h\rangle_P\).

The following theorem shows that,
in the one-sided misspecification regime, the isocalibrated DML estimator is
regular for the projection-based oracle parameter \(\Psi_0\).

\begin{theorem}[Behavior under local perturbations]
\label{theorem::regularity}
Assume Conditions~\ref{cond::linear}--\ref{cond::pathwise}
and~\ref{cond::DRlimits}--\ref{cond::variation}. If either
\(\overline{\mu}_0=\mu_0\) or \(\overline{\alpha}_0=\alpha_0\), then \(\chi_0\)
is the efficient influence function of \(\Psi_0\) at \(P_0\). Consequently,
\(\tau_n^*\) is regular, asymptotically linear, and efficient at \(P_0\) for the
projection-based oracle parameter \(\Psi_0\), with influence function
\(\chi_0\). In particular, under sampling from any \(n^{-1/2}\)-local
perturbation \(P_{0,t_n}\) along a regular parametric submodel through \(P_0\),
\[
n^{1/2}\{\tau_n^*-\Psi_0(P_{0,t_n})\}
\]
converges in distribution to a mean-zero normal random variable with variance
\(\sigma_0^2\), as defined in Theorem~\ref{theorem::DRinference}.
\end{theorem}

Theorem~\ref{theorem::regularity} concerns the oracle parameter \(\Psi_0\). The
next lemma identifies submodels on which this oracle parameter coincides with
the original target parameter \(\Psi\). On such submodels, regularity for
\(\Psi_0\) transfers directly to regularity for \(\Psi\).

\begin{lemma}[Agreement of the oracle and target parameters]
\label{lem::oracle-target-agreement}
If \(\overline{\mu}_0=\mu_0\), then \(\Psi_0(P)=\Psi(P)\) for every
\(P\in\mathcal{M}_{\mu_0}\), where
\[
\mathcal{M}_{\mu_0}
:=
\Bigl\{
P\in\mathcal{M}:\mu_P = g\circ \mu_0 \quad P\text{-a.s.}
\text{ for some } g:\mathbb{R}\to\mathbb{R}
\Bigr\}.
\]
Similarly, if \(\overline{\alpha}_0=\alpha_0\), then
\(\Psi_0(P)=\Psi(P)\) for every \(P\in\mathcal{M}_{\alpha_0}\), where
\[
\mathcal{M}_{\alpha_0}
:=
\Bigl\{
P\in\mathcal{M}:\alpha_P = f\circ \alpha_0 \quad P\text{-a.s.}
\text{ for some } f:\mathbb{R}\to\mathbb{R}
\Bigr\}.
\]
In particular, in either one-sided regime, \(\Psi_0(P_0)=\Psi(P_0)\).
\end{lemma}

Combining Lemma~\ref{lem::oracle-target-agreement} with
Theorem~\ref{theorem::regularity}, we conclude that \(\tau_n^*\) is regular
for the original target parameter \(\Psi\) over the submodel
\(\mathcal{M}_{\mu_0}\) when \(\overline{\mu}_0=\mu_0\), and over the submodel
\(\mathcal{M}_{\alpha_0}\) when \(\overline{\alpha}_0=\alpha_0\). In
particular, \(\tau_n^*\) is regular for \(\Psi\) over all local perturbations
that leave the consistently estimated nuisance function unchanged, in
agreement with Theorem~2 of \cite{dukes2024doubly}.

The preceding paragraph establishes regularity for \(\Psi\) only along
submodels where the oracle parameter \(\Psi_0\) and the original target
parameter \(\Psi\) agree. The following corollary characterizes what happens
under general local perturbations: inference on \(\Psi\) incurs a local
asymptotic bias, whose size is controlled by the misspecification of the
nuisance limits. The corollary is an immediate consequence of Theorem~7 of
\cite{van2023adaptive}.

\begin{corollary}[Asymptotic bias under local perturbations]
\label{cor::regularity}
Assume Conditions~\ref{cond::linear}--\ref{cond::pathwise}
and~\ref{cond::DRlimits}--\ref{cond::variation}, and suppose either
\(\overline{\mu}_0=\mu_0\) or \(\overline{\alpha}_0=\alpha_0\). Suppose that \(h\mapsto\phi_{0,h}\) in Condition~\ref{cond::pathwise} is Lipschitz in \(L^2(P_0)\): for some
\(L_\phi<\infty\),
\[
\|\phi_{0,h_1}-\phi_{0,h_2}\|_{P_0}
\le
L_\phi\|h_1-h_2\|_{P_0},
\qquad h_1,h_2\in\mathcal H .
\]
Under sampling from any \(n^{-1/2}\)-local perturbation \(P_{0,t_n}\) along a
regular parametric submodel through \(P_0\), with unit score
\(s \in L^2_0(P_0)\), \(\|s\|_{L^2(P_0)}=1\), and perturbation
\(t_n := t n^{-1/2}\) for fixed \(t \in \mathbb{R}\),
\[
n^{1/2}\{\tau_n^* - \Psi(P_{0,t_n})\}
\ \xrightarrow{d}\
N\!\left(b_0(t),\,\sigma_0^2\right),
\]
where \(b_0(t) := t \langle s, \chi_0 - D_0 \rangle\). Moreover,
\[
|b_0(t)| \lesssim |t| \,
\left\{
\|\overline{\mu}_0 - \mu_0\|_{P_0}
\vee
\|\overline{\alpha}_0 - \alpha_0\|_{P_0}
\right\},
\]
where the hidden constant depends only on \(L_\phi\) and the uniform bound in
Condition~\ref{cond::bound}.
\end{corollary}

Consequently, the local asymptotic bias under \(P_{0,t_n}\) equals \(b_0(t)\) and, uniformly over unit-score directions, is controlled by the misspecification of the nuisance limits: \[ |b_0(t)| \ \lesssim\ |t| \left\{ \|\overline{\mu}_0-\mu_0\|_{P_0} \vee \|\overline{\alpha}_0-\alpha_0\|_{P_0} \right\}. \] This bound vanishes when both nuisance limits are correctly specified. In the one-sided misspecification regimes considered here, it is small whenever the misspecified nuisance limit is close to the truth. Thus, even under local perturbations, mild nuisance misspecification may lead to only a small local asymptotic bias for inference on \(\Psi\). This also explains why, despite the pointwise robustness provided by calibration, one should still aim to estimate both nuisance functions as accurately as possible: reducing nuisance misspecification directly reduces the worst-case bound on the local asymptotic bias for the original target parameter.

 \section{Extensions of calibrated DML}
\subsection{Extension to mixed bias parameters}
 \label{sec::mixedbias}
In this section, we provide a brief overview of how our doubly robust inference framework can be extended to a general class of parameters with the mixed bias property, as introduced in \cite{rotnitzky2021characterization}. While our primary focus in this paper remains on the class of linear functionals discussed in Section \ref{section::setup}, it is important to emphasize that our technique and results can be readily generalized to parameters exhibiting the mixed bias property.

We focus on debiasing mixed bias remainders of the following form using isotonic calibration:
\begin{align*}
    \int w(z) \{\alpha_n(w,a) - \alpha_0(w,a)\}\{\gamma_n(w,a) - \gamma_0(w,a)\}  P_0(dz)
\end{align*}
where $z \mapsto w(z)$ is a known weighting function, and $\alpha_0 := \argmin_{\alpha} E_0[w(Z)\{\alpha(A,W)\}^2 - 2m_1(A,W, \alpha)]$ and $\gamma_0 := \argmin_{\gamma} E_0[w(Z)\{\gamma(A,W)\}^2 - 2m_2(A,W, \gamma)]$ are weighted Riesz representers for the linear functionals $\alpha \mapsto E_0[m_1(A,W, \alpha)]$ and $\gamma \mapsto E_0[m_2(A,W, \gamma)]$. Mixed bias parameters, as defined in \cite{rotnitzky2021characterization}, have remainders of this form. More generally, parameters may involve second-order remainders consisting of multiple such mixed bias terms, each of which can be linearized separately using the results of this section.

Given estimators $\alpha_n$ and $\gamma_n$ of $\alpha_0$ and $\gamma_0$, we propose using the isocalibrated nuisance estimators $\alpha_n^* = g_{n,1} \circ \alpha_n$ and $\gamma_n^* = g_{n,2} \circ \gamma_n$, where 
\begin{equation}
    \begin{aligned}
        g_{n,1} &\in \argmin_{g \in \mathcal{F}_{\text{iso}}} \sum_{i=1}^n w(Z_i) \big{\{}g(\alpha_n(A_i, W_i))\big{\}}^2 - 2m_1(Z_i, g \circ \alpha_n), \\
        g_{n,2} &\in \argmin_{g \in \mathcal{F}_{\text{iso}}} \sum_{i=1}^n w(Z_i) \big{\{}g(\gamma_n(A_i, W_i))\big{\}}^2 - 2m_2(Z_i, g \circ \gamma_n).
    \end{aligned}
\end{equation}
Notably, the first-order conditions of the minimizing solutions guarantee that
\begin{align*}
    \frac{1}{n} \sum_{i=1}^n \left[ w(Z_i) (g \circ \alpha_n^*)(W_i,A_i) \alpha_n^*(W_i, A_i) - m_1(W_i, A_i, g \circ \alpha_n^*)\right] &= 0, \\
    \frac{1}{n} \sum_{i=1}^n \left[ w(Z_i) (g \circ \gamma_n^*)(W_i,A_i) \gamma_n^*(W_i, A_i) - m_2(W_i, A_i, g \circ \gamma_n^*)\right] &= 0
\end{align*}
for all transformations $g: \mathbb{R} \rightarrow \mathbb{R}$. As in Algorithms \ref{alg:DR} and \ref{alg::bootstrap}, the initial nuisance estimators can be cross-fitted, and confidence intervals can be constructed using the bootstrap. 

Our proof techniques can be readily extended to establish that isotonic
calibration guarantees doubly robust asymptotic linearity of mixed bias
remainders. The key point is that the relevant argument is exactly the
Riesz-based decomposition in Theorem~\ref{theorem::DRbiasexpansion}, and in
particular the Riesz-based part of its proof. That proof treats the drift term
as a cross-product between an error in a Riesz representer and an error in a
second nuisance, then uses only three ingredients: the Riesz representation
identity, the empirical Riesz calibration equation \eqref{eqn::orthogonalRiesz},
and orthogonal projection identities. It does not use that the second nuisance
is an outcome regression or that it was constructed by minimizing a
least-squares loss. For a mixed bias remainder, one applies the same argument
with \(\mu_0\) replaced by the second Riesz representer \(\gamma_0\), with the
ordinary inner product replaced by the weighted inner product
\(P_0\{w(Z)\cdot\,\cdot\}\), and with \(\psi_0\) replaced by the corresponding
linear functional \(h\mapsto E_0\{m_1(Z,h)\}\) or
\(h\mapsto E_0\{m_2(Z,h)\}\). Thus the only substantive change is that the
least-squares calibration loss for the regression nuisance is replaced by the
appropriate weighted Riesz loss for the second representer. The resulting
expansion again expresses the mixed bias remainder as an empirical-process term
plus a higher-order projected cross-product remainder, controlled by analogues
of the partial-orthogonality and empirical-process conditions used in
Theorem~\ref{theorem::DRinference}.

To build intuition, consider the mixed bias remainder after calibration. Define
\[
    \langle f,h\rangle_w := P_0\{w(Z)f(A,W)h(A,W)\},\qquad
    \Delta_{\alpha,n}^*:=\alpha_n^*-\alpha_0,\qquad
    \Delta_{\gamma,n}^*:=\gamma_n^*-\gamma_0 .
\]
We use the decomposition induced by the calibration equation for
\(\alpha_n^*\); the corresponding decomposition based on \(\gamma_n^*\) is
symmetric. Let \(\Pi_{\alpha_n^*}^w\) denote the
\(\langle\cdot,\cdot\rangle_w\)-orthogonal projection onto functions of
\(\alpha_n^*(A,W)\), and let \(\Pi_{(\alpha_n^*,\alpha_0)}^w\) denote the
projection onto functions of \((\alpha_n^*(A,W),\alpha_0(A,W))\). Since
\(\Delta_{\alpha,n}^*\) is measurable with respect to
\((\alpha_n^*,\alpha_0)\), the weighted law of iterated expectations gives
\[
    \langle \Delta_{\alpha,n}^*,\Delta_{\gamma,n}^* \rangle_w
    =
    \left\langle
    \Delta_{\alpha,n}^*,
    \Pi_{(\alpha_n^*,\alpha_0)}^w\Delta_{\gamma,n}^*
    \right\rangle_w .
\]
Now define
\[
    r_{n,0}:=\Pi_{\alpha_n^*}^w(\gamma_0-\gamma_n^*)
            =-\Pi_{\alpha_n^*}^w\Delta_{\gamma,n}^* .
\]
Decomposing the projection around \(\Pi_{\alpha_n^*}^w\), as in the
Riesz-based proof of Theorem~\ref{theorem::DRbiasexpansion}, yields
\[
\begin{aligned}
    \langle \Delta_{\alpha,n}^*,\Delta_{\gamma,n}^* \rangle_w
    &=
    -\left\langle \Delta_{\alpha,n}^*, r_{n,0}\right\rangle_w  \\
    &\quad
    -
    \left\langle
    (\Pi_{(\alpha_n^*,\alpha_0)}^w-\Pi_{\alpha_n^*}^w)
    \Delta_{\gamma,n}^*,
    \alpha_0-\Pi_{\alpha_n^*}^w\alpha_0
    \right\rangle_w .
\end{aligned}
\]
Here the second term uses the fact that
\((\Pi_{(\alpha_n^*,\alpha_0)}^w-\Pi_{\alpha_n^*}^w)\Delta_{\gamma,n}^*\)
is orthogonal to every function of \(\alpha_n^*\). The first term is the leading bias component. By the weighted Riesz
representation property for \(\alpha_0\),
\[
    \left\langle \Delta_{\alpha,n}^*, r_{n,0}\right\rangle_w
    =
    P_0\{w(Z)\alpha_n^*(A,W)r_{n,0}(A,W)\}
    -
    P_0\{m_1(Z,r_{n,0})\}.
\]
Since \(r_{n,0}\) is a function of \(\alpha_n^*\), the calibration score
equation for \(\alpha_n^*\) turns this leading term into the empirical-process
fluctuation
\[
    -\left\langle \Delta_{\alpha,n}^*, r_{n,0}\right\rangle_w
    =
    (P_n-P_0)\{w(Z)\alpha_n^*(A,W)r_{n,0}(A,W)-m_1(Z,r_{n,0})\}.
\]
Consequently,
\[
\begin{aligned}
    \langle \Delta_{\alpha,n}^*,\Delta_{\gamma,n}^* \rangle_w
    &=
    (P_n-P_0)\{w(Z)\alpha_n^*(A,W)r_{n,0}(A,W)-m_1(Z,r_{n,0})\} \\
    &\quad
    -
    \left\langle
    (\Pi_{(\alpha_n^*,\alpha_0)}^w-\Pi_{\alpha_n^*}^w)
    \Delta_{\gamma,n}^*,
    \alpha_0-\Pi_{\alpha_n^*}^w\alpha_0
    \right\rangle_w .
\end{aligned}
\]
The first term is controlled as an empirical-process fluctuation. The second is
a higher-order cross-product remainder, directly analogous to the
corresponding remainder in the Riesz-based decomposition of
Theorem~\ref{theorem::DRbiasexpansion}.

\begin{example}[Calibrated DML for expected conditional covariance]
    Let $\pi_n$ and $m_n$ be estimators of $w \mapsto E_0[A \mid W = w]$ and $w \mapsto E_0[Y \mid W = w]$, respectively. The expected conditional covariance parameter 
\[
E_0\left[\{A - \pi_0(W)\}\{Y - m_0(W)\}\right] = E_0\left[\{A - \pi_0(W)\}^2\{\mu_0(1,W) - \mu_0(0,W)\}\right]
\]
arises in the context of estimating the ATE under treatment effect homogeneity. Doubly robust inference for this parameter was studied in \cite{dukes2024doubly} and \cite{bonvini2024doubly}. This parameter is a linear functional of the outcome regression, but the functional depends on the nuisance function $\pi_0$. In this case, a modification of our procedure yields improved properties. A doubly robust estimator of the expected covariance parameter is given by
\[
\frac{1}{n}\sum_{i=1}^n \{A_i - \pi_n(W_i)\}\{Y_i - m_n(W_i)\}.
\]
The second-order remainder of this estimator involves the cross-product term \citep{robinson1988root}:
\[
\langle \pi_n(W) - \pi_0(W), m_n - m_0(W)\rangle.
\]
This mixed bias term is exactly of the form addressed in Section~\ref{section::prelimDR}, involving an inner product between a regression function and a Riesz representer. The regression function $m_0$ corresponds to the least-squares loss $(z, m) \mapsto \{y - m(w)\}^2$, while the function $\pi_0$ is a Riesz representer of the linear functional $\alpha \mapsto E_0[A \alpha(W)]$, with Riesz loss corresponding to the usual least-squares loss $(z, \alpha) \mapsto \{a - \alpha(w)\}^2$.

Thus, a calibrated DML estimator of the expected covariance parameter is given by
\[
\frac{1}{n}\sum_{i=1}^n \{A_i - \pi_n^*(W_i)\}\{Y_i - m_n^*(W_i)\},
\]
where $\pi_n^* = g_n \circ \pi_n$ and $m_n^* = f_n \circ m_n$, with
\begin{align*}
    g_n &= \argmin_{g \in \mathcal{F}_{\text{iso}}} \sum_{i=1}^n \left\{ A_i - g(\pi_n(W_i))\right\}^2, \\
    f_n &= \argmin_{f \in \mathcal{F}_{\text{iso}}} \sum_{i=1}^n \left\{ Y_i - f(m_n(W_i))\right\}^2.
\end{align*}
Doubly robust asymptotic linearity of this estimator follows from conditions analogous to those in Section~\ref{section::theory}.
\end{example}

\subsection{A general template for higher-order debiasing}
\label{sec::generaltemplate}
The debiasing strategy of Section \ref{section::prelimDR} can be substantially
generalized, as the analysis relies only on empirical orthogonality
properties of the nuisance estimators and basic properties of
projections. In this section, we present a general debiasing strategy
that extends beyond this setting.

Our extension uses the following notation. Let $\Delta_{\mu,n}:=\mu_n-\mu_0$
and $\Delta_{\alpha,n}:=\alpha_n-\alpha_0$. Let
$\Theta_{\mu,n,0} \subset \mathcal{H}$ be a closed linear subspace that contains
the \emph{regression error} $\Delta_{\mu,n}$ (this space may depend on $n$), and
let $\Theta_{\mu,n} \subset \Theta_{\mu,n,0}$ be a lower-dimensional closed
\emph{approximating} subspace. We view $\Theta_{\mu,n,0}$ as a working model for
the directions in which $\mu_n$ may deviate from $\mu_0$, and $\Theta_{\mu,n}$
as the tractable subset of those directions on which we can enforce empirical
orthogonality. Similarly, let $\Theta_{\alpha,n,0} \subseteq \mathcal{H}$ be a
closed linear subspace that contains the \emph{Riesz error} $\Delta_{\alpha,n}$,
and let $\Theta_{\alpha,n} \subseteq \Theta_{\alpha,n,0}$ be a lower-dimensional
closed approximating subspace. We write $\Pi_{\mu,n,0}$, $\Pi_{\mu,n}$,
$\Pi_{\alpha,n,0}$, and $\Pi_{\alpha,n}$ for the $\|\cdot\|$-orthogonal
projections onto $\Theta_{\mu,n,0}$, $\Theta_{\mu,n}$, $\Theta_{\alpha,n,0}$,
and $\Theta_{\alpha,n}$, respectively; for example,
$\Pi_{\mu,n} f := \arg\min_{h \in \Theta_{\mu,n}} \|f-h\|$.

Two examples recover familiar constructions. First, the calibrated DML setup in Section \ref{sec::genstrat} is obtained by taking $\Theta_{\mu,n} := \overline{\{f \circ \mu_n : f \colon \mathbb{R} \to \mathbb{R}\}}$ and $\Theta_{\mu,n,0} := \overline{\{g \circ (\mu_n,\mu_0) : g \colon \mathbb{R}^2 \to \mathbb{R}\}}$, with $\Theta_{\alpha,n}$ and $\Theta_{\alpha,n,0}$ defined analogously; in this case, $\Theta_{\mu,n}$ corresponds to calibration directions available through post hoc one-dimensional transformations of $\mu_n$. Second, for a linear sieve given by a nested sequence $\mathcal{H}_1 \subseteq \mathcal{H}_2 \subseteq \cdots \subseteq \mathcal{H}_\infty \subseteq \mathcal{H}$, one may take $\Theta_{\mu,n} := \mathcal{H}_{k(n)}$ and $\Theta_{\mu,n,0} := \mathcal{H}_\infty$, where $k(n) \uparrow \infty$ as the sample size increases; here $\Theta_{\mu,n}$ is the finite-dimensional space used at sample size $n$, while $\Theta_{\mu,n,0}$ is the limiting sieve space containing the regression error. 

Our more general debiasing strategy constructs nuisance estimators $\mu_n$
and $\alpha_n$ that satisfy the empirical orthogonality (score) conditions
\begin{equation}
\frac{1}{n}\sum_{i=1}^n (\Pi_{\mu,n}\Delta_{\alpha,n})(A_i,W_i)
\{Y_i - \mu_n(A_i,W_i)\}
= 0 ,
\label{eqn::orthogonaloutcomesieve}
\end{equation}
\begin{equation}
\psi_n(\Pi_{\alpha,n}\Delta_{\mu,n})
- \frac{1}{n}\sum_{i=1}^n (\Pi_{\alpha,n}\Delta_{\mu,n})(A_i, W_i)\,
\alpha_n(A_i, W_i)
= 0 .
\label{eqn::orthogonalRieszsieve}
\end{equation}
In words, \eqref{eqn::orthogonaloutcomesieve} requires that the residuals of
$\mu_n$ are empirically orthogonal to the projection of the estimation
error $\Delta_{\alpha,n}$ onto the approximating subspace
$\Theta_{\mu,n}$ for $\mu_n - \mu_0$. If $\Theta_{\mu,n}$ happened to contain
$\Delta_{\alpha,n}$ itself, then this condition reduces to enforcing
\[
\frac{1}{n}\sum_{i=1}^n \Delta_{\alpha,n}(A_i,W_i)
\{Y_i - \mu_n(A_i,W_i)\} \approx 0,
\]
which sets the empirical estimate of the cross-product remainder
$\langle \Delta_{\alpha,n}, \Delta_{\mu,n} \rangle$ to zero.
If $\mu_n$ and $\alpha_n$ are sieve-type estimators that respectively
minimize the least-squares and Riesz losses over $\Theta_{\mu,n}$ and
$\Theta_{\alpha,n}$, then both conditions hold automatically by the
corresponding first-order optimality conditions. Otherwise, given initial
estimators $\mu_n^{(0)}$ and $\alpha_n^{(0)}$, these conditions can be
enforced post hoc by defining $\mu_n := \mu_n^{(0)} + h_{n,\mu}$ and
$\alpha_n := \alpha_n^{(0)} + h_{n,\alpha}$, where the adjustment terms
are obtained via the offset sieve problems
\begin{align*}
  h_{n,\mu} &:= \arg\min_{h \in \Theta_{\mu,n}}
  \sum_{i=1}^n \{Y_i - \mu_n^{(0)}(A_i, W_i) - h(A_i, W_i)\}^2,\\
  h_{n,\alpha} &:= \arg\min_{h \in \Theta_{\alpha,n}}
  \sum_{i=1}^n h(A_i, W_i)^2
  - 2\,\{\psi_n(h) - \langle \alpha_n^{(0)}, h \rangle_{P_n}\},
\end{align*}
where $\Theta_{\mu,n}$ and $\Theta_{\alpha,n}$ are taken to be
finite-dimensional and tuned to control overfitting. In high-dimensional
settings, one can add $\ell^1$ or $\ell^2$ regularization, in which case
the orthogonality equations typically hold only approximately, in view of
the KKT conditions for the minimizers. The resulting regularization error
must then be controlled, as in \cite{bradic2019minimax} and
\cite{smucler2019unifying}.  

In the following theorem, we denote
$\widetilde A_{n,0} : z \mapsto \widetilde{s}_{n,0}(a,w)\{y -
\mu_n(a,w)\}$,
$\widetilde B_{n,0} := \psi_0(\widetilde r_{n,0}) +
\phi_{\widetilde r_{n,0}} -
\widetilde r_{n,0}\alpha_n$, and
$\mathrm{Rem}_{\psi_n}(\widetilde r_{n,0}) :=
\psi_n(\widetilde r_{n,0}) - \psi_0(\widetilde r_{n,0}) -
(P_n - P_0)\phi_{\widetilde r_{n,0}}$, where $\widetilde s_{n,0} := \Pi_{\mu,n}(\alpha_0 - \alpha_n)$ and 
$\widetilde r_{n,0} := \Pi_{\alpha,n}(\mu_0 - \mu_n)$.

\begin{theorem}[Expansion of the cross-product remainder under general orthogonality] \label{theorem::DRbiasexpansionsieve}
Assume that the empirical functional $\psi_n$ is linear.
If $\mu_n$ satisfies \eqref{eqn::orthogonaloutcomesieve}, then we have the
regression-based decomposition:
\begin{align*}
\langle \alpha_0 - \alpha_n, \mu_n - \mu_0 \rangle
&= (P_n - P_0)\widetilde A_{n,0}
- \langle (\Pi_{\mu,n,0} - \Pi_{\mu,n})\Delta_{\alpha,n},
\,  (\Pi_{\mu,n,0} - \Pi_{\mu,n})\Delta_{\mu,n} \rangle.
\end{align*}
If $\alpha_n$ satisfies \eqref{eqn::orthogonalRieszsieve}, then we have the
Riesz-based decomposition:
\begin{align*}
\langle \alpha_0 - \alpha_n, \mu_n - \mu_0 \rangle
&= (P_n - P_0)\widetilde B_{n,0}
- \langle (\Pi_{\alpha,n,0} - \Pi_{\alpha,n})\Delta_{\mu,n},
\, (\Pi_{\alpha,n,0} - \Pi_{\alpha,n})\Delta_{\alpha,n}\rangle
+ \mathrm{Rem}_{\psi_n}(\widetilde r_{n,0}).
\end{align*}
\end{theorem}

The decompositions in Theorem~\ref{theorem::DRbiasexpansionsieve} are similar
to those in Theorem~\ref{theorem::DRbiasexpansion}. Each decomposition
consists of an empirical process term, which is asymptotically linear
under suitable conditions, and a higher-order cross-product remainder.
In the regression-based case, these terms are $(P_n - P_0)\widetilde
A_{n,0}$ and
$-\big\langle (\Pi_{\mu,n,0} - \Pi_{\mu,n})\Delta_{\alpha,n},\,
(\Pi_{\mu,n,0} - \Pi_{\mu,n})\Delta_{\mu,n} \big\rangle$,
respectively. By the Cauchy--Schwarz inequality,
\begin{equation}
\label{eqn::upperboundsieve}
\big|
\langle (\Pi_{\mu,n,0} - \Pi_{\mu,n})\Delta_{\alpha,n},\,
(\Pi_{\mu,n,0} - \Pi_{\mu,n})\Delta_{\mu,n} \rangle
\big|
\le
\|(\Pi_{\mu,n,0} - \Pi_{\mu,n})\Delta_{\alpha,n}\|\,
\|\Delta_{\mu,n} - \Pi_{\mu,n}\Delta_{\mu,n}\|.
\end{equation}
Consequently, the higher-order cross-product remainder is small if either the
regression error $\Delta_{\mu,n}$ is well approximated by $\Theta_{\mu,n}$, or
the projected Riesz error $\Pi_{\mu,n,0} \Delta_{\alpha,n}$ is
well approximated by $\Theta_{\mu,n}$. In the special case $\Theta_{\mu,n,0}=\mathcal{H}$, this bound reduces
to the product of approximation errors for the nuisance estimation errors,
namely
$\|\Delta_{\alpha,n} - \Pi_{\mu,n}\Delta_{\alpha,n}\|\,
\|\Delta_{\mu,n} - \Pi_{\mu,n}\Delta_{\mu,n}\|$.

To provide intuition, we interpret the bound in \eqref{eqn::upperboundsieve}
in the special case where $\Theta_{\mu,n} := \mathcal{H}_{k(n)}$ and
$\Theta_{\mu,n,0} := \mathcal{H}_\infty$ for a linear sieve
$\{\mathcal{H}_k\}_{k=1}^\infty$. In this setting, the right-hand side is
the product of two sieve approximation errors. The first term is small
when the projected Riesz error $\Pi_{\mu,n,0}\Delta_{\alpha,n}$ is
sufficiently smooth to be well approximated by the sieve space
$\Theta_{\mu,n}$, while the second term is bounded by
$\|\mu_n - \mu_0\|$ by standard projection properties. Theorem~\ref{theorem::DRbiasexpansionsieve}
can therefore be used to establish doubly robust asymptotic linearity of
sieve-based autoDML estimators that use sieve estimators for both the
Riesz representer and the outcome regression. As a special case, setting
either nuisance estimator to the zero function can be used to derive the classical
result that plug-in sieve estimators are asymptotically linear without
explicit debiasing. 

Rather than working with the projected errors, a sufficient---though
stronger than needed---condition for the higher-order cross-product
remainders to be negligible is that the spaces $\Theta_{\mu,n}$ and
$\Theta_{\alpha,n}$ provide good approximations to
$\Delta_{\alpha,n}$ and $\Delta_{\mu,n}$, respectively. From a practical
standpoint, however, directly choosing $\Theta_{\mu,n}$ to approximate
$\Delta_{\alpha,n}$ is often unrealistic, since $\Delta_{\alpha,n}$ is
typically high-dimensional and unknown. Calibrated DML avoids this curse of dimensionality by using $\mu_n$ and
$\alpha_n$ as low-dimensional summaries and defining $\Theta_{\mu,n}$
and $\Theta_{\alpha,n}$ as classes of one-dimensional functions composed
with them.

\subsection{Connection to sparsity-based doubly robust inference}
\label{appendix:relatedwork}

This section expands on the comparison in Section~\ref{sec:relatedwork}
between our calibration-based approach and sparsity-based methods for doubly
robust asymptotic linearity, especially \citet{liu2023root}.

When the nuisance functions are estimated by \(\ell_1\)-regularized empirical
risk minimization over linear classes, the Karush--Kuhn--Tucker (KKT)
conditions imply approximate empirical orthogonality over suitable linear
subspaces; see, for example, \citealp[Eq.~15]{tan2020model},
\citealp[Sec.~2.1]{bradic2019sparsity}, and
\citealp[Sec.~4.1]{smucler2019unifying}. Under additional approximation
conditions, typically requiring at least one nuisance error to be sufficiently
sparse, this orthogonality can yield doubly robust asymptotic linearity for
\(\ell_1\)-regularized autoDML estimators. \citet{liu2023root} show that this
type of \(\ell_1\)-regularized high-dimensional correction can be applied
post hoc to initial \(\ell_1\)-regularized learners. For the outcome
regression, their post-hoc correction can be viewed as enforcing a
high-dimensional orthogonality objective, often called
\emph{multiaccuracy}
\citep{kim2019multiaccuracy,deng2023happymap,roth2022uncertain}, by targeting
the moment conditions $E_{0}\left[(Y-\mu_n(A,W))\,f(A,W)\right]=0$ for all \(f\) in a rich class; in their setting, this is the linear span of the
design or basis functions used in the \(\ell_1\)-regularized learners.

In contrast, Section~\ref{sec::genstrat} shows that, for our purposes, it
suffices to target a single moment condition, namely $E_{0}\left[(Y-\mu_n(A,W))\,s_{n,0}(A,W)\right]=0,$ 
as in \eqref{eqn::orthogonaloutcome}. Calibrated DML
(Section~\ref{section::calibration}) achieves this by targeting the
one-dimensional calibration constraint $\mu_n(A,W)=E_0\left[Y \mid \mu_n(A,W)\right],$
or, equivalently, the one-dimensional family of orthogonality conditions
 $E_{0}\left[(Y-\mu_n(A,W))\,\theta\{\mu_n(A,W)\}\right]=0
$
for all \(\theta:\mathbb{R}\to\mathbb{R}\).

Taken together, these comparisons suggest that the high-dimensional generalized
linear model construction of \citet{liu2023root} can be viewed as a
sparsity-based analogue of our calibration-based post-hoc correction. Unlike our
approach, their method is built around first-stage \(\ell_1\)-regularized
estimators, a high-dimensional second-stage adjustment, and sparsity-based
conditions. By contrast, calibrated DML is learner-agnostic, uses a
one-dimensional adjustment, and is not tied to \(\ell_1\)-regularization or
sparsity assumptions. Rather than assuming approximate sparsity, we require the
asymptotic partial orthogonality condition in \ref{cond:ortho}.

An interesting open question is whether the first-stage estimators in
\citet{liu2023root} can be replaced by user-supplied machine learners, rather
than being restricted to \(\ell_1\)-regularized estimators in a
high-dimensional linear model. In Appendix~\ref{sec::generaltemplate}, we
extended our calibration framework and Theorem~\ref{theorem::DRbiasexpansion} to
more general orthogonality conditions and projection operators, including those
arising from regularized empirical risk minimization and sieve estimation. This
extension suggests one possible route toward such a generalization, with
assumptions formulated as sparsity conditions on first-stage nuisance
\emph{errors} rather than on the nuisance \emph{functions} themselves. More
broadly, it suggests that calibration and multiaccuracy, viewed as post-hoc
predictive-modeling techniques, can both be used to construct DRAL estimators.
These connections point toward a unified view of sparsity-based doubly robust
methods, including \citet{liu2023root}, our calibration-based approach, and
targeted-learning methods such as \citet{benkeser2017doubly}.

\section{Background on calibration}
\label{appendix::isocallit}

In this section, we provide background on calibration using binning and isotonic regression. For additional details, we refer to \citet{van2025generalized} for a general treatment of histogram binning and isotonic calibration; see also \citet{gupta2020distribution, gupta2021distribution} for histogram binning, \citet{van2023causal} and \citet{van_der_Laan2024stabilized} for applications of isotonic calibration in causal inference, and \citet{whitehouse2024orthogonal} for calibration with respect to general loss functions.

\subsection{Empirical calibration via histogram binning}

Let \((z,\nu) \mapsto \ell(z,\nu)\) be a loss function for some nuisance function \(\nu_0 \in \mathcal{H}\). An estimator \(\nu_n\) is said to be \emph{\(\ell\)--empirically calibrated} for \(\nu_0\) if it satisfies the self-consistency property
\begin{equation}
\frac{1}{n} \sum_{i=1}^n \ell(Z_i, \nu_n) = \min_{\theta} \frac{1}{n} \sum_{i=1}^n \ell(Z_i, \theta \circ \nu_n)\,, \label{eqn::ellcalibrated1}
\end{equation}
where the minimum is taken over all real-valued functions \(\theta: \mathbb{R} \to \mathbb{R}\). That is, the empirical risk of \(\nu_n\) cannot be improved through transformation by any one-dimensional mapping.  

When the loss $\ell$ is sufficiently smooth, the first-order equations characterizing the empirical risk minimization problem imply that $\ell$--empirical calibration is equivalent to requiring that, for each map $\theta: \mathbb{R} \rightarrow \mathbb{R}$, 
\begin{equation}
   \frac{1}{n}\sum_{i=1}^n \frac{\partial}{\partial\varepsilon}   \ell(Z_i, \nu_n + \varepsilon \theta \circ \nu_n) \Big |_{\varepsilon = 0}\,=\,0\,. \label{eqn::ellcalibrated2}
\end{equation} 

A simple, distribution-free method for calibrating nuisance estimators is \emph{histogram binning}, such as uniform-mass (quantile) binning \citep{gupta2020distribution, gupta2021distribution}. Given an estimated function \( \nu_n \), the calibration dataset \( \mathcal{C}_n = \{Z_i\}_{i=1}^n \), and a loss function \( \ell(z, \nu) \), the range of predictions \( \nu_n(Z_i) \) is partitioned into \( K \) disjoint bins \( \{B_k\}_{k=1}^K \) in an outcome-independent manner—for example, using quantiles of \( \{\nu_n(Z_i)\}_{i=1}^n \). The calibrated estimator \( \nu_n^* := \theta_n \circ \nu_n \) is then obtained by minimizing the empirical loss within each bin:
\[
\theta_n(t) := \argmin_{c \in \mathbb{R}} \sum_{i=1}^n \mathbbm{1}\{\nu_n(Z_i) \in B_{k(t)}\} \, \ell(Z_i, c),
\]
where \( k(t) \) denotes the index of the bin containing \( t \in \nu_n(\mathcal{Z}) \). For least-squares loss, this reduces to computing the average of the observed outcomes within each bin. Since any transformation of a predictor preserves its piecewise constant structure, the empirical risk cannot be reduced by further transforming its predictions, and hence it is \emph{empirically \( \ell \)-calibrated}.

While histogram binning guarantees empirical calibration, it does not ensure good out-of-sample performance. With a fine partition \( (K = n) \), the estimator may interpolate the data and overfit, leading to poor generalization. Conversely, with a coarse partition \( (K = 1) \), the estimator is well-calibrated but lacks predictive power, collapsing to a constant predictor. This trade-off highlights the importance of choosing \( K \) carefully: too few bins limit expressiveness, while too many increase variance and degrade calibration. Under appropriate choices of \( K \), histogram binning yields asymptotically calibrated estimators, with the conditional \( \ell^2 \)-calibration error satisfying \( \operatorname{Cal}_{\ell^2}(\nu_n^*) = O_p\left( \frac{K \log(n/K)}{n} \right) \) \citep{whitehouse2024orthogonal, van2025generalized}. Thus, tuning \( K \), for example via cross-validation, is essential to balance calibration and accuracy. In the next section, we adopt a data-adaptive alternative—\emph{isotonic calibration}—which flexibly enforces monotonicity while maintaining distribution-free guarantees \citep{zadrozny2001obtaining, niculescu2005predicting, van2023causal}.

\subsection{Empirical calibration via isotonic regression}

Isotonic calibration \citep{zadrozny2002transforming, niculescu2005obtaining} is a data-adaptive histogram binning method that learns the bins using isotonic regression, a nonparametric technique traditionally used for estimating monotone functions \citep{barlow1972isotonic, groeneboom1993isotonic}. Specifically, the bins are determined by minimizing an empirical mean least-squares criterion under the constraint that the calibrated predictor is a non-decreasing monotone transformation of the original predictor. Isotonic calibration is motivated by the heuristic that, for a good predictor \( z \mapsto \nu_n(z) \), the optimal transformation of the predictor should be approximately monotone as a function of \( \nu_n \). For example, for a perfect predictor, the optimal transformation should be the identity function. Even when this map is not approximately monotone, isotonic regression ensures that the quality of the initial predictor is not harmed, since the identity transformation—being monotone—can always be learned. Isotonic calibration is distribution-free—it does not rely on monotonicity assumptions—and, unlike traditional histogram binning, it is tuning-parameter-free and naturally preserves the mean least-squares of the original predictor (as the identity transformation is monotone) \citep{van2023causal}.

For clarity, we focus on the regression case where \( \ell \) denotes the least-squares loss. Formally, isotonic calibration takes a fitted nuisance estimator \( \nu_n \) and a calibration dataset \( \mathcal{C}_n \) to produce a calibrated model \( \nu_n^* := \theta_n \circ \nu_n \), where \( \theta_n: \mathbb{R} \rightarrow \mathbb{R} \) is an isotonic step function obtained by solving the optimization problem:
\begin{equation}
    \theta_n \in \argmin_{\theta \in \mathcal{F}_{\text{iso}}} \sum_{i=1}^n \left\{Y_i - \theta(\nu_n(W_i, A_i))\right\}^2,
\end{equation}
where \( \mathcal{F}_{\text{iso}} \) denotes the set of all univariate, piecewise constant functions that are monotonically nondecreasing. Following \citet{groeneboom1993isotonic}, we consider the unique c\`{a}dl\`{a}g piecewise constant solution to this problem, which has jumps only at observed values \( \{\nu_n(W_i, A_i): i \in [n]\} \). The first-order optimality conditions of this convex problem imply that the isotonic solution \( \theta_n \) acts as a binning calibrator over a data-adaptive set of bins determined by the jump points of the step function. Thus, isotonic calibration yields perfect empirical calibration. Specifically, for any transformation \( g: \mathbb{R} \rightarrow \mathbb{R} \), the perturbed step function \( \varepsilon \mapsto \theta_n + \varepsilon (g \circ \theta_n) \) remains isotonic for all sufficiently small \( \varepsilon \) such that \( |\varepsilon| \sup_{t \in \nu_n(\mathcal{Z})} |(g \circ \theta_n)(t)| \) is less than the maximum jump size of \( \theta_n \), given by \( \sup_{t \in \nu_n(\mathcal{Z})} |\theta_n(t) - \theta_n(t-)| \). Since \( \theta_n \) minimizes the empirical mean least-squares over all isotonic functions, it follows that for each \( g: \mathbb{R} \rightarrow \mathbb{R} \), the following condition holds:
\begin{align*}
    \frac{d}{d\varepsilon} \frac{1}{2} \sum_{i=1}^n \left\{Y_i - \theta_n(\nu_n(W_i, A_i)) - \varepsilon g(\theta_n(\nu_n(W_i, A_i)))\right\}^2 \Big|_{\varepsilon = 0} = \sum_{i=1}^n g(\nu_n^*(Z_i)) \{Y_i - \nu_n^*(Z_i)\} = 0.
\end{align*}
These orthogonality conditions are equivalent to perfect empirical calibration. In particular, by taking \( g(t) = 1\{t = \theta_n(\nu_n(z))\} \), we conclude that the isotonic calibrated predictor \( \nu_n^* \) is empirically calibrated.

\subsection{Benefits and trade-offs in empirical calibration}

In the context of debiased machine learning, empirical calibration is valuable not only for enabling doubly robust inference but also for improving estimator stability and predictive accuracy. In our framework, empirical calibration is defined relative to a loss function \( \ell(z, \nu) \) for a nuisance function \( \nu \in \mathcal{H} \), and requires that \( \nu_n \) satisfies the self-consistency condition
\[
\frac{1}{n} \sum_{i=1}^n \ell(Z_i, \nu_n) = \min_{\theta} \frac{1}{n} \sum_{i=1}^n \ell(Z_i, \theta \circ \nu_n),
\]
where the minimum is taken over all univariate transformations \( \theta: \mathbb{R} \to \mathbb{R} \). This condition ensures that the predictions \( \nu_n \) are optimal with respect to empirical risk and cannot be improved by recalibration.

In the case of least-squares loss \( \ell(z, \mu) = \{y - \mu(a, w)\}^2 \), calibration requires that \( \mu_n(a, w) \) equals the average outcome among individuals with the same predicted value. That is,
\[
\mu_n(a, w) = \frac{\sum_{i=1}^n 1\{ \mu_n(A_i, W_i) = \mu_n(a, w) \} Y_i}{\sum_{i=1}^n 1\{ \mu_n(A_i, W_i) = \mu_n(a, w) \}},
\]
ensuring that \( \mu_n \) does not systematically over- or under-predict outcomes. For the Riesz loss \( \ell(z, \alpha) = \alpha^2(a, w) - 2 \psi_n(\alpha) \), associated with estimating a counterfactual mean \( \psi_0 = E_0\{\mu_0(1, W)\} \), calibration of the Riesz representer \( \alpha_n \) implies approximate covariate balance across estimated inverse propensity strata. Specifically, the calibration condition
\[
\frac{1}{n} \sum_{i=1}^n A_i \alpha_n(1, W_i) f(\alpha_n(1, W_i)) = \frac{1}{n} \sum_{i=1}^n f(\alpha_n(1, W_i))
\]
must hold for all functions \( f \), ensuring that higher weights contribute meaningfully to balancing, rather than inflating variance \citep{deshpande2023calibrated, van_der_Laan2024stabilized}.

To ensure good performance of the empirically calibrated estimator, it is important that calibration holds not only on the observed data but also at the population level. Notably, perfect empirical calibration can be achieved trivially by overfitting the labels—for example, by interpolating the outcomes—so it is crucial to use calibration procedures that generalize well. Formally, a function \( \nu \) is said to be perfectly \emph{\( \ell \)-calibrated} with respect to a loss function \( \ell(z, \nu) \) if it minimizes the expected loss over all one-dimensional transformations of itself:
\[
E_P[\ell(Z, \nu)] = \inf_{\theta} E_P[\ell(Z, \theta \circ \nu)],
\]
where the infimum is taken over all functions \( \theta: \mathbb{R} \to \mathbb{R} \).  The notion of \( \ell \)-calibration generalizes common forms of calibration, such as regression and quantile calibration \citep{roth2022uncertain}, which correspond to the least-squares and pinball (hinge) losses, respectively \citep{deng2023happymap}. Related ideas have also been explored in the context of predictive modeling \citep{whitehouse2024orthogonal}.

\section{Implementation of calibrated DML}
\label{sec:code}

An \texttt{R} package \texttt{calibratedDML} as well as \texttt{Python} code implementing CDML can be found on GitHub at the following link: \href{https://github.com/Larsvanderlaan/calibratedDML}{https://github.com/Larsvanderlaan/calibratedDML}.

The following subsections provide \texttt{R} and \texttt{Python} code for calibrating inverse propensity weights and outcome regressions. These calibrated models can be directly used within DML frameworks to construct isocalibrated DML estimators of the average treatment effect. Here, we implement isotonic regression using \texttt{xgboost}, allowing control over the maximum tree depth and the minimum number of observations in each constant segment of the isotonic regression fit.

\subsection{R Code}

\begin{lstlisting}[style=compact, language=R]
# Function: isoreg_with_xgboost
# Purpose: Fits isotonic regression using XGBoost.
# Inputs:
#   - x: A vector or matrix of predictor variables.
#   - y: A vector of response variables.
#   - max_depth: Maximum depth of the trees in XGBoost (default = 15).
#   - min_child_weight: Minimum sum of instance weights (Hessian) 
#        needed in a child node (default = 20).
# Returns:
#   - A function that takes a new predictor variable x
#     and returns the model's predicted values.

isoreg_with_xgboost <- function(x, y, max_depth = 15, min_child_weight = 20) {
 
  # Create an XGBoost DMatrix object from the data
  data <- xgboost::xgb.DMatrix(data = as.matrix(x), label = as.vector(y))
  
  # Set parameters for the monotonic XGBoost model
  params <- list(max_depth = max_depth,
                 min_child_weight = min_child_weight,
                 monotone_constraints = 1,  # Enforce monotonic increase
                 eta = 1, gamma = 0,
                 lambda = 0)
  
  # Train the model with one boosting round
  iso_fit <- xgboost::xgb.train(params = params,
                                data = data, 
                                nrounds = 1)
  
  # Prediction function for new data
  fun <- function(x) {
    data_pred <- xgboost::xgb.DMatrix(data = as.matrix(x))
    pred <- predict(iso_fit, data_pred)
    return(pred)
  }
  return(fun)
}

# Function: calibrate_inverse_weights
# Purpose: Calibrates inverse weights using isotonic regression 
#          with XGBoost for two propensity scores.
# Inputs:
#   - A: Binary indicator variable.
#   - pi1: Estimated propensity score for treatment group A = 1.
#   - pi0: Estimated propensity score for control group A = 0.
# Returns:
#   - A list containing calibrated inverse weights for each group:
#       - alpha1_star: Calibrated inverse weights for A = 1.
#       - alpha0_star: Calibrated inverse weights for A = 0.

calibrate_inverse_weights <- function(A, pi1, pi0) {

  # Calibrate pi1 using monotonic XGBoost
  calibrator_pi1 <- isoreg_with_xgboost(pi1, A)
  pi1_star <- calibrator_pi1(pi1)
  
  # Set minimum truncation level for treated group
  c1 <- min(pi1_star[A == 1])
  pi1_star = pmax(pi1_star, c1)
  alpha1_star <- 1 / pi1_star
  
  # Calibrate pi0 using monotonic XGBoost
  calibrator_pi0 <- isoreg_with_xgboost(pi0, 1 - A)
  pi0_star <- calibrator_pi0(pi0)
  
  # Set minimum truncation level for control group
  c0 <- min(pi0_star[A == 0])
  pi0_star = pmax(pi0_star, c0)
  alpha0_star <- 1 / pi0_star
  
  # Return calibrated inverse weights for both groups
  return(list(alpha1_star = alpha1_star, alpha0_star = alpha0_star))
}

# Function: calibrate_outcome_regression
# Purpose: Calibrates outcome regression predictions using isotonic 
#          regression with XGBoost.
# Inputs:
#   - Y: Observed outcomes.
#   - mu1: Predicted outcome for treated group.
#   - mu0: Predicted outcome for control group.
#   - A: Binary indicator variable.
# Returns:
#   - A list containing calibrated predictions for each group:
#       - mu1_star: Calibrated predictions for A = 1.
#       - mu0_star: Calibrated predictions for A = 0.

calibrate_outcome_regression <- function(Y, mu1, mu0, A) {

  # Calibrate mu1 using monotonic XGBoost for treated group
  calibrator_mu1 <- isoreg_with_xgboost(mu1[A==1], Y[A==1])
  mu1_star <- calibrator_mu1(mu1)

  # Calibrate mu0 using monotonic XGBoost for control group
  calibrator_mu0 <- isoreg_with_xgboost(mu0[A==0], Y[A==0])
  mu0_star <- calibrator_mu0(mu0)

  # Return calibrated values for both groups
  return(list(mu1_star = mu1_star, mu0_star = mu0_star))
}

\end{lstlisting}

\subsection{Python code}

\begin{lstlisting}[style=compact, language=Python]

import xgboost as xgb
import numpy as np

def isoreg_with_xgboost(x, y, max_depth=15, min_child_weight=20):
    """
    Fits isotonic regression using XGBoost with monotonic constraints to ensure 
    non-decreasing predictions as the predictor variable increases.

    Args:
        x (np.array): A vector or matrix of predictor variables.
        y (np.array): A vector of response variables.
        max_depth (int, optional): Maximum depth of the trees in XGBoost. 
                                   Default is 15.
        min_child_weight (float, optional): Minimum sum of instance weights 
                                            needed in a child node. Default is 20.

    Returns:
        function: A prediction function that takes a new predictor variable x 
                  and returns the model's predicted values.
                  
    Example:
        >>> x = np.array([[1], [2], [3]])
        >>> y = np.array([1, 2, 3])
        >>> model = isoreg_with_xgboost(x, y)
        >>> model(np.array([[1.5], [2.5]]))
    """
    
    # Create an XGBoost DMatrix object from the data
    data = xgb.DMatrix(data=np.asarray(x), label=np.asarray(y))

    # Set parameters for the monotonic XGBoost model
    params = {
        'max_depth': max_depth,
        'min_child_weight': min_child_weight,
        'monotone_constraints': "(1)",  # Enforce monotonic increase
        'eta': 1,
        'gamma': 0,
        'lambda': 0
    }

    # Train the model with one boosting round
    iso_fit = xgb.train(params=params, dtrain=data, num_boost_round=1)

    # Prediction function for new data
    def predict_fn(x):
        """
        Predicts output for new input data using the trained isotonic regression model.
        
        Args:
            x (np.array): New predictor variables as a vector or matrix.
        
        Returns:
            np.array: Predicted values.
        """
        data_pred = xgb.DMatrix(data=np.asarray(x))
        pred = iso_fit.predict(data_pred)
        return pred

    return predict_fn

def calibrate_inverse_weights(A, pi1, pi0):
    """
    Calibrates inverse weights using isotonic regression with XGBoost for two propensity scores.

    Args:
        A (np.array): Binary indicator variable.
        pi1 (np.array): Propensity score for treatment group (A = 1).
        pi0 (np.array): Propensity score for control group (A = 0).

    Returns:
        dict: Contains calibrated inverse weights:
              - alpha1_star: Inverse weights for A = 1.
              - alpha0_star: Inverse weights for A = 0.
    """
    calibrator_pi1 = isoreg_with_xgboost(pi1, A)
    pi1_star = calibrator_pi1(pi1)
    c1 = np.min(pi1_star[A == 1])
    pi1_star = np.maximum(pi1_star, c1)
    alpha1_star = 1 / pi1_star

    calibrator_pi0 = isoreg_with_xgboost(pi0, 1 - A)
    pi0_star = calibrator_pi0(pi0)
    c0 = np.min(pi0_star[A == 0])
    pi0_star = np.maximum(pi0_star, c0)
    alpha0_star = 1 / pi0_star

    return {'alpha1_star': alpha1_star, 'alpha0_star': alpha0_star}

def calibrate_outcome_regression(Y, mu1, mu0, A):
    """
    Calibrates outcome regression predictions using isotonic regression with XGBoost.
    
    Args:
        Y (np.array): Observed outcomes.
        mu1 (np.array): Predicted outcome for the treated group (A = 1).
        mu0 (np.array): Predicted outcome for the control group (A = 0).
        A (np.array): Binary treatment indicator (1 for treatment, 0 for control).
    
    Returns:
        dict: Calibrated predictions for each group:
              - 'mu1_star': Calibrated predictions for A = 1.
              - 'mu0_star': Calibrated predictions for A = 0.
    """
    # Calibrate mu1 using treated group data (A = 1)
    calibrator_mu1 = isoreg_with_xgboost(mu1[A == 1], Y[A == 1])
    mu1_star = calibrator_mu1(mu1)  # Apply calibrator to all mu1 values

    # Calibrate mu0 using control group data (A = 0)
    calibrator_mu0 = isoreg_with_xgboost(mu0[A == 0], Y[A == 0])
    mu0_star = calibrator_mu0(mu0)  # Apply calibrator to all mu0 values

    return {'mu1_star': mu1_star, 'mu0_star': mu0_star}
    
\end{lstlisting}

\section{Proofs for Sections \ref{section::setup}, \ref{sec::genstrat}, and \ref{section::calibration}}
\label{Appendix::prelimproofs}

\subsection{Proof of Lemma \ref{lem:hellinger_continuity}}

\label{Appendix::hellinger_continuity}

\begin{proof}[Proof of Lemma \ref{lem:hellinger_continuity}]
Linearity of \(\psi_0\) follows from \ref{cond:expectlinear}. For \(\mu\ne0\),
applying \ref{cond:expectl2} to
\(\mu/\|\mu\|\) and using Cauchy--Schwarz give
\[
|\psi_0(\mu)|
\le \left(E_0[m(Z,\mu)^2]\right)^{1/2}
\le B_0 \|\mu\|,
\]
which verifies Condition~\ref{cond::linear}.
If \(H(P,P_0)<\eta\), then \ref{cond:expectdom} gives
\[
E_P[m(Z,\mu)^2]\le C_\eta E_0[m(Z,\mu)^2]\le C_\eta B_0^2\|\mu\|^2,
\]
so \(\psi_P\) is also a bounded linear functional.

We next verify Condition~\ref{cond::continuous}. Fix \(P\in\mathcal M\) such
that \(H(P,P_0)<\eta\), and let \(\nu\) be a common dominating measure for
\(P\) and \(P_0\), with densities \(p\) and \(p_0\). For fixed
\(\mu\in\mathcal H\), set \(f(z):=m(z,\mu)\). Then
\begin{align*}
E_P[f(Z)]-E_{P_0}[f(Z)]
&=
\int f(z)\{p(z)-p_0(z)\}\,d\nu(z)\\
&=
\int f(z)\{\sqrt{p(z)}-\sqrt{p_0(z)}\}\\
&\quad\times \{\sqrt{p(z)}+\sqrt{p_0(z)}\}\,d\nu(z).
\end{align*}
By Cauchy--Schwarz,
\begin{align*}
\bigl|E_P[f(Z)]-E_{P_0}[f(Z)]\bigr|
&\le
\left\{\int f^2(\sqrt p+\sqrt{p_0})^2\,d\nu\right\}^{1/2}\\
&\quad\times
\left\{\int(\sqrt p-\sqrt{p_0})^2\,d\nu\right\}^{1/2}.
\end{align*}
For the first factor, use \(2\sqrt{pp_0}\le p+p_0\) to obtain
\[
\int f^2(\sqrt p+\sqrt{p_0})^2\,d\nu
=
\int f^2\{p+p_0+2\sqrt{pp_0}\}\,d\nu
\le
2\int f^2(p+p_0)\,d\nu
=
2\{E_P[f(Z)^2]+E_{P_0}[f(Z)^2]\}.
\]
For the second factor,
\[
\left(\int(\sqrt p-\sqrt{p_0})^2\,d\nu\right)^{1/2}
=
\sqrt{2}\,H(P,P_0).
\]
Combining these bounds yields
\begin{align*}
\bigl|E_P\{m(Z,\mu)\}-E_{P_0}\{m(Z,\mu)\}\bigr|
&\le
\bigl(2\{E_P[f^2]+E_{P_0}[f^2]\}\bigr)^{1/2}\sqrt{2}\,H(P,P_0)\\
&=
2\,H(P,P_0)\,\bigl(E_P[f^2]+E_{P_0}[f^2]\bigr)^{1/2}.
\end{align*}
If \(\|\mu\|\le 1\), \ref{cond:expectl2} and \ref{cond:expectdom} bound the
last display by
\(2\sqrt{C_\eta+1}\,B_0 H(P,P_0)\). Taking the supremum over
\(\{\mu\in\mathcal H:\|\mu\|\le 1\}\) gives
\[
\|\psi_P-\psi_{P_0}\|_{\mathrm{op}}
\le 2\sqrt{C_\eta+1}\,B_0 H(P,P_0).
\]
Thus \(\|\psi_P-\psi_{P_0}\|_{\mathrm{op}}\to0\) whenever
\(H(P,P_0)\to0\), which verifies Condition~\ref{cond::continuous} for this
class of functionals.

It remains to verify Condition~\ref{cond::pathwise}. Fix
\(\mu\in\mathcal H\). The preceding bound implies \(m(\cdot,\mu)\in L^2(P_0)\).
For any regular parametric submodel \(t\mapsto P_t\) through \(P_0\) with score
\(s\in L^2_0(P_0)\),
\[
\left.\frac{d}{dt}\psi_{P_t}(\mu)\right|_{t=0}
= E_0[m(Z,\mu)s(Z)]
= E_0[\{m(Z,\mu)-\psi_0(\mu)\}s(Z)].
\]
Thus \(P\mapsto\psi_P(\mu)\) is pathwise differentiable at \(P_0\) with
influence function \(\phi_{0,\mu}=m(\cdot,\mu)-\psi_0(\mu)\). Since
\(\phi_{0,\mu}\in L^2_0(P_0)\), this is the efficient influence function in the
nonparametric model.
\end{proof}

\subsection{Proof of Theorem \ref{theorem::EIF}}

\begin{proof}[Proof of Theorem \ref{theorem::EIF}]
We first show that $(P,\mu) \mapsto \psi_P(\mu)$ is total pathwise
differentiable in the sense of Appendix A.4 of
\cite{luedtke2024simplifying}. Lemma S6 of
\cite{luedtke2024simplifying} reduces this to verifying that the total
operator $(P, \mu) \mapsto \psi_P(\mu)$ is partially differentiable in its
first argument and continuously partially differentiable in its second argument
at $(P_0, \mu_0)$. The former differentiability holds since, by
Condition~\ref{cond::pathwise}, $P \mapsto \psi_P(\mu_0)$ is pathwise
differentiable at $P_0$. For the latter, note that
$\mu \mapsto \psi_P(\mu)$ is a linear functional, so the partial derivative of
$(P, \mu) \mapsto \psi_P(\mu)$ in its second argument is
$\mu \mapsto \psi_P(\mu)$. Thus, continuous partial differentiability follows
from Condition~\ref{cond::continuous}, which gives continuity of
$P \mapsto \psi_P$ at $P_0$ when the domain is equipped with the Hellinger
distance and the range with the operator norm. By Lemma S6 in
\cite{luedtke2024simplifying}, the total pathwise derivative at
$(P_0, \mu_0)$ in the direction of
$(s,h) \in T_{\mathcal{M}}(P_0) \times \mathcal{H}$ is then given by
$\langle \phi_{0, \mu_0}, s \rangle + \langle \alpha_0, h\rangle $.

Under our assumptions, $P \mapsto \mu_P$ is pathwise differentiable at $P_0$ in a Hilbert-valued sense by Example 5 of \cite{luedtke2024one}, with pathwise derivative $d\mu_0(s)$ in the direction $s \in T_{\mathcal{M}}(P_0)$ given by $h \mapsto \langle h\{ \mathcal{I}_Y - \mu_0\} , s \rangle$, where $ \mathcal{I}_Y: o \mapsto y$. As a consequence, by the chain rule, $P \mapsto \psi_P(\mu_P)$ is pathwise differentiable at $P_0$ with pathwise derivative given by $\langle \phi_{0, \mu_0}, s \rangle + \langle \alpha_0\{\mathcal{I}_Y - \mu_0\}, s\rangle .$ It follows that the efficient influence function is $ \phi_{0, \mu_0} + D_{\mu_0, \alpha_0}$. 
\end{proof}

\subsection{Proofs for Theorem \ref{theorem::DRbiasexpansion}, Theorem \ref{theorem::DRbiasexpansionsieve}, and Lemma \ref{lemma::sufficientcoupling}}

\begin{proof}[Proof of Theorem \ref{theorem::DRbiasexpansion}]
    We begin with the proof of the Riesz-based decomposition, as the regression-based decomposition will follow as a special case.

By the law of iterated expectations, conditioning on
\((\alpha_n(A,W),\alpha_0(A,W))\) under \(P_0\) yields
\begin{align*}
\langle \alpha_0 - \alpha_n , \mu_n - \mu_0 \rangle
&=  \big\langle \alpha_0 - \alpha_n,\ \Pi_{(\alpha_n,\alpha_0)}(\mu_n-\mu_0)\big\rangle .
\end{align*}
We decompose the preceding display as
\begin{equation}
\label{eqn::DRexpansionproof}
\begin{aligned}
\langle \alpha_0 - \alpha_n , \mu_n - \mu_0 \rangle
&= \big\langle \alpha_0 - \alpha_n,\ \Pi_{\alpha_n}(\mu_n-\mu_0)\big\rangle
  + \big\langle (\Pi_{(\alpha_n,\alpha_0)}-\Pi_{\alpha_n})(\alpha_0-\alpha_n),\ 
          (\mu_n-\mu_0)-\Pi_{\alpha_n}(\mu_n-\mu_0)\big\rangle \\
&=  \big\langle \alpha_n - \alpha_0,\ r_{n,0}\big\rangle
   -\big\langle (\Pi_{(\alpha_n,\alpha_0)}-\Pi_{\alpha_n})\Delta_{\alpha,n},\ 
          \Delta_{\mu,n}-\Pi_{\alpha_n}\Delta_{\mu,n}\big\rangle,
\end{aligned}
\end{equation}
where $r_{n,0}  = - \Pi_{\alpha_n}(\mu_n-\mu_0)$.
By the Riesz representation property for \(\alpha_0\) in \eqref{eqn::reiszrep},
the first term on the right-hand side can be expressed as
\begin{align*}
\big\langle \alpha_n - \alpha_0,\, r_{n,0}\big\rangle
&= - \psi_0(r_{n,0})  + P_0\{\alpha_n r_{n,0}\} \\
&= -\psi_n(r_{n,0}) + P_0\{\alpha_n r_{n,0}\}
   - \psi_0(r_{n,0}) + \psi_n(r_{n,0}),
\end{align*}
where, in the second equality, we add and subtract \(\psi_n(r_{n,0})\). By the
orthogonality condition in \eqref{eqn::orthogonalRiesz},
\[
\psi_n(r_{n,0}) = P_n\{\alpha_n r_{n,0}\}.
\]
Hence,
\[
- \psi_n(r_{n,0}) + P_0\{\alpha_n r_{n,0}\}
= - (P_n-P_0)\{\alpha_n r_{n,0}\}.
\]
Therefore,
\begin{align*}
\big\langle \alpha_n - \alpha_0,\, r_{n,0}\big\rangle
&= (P_n-P_0)\{-\alpha_n r_{n,0}\} - \psi_0(r_{n,0}) + \psi_n(r_{n,0}) \\
&= (P_n-P_0)\{-\alpha_n r_{n,0} + \phi_{r_{n,0}}\}
   + \psi_n(r_{n,0}) - \psi_0(r_{n,0}) - P_n \phi_{r_{n,0}} \\
&= (P_n-P_0)B_{n,0} + Rem_{\psi_n}(r_{n,0}),
\end{align*}
where we add and subtract \(P_n \phi_{r_{n,0}}\) and use that
\(P_0 \phi_{r_{n,0}}=0\). Substituting this display into
\eqref{eqn::DRexpansionproof}, we obtain
\begin{equation}
\begin{aligned}
\langle \alpha_0 - \alpha_n , \mu_n - \mu_0 \rangle
&= (P_n-P_0)B_{n,0}
   + Rem_{\psi_n}(r_{n,0})
   - \big\langle (\Pi_{(\alpha_n,\alpha_0)}-\Pi_{\alpha_n})\Delta_{\alpha,n},\ 
          \Delta_{\mu,n}-\Pi_{\alpha_n}\Delta_{\mu,n}\big\rangle \\
&= (P_n-P_0)B_{n,0}
   + \big\langle \alpha_0 - \Pi_{\alpha_n}\alpha_0,\ 
          \Delta_{\mu,n}-\Pi_{\alpha_n}\Delta_{\mu,n}\big\rangle
   + Rem_{\psi_n}(r_{n,0}),
\end{aligned}
\end{equation}
where we use that
\[
(\Pi_{(\alpha_n,\alpha_0)}-\Pi_{\alpha_n})\Delta_{\alpha,n}
= \Pi_{\alpha_n}\alpha_0 - \alpha_0
= -(\alpha_0 - \Pi_{\alpha_n}\alpha_0).
\]
The Riesz-based decomposition now follows. If Condition~\ref{cond:orthoB} holds, then
\[
\big\langle (\Pi_{(\mu_n,\mu_0)}-\Pi_{\mu_n})\Delta_{\alpha,n},\ \mu_0-\Pi_{\mu_n}\mu_0 \big\rangle
= \gamma_{\alpha,0}(\mu_n,\alpha_n)\,\|\alpha_n-\alpha_0\|^2
= O_p\left(\|\alpha_n-\alpha_0\|^2\right).
\]
By the Cauchy--Schwarz inequality, we also have
\[
\big\langle (\Pi_{(\mu_n,\mu_0)}-\Pi_{\mu_n})\Delta_{\alpha,n},\ \mu_0-\Pi_{\mu_n}\mu_0 \big\rangle
\le \|(\Pi_{(\mu_n,\mu_0)}-\Pi_{\mu_n})\Delta_{\alpha,n}\|\,
   \|\mu_0-\Pi_{\mu_n}\mu_0\|.
\]
By nonexpansiveness of orthogonal projections,
\(\|(\Pi_{(\mu_n,\mu_0)}-\Pi_{\mu_n})\Delta_{\alpha,n}\|\le \|\Delta_{\alpha,n}\|\) and
\[
\|\mu_0-\Pi_{\mu_n}\mu_0\|
= \|\Delta_{\mu,n}-\Pi_{\mu_n}\Delta_{\mu,n}\|
\le \|\Delta_{\mu,n}\|.
\]
Hence,
\[
\big\langle (\Pi_{(\mu_n,\mu_0)}-\Pi_{\mu_n})\Delta_{\alpha,n},\ \mu_0-\Pi_{\mu_n}\mu_0 \big\rangle
\le \|\Delta_{\alpha,n}\|\,\|\Delta_{\mu,n}\|.
\]
Taking the minimum of these two valid bounds, we obtain
\[
\big\langle (\Pi_{(\alpha_n,\alpha_0)}-\Pi_{\alpha_n})\Delta_{\mu,n},\ \Delta_{\alpha,n} \big\rangle
= O_p\left(\|\alpha_n-\alpha_0\|^2 \wedge
\|\mu_n-\mu_0\|\,\|\Pi_{\alpha_n}\alpha_0-\alpha_0\|\right),
\]
as desired.

For the regression-based decomposition, we apply the Riesz-based decomposition with
\(\psi_0(\alpha) := E_0\{\alpha(A,W)\mu_0(A,W)\}\) and
\(\psi_n(\alpha) := \frac{1}{n}\sum_{i=1}^n \alpha(A_i,W_i)Y_i\) (see also the proof outline in Section \ref{sec::genstrat}). In particular,
\(\mu_0\) is the Riesz representer of the linear functional
\(\psi_0:\alpha \mapsto E_0\{\alpha(A,W)\mu_0(A,W)\}\). Moreover, the
orthogonality condition in \eqref{eqn::orthogonaloutcome} is equivalent to the
orthogonality condition \eqref{eqn::orthogonalRiesz} for the estimated linear
functional \(\psi_n\). For this choice of \(\psi_n\), we have
\(Rem_{\psi_n}(\alpha)=0\) with \(\phi_{\alpha}(z) := \alpha y -
\psi_0(\alpha)\). Under these definitions, \(B_0\) and \(B_{n,0}\) coincide with
\(A_0\) and \(A_{n,0}\), respectively. The regression-based decomposition now follows.

\end{proof}

\begin{proof}[Proof of Theorem \ref{theorem::DRbiasexpansionsieve}]
We use only the projection identities that underlie the proof of
Theorem~\ref{theorem::DRbiasexpansion}. If \(\Theta\subseteq\Theta_0\) are
closed linear subspaces with projections \(\Pi\) and \(\Pi_0\), then
\((\Pi_0-\Pi)h\) is orthogonal to \(\Theta\) for every \(h\in\mathcal H\). Also,
if \(g\in\Theta_0\), then \((\Pi_0-\Pi)g=g-\Pi g\).

We first prove the regression-based decomposition. Since
\(\Delta_{\mu,n}\in\Theta_{\mu,n,0}\),
\[
\langle \alpha_0-\alpha_n,\mu_n-\mu_0\rangle
=
-\langle \Pi_{\mu,n,0}\Delta_{\alpha,n},\Delta_{\mu,n}\rangle .
\]
Decomposing \(\Pi_{\mu,n,0}\Delta_{\alpha,n}\) into its projection onto
\(\Theta_{\mu,n}\) and its residual gives
\begin{align*}
\langle \alpha_0-\alpha_n,\mu_n-\mu_0\rangle
&=
-\langle \Pi_{\mu,n}\Delta_{\alpha,n},\Delta_{\mu,n}\rangle
-\langle (\Pi_{\mu,n,0}-\Pi_{\mu,n})\Delta_{\alpha,n},
        \Delta_{\mu,n}\rangle  \\
&=
-\langle \Pi_{\mu,n}\Delta_{\alpha,n},\Delta_{\mu,n}\rangle
-\langle (\Pi_{\mu,n,0}-\Pi_{\mu,n})\Delta_{\alpha,n},
        (\Pi_{\mu,n,0}-\Pi_{\mu,n})\Delta_{\mu,n}\rangle ,
\end{align*}
where the second equality uses that
\((\Pi_{\mu,n,0}-\Pi_{\mu,n})\Delta_{\alpha,n}\) is orthogonal to
\(\Theta_{\mu,n}\). Let
\(\widetilde s_{n,0}:=\Pi_{\mu,n}(\alpha_0-\alpha_n)
=-\Pi_{\mu,n}\Delta_{\alpha,n}\). Then
\[
P_0\widetilde A_{n,0}
=P_0\{\widetilde s_{n,0}(A,W)(Y-\mu_n(A,W))\}
=-\langle \widetilde s_{n,0},\Delta_{\mu,n}\rangle
=\langle \Pi_{\mu,n}\Delta_{\alpha,n},\Delta_{\mu,n}\rangle .
\]
The score equation \eqref{eqn::orthogonaloutcomesieve} implies
\(P_n\widetilde A_{n,0}=0\). Hence
\(-\langle \Pi_{\mu,n}\Delta_{\alpha,n},\Delta_{\mu,n}\rangle
=(P_n-P_0)\widetilde A_{n,0}\), which proves the first display.

We next prove the Riesz-based decomposition. Since
\(\Delta_{\alpha,n}\in\Theta_{\alpha,n,0}\),
\[
\langle \alpha_0-\alpha_n,\mu_n-\mu_0\rangle
=
-\langle \Delta_{\alpha,n},\Pi_{\alpha,n,0}\Delta_{\mu,n}\rangle .
\]
Using the same projection identity with
\(\Theta_{\alpha,n}\subseteq\Theta_{\alpha,n,0}\), we obtain
\begin{align*}
\langle \alpha_0-\alpha_n,\mu_n-\mu_0\rangle
&=
-\langle \Delta_{\alpha,n},\Pi_{\alpha,n}\Delta_{\mu,n}\rangle
-\langle (\Pi_{\alpha,n,0}-\Pi_{\alpha,n})\Delta_{\alpha,n},
        (\Pi_{\alpha,n,0}-\Pi_{\alpha,n})\Delta_{\mu,n}\rangle .
\end{align*}
Let
\(\widetilde r_{n,0}:=\Pi_{\alpha,n}(\mu_0-\mu_n)
=-\Pi_{\alpha,n}\Delta_{\mu,n}\). The first term is
\(\langle \Delta_{\alpha,n},\widetilde r_{n,0}\rangle\). By the Riesz
representation property for \(\alpha_0\) in \eqref{eqn::reiszrep},
\[
\langle \Delta_{\alpha,n},\widetilde r_{n,0}\rangle
=P_0\{\alpha_n\widetilde r_{n,0}\}-\psi_0(\widetilde r_{n,0}).
\]
Because \(\psi_n\) is linear, \eqref{eqn::orthogonalRieszsieve} applied to
\(\Pi_{\alpha,n}\Delta_{\mu,n}=-\widetilde r_{n,0}\) gives
\(\psi_n(\widetilde r_{n,0})
=P_n\{\alpha_n\widetilde r_{n,0}\}\). Therefore
\begin{align*}
\langle \Delta_{\alpha,n},\widetilde r_{n,0}\rangle
&=
-(P_n-P_0)\{\alpha_n\widetilde r_{n,0}\}
+\psi_n(\widetilde r_{n,0})-\psi_0(\widetilde r_{n,0}) \\
&=
(P_n-P_0)\widetilde B_{n,0}
+\mathrm{Rem}_{\psi_n}(\widetilde r_{n,0}),
\end{align*}
where \(P_0\phi_{\widetilde r_{n,0}}=0\) and the definitions of
\(\widetilde B_{n,0}\) and
\(\mathrm{Rem}_{\psi_n}(\widetilde r_{n,0})\) are used. Substitution into the
preceding projection decomposition proves the Riesz-based display.
\end{proof}

\begin{proof}[Proof of Lemma  \ref{lemma::sufficientcoupling}]
The proof that Conditions~\ref{cond:orthoA} and \ref{cond:orthoB} are implied by
\ref{cond:projectioncoupling1A} and \ref{cond:projectioncoupling1B},
respectively, follows directly from the Cauchy--Schwarz inequality applied to
the covariances in the numerators of \(\gamma_{\mu,0}(\mu_n,\alpha_n)\) and
\(\gamma_{\alpha,0}(\mu_n,\alpha_n)\). It therefore remains to show that
\ref{cond:projectioncoupling1A} and \ref{cond:projectioncoupling1B} are implied
by \ref{cond:projectioncoupling2A} and \ref{cond:projectioncoupling2B},
respectively. We prove that \ref{cond:projectioncoupling2A} implies
\ref{cond:projectioncoupling1A}; the remaining implication follows by symmetry.

Denote $X := (W,A)$. Let $g: \mathbb{R}^2 \rightarrow \mathbb{R}$ be a Lipschitz continuous function with constant $L > 0$.  By Lipschitz continuity, we have that
\begin{align*}
  &\left|  g(\mu_n(x), \mu_0(x)) - E[g(\mu_n(X), \mu_0(X))  |  \mu_n(X) = \mu_n(x)] \right|\\
  &\hspace{1in}= \left| E[ g(\mu_n(x), \mu_0(x))  - g(\mu_n(x), \mu_0(X))  |  \mu_n(X) = \mu_n(x)] \right|\\
  &\hspace{1in} \leq  E[ \left|g(\mu_n(x), \mu_0(x))  - g(\mu_n(x), \mu_0(X)) \right|  |  \mu_n(X) = \mu_n(x)]  \\
    &\hspace{1in} \leq L E[\left|\mu_0(x) - \mu_0(X) \right|  |  \mu_n(X) = \mu_n(x)]  .
\end{align*}
On the event $\{ \mu_n(X) = \mu_n(x)\}$, we know 
\begin{align*}
    |\mu_0(x) - \mu_0(X) | &\leq |\mu_0(x) - \mu_n(x)| + |\mu_n(x) - \mu_n(X)| + |\mu_0(X) - \mu_n(X)| \\
    & \leq |\mu_0(x) - \mu_n(x)| +  |\mu_0(X) - \mu_n(X)|.
\end{align*} 
Therefore, it holds that
\begin{align*}
    &\left|  g(\mu_n(x), \mu_0(x)) - E[g(\mu_n(X), \mu_0(X))  |  \mu_n(X) = \mu_n(x)] \right|\\
    &\hspace{1in}\lesssim E[\left|\mu_0(x) - \mu_0(X) \right|  \mid    \mu_n(X) = \mu_n(x)] \\ 
     &\hspace{1in} \lesssim E|\mu_0(x) - \mu_n(x)|     + E[ |\mu_0(X) - \mu_n(X)|  \mid   \mu_n(X) = \mu_n(x)] .
\end{align*}  

Now, for some function $f \in L^2(P_X)$, suppose that $(\mu_n(x), \mu_0(x)) \mapsto (\Pi_{(\mu_n, \mu_0)} f)(x)$ is Lipschitz continuous. Then, defining $g: (\widehat{m}, m) \mapsto E_0[f(X)  |  \mu_n(X) = \widehat{m}, \mu_0(X) = m, \mathcal{D}_n]$ and noting by the law of iterated expectation that $\Pi_{\mu_n} \Pi_{(\mu_n, \mu_0)} f = \Pi_{\mu_n} f$,
we obtain the following pointwise error bound:
$$\left| \Pi_{(\mu_n, \mu_0)} f - \Pi_{\mu_n} f \right| \lesssim \left|\mu_n - \mu_0 \right| + \Pi_{\mu_n}(|\mu_n - \mu_0|).$$
Since $\|\Pi_{\mu_n}(|\mu_n - \mu_0|)\| \leq \| \mu_n - \mu_0\| $ by the properties of projections, it follows that
$$\| \Pi_{(\mu_n, \mu_0)} f - \Pi_{\mu_n} f\| \lesssim  \| \mu_n - \mu_0 \|.$$
Taking $f$ to be $\Delta_{\alpha,n}$, we conclude that Condition \ref{cond:projectioncoupling2A} implies Condition \ref{cond:projectioncoupling1A}. By an identical argument, Condition \ref{cond:projectioncoupling2B} implies Condition \ref{cond:projectioncoupling1B}. The result then follows.

By symmetry of $\mu_n$ and $\mu_0$ in the Lipschitz continuity condition and the
preceding bound, we also have
\[
\| \Pi_{(\mu_n, \mu_0)} f - \Pi_{\mu_0} f\| \lesssim \| \mu_n - \mu_0 \|.
\]
Hence, by the triangle inequality,
\begin{align*}
\| \Pi_{\mu_0} f - \Pi_{\mu_n} f\|
&\leq \| \Pi_{(\mu_n, \mu_0)} f - \Pi_{\mu_0} f\|
  + \| \Pi_{(\mu_n, \mu_0)} f - \Pi_{\mu_n} f\| \\
&\lesssim \| \mu_n - \mu_0 \|.
\end{align*}

\end{proof}

\subsection{Proof of Lemma  \ref{theorem:empiricalcalibration}}

\begin{proof}[Proof of Lemma~\ref{theorem:empiricalcalibration}]
The orthogonality conditions follow directly from the first-order optimality
conditions of \eqref{eqn::calpropertyoutcome} and \eqref{eqn::calpropertyRiesz}
for the least-squares and Riesz losses. For the least-squares loss, note that
$\mu_n = \theta_n \circ \mu_n$, where
\[
\theta_n \in \argmin_{\theta}\sum_{i=1}^n \{Y_i - \theta(\mu_n(A_i, W_i))\}^2 .
\]
By the first-order optimality conditions for $\theta_n$, for each
$\theta:\mathbb{R}\to\mathbb{R}$,
\begin{align*}
0
&= \left.\frac{d}{d\varepsilon}\sum_{i=1}^n \Big\{Y_i-(\theta_n+\varepsilon\theta)\big(\mu_n(A_i,W_i)\big)\Big\}^2\right|_{\varepsilon=0} \\
&= 2\,\frac{1}{n}\sum_{i=1}^n \theta\big(\mu_n(A_i,W_i)\big)\{Y_i-\mu_n(A_i,W_i)\}.
\end{align*}
Hence, \eqref{eqn::calpropertyoutcome2} holds, and choosing $\theta$ such that
$\theta\circ\mu_n = r_{n,0}$, we conclude that $\mu_n$ satisfies
\eqref{eqn::orthogonaloutcome}.

The implication for the Riesz loss follows by an identical argument. Here we
use that, by \ref{cond::estboundedlinear}, $\psi_n$ is a bounded linear
functional on $L^2(P_n)$, and
\begin{align*}
\left.\frac{d}{d\varepsilon}\psi_n\left(Z_i,\alpha_n+\varepsilon\,\theta\circ\alpha_n\right)\right|_{\varepsilon=0}
&= \left.\frac{d}{d\varepsilon}\psi_n\left(Z_i,\varepsilon\,\theta\circ\alpha_n\right)\right|_{\varepsilon=0} \\
&= \left.\frac{d}{d\varepsilon}\,\varepsilon\,\psi_n\left(Z_i,\theta\circ\alpha_n\right)\right|_{\varepsilon=0} \\
&= \psi_n\left(Z_i,\theta\circ\alpha_n\right).
\end{align*}
See also Lemma~\ref{lemma::isoscores}, which establishes empirical calibration
for the isocalibrated nuisance estimators.
\end{proof}

\subsection{Proofs of Lemma \ref{lemma::negEPOutcome} and Lemma \ref{lemma::negEPRiesz}}
\label{appendix::proofEP}

\begin{proof}[Proof of Lemma \ref{lemma::negEPOutcome}]
Let $\mathcal{I}_Y: z \mapsto y$ be the coordinate projection onto the outcome variable.  By the triangle inequality, we have
    \begin{align*}
        \|A_{n,0} - A_0\|& \leq \| \Pi_{\mu_n}(\alpha_0 - \overline{\alpha}_0)\{\mathcal{I}_Y - \mu_0(z)\}  - \Pi_{\mu_0}(\alpha_0 - \overline{\alpha}_0)\{\mathcal{I}_Y - \mu_0(z)\}\| \\
        & \quad +  \| \Pi_{\mu_n}(\overline{\alpha}_0 - \alpha_n)\{\mathcal{I}_Y - \mu_n(z)\} - \Pi_{\mu_n}(\alpha_0 - \overline{\alpha}_0)\{\mathcal{I}_Y - \mu_0(z)\} \|.
    \end{align*}

    By boundedness of nuisances in \ref{controlEP::bounded}, consistency of $\mu_n$ for $\mu_0$, and the projection bound in \ref{controlEP::couplingmu}, the first term satisfies
    \begin{align*}
        \| \Pi_{\mu_n}(\alpha_0 - \overline{\alpha}_0)\{\mathcal{I}_Y - \mu_0(z)\}  - \Pi_{\mu_0}(\alpha_0 - \overline{\alpha}_0)\{\mathcal{I}_Y - \mu_0(z)\}\| &= \|\{ (\Pi_{\mu_n} - \Pi_{\mu_0})(\alpha_0 - \overline{\alpha}_0)\}\{\mathcal{I}_Y - \mu_0(z)\}\|\\
        &= O_p(\|(\Pi_{\mu_n} - \Pi_{\mu_0})(\alpha_0 - \overline{\alpha}_0)\|)\\
          &= O_p(\|\mu_n - \mu_0\|^{\beta})\\
            &=o_p(1).
          \end{align*}
 By boundedness of nuisances in \ref{controlEP::bounded} and nonexpansiveness of projections, the second term satisfies:
    \begin{align*}
     & \| \Pi_{\mu_n}(\overline{\alpha}_0 - \alpha_n)\{\mathcal{I}_Y - \mu_n(z)\} - \Pi_{\mu_n}(\alpha_0 - \overline{\alpha}_0)\{\mathcal{I}_Y - \mu_0(z)\} \|.\\
     & \quad  \lesssim   O_p(\|\mu_n - \mu_0\|) +  \| \{\Pi_{\mu_n} (\overline{\alpha}_0 - \alpha_n)\}\{\mathcal{I}_Y - \mu_0(z)\}  \|\\
      & \quad  \lesssim   O_p(\|\mu_n - \mu_0\|) +  O_p(\|\overline{\alpha}_0 - \alpha_n\|).
    \end{align*}
    Thus, by consistency of $\mu_n$ for $\mu_0$ and convergence of $\alpha_n$ to the misspecified limit $\overline{\alpha}_0$ in \ref{controlEP::couplingmu},
      \begin{align*}
    \| \Pi_{\mu_n}(\overline{\alpha}_0 - \alpha_n)\{\mathcal{I}_Y - \mu_n(z)\} - \Pi_{\mu_n}(\alpha_0 - \overline{\alpha}_0)\{\mathcal{I}_Y - \mu_0(z)\} \|
   &  = o_p(1).
    \end{align*}
    We conclude that
     \begin{align*}
        \|A_{n,0} - A_0\| = o_p(1).
    \end{align*}
By \ref{controlEP::donsker}, $A_{n,0}-A_0$ belongs, with probability tending to
one, to a fixed Donsker class. Hence, by asymptotic equicontinuity of the
empirical process on Donsker classes \citep{van1996weak} and the fact that
$\|A_{n,0}-A_0\|=o_p(1)$, we have $(P_n-P_0)(A_{n,0}-A_0)=o_p(n^{-1/2})$.

\end{proof}

\begin{proof}[Proof of Lemma \ref{lemma::negEPRiesz}]
By the triangle inequality and boundedness in \ref{controlEP::bounded},
\begin{align*}
\|\phi_{r_{n,0}} - \phi_{r_0} -  r_{n,0} \alpha_n + r_0 \alpha_0\|
& \lesssim \|\phi_{r_{n,0}} - \phi_{r_0} \| + \|r_{n,0} \alpha_n - r_0 \alpha_0\|\\
& \lesssim \|\phi_{r_{n,0}} - \phi_{r_0} \| + \|r_{n,0} - r_0\| + \|\alpha_n - \alpha_0\|.
\end{align*}
By \ref{controlEP::lip},
$\|\phi_{r_{n,0}} - \phi_{r_0} \|  = O_p(\|r_{n,0} - r_0 \| )$. Hence,
\[
\|\phi_{r_{n,0}} - \phi_{r_0} -  r_{n,0} \alpha_n + r_0 \alpha_0\|
= O_p(\|r_{n,0} - r_0\|) + O_p(\|\alpha_n - \alpha_0\|).
\]
By symmetry, applying \ref{controlEP::couplingriesz} and arguing exactly as in
the proof of Lemma~\ref{lemma::negEPOutcome}, we have $\|r_{n,0} - r_0\|=o_p(1)$
and therefore
$\|\phi_{r_{n,0}} - \phi_{r_0} -  r_{n,0} \alpha_n + r_0 \alpha_0\|=o_p(1)$.
By \ref{controlEP::donskerriesz}, $\phi_{r_{n,0}} - \phi_{r_0} -  r_{n,0} \alpha_n + r_0 \alpha_0$ belongs, with probability
tending to one, to a fixed Donsker class. Hence, by asymptotic equicontinuity
of the empirical process on Donsker classes \citep{van1996weak} and the
convergence established above, we have $(P_n-P_0)(B_{n,0}-B_0)=o_p(n^{-1/2})$.

\end{proof}

\section{Proofs of supporting theory and notation}

In this section, we introduce notation and collect technical lemmas used to
prove Theorems~\ref{theorem::DRinference} and~\ref{theorem::DRboot}.

\subsection{Notation}
\label{app:notationcross}
\paragraph{Cross-averaging and cross-fitting notation.} Let $J \in \mathbb{N}$ denote a fixed number of cross-fitting splits. Let
$\mathcal{D}_n^1,\mathcal{D}_n^2,\ldots,\mathcal{D}_n^J$ be a partition of the
available data $\mathcal{D}_n$ into $J$ subsets of approximately equal size,
with corresponding index sets $\mathcal{I}^1_n,\mathcal{I}^2_n,\ldots,\mathcal{I}^J_n$.
For each $i \in [n]$, let $j(i) \in [J]$ denote the index of the fold containing
observation $i$, so that $i \in \mathcal{I}_{n}^{j(i)}$. For a collection of fold-specific functions
$\nu_{n,\diamond} := \{\nu_{n,j}: j \in [J]\}$, define the fold-averaged operators
\[
\overline{P}_0\nu_{n,\diamond} := \frac{1}{J}\sum_{j=1}^J P_0 \nu_{n,j},
\qquad
\overline{P}_n\nu_{n,\diamond} := \frac{1}{J}\sum_{j=1}^J P_{n,j} \nu_{n,j}.
\]
This notation allows us to view the collections of calibrated cross-fitted
estimators $\alpha_{n,\diamond}^* := \{\alpha_{n,j}^*: j \in [J]\}$ and
$\mu_{n,\diamond}^* := \{\mu_{n,j}^*: j \in [J]\}$ as extended functions on
$\mathcal{A} \times \mathcal{W} \times [J]$. Specifically, let $j_{\mathrm{rv}} \in [J]$ be uniformly
distributed and independent of the data $\mathcal{D}_n$. Then
$\widetilde{Z} := (Z, j_{\mathrm{rv}}) \sim \overline{P}_0$, since $Z \sim P_0$
and $P(j_{\mathrm{rv}} = j) = 1/J$ for $j \in [J]$.

For a collection of fold-specific functions $f_{\diamond} := \{f_j : j \in [J]\}$
with each $f_j \in L^2(P_0)$, define the associated $L^2(\overline{P}_0)$ norm by
\[
\|f_{\diamond}\|_{\overline{P}_0} := \{\overline{P}_0(f_{\diamond}^2)\}^{1/2}.
\]
Under this notation, we may view $f_{\diamond}$ as a single function on
$\mathcal{Z}\times[J]$ and hence as an element of $L^2(\overline{P}_0)$. We define the cross-averaged functionals
\[
\overline{\psi}_n(f_{\diamond}) := \frac{1}{J}\sum_{j=1}^J \psi_{n,j}(f_{j}),
\qquad
\overline{\psi}_0(f_{\diamond}) := \frac{1}{J}\sum_{j=1}^J \psi_{0}(f_{j}).
\]
Let
$v_{\diamond} := \{v_j:\mathcal{A}\times\mathcal{W}\to\mathbb{R}^k\}_{j=1}^J$ be
feature maps, and define the fold-averaged projection operator
$\overline{\Pi}_{v_{\diamond}}:\mathcal{H}^J\to\mathcal{H}^J$ by
\[
(\overline{\Pi}_{v_{\diamond}} f_{\diamond})_j := \theta^\star \circ v_j,
\qquad
\theta^\star \in \arg\min_{\theta}\, \|f_{\diamond} - \theta \circ v_{\diamond}\|_{\overline{P}_0}^2,
\]
where the minimization is over measurable $\theta:\mathbb{R}^k\to\mathbb{R}$ such
that $\|\theta\circ v_{\diamond}\|_{\overline{P}_0}<\infty$. Thus,
$\overline{\Pi}_{v_{\diamond}} f_{\diamond}$ is the $L^2(\overline{P}_0)$
projection of $f_{\diamond}$ onto functions of $v_{\diamond}$, using a single map
$\theta$ shared across folds.

This notation is only a bookkeeping device for cross-fitting. Equivalently, one
may regard \(j_{\mathrm{rv}}\) as a uniformly distributed fold label, independent
of \(Z\), and view \(f_{\diamond}\) and \(v_{\diamond}\) as the single functions
\((z,j)\mapsto f_j(z)\) and \((z,j)\mapsto v_j(z)\) on the augmented space
\(\mathcal Z\times[J]\). Then \(\overline{P}_0\) is the law of
\((Z,j_{\mathrm{rv}})\), and \(\overline{\Pi}_{v_{\diamond}} f_{\diamond}\) is
the ordinary \(L^2(\overline{P}_0)\) projection of \(f_{\diamond}\) onto the
class of functions of \(v_{\diamond}\), with a common regression map
\(\theta\) pooled across folds. In particular, if \(J=1\), or if
\(f_j=f\) and \(v_j=v\) for all \(j\), then this notation reduces to the usual
one-sample notation: \(\overline{P}_0=P_0\),
\(\|f_{\diamond}\|_{\overline{P}_0}=\|f\|_{P_0}\), and
\(\overline{\Pi}_{v_{\diamond}} f_{\diamond}=\Pi_v f\). Thus, the
cross-averaged operators are the standard population operators applied on an
augmented probability space that keeps track of the fold label.

\paragraph{Entropy numbers and integrals.} To control empirical process remainders, we use maximal inequalities based on uniform entropy integrals. For a uniformly bounded function class $\mathcal{F}$, let $N(\epsilon,\mathcal{F},L_2(P))$ denote the $\epsilon$-covering number \citep{van1996weak} of $\mathcal{F}$ relative to the $L^2(P)$ metric, and define the uniform entropy integral of $\mathcal{F}$ by
\begin{equation*}
\mathcal{J}(\delta,\mathcal{F}) := \int_{0}^{\delta} \sup_{Q}\sqrt{\log N(\epsilon,\mathcal{F},L_2(Q))}\,d\epsilon\ ,
\end{equation*}
where the supremum is taken over all discrete probability distributions $Q$. In contrast to \citet{van1996weak}, we do not define the uniform entropy integral relative to an envelope function. We may do so because all function classes we consider are uniformly bounded; thus, any uniformly bounded envelope would change the uniform entropy integral in \citet{van1996weak} only by a multiplicative constant.

\paragraph{Function classes used to control empirical process remainders.}  We define relevant function classes that will be used to control empirical process remainders in our proofs. Let $M/2 < \infty$ be an upper bound on the nuisance functions and estimators in \ref{cond::estnuisbound}. Let $\mathcal{F}_{\text{iso}} \subset \{f:[-M,M]\rightarrow \mathbb{R} \mid \text{$f$ is monotone nondecreasing}\}$ consist of all univariate isotonic functions with supremum norm at most $M < \infty$. Let $M' < \infty$ be an upper bound on the total variation norms of the calibration functions $\{\theta_{n,j,\mu}, \theta_{n,j,\alpha}: j \in [J]\}$ in \ref{cond::variation}. Let $\mathcal{F}_{\text{TV}} \subset \{f:[-M,M]\rightarrow \mathbb{R} \mid \text{$f$ has finite total variation}\}$ consist of all univariate functions with total variation norm at most $3M'$ and supremum norm at most $M < \infty$.

For $j \in [J]$, define the function classes
\[
\mathcal{S}_{n,j} := \left\{f \circ \mu_{n,j} \,\middle|\, f \in \mathcal{F}_{\text{TV}} \cup \mathcal{F}_{\text{iso}} \right\}
\quad\text{and}\quad
\mathcal{R}_{n,j} := \left\{g \circ \alpha_{n,j} \,\middle|\, g \in \mathcal{F}_{\text{TV}} \cup \mathcal{F}_{\text{iso}} \right\}.
\]
By \ref{cond::variation} and \ref{cond::estnuisbound}, we will show that
$r_{n,j}^* = \Pi_{\mu_{n,j}^*}\Delta_{\alpha,n,j}$ and
$s_{n,j}^* = \Pi_{\alpha_{n,j}^*}\Delta_{\mu,n,j}$
can be expressed as one-dimensional transformations of $\mu_{n,j}$ and $\alpha_{n,j}$, respectively, whose total variation norms are bounded by $3M'$. Hence, with probability tending to one, $\mu_{n,j}^* = f_n \circ \mu_{n,j}$ and $s_{n,j}^*$ belong to $\mathcal{S}_{n,j}$. Similarly, $\alpha_{n,j}^* = g_n \circ \alpha_{n,j}$ and $r_{n,j}^*$ belong to $\mathcal{R}_{n,j}$. Note that $\mathcal{S}_{n,j}$ and $\mathcal{R}_{n,j}$ are fixed function classes conditional on the $j$-th training set $\mathcal{D}_n \backslash \mathcal{C}_j$.

\paragraph{Conventions for proofs.} In our proofs, we may, without loss of generality, treat the bounds in \ref{cond::estnuisbound} (and \ref{cond::estboundboot}) as holding almost surely rather than only with probability tending to one. This is justified because our results are stated in terms of stochastic order (e.g., $O_p$ and $o_p$), which is unaffected by restricting attention to events whose probabilities tend to one. Formally, we work on the event
\[
\mathcal{B}_{n} := \Bigl\{\|s_{n,j}^*\|_\infty \le 2M,\ \|r_{n,j}^*\|_\infty \le 2M,\ \|\alpha_{n,j}^*\|_\infty \le 2M,\ \|\mu_{n,j}^*\|_\infty \le 2M, \, \forall j \in [J]\Bigr\},
\]
which satisfies $P_0(\mathcal{B}_{n})\to 1$ by \ref{cond::estnuisbound}. Since all subsequent bounds are derived on $\mathcal{B}_{n}$, we suppress the indicator $1\{\mathcal{B}_{n}\}$ in the notation and interpret stochastic order statements (e.g., $O_p(\cdot)$, $o_p(\cdot)$) as holding under $P_0$ on events whose probabilities tend to one. For two quantities $x$ and $y$, we use the expression  $x \lesssim y$ to mean that $x$ is upper bounded by $y$ times a universal constant that may only depend on global constants that appear in \ref{cond::estnuisbound}. We denote by \( \mathcal{I}_Y \) the coordinate projection map \( z = (w, a, y) \mapsto y \); for example, \( P_0 \{\mathcal{I}_Y - \mu_0\} = E_0[Y - \mu_0(A,W)] \).

\subsection{Bias expansions for cross-fitted one-step estimators}

We begin by deriving a generic error expansion for cross-fitted one-step estimators, which will serve as the starting point for our analysis. We then derive a cross-fitted analogue of Theorem~\ref{theorem::DRbiasexpansion}, which we will use to establish DRAL for our isocalibrated one-step estimator.

Let $\alpha_{n,\diamond}=\{\alpha_{n,j}: j\in[J]\}$ and $\mu_{n,\diamond}=\{\mu_{n,j}: j\in[J]\}$ be arbitrary fold-specific estimators of $\alpha_0$ and $\mu_0$, respectively, which may be uncalibrated. Define the associated one-step debiased estimator by
\[
\tau_n := \frac{1}{J}\sum_{j=1}^J \psi_{n,j}(\mu_{n,j})
\;+\;
\frac{1}{n}\sum_{i=1}^n \alpha_{n,j(i)}(A_i,W_i)\{Y_i-\mu_{n,j(i)}(A_i,W_i)\}.
\]

The following expansion of the estimation error $\tau_n-\tau_0$ is a standard starting point for the analysis of doubly robust estimators.

\begin{lemma}[Standard doubly robust bias expansion]
\label{lemma::standardBiasExpansion}
Assume Conditions~\ref{cond::linear} and~\ref{cond::pathwise}. Then
    \begin{align*}
     \tau_n - \tau_0 = (\overline{P}_n - \overline{P}_0) \left\{\phi_{0, \mu_{n,\diamond}} + D_{\mu_{n, \diamond}, \alpha_{n, \diamond}} \right\} +  \langle \alpha_{n, \diamond} - \alpha_{0}, \mu_{0} -  \mu_{n, \diamond} \rangle_{\overline{P}_0} + Rem_{\mu_{n,\diamond}}(\psi_{n,\diamond}, \psi_0).
\end{align*}
where $Rem_{\mu_{n,\diamond}}(\psi_{n,\diamond}, \psi_0)  := \overline{\psi}_{n}(\mu_{n,\diamond}) - \overline{\psi}_0(\mu_{n,\diamond}) - \overline{P}_n \phi_{0, \mu_{n,\diamond}} $.
\end{lemma}
\begin{proof}
We can write
\[
\frac{1}{J}\sum_{j=1}^J \{\psi_{n,j}(\mu_{n,j}) - \psi_0(\mu_{n,j})\}
= (\overline{P}_n - \overline{P}_0)\phi_{0,\mu_{n,\diamond}}
+ Rem_{\mu_{n,\diamond}}(\psi_{n,\diamond},\psi_0),
\]
where
\[
Rem_{\mu_{n,\diamond}}(\psi_{n,\diamond},\psi_0)
:= \frac{1}{J}\sum_{j=1}^J \left\{\psi_{n,j}(\mu_{n,j}) - \psi_0(\mu_{n,j})
- (P_{n,j}-P_0)\phi_{0,\mu_{n,j}}\right\}.
\]
The estimation error of $\tau_n$ can be decomposed as
\begin{align*}
\tau_n - \tau_0
&= \frac{1}{J}\sum_{j=1}^J \psi_{n,j}(\mu_{n,j})
+ \overline{P}_n D_{\mu_{n,\diamond},\alpha_{n,\diamond}}
- \psi_0(\mu_0) \\
&= (\overline{P}_n - \overline{P}_0)\left\{\phi_{0,\mu_{n,\diamond}}
+ D_{\mu_{n,\diamond},\alpha_{n,\diamond}}\right\}
+ \overline{P}_0 D_{\mu_{n,\diamond},\alpha_{n,\diamond}}
+ \frac{1}{J}\sum_{j=1}^J \{\psi_0(\mu_{n,j})-\psi_0(\mu_0)\}
+ Rem_{\mu_{n,\diamond}}(\psi_{n,\diamond},\psi_0).
\end{align*}
By the Riesz representation $\psi_0(\cdot)=\langle \alpha_0,\cdot\rangle_{P_0}$, we have
$\psi_0(\mu_{n,j})-\psi_0(\mu_0)=\langle \alpha_0,\mu_{n,j}-\mu_0\rangle_{P_0}$.
Moreover, by the law of iterated expectations and $E_0[Y\mid A,W]=\mu_0(A,W)$,
\begin{align*}
\overline{P}_0 D_{\mu_{n,\diamond},\alpha_{n,\diamond}}
&= \frac{1}{J}\sum_{j=1}^J E_0\left[\alpha_{n,j}(A,W)\{Y-\mu_{n,j}(A,W)\}\right] \\
&= \frac{1}{J}\sum_{j=1}^J E_0\left[\alpha_{n,j}(A,W)\{\mu_0(A,W)-\mu_{n,j}(A,W)\}\right] \\
&= \langle \alpha_{n,\diamond}, \mu_0-\mu_{n,\diamond}\rangle_{\overline{P}_0}.
\end{align*}
Hence,
\[
\overline{P}_0 D_{\mu_{n,\diamond},\alpha_{n,\diamond}}
+ \frac{1}{J}\sum_{j=1}^J \{\psi_0(\mu_{n,j})-\psi_0(\mu_0)\}
= \langle \alpha_{n,\diamond}-\alpha_0,\mu_0-\mu_{n,\diamond}\rangle_{\overline{P}_0},
\]
which yields the claimed expansion.
\end{proof}

The next lemma gives regression- and Riesz-based decompositions that characterize
the leading-order behavior of the cross-product remainder. It is a cross-fitted
analogue of Theorem~\ref{theorem::DRbiasexpansion}, with explicit expressions
for the first-order bias terms, which vanish under suitable orthogonality
conditions on the cross-fitted nuisance estimators. The required orthogonality
conditions are
\[
\overline{P}_n\, s_{n,\diamond}\{\mathcal{I}_Y - \mu_{n,\diamond}\} = 0
\quad\text{and}\quad
\overline{\psi}_n(r_{n,\diamond}) - \overline{P}_n\bigl(r_{n,\diamond}\alpha_{n,\diamond}\bigr) = 0,
\]
where, for each $j \in [J]$,
$r_{n,j} := \bigl(\overline{\Pi}_{\alpha_{n,\diamond}}(\mu_0 - \mu_{n,\diamond})\bigr)_j$
and
$s_{n,j} := \bigl(\overline{\Pi}_{\mu_{n,\diamond}}(\alpha_0 - \alpha_{n,\diamond})\bigr)_j$.
We also recall that
$A_{n,j} \colon z \mapsto s_{n,j}(a,w)\{y - \mu_{n,j}(a,w)\}$ and
$B_{n,j} := \psi_0(r_0) + \phi_{0,r_{n,j}} - r_{n,j}\alpha_{n,j}$.
Finally, let $\Delta_{\alpha,n,j}:=\alpha_{n,j}-\alpha_0$ and
$\Delta_{\mu,n,j}:=\mu_{n,j}-\mu_0$.

\begin{lemma}[Expansion of cross-product remainder]
\label{lemma::outcomeRieszBiasDR}
Assume Conditions~\ref{cond::linear} and~\ref{cond::pathwise}. Then we have the regression-based decomposition:
\begin{align*}
\langle \alpha_0-\alpha_{n,\diamond},\;\mu_{n,\diamond}-\mu_0\rangle_{\overline{P}_0}
&=
-\overline{P}_n A_{n,\diamond} + (\overline{P}_n-\overline{P}_0)A_{n,\diamond} 
+\Big\langle (\Pi_{(\mu_{n,\diamond},\mu_0)}-\Pi_{\mu_{n,\diamond}})\Delta_{\alpha,n,\diamond},\;
\mu_0-\Pi_{\mu_{n,\diamond}}\mu_0 \Big\rangle_{\overline{P}_0}.
\end{align*}
Furthermore, we have the Riesz-based decomposition:
\begin{align*}
\langle \alpha_0-\alpha_{n,\diamond},\;\mu_{n,\diamond}-\mu_0\rangle_{\overline{P}_0}
&=
-\left\{\overline{\psi}_n(r_{n,\diamond})
-\overline{P}_n\bigl(r_{n,\diamond}\alpha_{n,\diamond}\bigr)\right\}
+ (\overline{P}_n-\overline{P}_0)B_{n,\diamond}
\\
&\quad
+\Big\langle (\Pi_{(\alpha_{n,\diamond},\alpha_0)}-\Pi_{\alpha_{n,\diamond}})\Delta_{\mu,n,\diamond},\;
\alpha_0-\Pi_{\alpha_{n,\diamond}}\alpha_0 \Big\rangle_{\overline{P}_0}
+ Rem_{\psi_{n,\diamond}}(r_{n,\diamond}),
\end{align*}
where $Rem_{\psi_{n,\diamond}}(r_{n,\diamond})
:=  \overline{\psi}_{n}(r_{n,\diamond})- \overline{\psi}_0(r_{n,\diamond})- \overline{P}_{n}\phi_{0, r_{n,\diamond}}.$
\end{lemma}
 \begin{proof} 

Our proof largely follows the proof of Theorem~\ref{theorem::DRbiasexpansion}
after a change of notation. Specifically, let $j_{\mathrm{rv}} \in [J]$ be uniformly
distributed and independent of the data $\mathcal{D}_n$. Then
$\widetilde{Z} := (Z, j_{\mathrm{rv}}) \sim \overline{P}_0$, since $Z \sim P_0$
and $P(j_{\mathrm{rv}} = j) = 1/J$ for $j \in [J]$. Our proof follows that of Theorem~\ref{theorem::DRbiasexpansion} applied with the extended data structure
$\widetilde{Z}$ with $P_0$ replaced by $\overline{P}_0$ and with
$\mu_n := \mu_{n,\diamond}$ and $\alpha_n := \alpha_{n,\diamond}$, where we view
$(w,a,j) \mapsto \mu_{n,j}(a,w)$ as a function on
$\mathcal{A}\times\mathcal{W}\times[J]$ (and similarly for $\alpha_{n,j}$).
Under this identification, the $P_0$-projection operator $\Pi_v$ is replaced by
the $\overline{P}_0$-projection operator $\overline{\Pi}_{v_{\diamond}}$.

We begin with the proof of the Riesz-based decomposition, as the regression-based decomposition will follow as a special case. Since $\alpha_0 - \alpha_{n,\diamond} \in L^2(\overline{P}_0)$ is a measurable
function of $(\alpha_{n,\diamond},\alpha_0)$, applying the projection
$\overline{\Pi}_{(\alpha_{n,\diamond},\,\alpha_0)}$ yields
\begin{align*}
\big\langle \alpha_0 - \alpha_{n,\diamond},\, \mu_{n,\diamond} - \mu_0 \big\rangle_{\overline{P}_0}
&=
\Big\langle \alpha_0 - \alpha_{n,\diamond},\,
\overline{\Pi}_{(\alpha_{n,\diamond},\,\alpha_0)}\Delta_{\mu,n,\diamond}
\Big\rangle_{\overline{P}_0}.
\end{align*}
We decompose the preceding display as
\begin{equation}
\label{eqn::DRexpansionproof::cross}
\begin{aligned}
\big\langle \alpha_0 - \alpha_{n,\diamond},\, \mu_{n,\diamond} - \mu_0 \big\rangle_{\overline{P}_0}
&=
\Big\langle \alpha_0 - \alpha_{n,\diamond},\,
\overline{\Pi}_{\alpha_{n,\diamond}}\Delta_{\mu,n,\diamond}\Big\rangle_{\overline{P}_0} \\
&\quad +
\Big\langle
\big(\overline{\Pi}_{(\alpha_{n,\diamond},\,\alpha_0)}-\overline{\Pi}_{\alpha_{n,\diamond}}\big)
(\alpha_0-\alpha_{n,\diamond}),\,
\Delta_{\mu,n,\diamond}-\overline{\Pi}_{\alpha_{n,\diamond}}\Delta_{\mu,n,\diamond}
\Big\rangle_{\overline{P}_0} \\
&=
\Big\langle \alpha_{n,\diamond} - \alpha_0,\, r_{n,\diamond}\Big\rangle_{\overline{P}_0}
-
\Big\langle
\big(\overline{\Pi}_{(\alpha_{n,\diamond},\,\alpha_0)}-\overline{\Pi}_{\alpha_{n,\diamond}}\big)
\Delta_{\alpha,n,\diamond},\,
\Delta_{\mu,n,\diamond}-\overline{\Pi}_{\alpha_{n,\diamond}}\Delta_{\mu,n,\diamond}
\Big\rangle_{\overline{P}_0},
\end{aligned}
\end{equation}
where $r_{n,\diamond} = -\,\overline{\Pi}_{\alpha_{n,\diamond}}\Delta_{\mu,n,\diamond}$.
By the Riesz representation property for $\alpha_0$ in \eqref{eqn::reiszrep}, the
first term on the right-hand side of \eqref{eqn::DRexpansionproof::cross} can be
written as
\begin{align*}
\big\langle \alpha_{n,\diamond} - \alpha_0,\, r_{n,\diamond}\big\rangle_{\overline{P}_0}
&=
-\psi_0(r_{n,\diamond}) + \overline{P}_0\!\left\{\alpha_{n,\diamond} r_{n,\diamond}\right\} \\
&=
-\overline{\psi}_n(r_{n,\diamond}) + \overline{P}_0\!\left\{\alpha_{n,\diamond} r_{n,\diamond}\right\}
-\psi_0(r_{n,\diamond}) + \overline{\psi}_n(r_{n,\diamond}),
\end{align*}
where, in the second equality, we add and subtract the cross-averaged functional
estimate
\[
\overline{\psi}_n(r_{n,\diamond}) := \frac{1}{J}\sum_{j=1}^J \psi_{n,j}(r_{n,j}).
\]
Rearranging gives
\begin{align*}
\big\langle \alpha_{n,\diamond} - \alpha_0,\, r_{n,\diamond}\big\rangle_{\overline{P}_0}
&=
-(\overline{P}_n-\overline{P}_0)\!\left\{\alpha_{n,\diamond} r_{n,\diamond}\right\}
+\left\{\overline{\psi}_n(r_{n,\diamond}) - \psi_0(r_{n,\diamond})\right\}
+\left\{\overline{P}_n\!\left(\alpha_{n,\diamond} r_{n,\diamond}\right)-\overline{\psi}_n(r_{n,\diamond})\right\}.
\end{align*}
Next, add and subtract $\overline{P}_n \phi_{r_{n,\diamond}}$ and use
$\overline{P}_0 \phi_{r_{n,\diamond}} = 0$ to obtain
\begin{align*}
\big\langle \alpha_{n,\diamond} - \alpha_0,\, r_{n,\diamond}\big\rangle_{\overline{P}_0}
&=
(\overline{P}_n-\overline{P}_0)\!\left\{-\alpha_{n,\diamond} r_{n,\diamond}
+\phi_{r_{n,\diamond}}\right\} \\
&\quad
+\left\{\overline{\psi}_n(r_{n,\diamond}) - \psi_0(r_{n,\diamond})
-(\overline{P}_n-\overline{P}_0)\phi_{r_{n,\diamond}}\right\} \\
&\quad
+\left\{\overline{P}_n\!\left(\alpha_{n,\diamond} r_{n,\diamond}\right)-\overline{\psi}_n(r_{n,\diamond})
-\overline{P}_n \phi_{r_{n,\diamond}}\right\}.
\end{align*}
Recalling
\[
B_{n,\diamond} := \psi_0(r_{n,\diamond} ) -\alpha_{n,\diamond} r_{n,\diamond} + \phi_{r_{n,\diamond}},
\qquad
Rem_{\psi_n}(r_{n,\diamond})
:= \overline{\psi}_n(r_{n,\diamond}) - \psi_0(r_{n,\diamond})
-(\overline{P}_n-\overline{P}_0)\phi_{r_{n,\diamond}},
\]
we may rewrite the preceding display as
\[
\big\langle \alpha_{n,\diamond} - \alpha_0,\, r_{n,\diamond}\big\rangle_{\overline{P}_0}
=
(\overline{P}_n-\overline{P}_0)B_{n,\diamond}
+ Rem_{\psi_n}(r_{n,\diamond})
+\left\{\overline{P}_n\!\left(\alpha_{n,\diamond} r_{n,\diamond}\right)-\overline{\psi}_n(r_{n,\diamond})
-\overline{P}_n \phi_{r_{n,\diamond}}\right\}.
\]
Substituting this display into \eqref{eqn::DRexpansionproof::cross} yields
\begin{equation}
\label{eqn::DRexpansionproof::cross2}
\begin{aligned}
\big\langle \alpha_0 - \alpha_{n,\diamond},\, \mu_{n,\diamond} - \mu_0 \big\rangle_{\overline{P}_0}
&=
(\overline{P}_n-\overline{P}_0)B_{n,\diamond}
+ Rem_{\psi_n}(r_{n,\diamond})
+\left\{\overline{P}_n\!\left(\alpha_{n,\diamond} r_{n,\diamond}\right)-\overline{\psi}_n(r_{n,\diamond})
-\overline{P}_n \phi_{r_{n,\diamond}}\right\} \\
&\quad
-\Big\langle
\big(\overline{\Pi}_{(\alpha_{n,\diamond},\,\alpha_0)}-\overline{\Pi}_{\alpha_{n,\diamond}}\big)\Delta_{\alpha,n,\diamond},\,
\Delta_{\mu,n,\diamond}-\overline{\Pi}_{\alpha_{n,\diamond}}\Delta_{\mu,n,\diamond}
\Big\rangle_{\overline{P}_0} \\
&=
(\overline{P}_n-\overline{P}_0)B_{n,\diamond}
+ Rem_{\psi_n}(r_{n,\diamond})
+\left\{\overline{P}_n\!\left(\alpha_{n,\diamond} r_{n,\diamond}\right)-\overline{\psi}_n(r_{n,\diamond})
-\overline{P}_n \phi_{r_{n,\diamond}}\right\} \\
&\quad
+\Big\langle \alpha_0 - \overline{\Pi}_{\alpha_{n,\diamond}}\alpha_0,\,
\Delta_{\mu,n,\diamond}-\overline{\Pi}_{\alpha_{n,\diamond}}\Delta_{\mu,n,\diamond}
\Big\rangle_{\overline{P}_0},
\end{aligned}
\end{equation}
where we use that
\[
\big(\overline{\Pi}_{(\alpha_{n,\diamond},\,\alpha_0)}-\overline{\Pi}_{\alpha_{n,\diamond}}\big)\Delta_{\alpha,n,\diamond}
=
\overline{\Pi}_{\alpha_{n,\diamond}}\alpha_0 - \alpha_0
=
-\big(\alpha_0 - \overline{\Pi}_{\alpha_{n,\diamond}}\alpha_0\big).
\]
The Riesz-based decomposition now follows.  

As in the proof of Theorem~\ref{theorem::DRbiasexpansion}, the regression-based
decomposition follows as a special case. For the regression-based decomposition,
we apply the Riesz-based decomposition with
\[
\psi_0(\alpha) := E_0\{\alpha(A,W)\mu_0(A,W)\}
\quad\text{and}\quad
\psi_{n,j}(\alpha) := E_{P_{n,j}}\{\alpha(A,W)Y\}.
\]
In particular,
$\mu_0$ is the Riesz representer of the linear functional
$\psi_0:\alpha \mapsto E_0\{\alpha(A,W)\mu_0(A,W)\}$.  For this
choice of $\psi_{n,j}$, we have $Rem_{\psi_n}(\alpha)=0$ with
$\widetilde{\phi}_{\alpha}(z) := \alpha(A,W)Y - \psi_0(\alpha)$. Under these
definitions, $B_0$ and $B_{n,\diamond}$ coincide with $A_0$ and $A_{n,\diamond}$,
respectively. 

\end{proof}

\subsection{Sufficient conditions for a negligible remainder: cross-fitted version of Lemma~\ref{lemma::sufficientcoupling}}
\label{appendix::sufficientCondsCross}

Recall
\begin{align*}
    \overline{\gamma}_{\mu,0}(\mu_{n, \diamond}^*, \alpha_{n, \diamond}^*)
    &=
    \left\langle
    \overline{\Pi}_{(\mu_{n,\diamond}^*,\, \mu_0)}\Delta_{\alpha,n,\diamond}
    - \overline{\Pi}_{\mu_{n,\diamond}^*}\Delta_{\alpha,n,\diamond}
    \,,\,
    \mu_0 - \overline{\Pi}_{\mu_{n,\diamond}^*}\mu_0
    \right\rangle_{\overline{P}_0},\\
    \overline{\gamma}_{\alpha,0}(\mu_{n, \diamond}^*, \alpha_{n, \diamond}^*)
    &=
    \left\langle
    \overline{\Pi}_{(\alpha_{n,\diamond}^*,\, \alpha_0)}\Delta_{\mu,n,\diamond}
    - \overline{\Pi}_{\alpha_{n,\diamond}^*}\Delta_{\mu,n,\diamond}
    \,,\,
    \alpha_0 - \overline{\Pi}_{\alpha_{n,\diamond}^*}\alpha_0
    \right\rangle_{\overline{P}_0}.
\end{align*}
We give sufficient conditions for the following requirement.
\begin{enumerate}[label=\bf{BB\arabic*)}, ref={BB\arabic*},series=sketch]
  \item \textit{Controlled partial orthogonality:} \label{cond:ortho:cross}
    \begin{enumerate}[label={\roman*)}, ref={\ref{cond:ortho:cross}\roman*}]\vspace{-.05in}
      \item \label{cond:orthoA:cross} \( \overline{\gamma}_{\mu,0}(\mu_{n, \diamond}^*, \alpha_{n, \diamond}^*) = O_p(  \|\mu_{n,\diamond}^* - \mu_0\|_{\overline{P}_0}^2)\),\vspace{-.05in}
      \item \label{cond:orthoB:cross} \( \overline{\gamma}_{\alpha,0}(\mu_{n, \diamond}^*, \alpha_{n, \diamond}^*) = O_p( \|\alpha_{n,\diamond}^* - \alpha_0\|_{\overline{P}_0}^2)\).
    \end{enumerate}
\end{enumerate}

The first sufficient condition is as follows.
\begin{enumerate}[label=\bf{\ref{cond:ortho:cross}$^{*}$}, ref={\ref{cond:ortho:cross}$^{*}$}]
\item \textit{Projection error coupling:} \label{cond:projectioncoupling1:cross}
  \begin{enumerate}[label={\roman*)}, ref={\ref{cond:projectioncoupling1:cross}\roman*}]\vspace{-.05in}
  \item \label{cond:projectioncoupling1A:cross}
  $\big\| \big(\overline{\Pi}_{(\mu_{n,\diamond}, \mu_0)} - \overline{\Pi}_{\mu_{n,\diamond}}\big)\Delta_{\alpha,n,\diamond}\big\|_{\overline{P}_0}
  = O_p\!\big( \|\mu_{n,\diamond}^* - \mu_0\|_{\overline{P}_0}\big)$\,;
  \vspace{-.05in}
  \item \label{cond:projectioncoupling1B:cross}
  $\big\| \big(\overline{\Pi}_{(\alpha_{n,\diamond}, \alpha_0)} - \overline{\Pi}_{\alpha_{n,\diamond}}\big)\Delta_{\mu,n,\diamond}\big\|_{\overline{P}_0}
  = O_p\!\big(\|\alpha_{n,\diamond}^* - \alpha_0\|_{\overline{P}_0}\big)$\,.
  \end{enumerate}
\end{enumerate}

The second sufficient condition is as follows.  In the condition, let \(j_{\mathrm{rv}}\) be uniformly distributed on \(\{1,2,\dots,J\}\), independently of \(\{Z_1,\dots,Z_n\}\). Let \(\overline{P}_0\) denote the joint distribution of \((Z,j_{\mathrm{rv}})\), and write \(\overline{E}_0 := E_{\overline{P}_0}\).

\begin{enumerate}[label=\bf{\ref{cond:ortho:cross}$^{**}$}, ref={\ref{cond:ortho:cross}$^{**}$}]
\item \textit{Lipschitz continuity:} Denoting \(\mathcal{D}_n := \{Z_1,\dots,Z_n\}\), assume that, for all sufficiently large \(n\), one of the following hold with \(P_0^n\)-probability one:
\label{cond:projectioncoupling2:cross}
\begin{enumerate}[
  label={\roman*)},
  ref=\ref{cond:projectioncoupling2:cross}\roman*,
  leftmargin=-10pt,
  labelsep=-0.7em,
  labelindent=0pt,
  align=left
]
\item \label{cond:projectioncoupling2A:cross}
 the map
\[
(\widehat m,m) \mapsto \overline{E}_0 \!\left\{\Delta_{\alpha,n,j_{\mathrm{rv}}}(A,W)\,\middle|\, 
\mu_{n,j_{\mathrm{rv}}}(A,W)=\widehat m,\ \mu_0(A,W)=m,\ \mathcal D_n\right\}
\]
is Lipschitz continuous with a uniformly bounded Lipschitz constant;

\item \label{cond:projectioncoupling2B:cross}
 the map
\[
(\widehat a,a) \mapsto \overline{E}_0\!\left\{\Delta_{\mu,n,j_{\mathrm{rv}}}(A,W)\,\middle|\, 
\alpha_{n,j_{\mathrm{rv}}}(A,W)=\widehat a,\ \alpha_0(A,W)=a,\ \mathcal D_n\right\}
\]
is Lipschitz continuous with a uniformly bounded Lipschitz constant.
\end{enumerate}
\end{enumerate}

\begin{lemma}[Sufficient conditions for \ref{cond:ortho:cross}]
\label{lemma::sufficientcoupling:cross}
Conditions \ref{cond:orthoA:cross} and \ref{cond:orthoB:cross} are implied by
\ref{cond:projectioncoupling1A:cross} and \ref{cond:projectioncoupling1B:cross},
respectively. Moreover, \ref{cond:projectioncoupling1A:cross} and
\ref{cond:projectioncoupling1B:cross} are implied by
\ref{cond:projectioncoupling2A:cross} and \ref{cond:projectioncoupling2B:cross},
respectively.
\end{lemma}
\begin{proof}
We prove the result by reduction to the non--cross-fitted version
Lemma~\ref{lemma::sufficientcoupling} via an extended data-structure construction.

Let $j_{\mathrm{rv}} \in [J]$ be uniformly distributed and independent of the
data $\mathcal{D}_n$, and define the extended observation
$\widetilde{Z} := (Z, j_{\mathrm{rv}})$. Then $\widetilde{Z}\sim \overline{P}_0$.
Moreover, we may view $\mu_{n,\diamond}^*,\alpha_{n,\diamond}^*$ as functions on
$\mathcal{A}\times\mathcal{W}\times[J]$ via
$(a,w,j)\mapsto \mu_{n,j}^*(a,w)$ and $(a,w,j)\mapsto \alpha_{n,j}^*(a,w)$, and
similarly for $\Delta_{\alpha,n,\diamond}$ and $\Delta_{\mu,n,\diamond}$.
Under this identification, the $\overline{P}_0$-projection
$\overline{\Pi}_{v_\diamond}$ coincides with the usual $L^2(\overline{P}_0)$
projection operator $\Pi_v$ applied to the corresponding function $v$ on the
extended space.

With this notation, $\gamma_{\mu,n,0}^*$ and $\gamma_{\alpha,n,0}^*$ are exactly
the quantities $\gamma_{\mu,n,0}$ and $\gamma_{\alpha,n,0}$ from the
non--cross-fitted setting in Section \ref{sec::genstrat}, but computed under $\overline{P}_0$ and with
$(\mu_n,\alpha_n)$ replaced by $(\mu_{n,\diamond}^*,\alpha_{n,\diamond}^*)$.
Likewise, Conditions \ref{cond:projectioncoupling1A:cross} and
\ref{cond:projectioncoupling1B:cross} (respectively,
\ref{cond:projectioncoupling2A:cross} and \ref{cond:projectioncoupling2B:cross})
are precisely the consistency conditions from
Lemma~\ref{lemma::sufficientcoupling}, again after replacing $P_0$ by
$\overline{P}_0$ and $(\mu_n,\alpha_n)$ by $(\mu_{n,\diamond}^*,\alpha_{n,\diamond}^*)$.
Therefore, applying Lemma~\ref{lemma::sufficientcoupling} on the extended space
(with \(P_0\) replaced by \(\overline{P}_0\)) yields that
\ref{cond:projectioncoupling1A:cross} implies \ref{cond:orthoA:cross}, and
\ref{cond:projectioncoupling1B:cross} implies \ref{cond:orthoB:cross}.

The second claim follows similarly from the second implication of
Lemma~\ref{lemma::sufficientcoupling}. In particular,
\ref{cond:projectioncoupling2A:cross} implies that the projection
\[
(\overline{\Pi}_{(\mu_{n,\diamond},\mu_0)}\Delta_{\alpha,n,\diamond})(A,W)
\]
is a Lipschitz transformation of
\((\mu_{n,\diamond}(A,W), \mu_0(A,W))\), and
\ref{cond:projectioncoupling2B:cross} implies that the projection
\[
(\overline{\Pi}_{(\alpha_{n,\diamond},\alpha_0)}\Delta_{\mu,n,\diamond})(A,W)
\]
is a Lipschitz transformation of
\((\alpha_{n,\diamond}(A,W), \alpha_0(A,W))\). Hence,
\ref{cond:projectioncoupling2A:cross} implies
\ref{cond:projectioncoupling1A:cross}, and
\ref{cond:projectioncoupling2B:cross} implies
\ref{cond:projectioncoupling1B:cross}. This completes the proof.
\end{proof}

\subsection{Entropy bounds and local maximal inequalities for empirical processes}

The following lemma controls the uniform entropy integral for the function classes $\mathcal{R}_{n,j}$ and $\mathcal{S}_{n,j}$, which contain $\mu_{n,j}^*$ and $\alpha_{n,j}^*$, respectively. 

\begin{lemma}
Under Conditions~\ref{cond::estnuisbound} and~\ref{cond::variation}, for each $j \in [J]$, we have $\mathcal{J}(\delta, \mathcal{R}_{n,j}) + \mathcal{J}(\delta, \mathcal{S}_{n,j})
\lesssim \sqrt{\delta}.$
\label{lemma::entropy}
\end{lemma}
\begin{proof}
We prove the claim for $\mathcal{S}_{n,j}$; the result for $\mathcal{R}_{n,j}$
follows by an identical argument. We have
\begin{align*}
\mathcal{J}(\delta,\mathcal{S}_{n,j})
&= \int_0^{\delta} \sup_Q \sqrt{N(\varepsilon,\mathcal{S}_{n,j},\norm{\,\cdot\,}_Q)}\,d\varepsilon \\
&= \int_0^{\delta} \sup_Q \sqrt{N\!\left(\varepsilon, \mathcal{F}_{\text{TV}} \cup \mathcal{F}_{\text{iso}},
\norm{\,\cdot\,}_{Q \circ \mu_{n,j}^{-1}}\right)}\, d\varepsilon \\
&\le \int_0^{\delta} \sup_Q \sqrt{N\!\left(\varepsilon, \mathcal{F}_{\text{TV}},
\norm{\,\cdot\,}_{Q \circ \mu_{n,j}^{-1}}\right)}\, d\varepsilon
+ \int_0^{\delta} \sup_Q \sqrt{N\!\left(\varepsilon, \mathcal{F}_{\text{iso}},
\norm{\,\cdot\,}_{Q \circ \mu_{n,j}^{-1}}\right)}\, d\varepsilon \\
&\le \mathcal{J}(\delta, \mathcal{F}_{\text{TV}}) + \mathcal{J}(\delta, \mathcal{F}_{\text{iso}}),
\end{align*}
where $Q \circ \mu_{n,j}^{-1}$ denotes the push-forward probability measure of
$\mu_{n,j}(W)$ conditional on the training sample
$\mathcal{D}_n \setminus \mathcal{C}^{(j)}$. By standard metric entropy bounds
for monotone functions,
$\mathcal{J}(\delta, \mathcal{F}_{\text{iso}}) \lesssim \sqrt{\delta}$
\citep{van1996weak}. Moreover, since any function of finite variation can be
written as the difference of two bounded monotone functions (e.g., Theorem~4 of \citet{royden1963real}), it follows that
\[
\mathcal{J}(\delta, \mathcal{F}_{\text{TV}})
\lesssim \mathcal{J}(\delta, \mathcal{F}_{\text{iso}})
\lesssim \sqrt{\delta}.
\]
 The result follows.
\end{proof}

The next lemma will be used to show that $r_{n,j}^*$ and $s_{n,j}^*$ are elements of $\mathcal{R}_{n,j} \cup \mathcal{S}_{n,j}$. 
 
\begin{lemma}
\label{lemma::TVnormBounded}
Let $\widehat{\eta}, \eta_0 \in L^2(P_0)$ and let $\widehat{\eta}^* := \widehat{\theta} \circ \widehat{\eta}$ for a isotonic (monotone nondecreasing) piece-wise constant transformation $ \widehat{\theta} : \mathbb{R} \rightarrow \mathbb{R}$ Suppose that there exists an $M < \infty$ such that the map $\widehat{\theta}_0 : \mathbb{R} \rightarrow \mathbb{R}$, defined such that $\widehat{\theta}_0(\widehat{\eta}(z)) := E_0[\eta_0(Z) \mid \widehat{\eta}(Z) = \widehat{\eta}(z), \mathcal{D}_n]$,  has a total variation norm that is almost surely bounded by $M$. Then, the map \(\widehat{\theta}_0^*\), defined such that \(\widehat{\theta}_0^*(\widehat{\eta}(z)) =  E_0[\eta_0(Z) \mid \widehat{\eta}^*(Z) = \widehat{\theta}(\widehat{\eta}(z)), \mathcal{D}_n]\), has its total variation, \(P_0\)-almost surely, bounded above by \(3 M\).
\end{lemma}
\begin{proof}
This proof is a direct modification of the argument used to establish Lemma~C.3
in \cite{van2023causal} (see also \cite{van_der_Laan2024stabilized}), rewritten to match the present statement and notation.

Define the sets $B_t:=\{z\in\mathbb R:\widehat{\theta}(z)=\widehat{\theta}(t)\}.$ Since $\widehat{\theta}$ is monotone nondecreasing, each $B_t$ is an interval of
the form $B_t=\{z\in\mathbb R: a(t)\le z<b(t)\}$ for some endpoints
$a(t)\le b(t)$ (possibly $a(t)=b(t)$). Let $\widehat{\theta}_0^*$ be as in the lemma statement. For any $z$,
\begin{align*}
\widehat{\theta}_0^*(\widehat{\eta}(z))
&=E_0\left[\eta_0(Z)\mid \widehat{\eta}^*(Z)=\widehat{\theta}(\widehat{\eta}(z)),\,\mathcal D_n\right] \\
&=E_0\left[\eta_0(Z)\mid \widehat{\theta}(\widehat{\eta}(Z))=\widehat{\theta}(\widehat{\eta}(z)),\,\mathcal D_n\right] \\
&=E_0\left[\eta_0(Z)\mid \widehat{\eta}(Z)\in B_{\widehat{\eta}(z)},\,\mathcal D_n\right].
\end{align*}
By the law of total expectation and the definition of $\widehat{\theta}_0$,
\begin{align*}
E_0\left[\eta_0(Z)\mid \widehat{\eta}(Z)\in B_{\widehat{\eta}(z)},\,\mathcal D_n\right]
&=
E_0\left[
E_0\left[\eta_0(Z)\mid \widehat{\eta}(Z),\,\mathcal D_n\right]
\;\middle|\;
\widehat{\eta}(Z)\in B_{\widehat{\eta}(z)},\,\mathcal D_n
\right] \\
&=
E_0\left[
\widehat{\theta}_0(\widehat{\eta}(Z))
\;\middle|\;
\widehat{\eta}(Z)\in B_{\widehat{\eta}(z)},\,\mathcal D_n
\right].
\end{align*}
Thus, $\widehat{\theta}_0^*$ is obtained by ``bin-averaging'' the function
$\widehat{\theta}_0$ over the partition induced by the level sets
$\{B_t:t\in\mathbb R\}$ of the monotone step map $\widehat{\theta}$.
Up to notation, this is exactly the setting of Lemma~C.3 in \cite{van2023causal},
which shows that such bin-averaging increases total variation by at most a
factor of $3$. For completeness, we prove this claim here.

Write the Jordan decomposition (Theorem~4, Section~5.2 of
\citealp{royden1963real}) as $\widehat{\theta}_0=\widehat{\theta}_0^{+}-\widehat{\theta}_0^{-},$
where $\widehat{\theta}_0^{+}$ and $\widehat{\theta}_0^{-}$ are bounded,
nondecreasing functions. Moreover, we may choose $\widehat{\theta}_0^{+}$ so
that $\widehat{\theta}_0^{+}(\infty)-\widehat{\theta}_0^{+}(-\infty)
=\|\widehat{\theta}_0\|_{TV}$. Since
$\|\widehat{\theta}_0^{-}\|_{TV}
=\|\widehat{\theta}_0-\widehat{\theta}_0^{+}\|_{TV}
\le \|\widehat{\theta}_0\|_{TV}+\|\widehat{\theta}_0^{+}\|_{TV}$,
it follows that $\|\widehat{\theta}_0^{-}\|_{TV}\le 2\|\widehat{\theta}_0\|_{TV}.$
Because $\widehat{\theta}$ is nondecreasing, by definition we have that if
$\widehat{\theta}(t_1)<\widehat{\theta}(t_2)$, then $x_1<x_2$ for any
$x_1\in B_{t_1}$ and $x_2\in B_{t_2}$. It follows that both maps
\[
t\mapsto
E_0\left[\widehat{\theta}_0^{+}(\widehat{\eta}(Z))
\mid \widehat{\eta}(Z)\in B_t,\,\mathcal D_n\right],
\qquad
t\mapsto
E_0\left[\widehat{\theta}_0^{-}(\widehat{\eta}(Z))
\mid \widehat{\eta}(Z)\in B_t,\,\mathcal D_n\right]
\]
are nondecreasing (and bounded). Hence, by Theorem~4 of
\citet{royden1963real}, their difference $\widehat{\theta}_0^*: t\mapsto
E_0\left[\widehat{\theta}_0(\widehat{\eta}(Z))
\mid \widehat{\eta}(Z)\in B_t,\,\mathcal D_n\right]$
is of bounded variation. Moreover, the total variation of a monotone function
equals the difference between its endpoint limits, and we have
\[
\essinf (\widehat{\theta}_0^{+}\circ \widehat{\eta})
\le
E_0\left[\widehat{\theta}_0^{+}(\widehat{\eta}(Z))
\mid \widehat{\eta}(Z)\in B_t,\,\mathcal D_n\right]
\le
\esssup (\widehat{\theta}_0^{+}\circ \widehat{\eta}),
\]
and similarly for $\widehat{\theta}_0^{-}\circ \widehat{\eta}$. Consequently,
the total variation norms of
$t\mapsto E_0[\widehat{\theta}_0^{+}(\widehat{\eta}(Z))\mid \widehat{\eta}(Z)\in B_t,\mathcal D_n]$
and
$t\mapsto E_0[\widehat{\theta}_0^{-}(\widehat{\eta}(Z))\mid \widehat{\eta}(Z)\in B_t,\mathcal D_n]$
are bounded by $\|\widehat{\theta}_0^{+}\|_{TV}$ and
$\|\widehat{\theta}_0^{-}\|_{TV}$, respectively. Using sublinearity of the
total variation norm, we conclude that the map $\widehat{\theta}_0^*: t\mapsto
E_0\left[\widehat{\theta}_0(\widehat{\eta}(Z))
\mid \widehat{\eta}(Z)\in B_t,\,\mathcal D_n\right]$
has total variation norm bounded by
\[
\|\widehat{\theta}_0^{+}\|_{TV}+\|\widehat{\theta}_0^{-}\|_{TV}
\le \|\widehat{\theta}_0\|_{TV}+2\|\widehat{\theta}_0\|_{TV}
=3\|\widehat{\theta}_0\|_{TV}.
\]
Since $\|\widehat{\theta}_0\|_{TV}\le M$ almost surely by assumption, it follows
that $\|\widehat{\theta}_0^*\|_{TV}\le 3M$ $P_0$-almost surely.

\end{proof}

The following lemma is a cross-averaged version of Lemma~\ref{lemma::TVnormBounded}, specialized to our isocalibrated cross-fitted nuisance estimators.

\begin{lemma}[Cross-averaged version of Lemma \ref{lemma::TVnormBounded}]
   Under \ref{cond::variation}, there exists an $M < \infty$ such that, for all $n \in \mathbb{N}$, the functions $\theta_{n,\mu}^*: \mathbb{R} \to \mathbb{R}$ and $\theta_{n,\alpha}^*: \mathbb{R} \to \mathbb{R}$, defined by $\theta_{n,\mu}^* \circ \mu_{n,\diamond}^* = \overline{\Pi}_{\mu_{n,\diamond}^*} \Delta_{\alpha,n,\diamond}$ and $\theta_{n,\alpha}^* \circ \alpha_{n,\diamond}^* = \overline{\Pi}_{\alpha_{n,\diamond}^*} \Delta_{\mu,n,\diamond}$, have total variation norms that are almost surely bounded by $3M$.
\label{lemma::TVnormBounded::cross}
\end{lemma}
\begin{proof}
  By symmetry, it suffices to prove the claim for
$\theta_{n,\mu}^* \circ \mu_{n,\diamond}^* = \overline{\Pi}_{\mu_{n,\diamond}^*}\Delta_{\alpha,n,\diamond}$.
For each $j \in [J]$, note that
\[
\{\overline{\Pi}_{\mu_{n,\diamond}^*} \Delta_{\alpha,n,\diamond}\}_j(z)
= E_{\overline{P}_0}\!\left[\Delta_{\alpha,n,j_{\mathrm{rv}}}(Z)\ \middle|\ \mu_{n,j_{\mathrm{rv}}}^*(Z) = \mu_{n,j}^*(z),\ \mathcal{D}_n\right],
\]
where the expectation is taken over $\widetilde{Z}=(Z,j_{\mathrm{rv}})\sim \overline{P}_0$, with
$j_{\mathrm{rv}}$ uniformly distributed on $[J]$ and independent of $Z$.

Define $\widehat{\eta}(\widetilde{Z}) := \mu_{n,j_{\mathrm{rv}}}(Z)$,
$\widehat{\eta}^*(\widetilde{Z}) := \mu_{n,j_{\mathrm{rv}}}^*(Z)$, and
$\eta_0(\widetilde{Z}) := \Delta_{\alpha,n,j_{\mathrm{rv}}}(Z)$.
By construction, $\widehat{\eta}^* = \widehat{\theta}\circ \widehat{\eta}$ for an isotonic,
piecewise-constant map $\widehat{\theta}$. Moreover, the map
\[
\widehat{\theta}_0(t)
:= E_{\overline{P}_0}\!\left[\eta_0(\widetilde{Z}) \ \middle|\ \widehat{\eta}(\widetilde{Z}) = t,\ \mathcal{D}_n\right]
= E_{\overline{P}_0}\!\left[\Delta_{\alpha,n,j_{\mathrm{rv}}}(Z)\ \middle|\ \mu_{n,j_{\mathrm{rv}}}(Z)=t,\ \mathcal{D}_n\right]
\]
has total variation norm almost surely bounded by $M$ under \ref{cond::variation}.
Therefore, applying Lemma~\ref{lemma::TVnormBounded} with expectations taken under $\overline{P}_0$ yields that
\[
\widehat{\theta}_0^*(t)
:= E_{\overline{P}_0}\!\left[\eta_0(\widetilde{Z}) \ \middle|\ \widehat{\eta}^*(\widetilde{Z}) = t,\ \mathcal{D}_n\right]
\]
has total variation norm almost surely bounded by $3M$. Setting $\theta_{n,\mu}^* := \widehat{\theta}_0^*$, we have
\[
\{\overline{\Pi}_{\mu_{n,\diamond}^*} \Delta_{\alpha,n,\diamond}\}_j(z)
= \theta_{n,\mu}^*\!\left(\mu_{n,j}^*(z)\right),
\]
and $\|\theta_{n,\mu}^*\|_{\mathrm{TV}} \le 3M$ almost surely, as claimed.
\end{proof}

The next lemma provides a local maximal inequality that we use to control the
empirical process remainders that appear in our analysis.

\begin{lemma}
\label{lemma::maximalIneq}
For each $j \in [J]$, let $\mathcal{H}_{n,j} = \mathcal{F}_{n,j}^4$ for a
function class $\mathcal{F}_{n,j}$ that is deterministic conditional on the
$j$th training set $\mathcal{D}_n \setminus \mathcal{C}^{(j)}$. Assume that
$\mathcal{J}(\delta, \mathcal{F}_{n,j}) \le t_n(\delta)$ for some concave map
$\delta \mapsto t_n(\delta)$ with $t_n(0)=0$ and such that
$\delta \mapsto t_n(\delta)/\delta$ is nonincreasing. For each $j \in [J]$, let
$T_{n,j}: \mathcal{Z} \times \mathbb{R}^4 \to \mathbb{R}$ be deterministic
conditional on $\mathcal{D}_n \setminus \mathcal{C}^{(j)}$ and uniformly bounded
by some fixed $B < \infty$. Assume further that there exists $L<\infty$ such
that, for each $z \in \mathcal{Z}$ and all $u,v \in \mathbb{R}^4$,
\[
\bigl|T_{n,j}(z,u)-T_{n,j}(z,v)\bigr| \le L \|u-v\|_2,
\]
i.e., the map $T_{n,j}(z,\cdot)$ is $L$-Lipschitz uniformly in $z$. Then, for all $\delta>0$,
\[
E_0^n \left[\sup_{h_{\diamond} \in \mathcal{H}_{n,\diamond}:
\|T_{n,\diamond}(\cdot, h_{\diamond}(\cdot))\|_{\overline{P}_0} \le \delta}
\left(\overline{P}_n - \overline{P}_0\right)
T_{n,\diamond}(\cdot, h_{\diamond}(\cdot))\right]
\lesssim n^{-1/2} t_n(\delta)\left(1 + \frac{t_n(\delta)}{\sqrt{n}\,\delta^2}\right).
\]
Furthermore, for any $\sqrt{n}\,\delta^2 > t_n(\delta)$,
\[
E_0^n \left[\sup_{h_{\diamond} \in \mathcal{H}_{n,\diamond}:
\|T_{n,\diamond}(\cdot, h_{\diamond}(\cdot))\|_{\overline{P}_0} \le \delta}
\left(\overline{P}_n^{\#} - \overline{P}_n\right)
T_{n,\diamond}(\cdot, h_{\diamond}(\cdot))\right]
\lesssim n^{-1/2} t_n(\delta)\left(1 + \frac{t_n(\delta)}{\sqrt{n}\,\delta^2}\right),
\]
where the implicit constant in ``$\lesssim$'' depends only on $B$, $L$, and the
constants appearing in the stated conditions.
\end{lemma}
\begin{proof}
To establish the first bound, it suffices by the triangle inequality to show that, for each $j \in [J]$,
\[
E_0^n \left[\sup_{h \in \mathcal{H}_{n,j}:\ \|T_{n,j}(\cdot, h(\cdot))\|\leq \delta}
\left(P_{n,j} - P_0\right) T_{n,j}(\cdot, h(\cdot))\right]
\;\lesssim\;
n^{-1/2}\, t_n(\delta)\left\{1 + \frac{t_n(\delta)}{\sqrt{n}\,\delta^2} \right\}.
\]
The desired bound then follows since
\[
E_0^n \left[\sup_{h_{\diamond} \in \mathcal{H}_{n,\diamond}:\ 
\|T_{n,\diamond}(\cdot, h_{\diamond}(\cdot))\|_{\overline{P}_0} \le \delta}
\left(\overline{P}_n - \overline{P}_0\right)
T_{n,\diamond}(\cdot, h_{\diamond}(\cdot))\right]
\;\lesssim\;
\max_{j \in [J]}
E_0^n \left[\sup_{h \in \mathcal{H}_{n,j}:\ \|T_{n,j}(\cdot, h(\cdot))\|\leq J\delta}
\left(P_{n,j} - P_0\right) T_{n,j}(\cdot, h(\cdot))\right],
\]
where we used that $\|T_{n,\diamond}(\cdot, h_{\diamond}(\cdot))\|_{\overline{P}_0} \le \delta$ implies
$\|T_{n,j}(\cdot, h(\cdot))\|\leq J\delta$ for each $j \in [J]$. Here, we also used that each fold has $\approx n/J$ folds with $J$ fixed.

To this end, let $\mathcal{G}_{n,j} := \{\,T_{n,j}(\cdot,h(\cdot)) : h \in \mathcal{H}_{n,j}\,\}.$
By Theorem~2.10.20 of \cite{van1996weak} on preservation of entropy integrals
under Lipschitz transformations and Lemma~\ref{lemma::entropy}, the class
$\mathcal{G}_{n,j}$ is uniformly bounded and satisfies
\[
\mathcal{J}(\delta, \mathcal{G}_{n,j}) \lesssim \mathcal{J}(\delta, \mathcal{H}_{n,j})
\lesssim 4\,\mathcal{J}(\delta, \mathcal{F}_{n,j}) \lesssim t_n(\delta).
\]
Hence, by Theorem~2.1 of \cite{vanderVaart2011local}, applied conditional on
$\mathcal{D}_n \setminus \mathcal{C}^{(j)}$ and then averaged over
$\mathcal{D}_n \setminus \mathcal{C}^{(j)}$ using the tower property,
\[
E_0^n \left[\sup_{h \in \mathcal{H}_{n,j}:\ \|T_{n,j}(\cdot, h(\cdot))\|\leq \delta}
\left(P_{n,j} - P_0\right) T_{n,j}(\cdot, h(\cdot))\right]
\;\lesssim\;
n^{-1/2}\, t_n(\delta)\left(1 + \frac{t_n(\delta)}{\sqrt{n}\,\delta^2} \right).
\]

We obtain the second bound (the bootstrap case) by adapting the proof of
Theorem~2.1 in \cite{vanderVaart2011local}. Throughout, we work conditional on
$\mathcal{D}_n \setminus \mathcal{C}^{(j)}$. Since the resulting bounds hold
with absolute constants that do not depend on $\mathcal{D}_n \setminus
\mathcal{C}^{(j)}$, averaging over $\mathcal{D}_n \setminus \mathcal{C}^{(j)}$
then yields the claimed result. To this end, note that $P_{n,j}^{\#}$ is the empirical distribution of $O(n/J)$ i.i.d.
samples from $P_{n,j}$. Define
the random radii
\[
\delta_n := \sup_{h \in \mathcal{H}_{n,j}:\ \|T_{n,j}(\cdot, h(\cdot))\|_{P_{0}} \leq \delta}
\| T_{n,j}(\cdot, h(\cdot))\|_{P_{n,j}}
\quad\text{and}\quad
\delta_n^\# := \sup_{h \in \mathcal{H}_{n,j}:\ \|T_{n,j}(\cdot, h(\cdot))\|_{P_{n,j}} \leq \delta_n}
\| T_{n,j}(\cdot, h(\cdot))\|_{P_{n,j}^\#}.
\]
By uniform boundedness of the class $\mathcal{G}_{n,j}$ and
Equation~(2.1) of \cite{vanderVaart2011local} (a consequence of Dudley's entropy
integral bound),
\begin{align*}
& E_0^n \left[\sup_{h \in \mathcal{H}_{n,j}:\ \|T_{n,j}(\cdot, h(\cdot))\|_{P_{0}} \leq J\delta}
\left(P_{n,j}^{\#} - P_{n,j}\right) T_{n,j}(\cdot, h(\cdot))  \,\middle|\,  P_{n,j}\right] \\
& \lesssim  n^{-1/2}\, E_0^n \left[ \mathcal{J}(\delta_n^\#, \mathcal{G}_{n,j})  \,\middle|\,  P_{n,j} \right] \\
& \lesssim  n^{-1/2}\, E_0^n \left[ t_n(\delta_n^\#)  \,\middle|\,  P_{n,j} \right].
\end{align*}
Since $\delta \mapsto t_n(\delta)$ is concave with $t_n(0)=0$ and
$\delta \mapsto t_n(\delta)/\delta$ is nonincreasing, the proof of
Theorem~2.1 of \cite{vanderVaart2011local} implies that
$\varepsilon \mapsto t_n(\sqrt{\varepsilon})$ is concave. Hence, by Jensen's
inequality,
\[
E_0^n \!\left[ t_n(\delta_n^\#) \,\middle|\, P_{n,j} \right]
\;\le\;
t_n\!\left(\widetilde{\delta}_n^{\#}\right),
\qquad
\widetilde{\delta}_n^{\#2} := E_0^n\!\left[\delta_n^{\#2} \,\middle|\, P_{n,j}\right].
\]
By the same argument and the tower property, averaging over $P_{n,j}$ yields
\[
E_0^n \!\left[ t_n(\delta_n^\#) \right]
\le
E_0^n \!\left[t_n\!\left(\widetilde{\delta}_n^{\#}\right)\right]
\le
t_n\!\left(\widetilde{\delta}_n\right),
\qquad
\widetilde{\delta}_n^{2} := E_0^n\!\left[\delta_n^{\#2} \right].
\]
Thus,
\begin{align}
\label{eqn::firstbound}
E_0^n \Bigg[
\sup_{\substack{h \in \mathcal{H}_{n,j}:\\ \|T_{n,j}(\cdot, h(\cdot))\|_{P_{0}} \leq J\delta}}
\left(P_{n,j}^{\#} - P_{n,j}\right) T_{n,j}(\cdot, h(\cdot))
\Bigg]
\lesssim
n^{-1/2} t_n\!\left(\widetilde{\delta}_n\right).
\end{align}

We now further bound the right hand side. Equation~(2.4) in the proof of Theorem~2.1 of \cite{vanderVaart2011local} shows that
\[
\widetilde{\delta}_n^{\#2} \lesssim \delta_n^2 + n^{-1/2} \, t_n(\widetilde{\delta}_n^{\#}).
\]
See also Theorem~2.2 of \cite{van2014uniform} for a related result.
Next, using again that $\varepsilon \mapsto t_n(\sqrt{\varepsilon})$ is concave and applying Jensen's inequality, we obtain
\begin{align*}
\widetilde{\delta}_n^2 = E_0^n\!\left[\widetilde{\delta}_n^{\#2}\right]
&\lesssim E_0^n\!\left[\delta_n^2\right] + n^{-1/2} E_0^n\!\left[t_n(\widetilde{\delta}_n^{\#})\right] \\
&\lesssim E_0^n\!\left[\delta_n^2\right] + n^{-1/2} t_n\!\left(\sqrt{E_0^n\!\left[\widetilde{\delta}_n^{\#2}\right]}\right)\\
&\lesssim E_0^n\!\left[\delta_n^2\right] + n^{-1/2} t_n\!\left(\widetilde{\delta}_n\right).
\end{align*}
By Theorem~2.2 of \cite{van2014uniform}, we have $E_0^n[\delta_n^2] \lesssim \delta^2$ for any $\delta>0$ satisfying $\sqrt{n}\,\delta^2 > t_n(\delta)$. For such a $\delta$, it follows that
\begin{align}
\label{eqn::secondbound}
\widetilde{\delta}_n^2
&\lesssim \delta^2 + n^{-1/2} t_n\!\left(\widetilde{\delta}_n\right).
\end{align}
 Finally, arguing as in the proof of Theorem~2.1 of \cite{vanderVaart2011local}, we apply Lemma~2.1 of that work to show that any $\widetilde{\delta}_n$ satisfying \eqref{eqn::secondbound} also satisfies
\[
t_n\!\left(\widetilde{\delta}_n\right)
\leq t_n(\delta)\left(1 + \frac{t_n(\delta)}{\sqrt{n}\,\delta^2}\right).
\]
Here we crucially use that $\delta \mapsto t_n(\delta)$ is concave with $t_n(0)=0$, and that $\delta \mapsto t_n(\delta)/\delta$ is nonincreasing. Thus, applying this bound to \eqref{eqn::secondbound}, we conclude that
\begin{align}
E_0^n \Bigg[
\sup_{\substack{h \in \mathcal{H}_{n,j}:\\ \|T_{n,j}(\cdot, h(\cdot))\|_{P_{0}} \leq \delta}}
\left(P_{n,j}^{\#} - P_{n,j}\right) T_{n,j}(\cdot, h(\cdot))
\,\Bigg|\, P_{n,j}
\Bigg]
\lesssim
n^{-1/2} t_n(\delta)\left(1 + \frac{t_n(\delta)}{\sqrt{n}\,\delta^2}\right),
\end{align}
as desired.

\end{proof}

Finally, the following lemma can be used to explicitly verify the locally
uniform asymptotic linearity conditions for estimated functionals, such as
Condition~\ref{cond::estboundedlinearcross}. We apply this lemma together with
Condition~\ref{cond::lipschitzisotonic} in the proof of
Theorem~\ref{theorem::DRboot}.

\begin{lemma}
\label{lemma::Lipschitzuniformentropy}
Suppose $\mathcal{A}$ is finite, and let $m: \mathcal{Z} \times \mathcal{H} \rightarrow \mathbb{R}$ be a functional that is linear in its second argument. Assume there exists $L \in (0,\infty)$ such that $|m(Z, h)| \leq L \max_{a \in \mathcal{A}} |h(a, W)|$ almost surely. Let $\nu \in \mathcal{H}$ be a fixed, uniformly bounded function, and let $\mathcal{F}$ be a class of functions from $\mathbb{R}$ to $\mathbb{R}$ with $\sup_{f \in \mathcal{F}} \|f\|_{\infty} \leq B$ and $\mathcal{J}(\infty, \mathcal{F}) < \infty$. Define the function classes $\mathcal{F}_{\nu} := \{f \circ \nu : f \in \mathcal{F}\}$ and $\mathcal{F}_{m,\nu} := \{m(\cdot, f) : f \in \mathcal{F}_{\nu}\}$. Then, there exists an $M < \infty$ depending only on $L$ such that, for all $\delta > 0$,
$\mathcal{J}(\delta,\mathcal{F}_{\nu} ) \leq M \mathcal{J}(\delta,\mathcal{F} ) $ and $\mathcal{J}(\delta,\mathcal{F}_{m,\nu} ) \leq M \mathcal{J}(\delta,\mathcal{F} ) $.
\end{lemma}
\begin{proof}
Since $|m(Z, h)| \leq L \max_{a \in \mathcal{A}} |h(a, W)|$ almost surely, we have
\[
N(\varepsilon, \mathcal{F}_{m,\nu}, L^2(Q)) \lesssim \sum_{a \in \mathcal{A}} N(\varepsilon, \mathcal{F}_{\nu}^{(a)}, L^2(Q)),
\]
where we define the function class $\mathcal{F}_{\nu}^{(a)} := \{w \mapsto f(a, w) : f \in \mathcal{F}_{\nu}\}$ and use the finiteness of $\mathcal{A}$. Hence,
\[
\mathcal{J}(\delta, \mathcal{F}_{m,\nu}) \lesssim \sum_{a \in \mathcal{A}} \mathcal{J}(\delta, \mathcal{F}_{\nu}^{(a)}).
\]
By the definition of $\mathcal{F}_{\nu}$, we have $\mathcal{F}_{\nu}^{(a)} = \{w \mapsto f(\nu(a, w)) : f \in \mathcal{F}\}$. Now, for all $a \in \mathcal{A}$,
\begin{align*}
\mathcal{J}(\delta, \mathcal{F}_{\nu}^{(a)}) 
&= \int_0^\delta \sup_Q \sqrt{N(\varepsilon, \mathcal{F}_{\nu}^{(a)}, L^2(Q))} \, d\varepsilon \\
&= \int_0^\delta \sup_{Q_{a,\nu}} \sqrt{N(\varepsilon, \mathcal{F}, L^2(Q_{a,\nu}))} \, d\varepsilon \\
&= \mathcal{J}(\delta, \mathcal{F}),
\end{align*}
where $Q_{a,\nu}$ denotes the pushforward measure of the random variable $\nu(a, W)$. It follows that $\mathcal{J}(\delta, \mathcal{F}_{m,\nu}) \lesssim \mathcal{J}(\delta, \mathcal{F})$. By an identical argument, we also have $\mathcal{J}(\delta, \mathcal{F}_{\nu}) \lesssim \mathcal{J}(\delta, \mathcal{F})$.
\end{proof}

\subsection{Dealing with empirical process remainders}
\label{appendix::dealingwithEP}
In this section, we provide technical lemmas establishing the asymptotic negligibility of certain empirical process remainders. These lemmas rely on the following result, which gives sufficient conditions for these remainders to be asymptotically negligible.

\begin{lemma}
\label{lemma::LipschitzEmpirical}
Assume \ref{cond::estnuisbound} and \ref{cond::variation}. Let $M$ be the
uniform bound on $Y$ and the nuisances in \ref{cond::estnuisbound}. Then, with
$P_0$-probability tending to one, $s_{n,j}^*$ and $r_{n,j}^*$ are uniformly
bounded by $M$ for each $j=1,\ldots,J$. For each $j \in [J]$, define
\[
\rho_{n,j}(z)
:=
T_{n,j}\bigl(z,\mu_{n,j}^*,s_{n,j}^*,\alpha_{n,j}^*,r_{n,j}^*\bigr),
\]
where $T_{n,j}:\mathcal{Z}\times[-2M,2M]^4\to\mathbb{R}$ 
satisfies the conditions of Lemma~\ref{lemma::maximalIneq}. Suppose that
$\|\rho_{n,\diamond}\|_{\overline{P}_0}=O_P(\varepsilon_n)$ for a deterministic
sequence $\varepsilon_n \to 0$. Then
\[
\bigl(\overline{P}_n-\overline{P}_0\bigr)\rho_{n,\diamond}
=
o_p(n^{-1/2}).
\]
Moreover, if $\overline{P}_n$, $\mu_{n,j}^*$, $s_{n,j}^*$, $\alpha_{n,j}^*$,
$r_{n,j}^*$, and $\rho_{n,j}$ are replaced by their bootstrap analogues
$\overline{P}_n^\#$, $\mu_{n,j}^\#$, $s_{n,j}^\#$, $\alpha_{n,j}^\#$,
$r_{n,j}^\#$, and $\rho_{n,j}^\#$, then
\[
\left|
\bigl(\overline{P}_n^\#-\overline{P}_n\bigr)\rho_{n,\diamond}^\#
\right|
+
\left|
\bigl(\overline{P}_n-\overline{P}_0\bigr)\rho_{n,\diamond}^\#
\right|
=
o_p(n^{-1/2})
\]
whenever $\|\rho_{n,\diamond}^\#\|_{\overline{P}_0}=O_P(\varepsilon_n).$
\end{lemma}
\begin{proof}
First, note that $s_{n,j}^*$ is bounded by the uniform bound for $\alpha_{n,j}^*-\alpha_0$, since conditional expectations preserve essential suprema and, by \ref{cond::estnuisbound}, $\|\alpha_{n,j}^*-\alpha_0\|_\infty \le 2M$ with $P_0$--probability tending to one. Similarly, $\|r_{n,j}^*\|_\infty \le 2M$ with $P_0$--probability tending to one, since $r_{n,j}^*$ is bounded by the uniform bound for $\mu_{n,j}^*-\mu_0$. Therefore, with $P_0$--probability tending to one, we have
\[
\bigl(z,\mu_{n,j}^*, s_{n,j}^*, \alpha_{n,j}^*, r_{n,j}^*\bigr)\in \mathcal{Z}\times [-2M,2M]^4,
\]
so $T_{n,j}\bigl(z,\mu_{n,j}^*, s_{n,j}^*, \alpha_{n,j}^*, r_{n,j}^*\bigr)$ is well-defined on this event. This establishes the first claim.

In what follows, we work on the event
\[
\mathcal{B}_{n,j} := \Bigl\{\|s_{n,j}^*\|_\infty \le 2M,\ \|r_{n,j}^*\|_\infty \le 2M,\ \|\alpha_{n,j}^*\|_\infty \le 2M,\ \|\mu_{n,j}^*\|_\infty \le 2M\Bigr\},
\]
which satisfies $P_0(\mathcal{B}_{n,j})\to 1$ by \ref{cond::estnuisbound}. Since all subsequent bounds are derived on $\mathcal{B}_{n,j}$, we suppress the indicator $1\{\mathcal{B}_{n,j}\}$ in the notation and interpret stochastic order statements (e.g., $O_P(\cdot)$, $o_P(\cdot)$) as holding under $P_0$ on events whose probabilities tend to one.

We apply Lemma~\ref{lemma::maximalIneq} with
$\mathcal{F}_{n,j} := \mathcal{S}_{n,j} \cup \mathcal{R}_{n,j}$, which contains
$\mu_{n,j}^*$, $s_{n,j}^*$, $\alpha_{n,j}^*$, and $r_{n,j}^*$ with probability
tending to one. This inclusion holds for $\mu_{n,j}^*$ and $\alpha_{n,j}^*$
because they are isotonic transformations of $\mu_{n,j}$ and $\alpha_{n,j}$,
respectively. Similarly, by Lemma~\ref{lemma::TVnormBounded::cross}, it holds for
$r_{n,j}^*$ and $s_{n,j}^*$ because they are bounded-variation transformations of
$\mu_{n,j}$ and $\alpha_{n,j}$, respectively. By Lemma~\ref{lemma::entropy},
\[
\mathcal{J}(\delta, \mathcal{F}_{n,j})
\;\lesssim\;
\mathcal{J}(\delta, \mathcal{R}_{n,j}) + \mathcal{J}(\delta, \mathcal{S}_{n,j})
\;\lesssim\;
t_n(\delta),
\qquad\text{where } t_n(\delta) := \sqrt{\delta}.
\]

Since $\|\rho_{n,\diamond}\|_{\overline{P}_0} = O_P(\varepsilon_n)$ with
$\varepsilon_n \to 0$, there exists a deterministic sequence $\delta_n \to 0$
such that $\|\rho_{n,\diamond}\|_{\overline{P}_0} \lesssim \delta_n$ with
probability tending to one. Let $A_n$ denote the event
$\{\|\rho_{n,\diamond}\|_{\overline{P}_0} \lesssim \delta_n\}$. On $A_n$,
\[
\bigl|(\overline{P}_n - \overline{P}_0)\rho_{n,\diamond}\bigr|
\le
\sup_{\substack{h_{\diamond} \in \mathcal{H}_{n,\diamond}:\\
\|T_{n,\diamond}(\cdot, h_{\diamond}(\cdot))\|_{\overline{P}_0} \le \delta_n}}
\bigl(\overline{P}_n - \overline{P}_0\bigr)\,
T_{n,\diamond}(\cdot, h_{\diamond}(\cdot)).
\]
Therefore, applying Lemma~\ref{lemma::maximalIneq} with $t_n(\delta)=\sqrt{\delta}$
yields
\begin{align*}
E_0^n\Bigg[
1(A_n)\,
\sup_{\substack{h_{\diamond} \in \mathcal{H}_{n,\diamond}:\\
\|T_{n,\diamond}(\cdot, h_{\diamond}(\cdot))\|_{\overline{P}_0} \le \delta_n}}
\bigl(\overline{P}_n - \overline{P}_0\bigr)\,
T_{n,\diamond}(\cdot, h_{\diamond}(\cdot))
\Bigg]
&\lesssim
n^{-1/2}\, t_n(\delta_n)\left\{1 + \frac{t_n(\delta_n)}{\sqrt{n}\,\delta_n^2}\right\}.
\end{align*}
Since $t_n(\delta)=\sqrt{\delta}$, we have
$t_n(\delta_n)\{1 + t_n(\delta_n)/(\sqrt{n}\,\delta_n^2)\}
= \sqrt{\delta_n} + (n^{-1/2})\delta_n^{-3/2} = o(1)$ for any deterministic
$\delta_n \downarrow 0$ satisfying $\sqrt{n}\,\delta_n^2 \to \infty$. Hence the
displayed expectation is $o(n^{-1/2})$, and Markov's inequality implies
\[
1(A_n)\, \bigl(\overline{P}_n - \overline{P}_0\bigr)\rho_{n,\diamond}
= o_p(n^{-1/2}).
\]
Because $P(A_n)\to 1$, it follows that
$(\overline{P}_n - \overline{P}_0)\rho_{n,\diamond} = o_p(n^{-1/2})$.

The bootstrap claims follow by the same argument, using the second part of
Lemma~\ref{lemma::maximalIneq} and noting that, conditional on $P_{n,j}$,
$P_{n,j}^{\#}$ is the empirical distribution of $n_j \asymp n/J$ i.i.d. draws
from $P_{n,j}$ (with $J$ fixed).
\end{proof}

The following lemma establishes consistency of $s_{n,j}^*$ and $r_{n,j}^*$ in the appropriate regimes of \ref{cond::DRlimits} , as well as consistency and rate preservation of the bootstrapped nuisance estimators.

\begin{lemma}[Consistency of nuisance estimators and rate preservation under the bootstrap]
\label{lemma::nuisanceconsistency}
    Under \ref{cond::DRlimits} and \ref{cond::mainprojectioncoupling1}, we have the following. Under \ref{cond::DRmisDRconsist::2} ($\overline{\mu}_0 = \mu_0$), $\|s_{n,j}^* - s_0\| = o_p(1)$, and under \ref{cond::DRmisDRconsist::1} ($\overline{\alpha}_0 = \alpha_0$), $\|r_{n,j}^* - r_0\| = o_p(1)$.
    
    Furthermore, under Conditions~\ref{cond::linear}, \ref{cond::estboundedlinearcross}, \ref{cond::estnuisbound}, and~\ref{cond::mainprojectioncoupling1::boot}--\ref{cond::lipschitzisotonic}, whenever one part of \ref{cond::DRmisDRconsist} holds for $\alpha_{n,\diamond}^{*}$ and $\mu_{n,\diamond}^{*}$, the same part holds for $\alpha_{n,\diamond}^{\#}$ and $\mu_{n,\diamond}^{\#}$. In particular, under \ref{cond::DRlimits}, we have $\|\alpha_{n,\diamond}^{\#} - \overline{\alpha}_0\|_{\overline{P}_0} = o_p(1)$ and $\|\mu_{n,\diamond}^{\#} - \overline{\mu}_0\|_{\overline{P}_0} = o_p(1)$. Moreover, under \ref{cond::DRmisDRconsist::2}, $\|s_{n,j}^{\#} - s_0\| = o_p(1)$, and under \ref{cond::DRmisDRconsist::1}, $1\{\overline{\alpha}_0 = \alpha_0\}\,\|r_{n,j}^{\#} - r_0\| = o_p(1)$.
\end{lemma}
\begin{proof}

First, assume case \ref{cond::DRmisDRconsist::2} in \ref{cond::DRlimits} ($\overline{\mu}_0 = \mu_0$). Then, the
second part of \ref{cond::mainprojectioncoupling1:ii} yields
\[
\|(\Pi_{\mu_{n,j}^*} - \Pi_{\mu_0})\Delta_{\alpha,n,j}\|
= o_p(1).
\]
Hence, by the triangle inequality,
\[
 \|s_{n,j}^* - s_0\|
\leq 1\{\overline{\mu}_0 = \mu_0\}\,\|(\Pi_{\mu_{n,j}^*} - \Pi_{\mu_0})\Delta_{\alpha,n,j}\|
  + 1\{\overline{\mu}_0 = \mu_0\}\,\|\Pi_{\mu_0}\{\alpha_{n,j}^* - \overline{\alpha}_0\}\|
= o_p(1),
\]
where we used the projection property
\[
\|\Pi_{\mu_0}\{\alpha_{n,j}^* - \overline{\alpha}_0\}\|
\leq \|\alpha_{n,j}^* - \overline{\alpha}_0\|
= o_p(1).
\]

Next, assume case \ref{cond::DRmisDRconsist::1} ($\overline{\alpha}_0 = \alpha_0$). A symmetric argument applying \ref{cond::mainprojectioncoupling1:i} shows that
\[
1\{\overline{\alpha}_0 = \alpha_0\}\,\|r_{n,j}^* - r_0\| = o_p(1).
\]

Under Conditions~\ref{cond::linear}, \ref{cond::estboundedlinearcross}, \ref{cond::estnuisbound}, \ref{cond::estboundboot}, and~\ref{cond::lipschitzisotonic}, the second part of Theorem~\ref{theorem::MSE} implies that
\begin{align*}
   \|\alpha_{n, \diamond}^{\#} - \overline{\alpha}_0 \|_{\overline{P}_0} &\leq   \|\alpha_{n,\diamond}^* - \overline{\alpha}_0\|_{\overline{P}_0} +  O_p(n^{-1/3}); \\
    \|\mu_{n,\diamond}^{\#} -   \overline{\mu}_0 \|_{\overline{P}_0} &\leq   \|\mu_{n,\diamond}^* -  \overline{\mu}_0\|_{\overline{P}_0} +  O_p(n^{-1/3}).
\end{align*}
Hence, by \ref{cond::DRlimits},
\begin{align*}
   \|\alpha_{n, \diamond}^{\#} - \overline{\alpha}_0 \|_{\overline{P}_0} =  o_p(1); \quad 
    \|\mu_{n,\diamond}^{\#} -   \overline{\mu}_0 \|_{\overline{P}_0} = o_p(1).
\end{align*}
Moreover, under \ref{cond::DRmisDRconsist::1}, we have $\|\alpha_{n, \diamond}^{\#} - \alpha_0 \|_{\overline{P}_0} = o_p(n^{-1/4})$, whereas under \ref{cond::DRmisDRconsist::2}, we have $\|\mu_{n, \diamond}^{\#} - \mu_0 \|_{\overline{P}_0} = o_p(n^{-1/4})$. Thus, whenever one part of \ref{cond::DRmisDRconsist} holds for $\alpha_{n, \diamond}^{*}$ and $\mu_{n, \diamond}^{*}$, the same part holds for $\alpha_{n, \diamond}^{\#}$ and $\mu_{n, \diamond}^{\#}$.

Finally, applying the same argument as in the nonbootstrap case, we obtain
$1\{\overline{\mu}_0 = \mu_0\}\,\|s_{n,j}^{\#} - s_0\| = o_p(1)$ and
$1\{\overline{\alpha}_0 = \alpha_0\}\,\|r_{n,j}^{\#} - r_0\| = o_p(1)$ under their respective cases in \ref{cond::DRlimits}, where we use the projection error bound in \ref{cond::mainprojectioncoupling1::boot}.

\end{proof}

\begin{lemma}[Negligible empirical process remainders]
Assume Conditions~\ref{cond::pathwise} and~\ref{cond::DRlimits}--\ref{cond::variation} hold. Then, it holds that
\[
(\overline{P}_n-\overline{P}_0) \{D_{\mu_{n,\diamond}^*, \alpha_{n,\diamond}^*} - D_{\overline{\mu}_0, \overline{\alpha}_0}\} = o_p(n^{-1/2}).
\]
Moreover, for the bootstrap estimator defined in Algorithm~\ref{alg::bootstrap}, if Conditions~\ref{cond::mainprojectioncoupling1::boot}--\ref{cond::lipschitzisotonic} also hold, then
\[
(\overline{P}_n^{\#}-\overline{P}_0) \{D_{\mu_{n,\diamond}^{\#}, \alpha_{n,\diamond}^{\#}} - D_{\overline{\mu}_0, \overline{\alpha}_0}\} = o_p(n^{-1/2}).
\]
   \label{lemma::firstEmpProcRemainders}  
 \end{lemma}
 \begin{proof}

Condition~\ref{cond::DRlimits} implies that
$\|\mu_{n,\diamond}^* - \overline{\mu}_0\|_{\overline{P}_0} = o_p(1)$ and
$\|\alpha_{n,\diamond}^* - \overline{\alpha}_0\|_{\overline{P}_0} = o_p(1)$.
Moreover, by the uniform boundedness in \ref{cond::estnuisbound}, the map
$(\mu,\alpha)\mapsto D_{\mu,\alpha}$ is Lipschitz as a function into
$L^2(\overline{P}_0)$. Hence,
\[
\|D_{\mu_{n,\diamond}^*, \alpha_{n,\diamond}^*}
- D_{\overline{\mu}_0, \overline{\alpha}_0}\|_{\overline{P}_0}
\lesssim
\|\mu_{n,\diamond}^* - \overline{\mu}_0\|_{\overline{P}_0}
+
\|\alpha_{n,\diamond}^* - \overline{\alpha}_0\|_{\overline{P}_0}
= o_p(1).
\]
Furthermore, for each $j\in[J]$ and $z=(y,a,w)\in\mathcal{Z}$,
\[
D_{\mu_{n,j}^*, \alpha_{n,j}^*}(z) - D_{\overline{\mu}_0, \overline{\alpha}_0}(z)
=
T_{n,j}\bigl(z,\mu_{n,j}^*, 0, \alpha_{n,j}^*, 0\bigr),
\]
where $T_{n,j}$ is the bounded Lipschitz transformation
\[
T_{n,j}(z,\mu,s,\alpha,r) := \alpha(a,w)\{y-\mu(a,w)\}.
\]
On the event in \ref{cond::estnuisbound}, we have
$(z,\mu(a,w),s(a,w),\alpha(a,w), r(a,w))\in \mathcal{Z} \times [-2M,2M]^4$, so $T_{n,j}$ is uniformly bounded. It is
also Lipschitz in $(\mu,\alpha)$ on this bounded set because $T_{n,j}$ is formed
from addition and multiplication, and multiplication is Lipschitz on bounded
domains. Therefore $T_{n,j}$ satisfies the conditions of
Lemma~\ref{lemma::LipschitzEmpirical}. Thus, by direct application of Lemma~\ref{lemma::LipschitzEmpirical}, we have that
\[
(\overline{P}_n-\overline{P}_0) \{D_{\mu_{n,\diamond}^*, \alpha_{n,\diamond}^*} - D_{\overline{\mu}_0, \overline{\alpha}_0}\} = o_p(n^{-1/2}).
\]

To prove the second part of this lemma, note that
\begin{align*}
(\overline{P}_n^{\#}-\overline{P}_0)\bigl\{D_{\mu_{n,\diamond}^{\#}, \alpha_{n,\diamond}^{\#}} - D_{\overline{\mu}_0, \overline{\alpha}_0}\bigr\}
&=
(\overline{P}_n^{\#}-\overline{P}_n)\bigl\{D_{\mu_{n,\diamond}^{\#}, \alpha_{n,\diamond}^{\#}} - D_{\overline{\mu}_0, \overline{\alpha}_0}\bigr\} \\
&\quad +
(\overline{P}_n-\overline{P}_0)\bigl\{D_{\mu_{n,\diamond}^{\#}, \alpha_{n,\diamond}^{\#}} - D_{\overline{\mu}_0, \overline{\alpha}_0}\bigr\}.
\end{align*}
By Lemma~\ref{lemma::nuisanceconsistency}, we have
$\|\alpha_{n,\diamond}^{\#} - \overline{\alpha}_0 \|_{\overline{P}_0} = o_p(1)$ and
$\|\mu_{n,\diamond}^{\#} - \overline{\mu}_0 \|_{\overline{P}_0} = o_p(1)$. Therefore,
arguing as before and using \ref{cond::estboundboot}, we obtain
\[
\|D_{\mu_{n,\diamond}^{\#}, \alpha_{n,\diamond}^{\#}} - D_{\overline{\mu}_0, \overline{\alpha}_0}\|_{\overline{P}_0}
= o_p(1).
\]
The second term on the right-hand side is $o_p(n^{-1/2})$ by the same argument
as above, applying Lemma~\ref{lemma::LipschitzEmpirical} to the random function
$z \mapsto D_{\mu_{n,\diamond}^{\#}, \alpha_{n,\diamond}^{\#}}(z) -
D_{\overline{\mu}_0, \overline{\alpha}_0}(z)$. The first term is also
$o_p(n^{-1/2})$ by an analogous argument, using the bootstrap version of
Lemma~\ref{lemma::LipschitzEmpirical}.

 \end{proof}

 For each $j \in [J]$, recall that $A_{n,j}^*: z \mapsto s_{n,j}^*(a,w)\{y- \mu_{n,j}^*(a,w)\}$ and $B_{n,j}^* := \psi_0(r_0) + \phi_{r_{n,j}^*}   - r_{n,j}^* \alpha_{n,j}^*$.  Also that  $Rem_{\psi_{n,\diamond}}(r_{n,\diamond}^*)
:= \frac{1}{J}\sum_{j=1}^J
\left\{\psi_{n,j}(r_{n,j}^*)-\psi_0(r_{n,j}^*)- P_{n,j}\phi_{0, r_{n,j}^*}\right\}.$  

\begin{lemma}[Negligible cross-product empirical process remainders]
\label{lemma::secondEmpProcRemainders}
Assume Conditions~\ref{cond::linear}--\ref{cond::pathwise} and~\ref{cond::DRlimits}--\ref{cond::variation}. If \ref{cond::DRmisDRconsist::2} ($\overline{\mu}_0=\mu_0$), then
\[
(\overline{P}_n - P_0)\{A_{n,\diamond}^* - A_0\} = o_p(n^{-1/2}),
\]
and if \ref{cond::DRmisDRconsist::1} ($\overline{\alpha}_0=\alpha_0$), then
\[
(\overline{P}_n - P_0)\{B_{n,\diamond}^* - B_0\} + Rem_{\psi_{n,\diamond}}(r_{n,\diamond}^*) = o_p(n^{-1/2}).
\]
Moreover, for the bootstrap estimator defined in Algorithm~\ref{alg::bootstrap}, if
Conditions~\ref{cond::mainprojectioncoupling1::boot}--\ref{cond::lipschitzisotonic} also hold, the same conclusions
hold with $\overline{P}_n$ replaced by $\overline{P}_n^{\#}$ and
$A_{n,\diamond}^*,B_{n,\diamond}^*$ replaced by $A_{n,\diamond}^{\#},
B_{n,\diamond}^{\#}$.
\end{lemma}
 \begin{proof}
The claim that
\[
(\overline{P}_n - P_0)\{A_{n,\diamond}^* - A_0\} = o_p(n^{-1/2})
\quad \text{under \ref{cond::DRmisDRconsist::2}}
\]
follows from a modification of the proof of Lemma~\ref{lemma::firstEmpProcRemainders}. Specifically, we replace $\alpha_{n,j}^*$ with $s_{n,j}^*$ in the definition of $T_{n,j}$ and then apply Lemma~\ref{lemma::LipschitzEmpirical}. This uses that
\[
\|s_{n,j}^* - s_0\| = o_p(1) \quad \text{under \ref{cond::DRmisDRconsist::2},}
\]
which follows from Lemma~\ref{lemma::nuisanceconsistency}.

Next, we establish the claim that
\[
(\overline{P}_n - P_0)\{B_{n,\diamond}^* - B_0\} = o_p(n^{-1/2})
\]
under \ref{cond::DRmisDRconsist::1}. Note that
\[
\frac{1}{J}\sum_{j=1}^J
\left\{\psi_{n,j}(r_{n,j}^*)-\psi_0(r_{n,j}^*)- P_{n,j}\phi_{0, r_{0}}\right\} = o_p(n^{-1/2})
\]
by \ref{cond::DRremainder}. Hence,
\begin{align*}
(\overline{P}_n - P_0)\{B_{n,\diamond}^* - B_0\} + Rem_{\psi_{n,\diamond}}(r_{n,\diamond}^*) 
&= \frac{1}{J}\sum_{j=1}^J
\left\{(P_{n,j} - P_0)\{\phi_{0,r_{n,j}^*} - \phi_{0,r_0} + r_0\alpha_0 - r_{n,j}^*\alpha_{n,j}^*\} \right\}\\
& \quad + \frac{1}{J}\sum_{j=1}^J
\left\{\psi_{n,j}(r_{n,j}^*)-\psi_0(r_{n,j}^*)- (P_{n,j} - P_0)\phi_{0, r_{n,j}^*}\right\} \\
&=\frac{1}{J}\sum_{j=1}^J
\left\{(P_{n,j} - P_0)\{r_0\alpha_0 - r_{n,j}^*\alpha_{n,j}^*\} \right\} \\
& \quad + \frac{1}{J}\sum_{j=1}^J
\left\{\psi_{n,j}(r_{n,j}^*)-\psi_0(r_{n,j}^*)- (P_{n,j} - P_0)\phi_{0, r_{0}}\right\} \\
&=\frac{1}{J}\sum_{j=1}^J
\left\{(P_{n,j} - P_0)\{r_0\alpha_0 - r_{n,j}^*\alpha_{n,j}^*\} \right\}  + o_p(n^{-1/2}),
\end{align*}
where we used that $(P_{n,j} - P_0)\phi_{0, r_{0}} =  P_{n,j} \phi_{0, r_{0}}$ since $\phi_{0, r_{0}}$ is mean zero. Arguing as in the proof of Lemma~\ref{lemma::firstEmpProcRemainders}, we may apply
Lemma~\ref{lemma::LipschitzEmpirical} to conclude that
\[
(\overline{P}_n-\overline{P}_0)\{r_0\alpha_0 - r_{n,j}^*\alpha_{n,j}^*\}
= o_p(n^{-1/2}),
\]
since, under \ref{cond::DRmisDRconsist::1} and \ref{cond::estnuisbound}, Lemma~\ref{lemma::nuisanceconsistency} implies that
\[
\|r_0\alpha_0 - r_{n,j}^*\alpha_{n,j}^*\|_{\overline{P}_0}
\lesssim
\|\alpha_0-\alpha_{n,j}^*\|_{\overline{P}_0} + \|r_0-r_{n,j}^*\|_{\overline{P}_0}
= o_p(1).
\]
The result now follows.

The bootstrap case follows along similar lines. Specifically, as in the proof of Lemma \ref{lemma::firstEmpProcRemainders} , we use the identity
\[
(\overline{P}_n^\#-\overline{P}_0)
=
(\overline{P}_n-\overline{P}_0)
+
(\overline{P}_n^\#-\overline{P}_n),
\]
where empirical process terms involving the first and second differences can be
handled using the first and second parts of Lemma~\ref{lemma::LipschitzEmpirical},
respectively. The bootstrap statements of Lemma~\ref{lemma::nuisanceconsistency}
show that
\[
\|r_{n,\diamond}^{\#} - r_0\|_{\overline{P}_0} = o_p(1)
\quad \text{under \ref{cond::DRmisDRconsist::1},}
\]
and
\[
\|s_{n,\diamond}^{\#} - s_0\|_{\overline{P}_0} = o_p(1)
\quad \text{under \ref{cond::DRmisDRconsist::2}.}
\]
To show that
\[
(\overline{P}_n^\# - P_0)\{B_{n,\diamond}^\# - B_0\} = o_p(n^{-1/2}),
\]
we use that
\[
\frac{1}{J}\sum_{j=1}^J
\left\{\psi_{n,j}(r_{n,j}^\#)-\psi_0(r_{n,j}^\#)- P_{n,j}\phi_{0, r_{0}}\right\}
\;+\;
\frac{1}{J}\sum_{j=1}^J (P_{n,j}- P_0)\{m(\cdot, r_{n,j}^\#) - m(\cdot, r_0)\}
= o_p(n^{-1/2})
\]
by Lemma~\ref{lemma::remainderlipschitzsmall} (proved next).

 \end{proof}

The following lemma can be used to explicitly verify the locally uniform
asymptotic linearity condition in \ref{cond::DRremainder} for the estimated
functional $\psi_n$ when $\psi_n(\mu) = \frac{1}{n}\sum_{i=1}^n m(Z_i,\mu)$ and
$\psi_0(\mu) = E_0\{m(Z,\mu)\}$, so that
$\varphi_{0,\mu} = m(\cdot,\mu) - \psi_0(\mu)$ (the setup of
Section~\ref{section::bootstrap}).

\begin{lemma}
Consider the setup of Section~\ref{section::bootstrap}, where
$\psi_0(h) = E_0\{m(Z,h)\}$. Assume Conditions~\ref{cond::linear}--\ref{cond::pathwise}, \ref{cond::DRlimits}--\ref{cond::variation}, and~\ref{cond::mainprojectioncoupling1::boot}--\ref{cond::lipschitzisotonic}.
Then, for $(f_{n,j}, f_0) \in \{(\mu_{n,j}, \overline{\mu}_0), (\mu_{n,j}^{\#}, \overline{\mu}_0)\}$,
\[
\frac{1}{J}\sum_{j=1}^J (P_{n,j}- P_{0}) \{m(\cdot, f_{n,j}) - m(\cdot, f_0)\}
= o_p(n^{-1/2})
\quad\text{and}\quad
\frac{1}{J}\sum_{j=1}^J (P_{n,j}^\#- P_{n,j}) \{m(\cdot, f_{n,j}) - m(\cdot, f_0)\}
= o_p(n^{-1/2}).
\]
The same conclusions hold for $(f_{n,j}, f_0) \in \{(r_{n,j}, r_0), (r_{n,j}^\#, r_0)\}$
under \ref{cond::DRmisDRconsist::1} (so that $\overline{\alpha}_0 = \alpha_0$).
\label{lemma::remainderlipschitzsmall}
\end{lemma}
\begin{proof}

Lemma~\ref{lemma::LipschitzEmpirical} is not directly applicable because, under
Condition~\ref{cond::lipschitzisotonic}, the map $\mu \mapsto m(\cdot,\mu)$ is
only pointwise Lipschitz in a sup-norm sense. Specifically, the condition assumes that
$\mathcal{A}$ is finite and that there exists a constant $L \in (0,\infty)$
such that, for all $\mu \in \mathcal{H}$,
\[
|m(Z,\mu)| \leq L \max_{a \in \mathcal{A}} |\mu(a,W)| \qquad P_0\text{-almost surely}.
\]
In what follows, we adapt the proof of Lemma~\ref{lemma::LipschitzEmpirical}.

By
\ref{cond::lipschitzisotonic}, we have $\|m(\cdot,\mu)\| \lesssim \|\mu\|$ for all
$\mu \in \mathcal{H}$. Therefore, by \ref{cond::DRlimits}, $\|m(\cdot, \mu_{n,j}^*) - m(\cdot, \overline{\mu}_0)\|
 \lesssim \|\mu_{n,j}^* - \overline{\mu}_0\|
  = o_p(1).$
Under \ref{cond::DRmisDRconsist::1}, by Lemma \ref{lemma::nuisanceconsistency}, we have $\|r_{n,j}^* - r_0\| = o_p(1).$
Hence, under \ref{cond::DRmisDRconsist::1}, $\|m(\cdot, r_{n,j}^*) - m(\cdot, r_0)\|
  \lesssim \|r_{n,j}^* - r_0\|
  = o_p(1).$
We conclude that, under \ref{cond::DRmisDRconsist::1},
\[
\|m(\cdot, \mu_{n,j}^*) - m(\cdot, \overline{\mu}_0)\| = o_p(1) \quad \text{and} \quad  \|m(\cdot, r_{n,j}^*) - m(\cdot, r_0)\| = o_p(1).
\]
A symmetric argument using \ref{cond::mainprojectioncoupling1::boot} also shows that
\[
\|m(\cdot, \mu_{n,j}^{\#}) - m(\cdot, \overline{\mu}_0)\| = o_p(1) \quad \text{and} \quad  \|m(\cdot, r_{n,j}^\#) - m(\cdot, r_0)\| = o_p(1).
\]

 Define the function class $\mathcal{F}_{m,n,j} := \{m(\cdot, f) - m(\cdot, g) : f \in \mathcal{R}_{n,j} \cup \mathcal{S}_{n,j}, g \in \{\overline{\mu}_0, r_0\}\}$. Note that $m(\cdot, \mu_{n,j}^*) - m(\cdot, \overline{\mu}_0)$, $m(\cdot, \mu_{n,j}^{\#}) - m(\cdot, \overline{\mu}_0)$, $m(\cdot, r_{n,j}^*) - m(\cdot, r_0)$, and $m(\cdot, r_{n,j}^\#) - m(\cdot, r_0)$ belong to $\mathcal{F}_{m,n,j}$ with probability tending to one. By Lemma~\ref{lemma::Lipschitzuniformentropy} and Lemma~\ref{lemma::entropy}, it holds that
\[
\mathcal{J}(\delta, \mathcal{F}_{m,n,j}) \lesssim \mathcal{J}(\delta, \mathcal{R}_{n,j} \cup \mathcal{S}_{n,j}) \lesssim \sqrt{\delta}.
\]
The proof of Lemma~\ref{lemma::LipschitzEmpirical} shows that $\mu_{n,j}^*$, $\alpha_{n,j}^*$, $r_{n,j}^*$, and $s_{n,j}^*$ lie in $\mathcal{R}_{n,j} \cup \mathcal{S}_{n,j}$ with probability tending to one. Consequently, the functions
$m(\cdot, \mu_{n,j}^*) - m(\cdot, \overline{\mu}_0)$,
$m(\cdot, \mu_{n,j}^{\#}) - m(\cdot, \overline{\mu}_0)$,
$m(\cdot, r_{n,j}^*) - m(\cdot, r_0)$, and
$m(\cdot, r_{n,j}^\#) - m(\cdot, r_0)$
belong to $\mathcal{F}_{m,n,j}$ with probability tending to one. 

To complete the proof, we apply the local maximal inequalities in Lemma \ref{lemma::maximalIneq} to each $\mathcal{F}_{m,n,j}$ and argue as in the proof of Lemma~\ref{lemma::LipschitzEmpirical} to obtain, for each $j \in [J]$,
\[
(P_{n,j}-P_0)\{m(\cdot,\mu_{n,j}^*)-m(\cdot,\overline{\mu}_0)\}
+(P_{n,j}-P_0)\{m(\cdot,r_{n,j}^*)-m(\cdot,r_0)\}
= o_p(n^{-1/2}),
\]
and
\[
(P_{n,j}^\#-P_{n,j})\{m(\cdot,\mu_{n,j}^{\#})-m(\cdot,\overline{\mu}_0)\}
+(P_{n,j}^\#-P_{n,j})\{m(\cdot,r_{n,j}^\#)-m(\cdot,r_0)\}
= o_p(n^{-1/2}).
\]
The remaining bounds follow from symmetric arguments.

\end{proof}

\section{Convergence rates for isotonic-calibrated nuisance estimators}
\label{appendix::calrates}

Let \(\alpha_{n,\diamond}^* := \{\alpha_{n,j}^*: j \in [J]\}\) and
\(\mu_{n,\diamond}^* := \{\mu_{n,j}^*: j \in [J]\}\) denote the isotonic-calibrated
nuisance estimators produced by Algorithm~\ref{alg:DR}. Let
\(\overline{\alpha}_0\) and \(\overline{\mu}_0\) be their respective
(possibly misspecified) limits, as in Condition~\ref{cond::DRlimits}. In this
section, we derive mean-squared error bounds for the calibrated estimators
\(\alpha_{n,j}^*\) and \(\mu_{n,j}^*\) relative to \(\alpha_0\) and \(\mu_0\).

We work in the setting where the linear functional \(\psi_0\) can be written
as \(\psi_0(\mu) = E_0\{m(Z,\mu)\}\) for a known map
\(m:\mathcal Z \times \mathcal H \to \mathbb R\), and we take \(\psi_n\) to be
the empirical plug-in estimator
\(\psi_n(\mu) = n^{-1}\sum_{i=1}^n m(Z_i,\mu)\). The arguments below extend to
more general estimators \(\psi_n\), provided suitable regularity conditions
hold for the associated second-order remainder.

We begin with a general result on convergence rates for isotonic-calibrated
estimators of Riesz representers of bounded linear functionals. This framework
includes both the outcome regression \(\mu_0\) and the Riesz representer
\(\alpha_0\) as special cases.
For \(z\in\mathcal Z\) and \(\nu\in\mathcal H\), define the Riesz loss \citep{chernozhukov2022automatic}
\[
\ell(z,\nu) := \nu^2(a,w) - 2\,m(z,\nu),
\]
where \(m\) is linear in its second argument. Let \(\{\nu_{n,j}: j\in[J]\}\) be
cross-fitted estimators of the population risk minimizer
\[
\nu_0 := \arg\min_{\nu\in\mathcal H} P_0\,\ell(\cdot,\nu).
\]
Write \(\nu_{n,\diamond} := \{\nu_{n,j}: j\in[J]\}\), and define the calibrated
cross-fitted estimators \(\nu_{n,\diamond}^* := h_n\circ \nu_{n,\diamond}\) and
\(\nu_{n,\diamond}^{\#} := h_n^{\#}\circ \nu_{n,\diamond}\), where the isotonic
regression maps \(h_n\), \(h_n^{\#}\), and \(h_{n,0}\) are given by
\begin{align*}
h_n
&:= \arg\min_{h\in\mathcal F_{\mathrm{iso}}}
\sum_{i=1}^n \ell\!\left(Z_i,\; h\circ \nu_{n,j(i)}\right),\\
h_n^{\#}
&:= \arg\min_{h\in\mathcal F_{\mathrm{iso}}}
\sum_{i=1}^n \ell\!\left(Z_i^{\#},\; h\circ \nu_{n,j(i)}\right),\\
h_{n,0}
&:= \arg\min_{h\in\mathcal F_{\mathrm{iso}}}
\sum_{j=1}^J E_0\!\left[\ell\!\left(Z,\; h\circ \nu_{n,j}\right)\right]
\;=\;
\arg\min_{h\in\mathcal F_{\mathrm{iso}}}
\bigl\|h\circ \nu_{n,\diamond} - \nu_0\bigr\|_{\overline P_0}.
\end{align*}

We impose the following conditions.
\begin{enumerate}[label=(F\arabic*), ref=F\arabic*]
\item \label{F1}
With probability tending to one, \(h_n\), \(h_n^{\#}\), and \(h_{n,0}\) lie in a
fixed, uniformly bounded subset of \(\mathcal F_{\mathrm{iso}}\).
\item \label{F2}
There exists \(L<\infty\) such that \(|E_0\{m(Z,h)\}|\le L\|h\|_{P_0}\) for all
\(h\in\mathcal H\).
\item \label{F3}
The action space \(\mathcal A\) is finite, and there exists \(L\in(0,\infty)\)
such that for all \(\mu\in\mathcal H\),
\(|m(Z,\mu)| \le L \max_{a\in\mathcal A}|\mu(a,W)|\) almost surely.
\end{enumerate}

 \begin{theorem}
  Assume \ref{F1}-\ref{F3}. Then, $ \|h_n \circ \nu_{n, \diamond} - h_{n,0} \circ \nu_{n, \diamond}\|_{\overline{P}_0} = O_p(n^{-1/3}) $. Hence,
\begin{align*}
 \|h_n \circ \nu_{n, \diamond} - \nu_0\|_{\overline{P}_0} & \leq \inf_{h \in \mathcal{F}_{iso}}\|h \circ \nu_{n,\diamond} -\nu_0\| +  O_p(n^{-1/3})
\end{align*}
Moreover, the bootstrap calibrated nuisance estimators satisfy, for any function $\overline{\nu}_0 \in L^2(P_0)$ and as $n \rightarrow \infty$, the following root mean square error bounds:
\begin{align*}
   \|h_n^{\#} \circ \nu_{n, \diamond} - \overline{\nu}_0 \|_{\overline{P}_0} &\leq   \|h_n \circ \nu_{n, \diamond} - \overline{\nu}_0\|_{\overline{P}_0} +  O_p(n^{-1/3}).
\end{align*}\label{theorem::MSEgeneral}
\end{theorem}

\begin{proof}
Let $\widetilde{\mathcal{F}}_{iso}$ be a uniformly bounded subset of $\mathcal{F}_{iso}$ that contains $h_n$, $h_n^{\#}$, and $h_{n,0}$ with probability tending to one (which exists by \ref{F1}). Define $\mathcal{F}_{n,j} := \{f \circ \nu_{n,j}: f \in \widetilde{\mathcal{F}}_{iso}\}$. Since $h_n$ and $h_{n,0}$ fall in $\mathcal{F}_{iso}$ with probability tending to one by \ref{F1}, it holds that $\nu_{n,j}^* = h_n \circ \nu_{n,j}$ and $h_{n,0} \circ \nu_{n,j}$ fall in $\mathcal{F}_{n,j}$ with probability tending to one.

We first show that the population risk induced by the Riesz loss is quadratic and strongly convex. By
Assumption~\ref{F2}, the map \(\nu \mapsto P_0 m(\cdot,\nu)\) is a bounded linear
functional on \(\mathcal H\). Moreover, \(\ell\) is the corresponding Riesz loss
\citep{chernozhukov2022automatic}, so the population minimizer
\(\nu_0 := \arg\min_{\nu\in\mathcal H} P_0\ell(\cdot,\nu)\) is its (nonparametric)
Riesz representer. In particular, the Riesz representation property yields
\(P_0 m(\cdot,\nu) = P_0(\nu\nu_0)\) for all \(\nu\in\mathcal H\), and hence
\[
P_0\ell(\cdot,\nu)
= P_0\{\nu^2 - 2 m(\cdot,\nu)\}
= P_0\{\nu^2 - 2 \nu\nu_0\}.
\]
Now fix a uniformly bounded, closed, convex class \(\mathcal F \subseteq \mathcal H\),
and let \(\nu_{0,\mathcal F} := \arg\min_{\nu\in\mathcal F} P_0\ell(\cdot,\nu)\).
Since \(\nu \mapsto P_0\ell(\cdot,\nu)\) is differentiable and convex, the
first-order optimality condition at \(\nu_{0,\mathcal F}\) gives, for all
\(\nu\in\mathcal F\),
\[
\left.\frac{d}{dt}\,P_0\ell\bigl(\cdot,\nu_{0,\mathcal F}+t(\nu-\nu_{0,\mathcal F})\bigr)\right|_{t=0}
= 2\,P_0\!\left\{(\nu_{0,\mathcal F}-\nu_0)(\nu-\nu_{0,\mathcal F})\right\}
\ge 0,
\]
equivalently \(P_0\{(\nu_{0,\mathcal F}-\nu_0)(\nu-\nu_{0,\mathcal F})\}\ge 0\).
Using the quadratic form above, we then expand the excess risk:
\begin{align*}
P_0\{\ell(\cdot,\nu)-\ell(\cdot,\nu_{0,\mathcal F})\}
&= P_0\!\left\{(\nu-\nu_{0,\mathcal F})(\nu+\nu_{0,\mathcal F}-2\nu_0)\right\} \\
&= \|\nu-\nu_{0,\mathcal F}\|_{P_0}^2
+ 2\,P_0\!\left\{(\nu_{0,\mathcal F}-\nu_0)(\nu-\nu_{0,\mathcal F})\right\} \\
&\ge \|\nu-\nu_{0,\mathcal F}\|_{P_0}^2,
\end{align*}
which establishes the strong-convexity inequality
\[
\|\nu-\nu_{0,\mathcal F}\|_{P_0}^2
\le P_0\{\ell(\cdot,\nu)-\ell(\cdot,\nu_{0,\mathcal F})\}.
\]

We prove the convergence rates for the bootstrap estimators, since the rates for
the non-bootstrap estimators follow by similar arguments. To this end, for
$j \in [J]$, define the random rate
$\delta_{n,j}^{\#} := \|h_n^{\#} \circ \nu_{n,j} - h_{n,0} \circ \nu_{n,j}\|_{P_0}$
and let $(\delta_n^{\#})^2 := \frac{1}{J}\sum_{j=1}^J (\delta_{n,j}^{\#})^2 = \|h_n^{\#} \circ \nu_{n,\diamond} - h_{n,0} \circ \nu_{n,\diamond}\|_{\overline{P}_0}^2$.
By convexity of $\mathcal{F}_{iso}$, the quadratic excess risk bound of the
previous paragraph, and the fact that $h_{n,0}$ is a population risk minimizer,
we have
\[
(\delta_n^{\#})^2 \lesssim
\overline{P}_0 \ell(\cdot, h_n^{\#} \circ \nu_{n,\diamond})
- \overline{P}_0 \ell(\cdot, h_{n,0} \circ \nu_{n,\diamond}).
\]
Furthermore, since $h_n^{\#} \circ \nu_{n,\diamond}$ is the bootstrap empirical
risk minimizer, it holds that
\[
\overline{P}_n^{\#}\!\left\{\ell(\cdot, h_n^{\#} \circ \nu_{n,\diamond})
- \ell(\cdot, h_{n,0} \circ \nu_{n,\diamond}) \right\} \le 0 .
\]
Combining the above two displays, we obtain
\begin{align*}
\overline{P}_0 \ell(\cdot, h_n^{\#} \circ \nu_{n,\diamond})
- \overline{P}_0 \ell(\cdot, h_{n,0} \circ \nu_{n,\diamond})
&=
(\overline{P}_0-\overline{P}_n^{\#}) \ell(\cdot, h_n^{\#} \circ \nu_{n,\diamond}) \\
&\quad
+ \overline{P}_n^{\#}\!\left\{\ell(\cdot, h_n^{\#} \circ \nu_{n,\diamond})
- \ell(\cdot, h_{n,0} \circ \nu_{n,\diamond}) \right\} \\
&\quad
- (\overline{P}_0-\overline{P}_n^{\#}) \ell(\cdot, h_{n,0} \circ \nu_{n,\diamond}) \\
&\le
(\overline{P}_0-\overline{P}_n^{\#}) \left\{\ell(\cdot, h_n^{\#} \circ \nu_{n,\diamond})
- \ell(\cdot, h_{n,0} \circ \nu_{n,\diamond}) \right\},
\end{align*}
and therefore
\begin{align*}
(\delta_n^{\#})^2
&\le
(\overline{P}_0-\overline{P}_n^{\#}) \left\{\ell(\cdot, h_n^{\#} \circ \nu_{n,\diamond})
- \ell(\cdot, h_{n,0} \circ \nu_{n,\diamond}) \right\} \nonumber \\
&\le
(\overline{P}_0-\overline{P}_n) \left\{\ell(\cdot, h_n^{\#} \circ \nu_{n,\diamond})
- \ell(\cdot, h_{n,0} \circ \nu_{n,\diamond}) \right\} \nonumber \\
&\quad
+ (\overline{P}_n-\overline{P}_n^{\#}) \left\{\ell(\cdot, h_n^{\#} \circ \nu_{n,\diamond})
- \ell(\cdot, h_{n,0} \circ \nu_{n,\diamond}) \right\} \nonumber \\
&\le
\frac{1}{J}\sum_{j=1}^J \left[\gamma_{n,j}(\delta_{n,j}^{\#})
+ \gamma_{n,j}^{\#}(\delta_{n,j}^{\#})\right], \nonumber
\end{align*}
where we apply the triangle inequality and define
\begin{align*}
\gamma_{n,j}(\delta)
&:= \sup_{\nu_1, \nu_2 \in \mathcal{F}_{n,j}:\ \| \nu_1 - \nu_2\|\le \delta }
\left|({P}_0-{P}_{n,j}) \left\{\ell(\cdot, \nu_1) - \ell(\cdot, \nu_2) \right\} \right|,\\
\gamma_{n,j}^{\#}(\delta)
&:= \sup_{\nu_1, \nu_2 \in \mathcal{F}_{n,j}:\ \| \nu_1 - \nu_2\|\le \delta }
\left|(P_{n,j} - P_{n,j}^{\#}) \left\{\ell(\cdot, \nu_1) - \ell(\cdot, \nu_2) \right\} \right|.
\end{align*}
Moreover, since $\max_{j \in [J]} \delta_{n,j}^{\#} \le J \delta_{n}^{\#}$, we have
\begin{align}
(\delta_n^{\#})^2
&\le \frac{1}{J}\sum_{j=1}^J \left[\gamma_{n,j}(J \delta_{n}^{\#})
+ \gamma_{n,j}^{\#}(J \delta_{n}^{\#})\right].
\label{eqn::basicineq}
\end{align}

By Assumption~\ref{F3} and the boundedness of $\mathcal{F}_{n,j}$, the function
class \(\{\ell(\cdot, \nu_1) - \ell(\cdot, \nu_2) : \nu_1, \nu_2 \in \mathcal{F}_{n,j}\}\)
has a controlled uniform entropy integral, conditional on the training data for
the nuisance estimators (so that the class is fixed given this training data).
In particular, by preservation properties of the uniform entropy integral
\citep{van1996weak}, the product class $\mathcal{F}_{n,j} \times \mathcal{F}_{n,j}$
has a uniform entropy integral bounded, up to a constant, by
\(\mathcal{J}(\delta, \mathcal{F}_{\nu_{n,j}, \mathrm{iso}}) + \mathcal{J}(\delta, \mathcal{F}_{\nu_{n,j}, \mathrm{iso}})\)
for each $j \in [J]$. This quantity is itself, up to a constant, bounded above by
\(\mathcal{J}(\delta, \mathcal{F}_{\mathrm{iso}})\) (see, e.g., the proof of
Lemma~\ref{lemma::entropy}). Therefore, applying Lemma~\ref{lemma::Lipschitzuniformentropy}
and following an argument identical to the proof of Lemma~\ref{lemma::maximalIneq}
with \( t_n(\delta) := \sqrt{\delta} \), we obtain, for any fixed \( \delta > 0 \),
conditional on the training sample
\( \mathcal{T}_{n,j} := \mathcal{D}_n \setminus \mathcal{C}^{(j)} \),
\begin{equation}
\label{eqn::proofermrate}
\begin{aligned}
E_0^n\!\left[ \gamma_{n,j}(\delta) \mid \mathcal{T}_{n,j}\right]
&\lesssim n^{-\frac{1}{2}} \sqrt{\delta}
\left(1 + \frac{\sqrt{\delta}}{\sqrt{n}\,\delta^2} \right), \\
E_0^n\!\left[ \gamma_{n,j}^{\#}(\delta) \mid \mathcal{T}_{n,j}\right]
&\lesssim n^{-\frac{1}{2}} \sqrt{\delta}
\left(1 + \frac{\sqrt{\delta}}{\sqrt{n}\,\delta^2} \right),
\end{aligned}
\end{equation}
where the bound on the right-hand side is deterministic with ``$\lesssim$'' holding
for a fixed constant.

 We are now ready to obtain a convergence rate bound for $\delta_n^{\#} = \|h_n^{\#} \circ \nu_{n,\diamond} - h_{n,0} \circ \nu_{n,\diamond}\|_{\overline{P}_0} $. To do so, we use a peeling argument similar to the proof of Theorem 3.2.5 in \citet{van1996weak}.  Proceeding with the proof and letting $\varepsilon_n \in \mathbb{R}$ be a deterministic rate to be set later and using that \eqref{eqn::basicineq} holds with probability one
\begin{align*}
P\!\left(2^{k+1}\varepsilon_n \ge \delta_n^{\#} \ge 2^k\varepsilon_n\right)
&=
P\!\left(
2^{k+1}\varepsilon_n \ge \delta_n^{\#} \ge 2^k\varepsilon_n,\;
(\delta_n^{\#})^2
\le
\frac{1}{J}\sum_{j=1}^J\Bigl[\gamma_{n,j}(J\delta_n^{\#})+\gamma_{n,j}^{\#}(J\delta_n^{\#})\Bigr]
\right)\\
&\le
P\!\left(
2^{k+1}\varepsilon_n \ge \delta_n^{\#} \ge 2^k\varepsilon_n,\;
2^{2k}\varepsilon_n^2
\le
\frac{1}{J}\sum_{j=1}^J\Bigl[\gamma_{n,j}(J2^{k+1}\varepsilon_n)+\gamma_{n,j}^{\#}(J2^{k+1}\varepsilon_n)\Bigr]
\right)\\
&\le
P\!\left(
2^{2k}\varepsilon_n^2
\le
\frac{1}{J}\sum_{j=1}^J\Bigl[\gamma_{n,j}(J2^{k+1}\varepsilon_n)+\gamma_{n,j}^{\#}(J2^{k+1}\varepsilon_n)\Bigr]
\right)\\
&\le
\sum_{j=1}^J
E_0^n\!\left[
P\!\left(
2^{2k}\varepsilon_n^2
\le
\gamma_{n,j}(J2^{k+1}\varepsilon_n)+\gamma_{n,j}^{\#}(J2^{k+1}\varepsilon_n)
\;\middle|\; \mathcal T_{n,j}
\right)
\right]\\
&\le
\sum_{j=1}^J
E_0^n\!\left[
\frac{
E_0^n\!\left[\gamma_{n,j}(J2^{k+1}\varepsilon_n)\mid \mathcal T_{n,j}\right]
+
E_0^n\!\left[\gamma_{n,j}^{\#}(J2^{k+1}\varepsilon_n)\mid \mathcal T_{n,j}\right]
}{
2^{2k}\varepsilon_n^2
}
\right].
\end{align*}
Taking \(\varepsilon_n := n^{-1/3}\) above and applying \eqref{eqn::proofermrate},
we can show, for some constant \(C'>0\), that
\[
P\!\left(2^{k+1}n^{-1/3} \ge \delta_n^{\#} \ge 2^{k}n^{-1/3}\right)
\lesssim
\frac{ n^{-1/2}\sqrt{2^{k+1}\varepsilon_n}\Bigl(1+\frac{\sqrt{2^{k+1}\varepsilon_n}}{\sqrt{n}\,\varepsilon_n^2}\Bigr)}{2^{2k}\varepsilon_n^2}
\le C'\,2^{-k}.
\]
Therefore, for any \(K>0\),
\[
P\!\left(\delta_n^{\#} \ge 2^{K} n^{-1/3}\right)
=
\sum_{k=K}^{\infty}
P\!\left(2^{k+1}n^{-1/3} \ge \delta_n^{\#} \ge 2^{k}n^{-1/3}\right)
\le
C'\sum_{k=K}^{\infty}2^{-k},
\]
and \(\sum_{k=K}^{\infty}2^{-k}\to 0\) as \(K\to\infty\). Hence, for every
\(\varepsilon>0\), there exists \(K>0\) such that
\(P(\delta_n^{\#} \ge 2^{K}n^{-1/3}) \le \varepsilon\). By definition,
\(\delta_n^{\#}=O_p(n^{-1/3})\), and therefore
\[
\|h_n^{\#}\circ\nu_{n,\diamond}-h_{n,0}\circ\nu_{n,\diamond}\|_{\overline P_0}^2
=O_p(n^{-2/3}).
\]
An analogous argument for the non-bootstrap case yields
\[
\|h_n\circ\nu_{n,\diamond}-h_{n,0}\circ\nu_{n,\diamond}\|_{\overline P_0}^2
=O_p(n^{-2/3}).
\]
By the triangle inequality,
\[
\|h_n^{\#}\circ\nu_{n,\diamond}-h_n\circ\nu_{n,\diamond}\|_{\overline P_0}
\le
\|h_n^{\#}\circ\nu_{n,\diamond}-h_{n,0}\circ\nu_{n,\diamond}\|_{\overline P_0}
+
\|h_n\circ\nu_{n,\diamond}-h_{n,0}\circ\nu_{n,\diamond}\|_{\overline P_0}
=O_p(n^{-1/3}),
\]
and hence
\[
\|h_n^{\#}\circ\nu_{n,\diamond}-h_n\circ\nu_{n,\diamond}\|_{\overline P_0}^2
=O_p(n^{-2/3}).
\]
Therefore, for any function \(\overline{\nu}_0\),
\begin{align*}
\|h_n^{\#}\circ\nu_{n,\diamond}-\overline{\nu}_0\|_{\overline P_0}
&\le
\|h_n^{\#}\circ\nu_{n,\diamond}-h_n\circ\nu_{n,\diamond}\|_{\overline P_0}
+
\|h_n\circ\nu_{n,\diamond}-\overline{\nu}_0\|_{\overline P_0}\\
&\le
O_p(n^{-1/3})
+
\|h_n\circ\nu_{n,\diamond}-\overline{\nu}_0\|_{\overline P_0}.
\end{align*}

\end{proof}

The following theorem is a direct application of Theorem \ref{theorem::MSEgeneral}.
 In the following, we define the population minimizers corresponding to the calibrators $f_n$ and $g_n$ as:
\begin{align*}
   f_{n, 0} &:= \argmin_{f \in \mathcal{F}_{iso}} \sum_{j=1}^J \|\mu_0 - f \circ \mu_{n,j}  \|_{P_0}^2;\\
    g_{n, 0} &:= \argmin_{g \in \mathcal{F}_{iso}}  \sum_{j=1}^J \left\{\|g \circ \alpha_{n,j} \|_{P_0}^2 - 2 \psi_0(g \circ \alpha_{n,j})\right\}.
\end{align*}

\begin{theorem}
\label{theorem::MSE}
Suppose that Conditions~\ref{cond::linear}, \ref{cond::estboundedlinearcross}, \ref{cond::estnuisbound}, and~\ref{cond::lipschitzisotonic} hold.
 Then, $ \|\alpha_{n,\diamond}^* - g_{n,0} \circ \alpha_{n,\diamond}\|_{\overline{P}_0} = O_p(n^{-1/3}) $ and $ \|\mu_{n,\diamond}^* - f_{n,0} \circ \mu_{n,\diamond}\|_{\overline{P}_0} = O_p(n^{-1/3})$. Hence,
\begin{align*}
 \|\alpha_{n,\diamond}^* - \alpha_0\|_{\overline{P}_0} & \leq \min_{g \in \mathcal{F}_{iso}}\|g \circ \alpha_{n,\diamond} -\alpha_0\| +  O_p(n^{-1/3}); \\
 \|\mu_{n,\diamond}^* - \mu_0\|_{\overline{P}_0} &\leq  \min_{f \in \mathcal{F}_{iso}}\|f \circ \mu_{n,\diamond} -\mu_0\| +  O_p(n^{-1/3}).
\end{align*}
Additionally, assume Condition~\ref{cond::estboundboot} holds. Then, the bootstrap calibrated nuisance estimators satisfy, for any functions $\overline{\alpha}_0, \overline{\mu}_0 \in L^2(P_0)$ and as $n \rightarrow \infty$, the following root mean square error bounds:
\begin{align*}
   \|\alpha_{n,\diamond}^{\#} - \overline{\alpha}_0 \|_{\overline{P}_0} &\leq   \|\alpha_{n,\diamond}^* - \overline{\alpha}_0\|_{\overline{P}_0} +  O_p(n^{-1/3}); \\
    \|\mu_{n,\diamond}^{\#} -   \overline{\mu}_0 \|_{\overline{P}_0} &\leq   \|\mu_{n,\diamond}^* -  \overline{\mu}_0\|_{\overline{P}_0} +  O_p(n^{-1/3}).
\end{align*}

\end{theorem}
\begin{proof}
  The proof follows from two direct applications of Theorem~\ref{theorem::MSEgeneral}. To obtain the claimed rates for \( \alpha_{n,\diamond}^* \), we take 
\[
\ell(z, \nu) := \nu^2(a,w) - 2m(z,\nu), \quad h_n := g_n, \quad h_{n,0} := g_{n,0}, \quad h_n^{\#} := g_n^{\#}, \quad \nu_0 := \alpha_0, \quad \nu_{n,j} := \alpha_{n,j}.
\] 
Similarly, to obtain the rates for \( \mu_{n,\diamond}^* \), we take 
\[
\ell(z, \nu) := \nu^2(a,w) - 2y\nu(a,w), \quad h_n := f_n, \quad h_{n,0} := f_{n,0}, \quad h_n^{\#} := f_n^{\#}, \quad \nu_0 := \mu_0, \quad \nu_{n,j} := \mu_{n,j}.
\]

In both cases, Condition~\ref{F1} holds by Conditions~\ref{cond::bound} and~\ref{cond::estboundboot}. For the least-squares loss, Condition~\ref{F2} holds trivially, while for the Riesz loss, it holds by Condition~\ref{cond::linear}. For the least-squares loss, Condition~\ref{F3} holds by the pointwise Lipschitz continuity of the map \( m \mapsto m^2 - 2ym \) and the uniform boundedness of the outcome \( Y \). For the Riesz loss, Condition~\ref{F3} holds by \ref{cond::lipschitzisotonic}.

\end{proof}

 \section{Proofs of main results}

\subsection{Proof of Theorem~\ref{theorem::DRinference}}

Let \(\alpha_{n,\diamond}^* := \{\alpha_{n,j}^*: j \in [J]\}\) and
\(\mu_{n,\diamond}^* := \{\mu_{n,j}^*: j \in [J]\}\) denote the isotonic-calibrated
nuisance estimators produced by Algorithm~\ref{alg:DR}.  

The following lemma shows that isotonic calibration of the cross-fitted nuisance
estimators ensures empirical calibration for their respective loss functions.
Consequently, the calibrated nuisance estimators satisfy cross-fitted
orthogonality conditions, and the leading bias terms in the expansion of the
cross-product remainder in Lemma~\ref{lemma::outcomeRieszBiasDR} vanish.

 \begin{lemma}[Empirical calibration via isotonic calibration]
 \label{lemma::isoscores}
 Assume \ref{cond::estboundedlinearcross}. Then, for all bounded measurable $\theta_1, \theta_2: \mathbb{R} \rightarrow \mathbb{R}$, it holds that
     \begin{align*}
\frac{1}{n}\sum_{i=1}^n  (\theta_1 \circ \mu_{n,j(i)}^*)(A_i, W_i) \left\{Y_i - \mu_{n,j(i)}^*(A_i, W_i) \right\} &= 0;\\
    \frac{1}{n}\sum_{i=1}^n \left\{  (\theta_2 \circ \alpha_{n,j(i)}^*)(A_i, W_i)\alpha_{n,j(i)}^*(A_i, W_i) - \psi_{n,j(i)}(\theta_2 \circ \alpha_{n,j(i)}^*) \right\}&= 0.         
     \end{align*} 
 \end{lemma}
 As a consequence,
      \begin{align*}
\frac{1}{n}\sum_{i=1}^n  s_{n,j(i)}^*(A_i, W_i) \left\{Y_i - \mu_{n,j(i)}^*(A_i, W_i) \right\} &= 0;\\
    \frac{1}{n}\sum_{i=1}^n \left\{ r_{n,j(i)}^*(A_i, W_i)\alpha_{n,j(i)}^*(A_i, W_i) - \psi_{n,j(i)}(r_{n,j(i)}^*) \right\}&= 0.         
     \end{align*} 
 \begin{proof}

The second part of the lemma follows from the first part because, by definition of the projection,
each $s_{n,j}^*$ can be written as $s_{n,j}^* = \theta \circ \mu_{n,j}^*$ for a
common map $\theta$ across $j \in [J]$, where
$s_{n,j}^* = (\overline{\Pi}_{\mu_{n,\diamond}^*}(\alpha_0 - \alpha_{n,\diamond}))_j$.
Similarly, by definition of the projection, each $r_{n,j}^*$ can be written as
$r_{n,j}^* = \vartheta \circ \alpha_{n,j}^*$ for a common map $\vartheta$, where
$r_{n,j}^* = (\overline{\Pi}_{\alpha_{n,\diamond}^*}(\mu_0 - \mu_{n,\diamond}))_j$.

The proof of this lemma follows from a simplification of the proof of Lemma~4 in
\cite{van2023causal}. We give the proof for \(\alpha_{n,\diamond}^*\), as the
argument for \(\mu_{n,\diamond}^*\) follows along similar lines. Recall that
\(\alpha_{n,\diamond}^* = g_n \circ \alpha_{n,\diamond}\), where \(g_n \in
\mathcal{F}_{\mathrm{iso}}\) is the empirical risk minimizer defined in
Algorithm~\ref{alg:DR}.  The solution \(g_n\) is unique only in \(L^2(P_n)\).
Following \citet{groeneboom1993isotonic} and without loss of generality, we take
\(g_n\) to be the unique c\`adl\`ag piecewise-constant solution of the isotonic
regression problem that can jump only at the observed values
\(\{\alpha_{n,j(i)}(A_i,W_i): i \in [n]\}\).

    By Condition~\ref{cond::estboundedlinearcross}, the empirical risk used to define the isotonic solution \( g_n \) is strongly convex and continuous on \( \mathcal{F}_{\text{iso}} \) with respect to the empirical \( L^2 \) norm \( \left\{\sum_{i=1}^n \left[(g \circ \alpha_{n,j(i)})(A_i, W_i)\right]^2\right\}^{1/2} \). By Theorem~5.5 of \cite{alexanderian2019optimization}, a strongly convex and continuous functional defined on a closed and bounded convex subset of a Hilbert space admits a unique minimizing solution on the set $\{\alpha_{n,j(i)}(A_i, W_i): i \in [n]\}$. Consequently, the solution \( g_n \) is uniquely defined on the set \( \{\alpha_{n,j(i)}(A_i, W_i) : 1 \leq i \leq n\} \), and, by assumption, it is the uniquely defined piecewise constant solution with jumps occurring only at these values.

    It holds that $g_n$ is piecewise constant. We claim that, for any function $h: \mathbb{R} \rightarrow \mathbb{R}$ and all $\varepsilon \in \mathbb{R}$ sufficiently close to zero, the perturbed function $g_n + \varepsilon h \circ g_n$ belongs to $\mathcal{F}_{iso}$. To see this, observe that $g_n$ is constant on some interval of its domain if and only if the same holds for $h \circ g_n$. Hence, the function $g_n + \varepsilon h \circ g_n$ is piecewise constant and jumps at the same points as $g_n$. Moreover, by taking $\varepsilon$ sufficiently small, we can ensure that the maximal jump change $\varepsilon \sup_{i \in [n]} |(h \circ g_n)(\mu_{n,j(i)}(A_i,W_i))|$ of $h \circ g_n$ is strictly smaller than the maximal jump change of $g_n$. Therefore, for this choice of $\varepsilon$, the perturbed function $g_n + \varepsilon h \circ g_n \in \mathcal{F}_{iso}$ must be nondecreasing and, therefore, an element of $\mathcal{F}_{iso}$. The claim now follows.

    Proceeding with the proof of the lemma, we denote $\xi_n(\varepsilon)  := g_n + \varepsilon (h \circ g_n)$ and $\xi_n(\varepsilon, \alpha_{n,j}) := g_n \circ \alpha_{n,j} + \varepsilon (h \circ g_n)(\alpha_{n,j})$ and, in a slight abuse of notation, let
    $$R_n(\xi_n(\varepsilon)) := \sum_{i=1}^n\left\{ \xi_n(\varepsilon, \alpha_{n,j(i)})^2(A_i, W_i) - 2 \psi_{n,j(i)}(\xi_n(\varepsilon, \alpha_{n,j(i)}))\right\}.$$
    Now, because \(g_n\) minimises \(g \mapsto R_n(g)\) over \(\mathcal{F}_{\mathrm{iso}}\) and \(g_n + \varepsilon h \circ g_n \in \mathcal{F}_{\mathrm{iso}}\) for \(\varepsilon\) sufficiently close to \(0\), \(g_n\) also minimizes \(g \mapsto R_n(g)\) over the one-dimensional submodel \(\varepsilon \mapsto g_n + \varepsilon h \circ g_n\) in a neighborhood of \(\varepsilon = 0\). Thus, the first-order optimality condition along this submodel implies that
\[
\frac{d}{d\varepsilon}[R_n(\xi_n(\varepsilon)) - R_n(g_n)]\Big|_{\varepsilon = 0} = 0.
\]
    Computing the derivative, using linearity of the functional $\psi_{n,\diamond}$ and that $h \circ g_n \circ \alpha_{n,\diamond} = h \circ \alpha_{n,\diamond}^*$, this further implies:
     \[
 \frac{1}{n}\sum_{i=1}^n \left\{  (h \circ \alpha_{n,j(i)}^*)(A_i, W_i)\alpha_{n,j(i)}^*(A_i, W_i) - \psi_{n,j(i)}(h \circ \alpha_{n,j(i)}^*) \right\}= 0. 
\] 
The result for $\alpha_{n,\diamond}^*$ now follows, noting that $h$ was an arbitrary bounded function.

\end{proof}

 \begin{proof}[Proof of Theorem \ref{theorem::DRinference}]

\noindent \textbf{Outline.} Our proof proceeds in three steps. First, we show that the estimation error
$\tau_n^* - \tau_0$ admits the decomposition
\[
\tau_n^* - \tau_0
=
(P_n - P_0)\{\phi_{0,\overline{\mu}_0}+D_{\overline{\mu}_0,\overline{\alpha}_0}\}
+ o_p(n^{-1/2})
\;+\;
\langle \alpha_{n,\diamond}^*-\alpha_0,\ \mu_0-\mu_{n,\diamond}^* \rangle_{\overline{P}_0},
\]
that is, an asymptotically linear term plus the usual cross-product remainder.
Second, we use the orthogonality properties implied by nuisance calibration to
control this cross-product remainder under each of the two cases in
Condition~\ref{cond::DRmisDRconsist}. Specifically, we show that
\[
\langle \alpha_{n,\diamond}^*-\alpha_0,\ \mu_0-\mu_{n,\diamond}^* \rangle_{\overline{P}_0}
=
1(\overline{\alpha}_0 \neq \alpha_0)\, P_n A_0
+ 1(\overline{\mu}_0 \neq \mu_0)\, P_n B_0
+ o_p(n^{-1/2}).
\]
Finally, combining these steps establishes that $\tau_n^*$ is DRAL with influence
function $\chi_0$.

\paragraph{Starting point: decomposition of the estimation error into an asymptotically linear term and a cross-product drift term.}  By Lemma~\ref{lemma::standardBiasExpansion}, we have the standard doubly robust expansion
\begin{align}
\label{eqn::standardDR}
\tau_n^* - \tau_0
&= (\overline{P}_n - \overline{P}_0)\left\{\phi_{0,\mu_{n,\diamond}^*} + D_{\mu_{n,\diamond}^*, \alpha_{n,\diamond}^*}\right\}
+ \langle \alpha_{n,\diamond}^* - \alpha_0,\ \mu_0 - \mu_{n,\diamond}^* \rangle_{\overline{P}_0}
+ Rem_{\mu_{n,\diamond}^*}(\psi_{n,\diamond}, \psi_0),
\end{align}
where $Rem_{\mu_{n,\diamond}^*}(\psi_{n,\diamond}, \psi_0)
:= \frac{1}{J}\sum_{j=1}^J \left\{\psi_{n,j}(\mu_{n,j}^*) - \psi_0(\mu_{n,j}^*) - P_{n,j}\phi_{0,\mu_{n,j}^*}\right\}.$
By the locally uniform asymptotic linearity condition in \ref{cond::DRremainder}, we have
\begin{align*}
Rem_{\mu_{n,\diamond}^*}(\psi_{n,\diamond}, \psi_0)
&= (\overline{P}_n - \overline{P}_0)\left\{ \phi_{0,\overline{\mu}_0} - \phi_{0,\mu_{n,\diamond}^*} \right\} +  \frac{1}{J}\sum_{j=1}^J \left\{\psi_{n,j}(\mu_{n,j}^*) - \psi_0(\mu_{n,j}^*) - P_{n,j}\phi_{0,\overline{\mu}_0}\right\}.\\
&= (\overline{P}_n - \overline{P}_0)\left\{\phi_{0,\overline{\mu}_0} - \phi_{0,\mu_{n,\diamond}^*}\right\} + o_p(n^{-1/2}),
\end{align*}
where we used that $\overline{P}_0 \phi_{0,\mu_{n,\diamond}^*} = \overline{P}_0 \phi_{0,\overline{\mu}_0} = 0$, as influence functions are mean zero. Substituting the preceding display into \eqref{eqn::standardDR} and collecting like terms, we obtain
\begin{align}
\label{eqn::standardDR2}
\tau_n^* - \tau_0
&= (\overline{P}_n - \overline{P}_0)\left\{\phi_{0,\overline{\mu}_0} + D_{\mu_{n,\diamond}^*, \alpha_{n,\diamond}^*}\right\}
+ \langle \alpha_{n,\diamond}^* - \alpha_0,\ \mu_0 - \mu_{n,\diamond}^* \rangle_{\overline{P}_0}
+ o_p(n^{-1/2}).
\end{align}
By Lemma~\ref{lemma::firstEmpProcRemainders}, $(\overline{P}_n - \overline{P}_0)\left\{D_{\mu_{n,\diamond}^*, \alpha_{n,\diamond}^*} - D_{\overline{\mu}_0, \overline{\alpha}_0}\right\}
= o_p(n^{-1/2}).$
Hence,
\begin{align*}
(\overline{P}_n - \overline{P}_0)D_{\mu_{n,\diamond}^*, \alpha_{n,\diamond}^*}
&= (\overline{P}_n - \overline{P}_0)D_{\overline{\mu}_0, \overline{\alpha}_0}
+ (\overline{P}_n - \overline{P}_0)\left\{D_{\mu_{n,\diamond}^*, \alpha_{n,\diamond}^*} - D_{\overline{\mu}_0, \overline{\alpha}_0}\right\} \\
&= (\overline{P}_n - \overline{P}_0)D_{\overline{\mu}_0, \overline{\alpha}_0} + o_p(n^{-1/2}).
\end{align*}
Substituting the above into \eqref{eqn::standardDR2}, we obtain
\begin{align}
\label{eqn::standardDR3}
\tau_n^* - \tau_0
&= (P_n - P_0) \left\{\phi_{0,\overline{\mu}_0} + D_{\overline{\mu}_0, \overline{\alpha}_0}\right\}
+ \langle \alpha_{n,\diamond}^* - \alpha_0,\ \mu_0 - \mu_{n,\diamond}^* \rangle_{\overline{P}_0}
+ o_p(n^{-1/2}).
\end{align}
The above expansion is the starting point for our DRAL analysis. In what follows, we control the cross-product remainder $ \langle \alpha_{n,\diamond}^* - \alpha_0,\ \mu_0 - \mu_{n,\diamond}^* \rangle_{\overline{P}_0}$ separately under the two cases in \ref{cond::DRlimits}.

\noindent \paragraph{Establishing asymptotic linearity when Riesz representer is estimated well.} Suppose the theorem is applied under \ref{cond::DRmisDRconsist::1} ($\overline{\alpha}_0 = \alpha_0$), so that $\|\alpha_{n,j}^* - \alpha_0 \| = o_p(n^{-1/4})$ and $\|\mu_{n,j}^* - \overline{\mu}_0\| = o_p(1)$ for $j=1,\ldots,J$. Applying the second part of Lemma~\ref{lemma::outcomeRieszBiasDR}, we have the Riesz-based decomposition:
\begin{equation}
\begin{aligned}
\label{eqn::crossriesz1}
\big\langle \alpha_0 - \alpha_{n,\diamond}^*,\ \mu_{n,\diamond}^* - \mu_0 \big\rangle_{\overline{P}_0}
&=
-\left\{\overline{\psi}_n(r_{n, \diamond}^*)
-\overline{P}_n\bigl(r_{n,\diamond}^*\,\alpha_{n,\diamond}^*\bigr)\right\}
\\
&\quad
+\Big\langle \bigl(\overline{\Pi}_{(\alpha_{n,\diamond}^*,\alpha_0)}-\overline{\Pi}_{\alpha_{n,\diamond}^*}\bigr)\Delta_{\mu,n,\diamond},\;
\alpha_0-\overline{\Pi}_{\alpha_{n,\diamond}^*}\alpha_0 \Big\rangle_{\overline{P}_0}
+ (\overline{P}_n-\overline{P}_0)B_{n,\diamond}^* + Rem_{\psi_{n,\diamond}}(r_{n,\diamond}^*),
\end{aligned}
\end{equation}
where $Rem_{\psi_{n,\diamond}}(r_{n,\diamond}^*)
:= \frac{1}{J}\sum_{j=1}^J
\left\{\psi_{n,j}(r_{n,j}^*)-\psi_0(r_{n,j}^*)- P_{n,j}\phi_{0, r_{n,j}^*}\right\}.$
By the second part of Lemma~\ref{lemma::secondEmpProcRemainders}, we have
\[
(\overline{P}_n - \overline{P}_0)\{B_{n,\diamond}^* - B_0\} + Rem_{\psi_{n,\diamond}}(r_{n,\diamond}^*)
= o_p(n^{-1/2}).
\]
Hence,
\[
(\overline{P}_n-\overline{P}_0)B_{n,\diamond}^* + Rem_{\psi_{n,\diamond}}(r_{n,\diamond}^*)
= (\overline{P}_n-\overline{P}_0) B_0 + o_p(n^{-1/2}).
\]
Substituting the above into \eqref{eqn::crossriesz1} and using that $\overline{P}_0 B_0 = 0$, we obtain
\begin{equation}
\begin{aligned}
\label{eqn::crossriesz2}
\big\langle \alpha_0 - \alpha_{n,\diamond}^*,\ \mu_{n,\diamond}^* - \mu_0 \big\rangle_{\overline{P}_0}
&=
-\left\{\overline{\psi}_n(r_{n, \diamond}^*)
-\overline{P}_n\bigl(r_{n,\diamond}^*\,\alpha_{n,\diamond}^*\bigr)\right\}
\\
&\quad
+\Big\langle \bigl(\overline{\Pi}_{(\alpha_{n,\diamond}^*,\alpha_0)}-\overline{\Pi}_{\alpha_{n,\diamond}^*}\bigr)\Delta_{\mu,n,\diamond},\;
\alpha_0-\overline{\Pi}_{\alpha_{n,\diamond}^*}\alpha_0 \Big\rangle_{\overline{P}_0}
+  \overline{P}_n  B_0  + o_p(n^{-1/2}).
\end{aligned}
\end{equation}
Since $\alpha_{n,\diamond}^*$ is calibrated using isotonic regression, the second part of Lemma~\ref{lemma::isoscores} implies that
\[
-\left\{\overline{\psi}_n(r_{n, \diamond}^*)
-\overline{P}_n\bigl(r_{n,\diamond}^*\,\alpha_{n,\diamond}^*\bigr)\right\} = 0.
\]
Thus, the first term in \eqref{eqn::crossriesz2} vanishes, and we conclude that
\begin{equation}
\begin{aligned}
\label{eqn::crossriesz3}
\big\langle \alpha_0 - \alpha_{n,\diamond}^*,\ \mu_{n,\diamond}^* - \mu_0 \big\rangle_{\overline{P}_0}
&= \Big\langle \bigl(\overline{\Pi}_{(\alpha_{n,\diamond}^*,\alpha_0)}-\overline{\Pi}_{\alpha_{n,\diamond}^*}\bigr)\Delta_{\mu,n,\diamond},\;
\alpha_0-\overline{\Pi}_{\alpha_{n,\diamond}^*}\alpha_0 \Big\rangle_{\overline{P}_0}
+  \overline{P}_n B_0  +  o_p(n^{-1/2}).
\end{aligned}
\end{equation}
It remains to control the final term. By \ref{cond::mainprojectioncoupling1:i}, which holds under \ref{cond::DRmisDRconsist::1}, we have that
\begin{align*}
\Big\langle \bigl(\overline{\Pi}_{(\alpha_{n,\diamond}^*,\alpha_0)}-\overline{\Pi}_{\alpha_{n,\diamond}^*}\bigr)\Delta_{\mu,n,\diamond},\;
\alpha_0-\overline{\Pi}_{\alpha_{n,\diamond}^*}\alpha_0 \Big\rangle_{\overline{P}_0}
= O_p\!\left(\|\alpha_{n,\diamond}^* -\alpha_0\|_{\overline{P}_0}^2\right).
\end{align*}
Moreover, by \ref{cond::mainprojectioncoupling1:i}, we have
$\|\alpha_{n,\diamond}^* -\alpha_0\|_{\overline{P}_0}^2 = o_p(n^{-1/2})$. Hence,
\begin{align*}
\Big\langle \bigl(\overline{\Pi}_{(\alpha_{n,\diamond}^*,\alpha_0)}-\overline{\Pi}_{\alpha_{n,\diamond}^*}\bigr)\Delta_{\mu,n,\diamond},\;
\alpha_0-\overline{\Pi}_{\alpha_{n,\diamond}^*}\alpha_0 \Big\rangle_{\overline{P}_0}
= o_p(n^{-1/2}).
\end{align*}
Substituting the above into \eqref{eqn::crossriesz3}, we conclude that
\begin{equation}
\begin{aligned}
\label{eqn::crossriesz4}
\big\langle \alpha_0 - \alpha_{n,\diamond}^*,\ \mu_{n,\diamond}^* - \mu_0 \big\rangle_{\overline{P}_0}
&= P_n B_0 + o_p(n^{-1/2}).
\end{aligned}
\end{equation}
Substituting the above into \eqref{eqn::standardDR3}, we obtain
\begin{align*}
\tau_n^* - \tau_0
&= (P_n - P_0) \left\{\phi_{0,\overline{\mu}_0} + D_{\overline{\mu}_0, \overline{\alpha}_0}\right\}
+ P_n B_0 + o_p(n^{-1/2}) \\
&= P_n\{B_0 + \phi_{0,\overline{\mu}_0}\} + (P_n - P_0) D_{\overline{\mu}_0, \overline{\alpha}_0} + o_p(n^{-1/2}) \\
&= P_n\{B_0 + \phi_{0,\overline{\mu}_0} + D_{\overline{\mu}_0, \overline{\alpha}_0}\}
- P_0 D_{\overline{\mu}_0, \overline{\alpha}_0}
+ o_p(n^{-1/2}).
\end{align*}
By the Riesz representation property and the fact that $\overline{\alpha}_0=\alpha_0$, we have
\[
P_0 D_{\overline{\mu}_0, \overline{\alpha}_0}
= P_0\!\left[\alpha_0\{\mu_0 - \overline{\mu}_0\}\right]
= \psi_0(\mu_0 - \overline{\mu}_0).
\]
Hence,
\begin{align*}
\tau_n^* - \tau_0
&= P_n\{B_0 + \phi_{0,\overline{\mu}_0} + D_{\overline{\mu}_0, \overline{\alpha}_0} - \psi_0(\mu_0 - \overline{\mu}_0)\}
+ o_p(n^{-1/2})\\
&= P_n \chi_0
+ o_p(n^{-1/2}),
\end{align*}
as desired.

\noindent \paragraph{Establishing asymptotic linearity when outcome regression is estimated well.} Suppose the theorem is applied under \ref{cond::DRmisDRconsist::2} ($\overline{\mu}_0 = \mu_0$), so that $\|\mu_{n,j}^* - \mu_0 \| = o_p(n^{-1/4})$ and $\|\alpha_{n,j}^* - \overline{\alpha}_0\| = o_p(1)$ for $j=1,\ldots,J$. Applying the first part of Lemma~\ref{lemma::outcomeRieszBiasDR}, we have the regression-based decomposition:
\begin{align}
\label{eqn::crossoutcome1}
\big\langle \alpha_0-\alpha_{n,\diamond}^*,\;\mu_{n,\diamond}^*-\mu_0\big\rangle_{\overline{P}_0}
&=
-\overline{P}_n A_{n,\diamond}^* + (\overline{P}_n-\overline{P}_0)A_{n,\diamond}^*
+\Big\langle (\overline{\Pi}_{(\mu_{n,\diamond}^*,\mu_0)}-\overline{\Pi}_{\mu_{n,\diamond}^*})\Delta_{\alpha,n,\diamond},\;
\mu_0-\overline{\Pi}_{\mu_{n,\diamond}^*}\mu_0 \Big\rangle_{\overline{P}_0}.
\end{align}
By the first part of Lemma~\ref{lemma::secondEmpProcRemainders}, we have
$(\overline{P}_n-\overline{P}_0)\{A_{n,\diamond}^* - A_0\} = o_p(n^{-1/2})$ and, hence,
\begin{align*}
(\overline{P}_n-\overline{P}_0)A_{n,\diamond}^*
&= (\overline{P}_n-\overline{P}_0)A_0 + (\overline{P}_n-\overline{P}_0)\{A_{n,\diamond}^* - A_0\} \\
&= P_n A_0 + o_p(n^{-1/2}),
\end{align*}
where we used that $\overline{P}_0 A_0 = 0$. Substituting the above into \eqref{eqn::crossoutcome1}, we find that
\begin{equation}
    \label{eqn::crossoutcome2}
\begin{aligned}
\big\langle \alpha_0-\alpha_{n,\diamond}^*,\;\mu_{n,\diamond}^*-\mu_0\big\rangle_{\overline{P}_0}
&=
-\overline{P}_n A_{n,\diamond}^* + \overline{P}_n A_0  + o_p(n^{-1/2})
\\
& \quad +\Big\langle (\overline{\Pi}_{(\mu_{n,\diamond}^*,\mu_0)}-\overline{\Pi}_{\mu_{n,\diamond}^*})\Delta_{\alpha,n,\diamond},\;
\mu_0-\overline{\Pi}_{\mu_{n,\diamond}^*}\mu_0 \Big\rangle_{\overline{P}_0}.
\end{aligned}
\end{equation}
Since $\mu_{n,\diamond}^*$ is calibrated using isotonic regression, the first part of Lemma~\ref{lemma::isoscores} establishes that
\[
\overline{P}_n A_{n,\diamond}^*  = \frac{1}{n}\sum_{i=1}^n  s_{n,j(i)}^*(A_i, W_i) \left\{Y_i - \mu_{n,j(i)}^*(A_i, W_i) \right\} = 0.
\]
Thus, \eqref{eqn::crossoutcome2} implies
\begin{align}
\label{eqn::crossoutcome3}
\big\langle \alpha_0-\alpha_{n,\diamond}^*,\;\mu_{n,\diamond}^*-\mu_0\big\rangle_{\overline{P}_0}
&= P_n A_0  + o_p(n^{-1/2})
+\Big\langle (\overline{\Pi}_{(\mu_{n,\diamond}^*,\mu_0)}-\overline{\Pi}_{\mu_{n,\diamond}^*})\Delta_{\alpha,n,\diamond},\;
\mu_0-\overline{\Pi}_{\mu_{n,\diamond}^*}\mu_0 \Big\rangle_{\overline{P}_0}.
\end{align}
It remains to control the final term. By \ref{cond::mainprojectioncoupling1:ii}, which holds under \ref{cond::DRmisDRconsist::2}, we have that
\begin{align*}
\Big\langle (\overline{\Pi}_{(\mu_{n,\diamond}^*,\mu_0)}-\overline{\Pi}_{\mu_{n,\diamond}^*})\Delta_{\alpha,n,\diamond},\;
\mu_0-\overline{\Pi}_{\mu_{n,\diamond}^*}\mu_0 \Big\rangle_{\overline{P}_0}
= O_p\!\left(\|\mu_{n,\diamond}^* -\mu_0\|_{\overline{P}_0}^2\right).
\end{align*}
Moreover, by \ref{cond::mainprojectioncoupling1:ii}, we have
$\|\mu_{n,\diamond}^* -\mu_0\|_{\overline{P}_0}^2 = o_p(n^{-1/2})$. Hence,
\begin{align*}
\Big\langle (\overline{\Pi}_{(\mu_{n,\diamond}^*,\mu_0)}-\overline{\Pi}_{\mu_{n,\diamond}^*})\Delta_{\alpha,n,\diamond},\;
\mu_0-\overline{\Pi}_{\mu_{n,\diamond}^*}\mu_0 \Big\rangle_{\overline{P}_0}
= o_p(n^{-1/2}).
\end{align*}
Substituting the above into \eqref{eqn::crossoutcome3}, we conclude that
\begin{equation}
\begin{aligned}
\label{eqn::crossoutcome4}
\big\langle \alpha_0 - \alpha_{n,\diamond}^*,\ \mu_{n,\diamond}^* - \mu_0 \big\rangle_{\overline{P}_0}
&= P_n A_0 + o_p(n^{-1/2}).
\end{aligned}
\end{equation}
Substituting the above into \eqref{eqn::standardDR3}, we obtain
\begin{align*}
\tau_n^* - \tau_0
&= (P_n - P_0) \left\{\phi_{0,\overline{\mu}_0} + D_{\overline{\mu}_0, \overline{\alpha}_0}\right\}
+ P_n A_0 + o_p(n^{-1/2}) \\
&= P_n \left\{\phi_{0,\overline{\mu}_0} + D_{\overline{\mu}_0, \overline{\alpha}_0} + A_0\right\}
+ o_p(n^{-1/2})\\
&= P_n \chi_0
+ o_p(n^{-1/2}),
\end{align*}
where we used that $P_0 D_{\overline{\mu}_0, \overline{\alpha}_0} = 0$ when $\overline{\mu}_0 = \mu_0$. The claim follows.

\paragraph{Completing the proof.}
We have now shown that if either of the two conditions in \ref{cond::DRmisDRconsist}
holds, then
\begin{align*}
\tau_n^* - \tau_0 = P_n \chi_0 + o_p(n^{-1/2}).
\end{align*}
Hence, $\tau_n^*$ is DRAL. If both nuisance functions are consistently estimated
($\overline{\alpha}_0 = \alpha_0$ and $\overline{\mu}_0 = \mu_0$), then the above
display reduces to
\begin{align*}
\tau_n^* - \tau_0 = P_n D_0 + o_p(n^{-1/2}),
\end{align*}
since $\chi_0 = D_0$ in this case. By Theorem~\ref{theorem::EIF}, this implies
that $\tau_n^*$ is asymptotically linear with influence function equal to the
efficient influence function of $\Psi$ under the nonparametric model
$\mathcal{M}$. By the characterization of efficient estimators in
\citet[Lemma~25.23]{van2000asymptotic}, a regular estimator is
asymptotically efficient only if it is asymptotically linear with influence
function equal to the efficient influence function; see also \cite{bickel1993efficient}. Hence, $\tau_n^*$ is a
regular and efficient estimator of $\tau_0$ under the nonparametric model
$\mathcal{M}$.

 \end{proof}

\subsection{Proof of Theorem \ref{theorem::DRboot}}

\begin{proof}[Proof of Theorem \ref{theorem::DRboot}]
\textbf{Starting point.} By Lemma~\ref{lemma::standardBiasExpansion}, it holds that
\begin{align*}
\tau_n^{\#} - \tau_0
&= (\overline{P}_n^{\#} - \overline{P}_0)
\left\{\phi_{0, \mu_{n,\diamond}^{\#}} + D_{\mu_{n,\diamond}^{\#}, \alpha_{n,\diamond}^{\#}} \right\}
+ \big\langle \alpha_{n,\diamond}^{\#} - \alpha_0,\; \mu_0 - \mu_{n,\diamond}^{\#} \big\rangle_{\overline{P}_0}
+ Rem_{\mu_{n,\diamond}^{\#}}(\psi_{n,\diamond}, \psi_0) \\
&= (\overline{P}_n^{\#} - \overline{P}_0)
\left\{ m(\cdot, \mu_{n,\diamond}^{\#}) + D_{\mu_{n,\diamond}^{\#}, \alpha_{n,\diamond}^{\#}} \right\}
+ \big\langle \alpha_{n,\diamond}^{\#} - \alpha_0,\; \mu_0 - \mu_{n,\diamond}^{\#} \big\rangle_{\overline{P}_0},
\end{align*}
where we used that
$\phi_{0, \mu_{n,\diamond}^{\#}} = m(\cdot, \mu_{n,\diamond}^{\#}) - \overline{P}_0 m(\cdot, \mu_{n,\diamond}^{\#})$
and that $Rem_{\mu_{n,\diamond}^{\#}}(\psi_{n,\diamond}, \psi_0) = 0$ in the setup of
Section~\ref{section::bootstrap}.
Arguing as in the proof of Theorem~\ref{theorem::DRinference}, we may apply the second part of
Lemma~\ref{lemma::firstEmpProcRemainders} to obtain
\begin{align*}
(\overline{P}_n^{\#} - \overline{P}_0)
\left\{ m(\cdot, \mu_{n,\diamond}^{\#}) + D_{\mu_{n,\diamond}^{\#}, \alpha_{n,\diamond}^{\#}} \right\}
&=
(\overline{P}_n^{\#} - \overline{P}_0)
\left\{ m(\cdot, \overline{\mu}_0) + D_{\overline{\mu}_0, \overline{\alpha}_0} \right\}
+ o_p(n^{-1/2}).
\end{align*}

Conditions~\ref{cond::linear}, \ref{cond::estboundedlinearcross}, \ref{cond::estnuisbound}, \ref{cond::estboundboot}, and~\ref{cond::lipschitzisotonic} are included in the current theorem's assumptions. Hence, we may apply Theorem~\ref{theorem::MSE} to conclude that
\begin{align*}
\|\alpha_{n,\diamond}^{\#} - \alpha_0\|_{\overline{P}_0}
&= O_p\!\left(\|\alpha_{n,\diamond}^{*} - \alpha_0\|_{\overline{P}_0}\right) + O_p(n^{-1/3}),\\
\|\mu_{n,\diamond}^{\#} - \mu_0\|_{\overline{P}_0}
&= O_p\!\left(\|\mu_{n,\diamond}^{*} - \mu_0\|_{\overline{P}_0}\right) + O_p(n^{-1/3}).
\end{align*}
Furthermore, by Lemma~\ref{lemma::nuisanceconsistency}, under
Conditions~\ref{cond::linear}, \ref{cond::estboundedlinearcross}, \ref{cond::estnuisbound}, and~\ref{cond::mainprojectioncoupling1::boot}--\ref{cond::lipschitzisotonic},
whenever one part of \ref{cond::DRmisDRconsist} holds for $\alpha_{n,\diamond}^{*}$ and $\mu_{n,\diamond}^{*}$,
the same part holds for $\alpha_{n,\diamond}^{\#}$ and $\mu_{n,\diamond}^{\#}$. In particular, under
\ref{cond::DRlimits}, we have $\|\alpha_{n,\diamond}^{\#} - \overline{\alpha}_0\|_{\overline{P}_0} = o_p(1)$ and
$\|\mu_{n,\diamond}^{\#} - \overline{\mu}_0\|_{\overline{P}_0} = o_p(1)$. Moreover, under
\ref{cond::DRmisDRconsist::2}, $\|s_{n,j}^{\#} - s_0\| = o_p(1)$, and under \ref{cond::DRmisDRconsist::1},
$1\{\overline{\alpha}_0 = \alpha_0\}\,\|r_{n,j}^{\#} - r_0\| = o_p(1)$. Consequently, under \eqref{cond::mainprojectioncoupling1::boot}, we can apply the proof of
Theorem~\ref{theorem::DRinference} with the original dataset $\mathcal{D}_n$ replaced by the bootstrap data $\mathcal{D}_n^\#$ and nuisance estimators replaced by their bootstrap counterparts.

We now establish asymptotic linearity of the cross-product remainder in the two cases.

 \noindent \paragraph{Establishing asymptotic linearity when the outcome regression is estimated well.}
Suppose the theorem is applied under \ref{cond::DRmisDRconsist::2} ($\overline{\mu}_0 = \mu_0$), so that
$\|\mu_{n,j}^* - \mu_0 \| = o_p(n^{-1/4})$ and $\|\alpha_{n,j}^* - \overline{\alpha}_0\| = o_p(1)$ for
$j=1,\ldots,J$. The first part of Lemma~\ref{lemma::secondEmpProcRemainders} implies that
\[
(\overline{P}_n^\# - P_0)\{A_{n,\diamond}^\# - A_0\} = o_p(n^{-1/2}).
\]
Applying the regression-based decomposition in Lemma~\ref{lemma::outcomeRieszBiasDR} to the bootstrap nuisance
estimators, and arguing as in the proof of Theorem~\ref{theorem::DRinference} using the partial orthogonality and
projection error bounds in \ref{cond::mainprojectioncoupling1::boot} together with the first part of
Lemma~\ref{lemma::secondEmpProcRemainders}, we obtain
\begin{align*}
\big\langle \alpha_{n,\diamond}^{\#} - \alpha_0,\; \mu_0 - \mu_{n,\diamond}^{\#} \big\rangle_{\overline{P}_0}
&=
1(\overline{\alpha}_0 \neq \alpha_0)\,(\overline{P}_n^{\#} - P_0)A_0
+ O_p\!\left(\|\mu_{n,\diamond}^{\#} - \mu_0\|_{\overline{P}_0}^2\right).
\end{align*}
By Theorem~\ref{theorem::MSE}, we have
$\|\mu_{n,\diamond}^{\#} - \mu_0\|_{\overline{P}_0}^2
= O_p(n^{-2/3}) + \|\mu_{n,\diamond}^{*} - \mu_0\|_{\overline{P}_0}^2$,
and thus, by \ref{cond::DRmisDRconsist}, it follows that
$\|\mu_{n,\diamond}^{\#} - \mu_0\|_{\overline{P}_0}^2 = o_p(n^{-1/2})$.
We conclude that
\begin{align*}
\big\langle \alpha_{n,\diamond}^{\#} - \alpha_0,\; \mu_0 - \mu_{n,\diamond}^{\#} \big\rangle_{\overline{P}_0}
&=
1(\overline{\alpha}_0 \neq \alpha_0)\,(\overline{P}_n^{\#} - P_0)A_0
+ o_p(n^{-1/2}).
\end{align*}

\noindent \paragraph{Establishing asymptotic linearity when the Riesz representer is estimated well.} Suppose the theorem is applied under \ref{cond::DRmisDRconsist::1} ($\overline{\alpha}_0 = \alpha_0$), so that $\|\alpha_{n,j}^* - \alpha_0 \| = o_p(n^{-1/4})$ and $\|\mu_{n,j}^* - \overline{\mu}_0\| = o_p(1)$ for $j=1,\ldots,J$. By the second part of Lemma~\ref{lemma::secondEmpProcRemainders} applied in the bootstrap setting, we have
\[
(\overline{P}_n^\# - P_0)\{B_{n,\diamond}^\# - B_0\}
+ \frac{1}{J}\sum_{j=1}^J
\left\{\psi_{n,j}(r_{n,j}^\#)-\psi_0(r_{n,j}^\#)- P_{n,j}\phi_{0, r_{n,j}^\#}\right\}
= o_p(n^{-1/2}).
\]
Applying the Riesz-based decomposition in Lemma~\ref{lemma::outcomeRieszBiasDR} to the bootstrap nuisance
estimators, and arguing as in the proof of Theorem~\ref{theorem::DRinference} using the partial orthogonality and
projection error bounds in \ref{cond::mainprojectioncoupling1::boot} together with the second part of
Lemma~\ref{lemma::secondEmpProcRemainders}, we obtain
   \begin{align*}
      \langle \alpha_{n, \diamond}^{\#} - \alpha_{0}, \mu_{0} -  \mu_{n, \diamond}^{\#} \rangle_{\overline{P}_0}& =  1(\overline{\mu}_0 \neq \mu_0) (\overline{P}_n^{\#} - P_0) B_0  +  O_p\left(\|\alpha_{n, \diamond}^{\#}  - \alpha_0\|_{\overline{P}_0}^2 \right).
    \end{align*} 
By Theorem~\ref{theorem::MSE}, it holds that
    \[
    \|\alpha_{n, \diamond}^{\#}  - \alpha_0\|_{\overline{P}_0}^2 = O_p(n^{-2/3}) + \|\alpha_{n, \diamond}^{*}  - \alpha_0\|_{\overline{P}_0}^2,
    \]
    and, thus, by \ref{cond::DRmisDRconsist},
    $\|\alpha_{n, \diamond}^{\#}  - \alpha_0\|_{\overline{P}_0}^2 = o_p(n^{-1/2})$.
We conclude that 
   \begin{align*}
      \langle \alpha_{n, \diamond}^{\#} - \alpha_{0}, \mu_{0} -  \mu_{n, \diamond}^{\#} \rangle_{\overline{P}_0}& = 1(\overline{\mu}_0 \neq \mu_0) (\overline{P}_n^{\#} - P_0) B_0  +  o_p(n^{-1/2}).
    \end{align*} 

\noindent \textbf{Establishing sample-conditional DRAL of $\tau_n^{\#} - \tau_n^{*}$.}
Combining the two cases treated above, we obtain
\begin{align*}
\big\langle \alpha_{n,\diamond}^{\#} - \alpha_{0},\ \mu_{0} - \mu_{n,\diamond}^{\#} \big\rangle_{\overline{P}_0}
&=
1(\overline{\alpha}_0 \neq \alpha_0)\, (\overline{P}_n^{\#} - P_0) A_0
+ 1(\overline{\mu}_0 \neq \mu_0)\, (\overline{P}_n^{\#} - P_0) B_0
+ o_p(n^{-1/2}).
\end{align*}
Hence,
\begin{align*}
\tau_n^{\#} - \tau_0
&=
\overline{P}_n^{\#}\chi_0 + o_p(n^{-1/2}).
\end{align*}
By Theorem~\ref{theorem::DRinference}, we also have
\[
\tau_n^{*} - \tau_0 = (\overline{P}_n - \overline{P}_0)\chi_0 + o_p(n^{-1/2}),
\]
and therefore
\[
\tau_n^{\#} - \tau_n^{*}
= (\overline{P}_n^{\#} - \overline{P}_n)\chi_0 + o_p(n^{-1/2}).
\]

\noindent \textbf{Establishing weak convergence of bootstrap law.} Let $n_j$ denote the number of observations in $\mathcal{C}^{(j)}$ (so that $P_{n,j}$ is the empirical measure based on $n_j$ samples). By assumption, the splits are approximately equal, in the sense that $J n_j/n \to 1$. For each $j \in [J]$, the bootstrap CLT \citep{van1996weak} yields
\[
\sqrt{n_j}\,(P_{n,j}^{\#}-P_{n,j})\chi_0 \,\big|\, P_{n,j}
\ \rightsquigarrow\ N\!\left(0,\, P_0\chi_0^2\right).
\]
Since the bootstrap resamples are generated independently across folds conditional on the data, the same limit holds conditional on the full collection $\{P_{n,j}: j \in [J]\}$:
\[
\sqrt{n_j}\,(P_{n,j}^{\#}-P_{n,j})\chi_0 \,\big|\, \{P_{n,j}: j \in [J]\}
\ \rightsquigarrow\ N\!\left(0,\, P_0\chi_0^2\right).
\]
Moreover,
\[
(\overline{P}_n^{\#}-\overline{P}_n)\chi_0
= \frac{1}{J}\sum_{j=1}^J (P_{n,j}^{\#}-P_{n,j})\chi_0,
\]
and $J n_j/n \to 1$ implies $\sqrt{n}/J = \sqrt{n_j}\,(1+o(1))$. Therefore, by Slutsky's lemma,
\[
\sqrt{n}\,(\overline{P}_n^{\#}-\overline{P}_n)\chi_0 \,\big|\, \{P_{n,j}: j \in [J]\}
\ \rightsquigarrow\ N\!\left(0,\, P_0\chi_0^2\right).
\]

Consequently, the conditional law of $\sqrt{n}\,(\tau_n^{\#}-\tau_n^{*})$ given $\{P_{n,j}: j \in [J]\}$
converges in probability to the law of $N(0, P_0\chi_0^2)$. Indeed, writing
\[
\sqrt{n}\,(\tau_n^{\#}-\tau_n^{*})
=
\sqrt{n}\,(\overline{P}_n^{\#}-\overline{P}_n)\chi_0 + \sqrt{n}\,R_n,
\]
our preceding argument shows that
\[
\mathcal{L}\!\left(\sqrt{n}\,(\overline{P}_n^{\#}-\overline{P}_n)\chi_0 \,\middle|\, \{P_{n,j}: j \in [J]\}\right)
\Rightarrow N(0,P_0\chi_0^2)
\quad\text{in }P_0\text{-probability}.
\]
Moreover, since $\sqrt{n}\,R_n=o_P(1)$ under the joint law of $(P_n,P_n^\#)$, we have
$P(|\sqrt{n}\,R_n|>\eta)\to 0$ for every $\eta>0$. Fix $\eta,\delta>0$ and define
\[
Q_n \;:=\; P^\#\!\left(|\sqrt{n}\,R_n|>\eta \,\middle|\, \{P_{n,j}: j \in [J]\}\right)\in[0,1].
\]
By the tower property,
\[
E_0[Q_n] \;=\; P\!\left(|\sqrt{n}\,R_n|>\eta\right),
\]
where $E_0$ denotes expectation with respect to the data-generating law and $P$ denotes the joint law of $(P_n,P_n^\#)$.
Therefore, by Markov's inequality,
\[
P_0(Q_n>\delta)
\;\le\;
\frac{E_0[Q_n]}{\delta}
\;=\;
\frac{P\!\left(|\sqrt{n}\,R_n|>\eta\right)}{\delta}
\;\to\;0.
\]
Thus, $\sqrt{n}\,R_n=o_{P^\#}(1)$ in $P_0$-probability, in the sense that for every $\eta,\delta>0$,
\[
P_0\!\left(
\Pr\!\left(|\sqrt{n}\,R_n|>\eta \,\middle|\, \{P_{n,j}: j \in [J]\}\right)>\delta
\right)\to 0.
\]
By Slutsky's lemma for conditional laws, it follows that
\[
\mathcal{L}\!\left(\sqrt{n}\,(\tau_n^{\#}-\tau_n^{*}) \,\middle|\, \{P_{n,j}: j \in [J]\}\right)
\Rightarrow N(0,P_0\chi_0^2)
\quad\text{in }P_0\text{-probability}.
\]
That is, the law $\mathcal{L}(\sqrt{n}\,(\tau_n^{\#}-\tau_n^{*}) \mid \{P_{n,j}: j \in [J]\})$
converges weakly in probability to the law of $N(0, P_0\chi_0^2)$, in the sense that for every bounded Lipschitz function
$f:\mathbb{R}\to\mathbb{R}$,
\[
E\!\left[f\!\left(\sqrt{n}\,(\tau_n^{\#}-\tau_n^{*})\right)\,\middle|\, \{P_{n,j}: j \in [J]\}\right]
\;\xrightarrow{P_0}\;
E\!\left[f(Z)\right],
\qquad Z\sim N(0,P_0\chi_0^2).
\]

\end{proof}

\section{Proofs for Appendix \ref{sec::irregular}}
\label{appendix::regularproof}

\begin{proof}[Proof of Lemma \ref{lem::oracle-target-agreement}]

Define
\[
\mathcal H_{\mu_0,P}
:=
\{g\circ \mu_0:g:\mathbb R\to\mathbb R\}\cap L^2(P),
\qquad
\mathcal H_{\alpha_0,P}
:=
\{f\circ \alpha_0:f:\mathbb R\to\mathbb R\}\cap L^2(P).
\]

If \((\overline{\alpha}_0,\overline{\mu}_0)=(\alpha_0,\mu_0)\), then
\(\Psi_0=\Psi\) by definition, and the lemma is immediate. We therefore
consider the strict one-sided regimes.

We first consider the case \(\overline{\mu}_0=\mu_0\). By the normal equations
for the \(L^2(P)\)-projection of \(\alpha_P\) onto the affine space
\(\overline{\alpha}_0+\mathcal H_{\mu_0,P}\),
\[
\alpha_P-\overline{\Pi}_{\mu_0,P}\alpha_P
\perp
\mathcal H_{\mu_0,P}.
\]
Hence, if \(\mu_P\in\mathcal H_{\mu_0,P}\), then
\[
\Psi_0(P)
=
\overline{\psi}_{\mu_0,P}(\mu_P)
=
\langle \overline{\Pi}_{\mu_0,P}\alpha_P,\mu_P\rangle_P
=
\langle \alpha_P,\mu_P\rangle_P
=
\Psi(P).
\]

The case \(\overline{\alpha}_0=\alpha_0\) is analogous. The normal equations
for the \(L^2(P)\)-projection of \(\mu_P\) onto the affine space
\(\overline{\mu}_0+\mathcal H_{\alpha_0,P}\) give
\[
\mu_P-\overline{\Pi}_{\alpha_0,P}\mu_P
\perp
\mathcal H_{\alpha_0,P}.
\]
Thus, if \(\alpha_P\in\mathcal H_{\alpha_0,P}\), then
\[
\Psi_0(P)
=
\psi_P(\overline{\Pi}_{\alpha_0,P}\mu_P)
=
\langle \alpha_P,\overline{\Pi}_{\alpha_0,P}\mu_P\rangle_P
=
\langle \alpha_P,\mu_P\rangle_P
=
\Psi(P).
\]
Finally, at \(P_0\), we have
\(\mu_0=\operatorname{id}\circ\mu_0\in\mathcal H_{\mu_0,P_0}\) and
\(\alpha_0=\operatorname{id}\circ\alpha_0\in\mathcal H_{\alpha_0,P_0}\), so the
corresponding agreement statement gives \(\Psi_0(P_0)=\Psi(P_0)\) in either
one-sided regime.
\end{proof}

\begin{proof}[Proof of Theorem \ref{theorem::regularity}]
If \((\overline{\alpha}_0,\overline{\mu}_0)=(\alpha_0,\mu_0)\), then
\(\Psi_0=\Psi\) by definition. Theorem~\ref{theorem::EIF} identifies
\(\chi_0\) with the efficient influence function, and
Theorem~\ref{theorem::DRinference} gives the asymptotic linear expansion of
\(\tau_n^*\) with this influence function. It remains to consider the two
strict one-sided regimes.

First suppose that \(\overline{\alpha}_0=\alpha_0\) and
\(\overline{\mu}_0\neq\mu_0\). For \(P\in\mathcal M\), define the reduced
regression
\[
    r_P(a,w)
    :=
    E_P\!\left\{
        Y-\overline{\mu}_0(A,W)
        \mid \alpha_0(A,W)=\alpha_0(a,w)
    \right\}.
\]
The normal equations for conditional expectation give
\begin{equation}
\label{eqn::irregular-r-projection}
    \overline{\Pi}_{\alpha_0,P}\mu_P
    =
    \overline{\mu}_0+r_P,
    \qquad
    r_{P_0}=r_0 .
\end{equation}
By \eqref{eqn::irregular-r-projection},
\begin{equation}
\label{eqn::irregular-r-oracle}
    \Psi_0(P)
    =
    \psi_P(\overline{\mu}_0)+\psi_P(r_P).
\end{equation}
The map \(P\mapsto\psi_P(\overline{\mu}_0)\) has influence function
\(\phi_{0,\overline{\mu}_0}\) by Condition~\ref{cond::pathwise}. For the map
\(P\mapsto\psi_P(r_P)\) in \eqref{eqn::irregular-r-oracle}, Example~5 of
\citet{luedtke2024one} and Lemma~S6 of \citet{luedtke2024simplifying} verify
the differentiability and chain-rule hypotheses of
Theorem~\ref{theorem::EIF}. Applying Theorem~\ref{theorem::EIF} with
pseudo-outcome
\(Y-\overline{\mu}_0(A,W)\), scalar summary \(\alpha_0(A,W)\), and linear
functional \(h\mapsto \psi_P(h)\) yields the influence function
\begin{equation}
\label{eqn::irregular-rterm-if}
    \phi_{0,r_0}
    +
    \alpha_0\{\mathcal{I}_Y-\overline{\mu}_0-r_0\}.
\end{equation}
Here \(\alpha_0=\operatorname{id}\circ\alpha_0\) is the Riesz representer of
\(h\mapsto\psi_0(h)\) on functions of \(\alpha_0\). Combining
\eqref{eqn::irregular-r-oracle}, the influence function
\(\phi_{0,\overline{\mu}_0}\) of \(P\mapsto\psi_P(\overline{\mu}_0)\), and
\eqref{eqn::irregular-rterm-if}, the efficient influence function of
\(\Psi_0\) at \(P_0\) in this regime is
\begin{equation}
\label{eqn::irregular-alpha-correct-if}
    \phi_{0,\overline{\mu}_0}
    +\phi_{0,r_0}
    +
    \alpha_0\{\mathcal{I}_Y-\overline{\mu}_0-r_0\}.
\end{equation}
To compare \eqref{eqn::irregular-alpha-correct-if} with \(\chi_0\), note that
\(h\mapsto\phi_{0,h}\) is linear: differentiating
\(\psi_P(c h_1+h_2)=c\psi_P(h_1)+\psi_P(h_2)\) at \(P_0\) gives
\(\phi_{0,ch_1+h_2}=c\phi_{0,h_1}+\phi_{0,h_2}\). Moreover, the
normal equation underlying \eqref{eqn::irregular-r-projection} gives
\begin{equation}
\label{eqn::irregular-r-orthogonality}
    \psi_0(\mu_0-\overline{\mu}_0-r_0)
    =
    \langle \alpha_0,\mu_0-\overline{\mu}_0-r_0\rangle
    =
    0 .
\end{equation}
Using the definition of \(\chi_0\) with \(A_0\equiv0\),
\(B_0=\phi_{0,r_0}+\psi_0(r_0)-r_0\alpha_0\), linearity of
\(h\mapsto\phi_{0,h}\), and \eqref{eqn::irregular-r-orthogonality} gives
\eqref{eqn::irregular-alpha-correct-if}.

Next suppose that \(\overline{\alpha}_0\neq\alpha_0\) and
\(\overline{\mu}_0=\mu_0\). Let
\[
    q_P(a,w)
    :=
    E_P\!\left\{Y\mid \mu_0(A,W)=\mu_0(a,w)\right\}.
\]
Then \(q_{P_0}=\mu_0\). Write
\[
    a_P:=\overline{\Pi}_{\mu_0,P}\alpha_P .
\]
Since \(a_P-\overline{\alpha}_0\) is a function of \(\mu_0\), the normal
equations give
\begin{equation}
\label{eqn::irregular-q-normal}
P\{(a_P-\overline{\alpha}_0)Y\}
=P\{(a_P-\overline{\alpha}_0)q_P\},
\end{equation}
and \(a_P-\overline{\alpha}_0\) is the Riesz representer, on functions of
\(\mu_0\), of \(h\mapsto\psi_P(h)-P\{\overline{\alpha}_0h\}\). Therefore, by
the definition of \(\Psi_0\) in this regime,
\begin{equation}
\label{eqn::irregular-q-oracle}
\begin{aligned}
    \Psi_0(P)
    &=
    P\{a_PY\}
    \\
    &=
    P\{\overline{\alpha}_0Y\}
    +\psi_P(q_P)
    -P\{\overline{\alpha}_0 q_P\},
\end{aligned}
\end{equation}
where the second equality uses \eqref{eqn::irregular-q-normal} and the
preceding Riesz-representation identity.

The influence function of \(P\{\overline{\alpha}_0Y\}\) is
\begin{equation}
\label{eqn::irregular-y-if}
\overline{\alpha}_0Y-P_0(\overline{\alpha}_0\mu_0).
\end{equation}
Example~5 of \citet{luedtke2024one} and Lemma~S6 of
\citet{luedtke2024simplifying} verify the differentiability and chain-rule
hypotheses required for Theorem~\ref{theorem::EIF}. Applying
Theorem~\ref{theorem::EIF} with scalar summary \(\mu_0(A,W)\) and linear
functional \(h\mapsto \psi_P(h)\) gives the
influence function
\begin{equation}
\label{eqn::irregular-q-psi-if}
    \phi_{0,\mu_0}
    +\Pi_{\mu_0}\alpha_0\{\mathcal{I}_Y-\mu_0\}
\end{equation}
for \(P\mapsto\psi_P(q_P)\). Applying Theorem~\ref{theorem::EIF} with scalar
summary \(\mu_0(A,W)\) and linear functional
\(h\mapsto P\{\overline{\alpha}_0h\}\) gives the influence function
\begin{equation}
\label{eqn::irregular-q-alpha-if}
    \overline{\alpha}_0\mu_0-P_0(\overline{\alpha}_0\mu_0)
    +\Pi_{\mu_0}\overline{\alpha}_0\{\mathcal{I}_Y-\mu_0\}.
\end{equation}
Combining \eqref{eqn::irregular-q-oracle},
\eqref{eqn::irregular-y-if}, \eqref{eqn::irregular-q-psi-if}, and
\eqref{eqn::irregular-q-alpha-if} yields
\begin{equation}
\label{eqn::irregular-mu-correct-if}
    \phi_{0,\mu_0}
    +
    \left\{
        \overline{\alpha}_0
        +\Pi_{\mu_0}(\alpha_0-\overline{\alpha}_0)
    \right\}
    \{\mathcal{I}_Y-\mu_0\}.
\end{equation}
Since \(s_0=\Pi_{\mu_0}(\alpha_0-\overline{\alpha}_0)\),
\eqref{eqn::irregular-mu-correct-if} equals
\begin{equation}
\label{eqn::irregular-mu-correct-chi}
    \phi_{0,\mu_0}
    +(\overline{\alpha}_0+s_0)\{\mathcal{I}_Y-\mu_0\},
\end{equation}
which equals \(\chi_0\) in this one-sided regime because
\(\overline{\mu}_0=\mu_0\) implies
\(B_0+\psi_0(\overline{\mu}_0-\mu_0)\equiv0\), while
\(A_0=s_0\{\mathcal{I}_Y-\mu_0\}\).

Equations \eqref{eqn::irregular-alpha-correct-if} and
\eqref{eqn::irregular-mu-correct-chi} identify \(\chi_0\) as the efficient
influence function of \(\Psi_0\) at \(P_0\) in the two strict one-sided
regimes. In the doubly consistent case, Theorem~\ref{theorem::EIF} identifies
\(\chi_0\) with the efficient influence function of \(\Psi_0=\Psi\). Since
Lemma~\ref{lem::oracle-target-agreement} gives
\(\Psi_0(P_0)=\Psi(P_0)=\tau_0\), Theorem~\ref{theorem::DRinference} gives
\[
\tau_n^*-\Psi_0(P_0)=P_n\chi_0+o_{P_0}(n^{-1/2}).
\]
Thus \(\tau_n^*\) is \(P_0\)-asymptotically linear with the efficient influence
function of \(\Psi_0\). The standard characterization of efficient estimators
\citep[Lemma~25.23]{van2000asymptotic} implies that \(\tau_n^*\) is regular and
efficient for \(\Psi_0\) at \(P_0\). Hence, for every regular parametric
submodel through \(P_0\) and every fixed \(t\), with \(t_n:=t n^{-1/2}\), under
sampling from \(P_{0,t_n}\),
\[
n^{1/2}\{\tau_n^*-\Psi_0(P_{0,t_n})\}
\rightsquigarrow
N(0,P_0\chi_0^2)
=
N(0,\sigma_0^2),
\]
which is the stated local limit.
\end{proof}

 \begin{proof}[Proof of Corollary~\ref{cor::regularity}]
 
We adapt the proof of Theorem 7 in \cite{van2023adaptive}. 
By Theorem~\ref{theorem::regularity}, \(\chi_0\) is the efficient influence
function of \(\Psi_0\) at \(P_0\). By Theorem~\ref{theorem::EIF}, \(D_0\) is the
efficient influence function of \(\Psi\) at \(P_0\). Since
\(\Psi_0(P_0)=\Psi(P_0)\), pathwise differentiability along
\(u\mapsto P_{0,u}\) gives
\[
\Psi_0(P_{0,t_n})-\Psi(P_{0,t_n})
=
t n^{-1/2}\langle s,\chi_0-D_0\rangle
+
o(n^{-1/2}).
\]
Therefore,
\[
n^{1/2}\{\tau_n^*-\Psi(P_{0,t_n})\}
=
n^{1/2}\{\tau_n^*-\Psi_0(P_{0,t_n})\}
+
t\langle s,\chi_0-D_0\rangle
+
o(1).
\]
Theorem~\ref{theorem::regularity} and Slutsky's theorem imply the stated
limiting distribution. The first bound follows from Cauchy--Schwarz.

It remains to bound \(\|\chi_0-D_0\|_{P_0}\). First suppose
\(\overline{\mu}_0=\mu_0\). Then
\(B_0+\psi_0(\overline{\mu}_0-\mu_0)\equiv0\), and
\[
\chi_0-D_0
=
\{\overline{\alpha}_0-\alpha_0+s_0\}\{Y-\mu_0\}.
\]
Let \(\eta_\alpha:=\alpha_0-\overline{\alpha}_0\). Since
\(s_0=\Pi_{\mu_0}\eta_\alpha\), Condition~\ref{cond::bound} and
nonexpansiveness of orthogonal projection give
\[
\|\chi_0-D_0\|_{P_0}
\lesssim
\|\Pi_{\mu_0}\eta_\alpha-\eta_\alpha\|_{P_0}
\le
\|\eta_\alpha\|_{P_0}.
\]

Next suppose \(\overline{\alpha}_0=\alpha_0\). Let
\(\eta_\mu:=\mu_0-\overline{\mu}_0\). Then \(A_0\equiv0\) and
\(r_0=\Pi_{\alpha_0}\eta_\mu\), so
\[
\chi_0-D_0
=
\phi_{0,\overline{\mu}_0}-\phi_{0,\mu_0}
+\phi_{0,r_0}
+\psi_0(r_0)-\psi_0(\eta_\mu)
+\alpha_0(\eta_\mu-r_0).
\]
Because \(P\mapsto\psi_P(0)\) is identically zero, its efficient influence
function satisfies \(\phi_{0,0}=0\). Hence, by the Lipschitz condition,
Condition~\ref{cond::bound}, the Riesz representation of \(\psi_0\), and
nonexpansiveness of orthogonal projection,
\[
\|\chi_0-D_0\|_{P_0}
\lesssim
\|\eta_\mu\|_{P_0}
+
\|r_0\|_{P_0}
+
\|\eta_\mu-r_0\|_{P_0}
\lesssim
\|\eta_\mu\|_{P_0}.
\]
Combining the two cases gives the claimed nuisance-misspecification bound.
\end{proof}

\section{Additional details on numerical experiments}

\subsection{Complete results for experiment 1}

\label{appendix::exp1}

\begin{figure}[H]
     \centering 
\includegraphics[width=0.5\linewidth]{newplots/legend_only.pdf}

    \includegraphics[width=\linewidth]{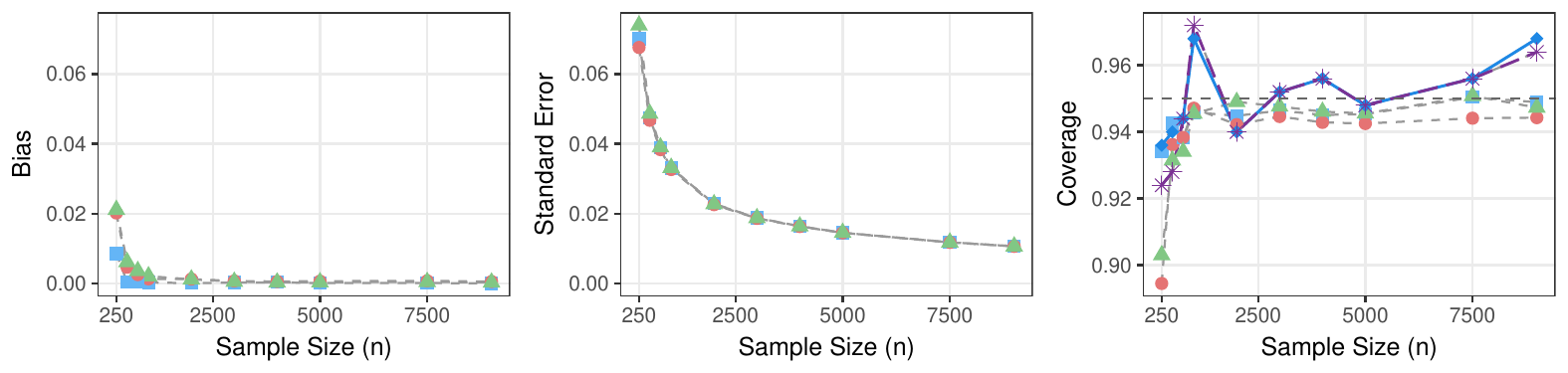}
    \caption{Empirical relative bias, standard error, and 95\% confidence interval coverage for isocalibrated DML (IC-DML), DR-TMLE and AIPW estimators under both doubly consistent estimation of the outcome regression and treatment mechanism. Both propensity score and outcome regression are estimated consistently.}
       \label{fig::exp1::both}   

\end{figure}

\subsection{Additional details for Experiment 2}
\label{appendix::exp2_holder}

This appendix gives implementation details for Experiment~2. The covariates are
\(W=(W_1,\ldots,W_d)\), with independent coordinates drawn from \(U[-1,1]\).
For a vector \(v=(v_1,\ldots,v_n)\), let \(\operatorname{Std}_n(v)\) denote
empirical centering and scaling to mean zero and standard deviation one within
the simulated dataset. Define
\[
  a_j=\frac{j^{-3/4}}{\left(\sum_{\ell=1}^d \ell^{-3/2}\right)^{1/2}} .
\]
The simulation uses two standardized scores. The first is an additive smooth
score,
\[
  s_f(W)=\operatorname{Std}_n\left[
    \sum_{j=1}^d a_j\{\sin(\pi W_j)+0.5W_j^2+0.25\cos(2\pi W_j)\}
  \right].
\]
The second is a rough score. It includes the univariate terms
\[
  r_j(W)=\operatorname{sign}(W_j)|W_j|^{1/2}+0.25\sin(3\pi W_j).
\]
For \(d>1\), let \(\mathcal I_d=\{1,\ldots,\min(d,4)\}\). The rough score also
includes
\[
\begin{aligned}
&\sum_{j<k,\ j,k\in\mathcal I_d}\sqrt{a_ja_k}
  \{\sin(\pi W_jW_k)+0.5\cos(\pi(W_j-W_k))\}\\
&\quad
  +0.45\prod_{j\in\mathcal I_d}\tanh(1.5W_j)
  +\sum_{j\in\mathcal I_d}a_j\{0.35\,1(W_j>0.2)-0.25\,1(W_j<-0.45)\}.
\end{aligned}
\]
When \(d=1\), the pairwise and product terms are omitted. Let \(s_s(W)\) denote
the standardized rough score. The propensity score depends on a mixture of the additive and rough scores:
\[
  \eta_\pi(W)=\operatorname{Std}_n\{0.65s_f(W)+\sqrt{1-0.65^2}s_s(W)\}.
\]
The propensity score is
\[
  \pi_0(W)
  =
  \operatorname{clip}_{[0.03,0.97]}
  \{\operatorname{expit}(-0.2+1.15\eta_\pi(W))\}.
\]
Treatment is generated as
\[
  A\mid W \sim \operatorname{Bernoulli}\{\pi_0(W)\}.
\]
The outcome logits are additive. Let
\[
  s_\tau(W)=
  \operatorname{Std}_n\{0.75s_f(W)+0.25\sin(\pi W_1)\}.
\]
The conditional outcome means are
\[
  \mu_0(0,W)=\operatorname{expit}(-0.6+0.9s_f(W)),
\]
and
\[
  \mu_0(1,W)=
  \operatorname{expit}(-0.6+0.9s_f(W)+0.35+0.15s_\tau(W)).
\]
The outcome is generated as
\[
  Y\mid A,W \sim \operatorname{Bernoulli}\{\mu_0(A,W)\}.
\]
The estimand is the ATE,
\[
  E\{\mu_0(1,W)-\mu_0(0,W)\}.
\]

The nuisance estimators are trained using 5-fold cross-fitting. Outcome regressions are estimated separately by treatment
arm with depth-one \texttt{xgboost} using singleton interaction constraints,
learning rate 0.1, up to 5000 boosting rounds, early stopping after 100 rounds,
and a 20\% validation split for selecting number of rounds. The propensity score is estimated with
\texttt{ranger} probability forests using 500 trees, maximum depth 8, and
\(mtry=\lfloor\sqrt d\rfloor\). Calibrated DML uses isotonic calibration, and
the displayed confidence interval is the level-set corrected Wald interval.

\subsection{Complete results for experiment 3}
\label{appendix::simRealFull}

\begin{table}[ht]
\centering
\scriptsize
\setlength{\tabcolsep}{3pt}
\renewcommand{\arraystretch}{0.82}
\caption{Full semi-synthetic benchmark summary for AIPW and IC-DML.}
\label{fig::exp2full}
\begin{tabular}{llrrrr}
\hline
Dataset & Method & Bias & RMSE & Coverage & Width \\
\hline
ACIC-2017 (17) & AIPW & 0.0195 & 0.0889 & 0.98 & 0.0462 \\
ACIC-2017 (17) & IC-DML & 0.0239 & 0.0929 & 0.97 & 0.0467 \\
ACIC-2017 (18) & AIPW & 0.371 & 1.01 & 0.21 & 0.0948 \\
ACIC-2017 (18) & IC-DML & 0.169 & 0.607 & 0.71 & 0.202 \\
ACIC-2017 (19) & AIPW & 0.0425 & 0.426 & 0.96 & 0.213 \\
ACIC-2017 (19) & IC-DML & 0.0428 & 0.426 & 0.96 & 0.213 \\
ACIC-2017 (20) & AIPW & 1.82 & 2.16 & 0.29 & 0.319 \\
ACIC-2017 (20) & IC-DML & 0.534 & 1.48 & 0.89 & 0.652 \\
ACIC-2017 (21) & AIPW & 0.00292 & 0.0166 & 1.00 & 0.097 \\
ACIC-2017 (21) & IC-DML & 0.00354 & 0.0176 & 1.00 & 0.0971 \\
ACIC-2017 (22) & AIPW & 0.0395 & 0.147 & 0.40 & 0.135 \\
ACIC-2017 (22) & IC-DML & 0.0216 & 0.109 & 0.80 & 0.242 \\
ACIC-2017 (23) & AIPW & 0.00758 & 0.0789 & 0.97 & 0.252 \\
ACIC-2017 (23) & IC-DML & 0.00677 & 0.079 & 0.97 & 0.251 \\
ACIC-2017 (24) & AIPW & 0.316 & 0.37 & 0.30 & 0.355 \\
ACIC-2017 (24) & IC-DML & 0.087 & 0.269 & 0.90 & 0.709 \\
ACIC-2018 (1000) & AIPW &  292 & 4310 & 0.59 &  568 \\
ACIC-2018 (1000) & IC-DML &  361 & 4530 & 0.66 & 1460 \\
ACIC-2018 (10000) & AIPW & 1.47 & 16.7 & 0.63 & 1.75 \\
ACIC-2018 (10000) & IC-DML &  1.5 & 16.7 & 0.73 & 5.37 \\
ACIC-2018 (2500) & AIPW & 78.2 & 1270 & 0.58 &  352 \\
ACIC-2018 (2500) & IC-DML & 2450 & 25900 & 0.65 & 10300 \\
ACIC-2018 (5000) & AIPW &  501 & 4130 & 0.57 &  352 \\
ACIC-2018 (5000) & IC-DML &  545 & 4390 & 0.63 &  758 \\
IHDP & AIPW & 0.129 & 0.461 & 0.58 & 1.02 \\
IHDP & IC-DML & 0.132 & 0.47 & 0.55 & 1.05 \\
Lalonde CPS & AIPW & 0.105 & 0.193 & 0.14 &  464 \\
Lalonde CPS & IC-DML & 0.0696 & 0.33 & 0.65 & 5990 \\
Lalonde PSID & AIPW & 0.0424 & 0.19 & 0.48 & 1670 \\
Lalonde PSID & IC-DML & 0.0236 & 0.432 & 0.83 & 8600 \\
Twins & AIPW & 0.0603 & 0.112 & 0.96 & 0.0313 \\
Twins & IC-DML & 0.0565 & 0.11 & 0.96 & 0.0312 \\
\hline
\end{tabular}
\end{table}

\singlespacing

\singlespacing
\bibliography{ref}
\end{document}